\documentclass[10pt,twocolumn,letterpaper]{article}

\usepackage[pagenumbers]{cvpr} 

\usepackage{graphicx}
\usepackage{amsmath}
\usepackage{amssymb}
\usepackage{booktabs}

\usepackage[pagebackref,breaklinks,colorlinks]{hyperref}

\usepackage[capitalize]{cleveref}
\crefname{section}{Sec.}{Secs.}
\Crefname{section}{Section}{Sections}
\Crefname{table}{Table}{Tables}
\crefname{table}{Tab.}{Tabs.}
\usepackage{dsfont}

\iftrue
\newcommand{\berivan}[1]{{\color{red}\textbf{BI}: #1}}
\fi

\def\cvprPaperID{4397} 
\def\confName{CVPR}
\def\confYear{2022}

\begin{document}



\newfont{\bbb}{msbm10 scaled 700}
\newcommand{\CCC}{\mbox{\bbb C}}

\newfont{\bb}{msbm10 scaled 1100}
\newcommand{\CC}{\mbox{\bb C}}
\newcommand{\RR}{\mbox{\bb R}}
\newcommand{\QQ}{\mbox{\bb Q}}
\newcommand{\ZZ}{\mbox{\bb Z}}
\newcommand{\FF}{\mbox{\bb F}}
\newcommand{\GG}{\mbox{\bb G}}
\newcommand{\EE}{\mbox{\bb E}}
\newcommand{\NN}{\mbox{\bb N}}
\newcommand{\KK}{\mbox{\bb K}}


\newcommand{\av}{{\bf a}}
\newcommand{\bv}{{\bf b}}
\newcommand{\cv}{{\bf c}}
\newcommand{\dv}{{\bf d}}
\newcommand{\ev}{{\bf e}}
\newcommand{\fv}{{\bf f}}
\newcommand{\gv}{{\bf g}}
\newcommand{\hv}{{\bf h}}
\newcommand{\iv}{{\bf i}}
\newcommand{\jv}{{\bf j}}
\newcommand{\kv}{{\bf k}}
\newcommand{\lv}{{\bf l}}
\newcommand{\mv}{{\bf m}}
\newcommand{\nv}{{\bf n}}
\newcommand{\ov}{{\bf o}}
\newcommand{\pv}{{\bf p}}
\newcommand{\qv}{{\bf q}}
\newcommand{\rv}{{\bf r}}
\newcommand{\sv}{{\bf s}}
\newcommand{\tv}{{\bf t}}
\newcommand{\uv}{{\bf u}}
\newcommand{\wv}{{\bf w}}
\newcommand{\vv}{{\bf v}}
\newcommand{\xv}{{\bf x}}
\newcommand{\yv}{{\bf y}}
\newcommand{\zv}{{\bf z}}
\newcommand{\zerov}{{\bf 0}}
\newcommand{\onev}{{\bf 1}}


\newcommand{\Am}{{\bf A}}
\newcommand{\Bm}{{\bf B}}
\newcommand{\Cm}{{\bf C}}
\newcommand{\Dm}{{\bf D}}
\newcommand{\Em}{{\bf E}}
\newcommand{\Fm}{{\bf F}}
\newcommand{\Gm}{{\bf G}}
\newcommand{\Hm}{{\bf H}}
\newcommand{\Id}{{\bf I}}
\newcommand{\Jm}{{\bf J}}
\newcommand{\Km}{{\bf K}}
\newcommand{\Lm}{{\bf L}}
\newcommand{\Mm}{{\bf M}}
\newcommand{\Nm}{{\bf N}}
\newcommand{\Om}{{\bf O}}
\newcommand{\Pm}{{\bf P}}
\newcommand{\Qm}{{\bf Q}}
\newcommand{\Rm}{{\bf R}}
\newcommand{\Sm}{{\bf S}}
\newcommand{\Tm}{{\bf T}}
\newcommand{\Um}{{\bf U}}
\newcommand{\Wm}{{\bf W}}
\newcommand{\Vm}{{\bf V}}
\newcommand{\Xm}{{\bf X}}
\newcommand{\Ym}{{\bf Y}}
\newcommand{\Zm}{{\bf Z}}
\newcommand{\Lam}{{\bf \Lambda}}

\newcommand{\Ac}{{\cal A}}
\newcommand{\Bc}{{\cal B}}
\newcommand{\Cc}{{\cal C}}
\newcommand{\Dc}{{\cal D}}
\newcommand{\Ec}{{\cal E}}
\newcommand{\Fc}{{\cal F}}
\newcommand{\Gc}{{\cal G}}
\newcommand{\Hc}{{\cal H}}
\newcommand{\Ic}{{\cal I}}
\newcommand{\Jc}{{\cal J}}
\newcommand{\Kc}{{\cal K}}
\newcommand{\Lc}{{\cal L}}
\newcommand{\Mc}{{\cal M}}
\newcommand{\Nc}{{\cal N}}
\newcommand{\Oc}{{\cal O}}
\newcommand{\Pc}{{\cal P}}
\newcommand{\Qc}{{\cal Q}}
\newcommand{\Rc}{{\cal R}}
\newcommand{\Sc}{{\cal S}}
\newcommand{\Tc}{{\cal T}}
\newcommand{\Uc}{{\cal U}}
\newcommand{\Wc}{{\cal W}}
\newcommand{\Vc}{{\cal V}}
\newcommand{\Xc}{{\cal X}}
\newcommand{\Yc}{{\cal Y}}
\newcommand{\Zc}{{\cal Z}}
\newcommand{\Ecb}{\bf {\cal E}}
\newcommand{\Ocb}{\bf {\cal O}}
\newcommand{\Lcb}{{\bm {\mathcal L}}}


\newcommand{\alphav}{\hbox{\boldmath$\alpha$}}
\newcommand{\betav}{\hbox{\boldmath$\beta$}}
\newcommand{\gammav}{\hbox{\boldmath$\gamma$}}
\newcommand{\deltav}{\hbox{\boldmath$\delta$}}
\newcommand{\etav}{\hbox{\boldmath$\eta$}}
\newcommand{\lambdav}{\hbox{\boldmath$\lambda$}}
\newcommand{\epsilonv}{\hbox{\boldmath$\epsilon$}}
\newcommand{\nuv}{\hbox{\boldmath$\nu$}}
\newcommand{\muv}{\hbox{\boldmath$\mu$}}
\newcommand{\zetav}{\hbox{\boldmath$\zeta$}}
\newcommand{\phiv}{\hbox{\boldmath$\phi$}}
\newcommand{\psiv}{\hbox{\boldmath$\psi$}}
\newcommand{\thetav}{\hbox{\boldmath$\theta$}}
\newcommand{\tauv}{\hbox{\boldmath$\tau$}}
\newcommand{\omegav}{\hbox{\boldmath$\omega$}}
\newcommand{\xiv}{\hbox{\boldmath$\xi$}}
\newcommand{\sigmav}{\hbox{\boldmath$\sigma$}}
\newcommand{\piv}{\hbox{\boldmath$\pi$}}
\newcommand{\rhov}{\hbox{\boldmath$\rho$}}

\newcommand{\Gammam}{\hbox{\boldmath$\Gamma$}}
\newcommand{\Lambdam}{\hbox{\boldmath$\Lambda$}}
\newcommand{\Deltam}{\hbox{\boldmath$\Delta$}}
\newcommand{\Sigmam}{\hbox{\boldmath$\Sigma$}}
\newcommand{\Phim}{\hbox{\boldmath$\Phi$}}
\newcommand{\Pim}{\hbox{\boldmath$\Pi$}}
\newcommand{\Psim}{\hbox{\boldmath$\Psi$}}
\newcommand{\Thetam}{\hbox{\boldmath$\Theta$}}
\newcommand{\Omegam}{\hbox{\boldmath$\Omega$}}
\newcommand{\Xim}{\hbox{\boldmath$\Xi$}}






\newcommand{\indicator}{\mathds{1}}

\title{LVAC: Learned Volumetric Attribute Compression for Point Clouds using Coordinate Based Networks}

\author{Berivan Isik\thanks{Work done while the first author was an intern at Google.}\\
Stanford University\\
{\tt\small berivan.isik@stanford.edu} 
\and
Philip A.\ Chou\\
Google\\
{\tt\small philchou@google.com}
\and
Sung Jin Hwang\\
Google\\
{\tt\small sjhwang@google.com}
\and
Nick Johnston\\
Google\\
{\tt\small nickj@google.com}
\and
George Toderici\\
Google\\
{\tt\small gtoderici@google.com}
}
\maketitle

\begin{abstract}
We consider the attributes of a point cloud as samples of a vector-valued volumetric function at discrete positions.  To compress the attributes given the positions, we compress the parameters of the volumetric function.  We model the volumetric function by tiling space into blocks, and representing the function over each block by shifts of a coordinate-based, or implicit, neural network.  Inputs to the network include both spatial coordinates and a latent vector per block.  We represent the latent vectors using coefficients of the region-adaptive hierarchical transform (RAHT) used in the MPEG geometry-based point cloud codec G-PCC.  The coefficients, which are highly compressible, are rate-distortion optimized by back-propagation through a rate-distortion Lagrangian loss in an auto-decoder configuration.  The result outperforms RAHT by 2--4 dB. This is the first work to compress volumetric functions represented by local coordinate-based neural networks.  As such, we expect it to be applicable beyond point clouds, for example to compression of high-resolution neural radiance fields.

\end{abstract}
\section{Introduction}
\label{sec:intro}

Our work addresses the problem of 3D point cloud attribute compression, using coordinate-based neural networks. Point clouds are a fundamental data type underlying 3D sampling and hence play a critical role in applications such as mapping and navigation, virtual and augmented reality, telepresence, and cultural heritage preservation, which rely on sampled 3D data \cite{SunEtAl:20,ParkCH:19,mekuria_2016,Pierdicca_2020}.  Given the volume of data in such applications, compression is important for both storage and communication.  Indeed, standards for point cloud compression are underway in both MPEG and JPEG \cite{Schwarz:18,JangEtAl:19,GraziosiEtAl:20,JPEG_Pleno_PC_CFE:20}.

\begin{figure}[t]
  \centering
   \includegraphics[height=2.0in]{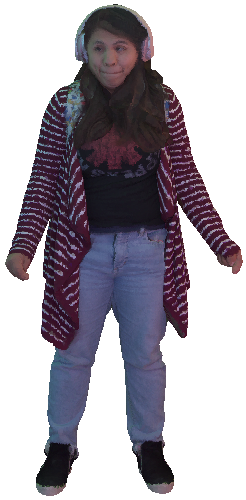}
   \includegraphics[height=2.0in]{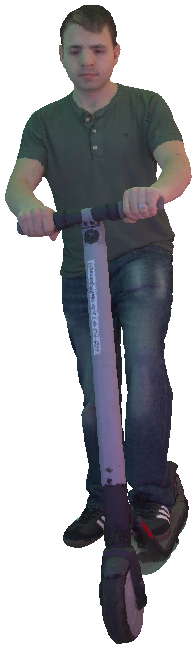}
   \includegraphics[height=2.0in]{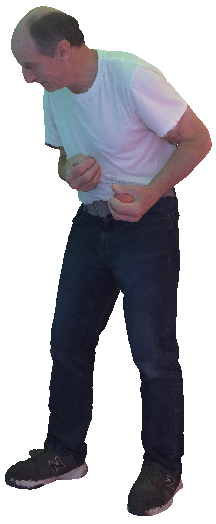}
   \includegraphics[height=2.0in]{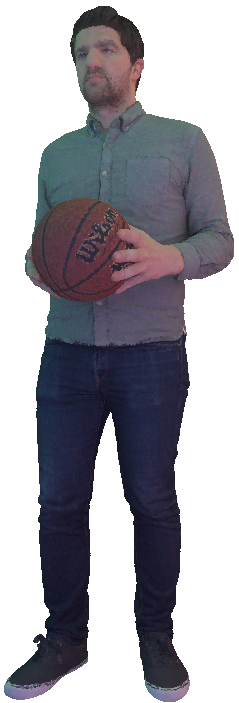}

   \caption{Point clouds {\em rock}, {\em scooter}, {\em juggling}, and {\em basketball}.}
   \label{fig:pointclouds}
   \vspace{-0.1in}
\end{figure}

\begin{figure}[t]
    \centering
    \includegraphics[width=0.8\linewidth]{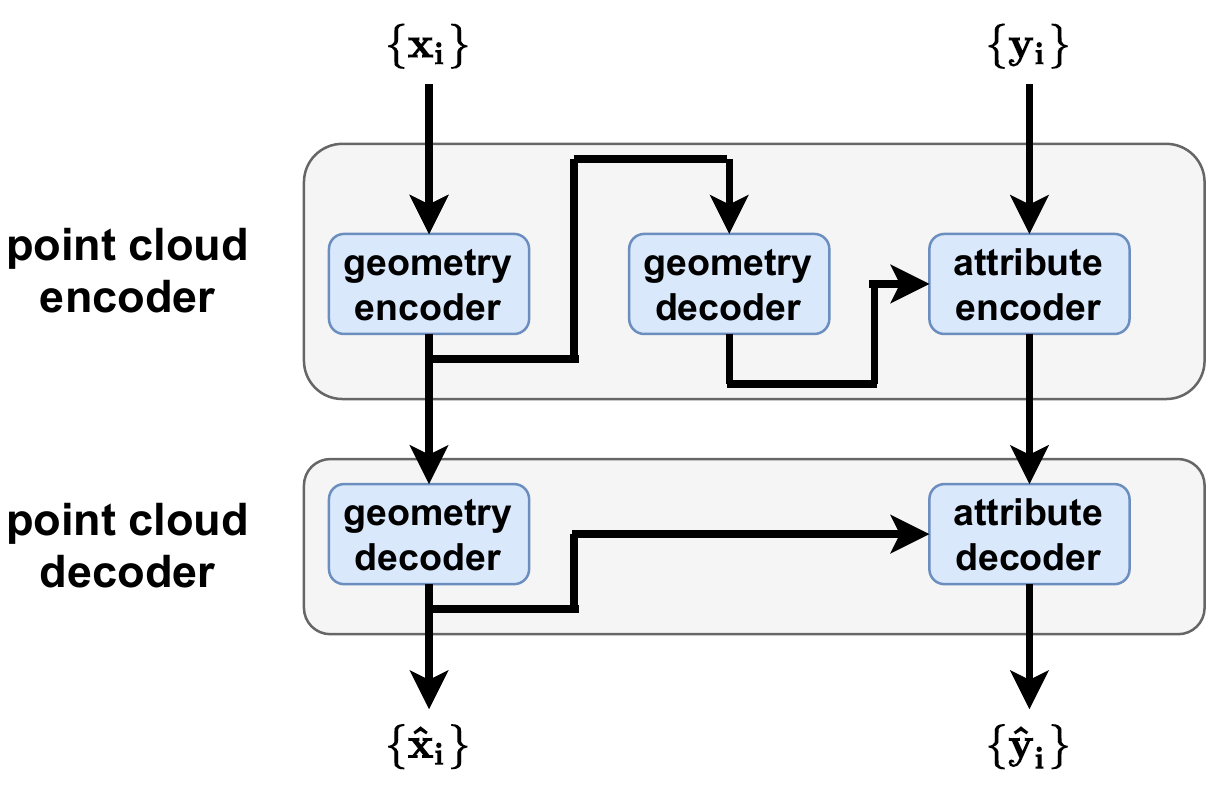}
    \caption{Point cloud codec, consisting of a geometry encoder and decoder, and an attribute encoder and decoder conditioned on the decoded geometry.}
    \label{fig:encoder_decoder}
    \vspace{-0.1in}
\end{figure}

3D point clouds, such as those shown in \cref{fig:pointclouds}, each consist of a set of points $\{(\xv_i,\yv_i)\}$, where $\xv_i$ is the 3D position of the $i$th point and $\yv_i$ is a vector of attributes associated with the point.  Attributes typically include color components, e.g., RGB, but may alternatively include reflectance, normals, transparency, density, spherical harmonics, and so forth.  Commonly (e.g., \cite{zhang_icip_2014,ThanouCF16,QuerozC:16,QueirozC:17b,PavezCQO:18,CohenTV16,ChouKK:20,KrivokucaCK:20,Schwarz:18}), point cloud compression is broken into two steps: compression of the point cloud positions, called the {\em geometry}, and compression of the point cloud {\em attributes}.  Compression of the attributes is conditioned on the decoded geometry, as illustrated in \cref{fig:encoder_decoder}.  It is important to note that this conditioning is crucial in achieving good compression.  This will become one of the themes of this paper.

Following successful application of neural networks in image compression \cite{ToOMHwViMi16,BaLaSi16a,ToViJoHwMi17,BaLaSi17,Ba18,BaMiSiHwJo18,MiBaTo18,BalleEtAl:20,mentzer2020high,hu2021learning}, neural networks have been used successfully for point cloud geometry compression, demonstrating significant gains over traditional techniques \cite{YanSLLLL:19,quach2019learning,GuardaRP:19a,GuardaRP:19b,guarda2020deep,tang2020deep,Quach2020ImprovedDP}.  However, the same cannot be said for point cloud attribute compression.  To our knowledge, our work is among the first to use neural networks for point cloud attribute compression.  Previous attempts may have been be hindered by the inability to properly condition the attribute compression on the decoded geometry, thus leading to poor results.  In our work, we show that proper conditioning improves attribute compression performance by over 30\% reduction in BD-Rate.  This results in a gain of 2--4 dB over region-adaptive linear transform (RAHT) coding, which is used in the ``geometry-based'' point cloud compression standard MPEG G-PCC.

Although learned image compression systems have been based on convolutional neural networks (CNNs), in this work we employ what have come to be called {\em coordinate based networks} (CBNs), also called {\em implicit networks}.  (See \cite{tancik2021learned, czerkawski2021neural} and the references in \cref{sec:related}.)  A CBN is a network, such as a multilayer perceptron (MLP), whose inputs include the coordinates of the spatial domain of interest, e.g., $\xv\in\RR^3$.  Thus a CBN can directly represent a nonlinear function of the spatial coordinates $\xv$, possibly indexed with a latent or feature vector $\zv$, as $\yv=f_\theta(\xv)$ or $\yv=f_\theta(\xv;\zv)$.  CBNs have recently come to the fore in accurately representing geometry and spatial phenomena such as radiance fields.
However, while there has been an explosion of work using CBNs for {\em representing} specific objects and scenes, none of that work focuses on {\em compressing} those representations.  (Two exceptions are \cite{BirdBSC:21,isik2021neural}, which apply model compression to the CBNs themselves.)  Good lossy compression is nontrivial, and must make the optimal trade-off between the fidelity of the reconstruction and the number of bits used in its binary representation.  We show that na\"{\i}ve scalar quantization and entropy coding of the parameters $\theta$ and/or latent vectors $\zv$ lead to very poor results, and that superior results can be achieved by proper normalization prior to uniform scalar quantization.  This normalization amounts to using different quantization step sizes, or different numbers of bits, for different latent vectors --- depending on the geometry.  In addition, the entropy model and CBN must be jointly trained to minimize a loss function that penalizes not only large distortion (or error) but large bit rate as well.

Our main contributions include the following:
\begin{itemize}
    \item We are among the first to compress point cloud {\em attributes} using neural networks.  Our solution allows the network to interpolate the reconstructed attributes continuously across space, and offers a 2--4 dB improvement over our linear baseline, RAHT with adaptive Run-Length Golomb-Rice (RLGR) entropy coding. Note that RAHT is the transform used in the latest MPEG G-PCC standard.
    \item We are the first to {\em compress volumetric functions} modeled by {\em local coordinate based networks}, by training with the rate-distortion Lagrangian as the loss function (as in image compression), thereby offering scalable, high fidelity reconstructions at low bit rates.  We show that na\"{\i}ve uniform scalar quantization and entropy coding leads to poor results, and we show formulas for normalizing the coefficients to achieve over a 30\% reduction in bit rate.
\end{itemize}

\Cref{sec:related} covers related work, \cref{sec:framework} details our Learned Volumetric Attribute Compression (LVAC) framework, \cref{sec:experiments} reports experimental results, and \cref{sec:conclusion} discusses and concludes.

\section{Related Work}
\label{sec:related}

\subsection{Coordinate Based Networks}

Early work that used coordinate based networks \cite{ParkFSNL:19, MeschederONNG:19,sitzmann2020metasdf}, exemplified by DeepSDF \cite{ParkFSNL:19}, focused on representing geometry {\em implicitly}, for example as the $c$-level set $\{\xv:c=f_\theta(\xv;\zv)\}\subset\RR^3$ of a function $f_\theta:\RR^3\times\RR^C\rightarrow\RR$ modeled by a neural network, where $\zv\in\RR^C$ is a global latent vector.  As a result such networks were called ``implicit'' networks.
Much of this work focused on auto-decoder architectures, in which the latent vector $\zv$ was determined for each instance by back propagation through the loss function.
The loss function $L(\theta,\zv)$ measured a pointwise error between samples $f_\theta(\xv_i;\zv)$ of the network and samples $f(\xv_i)$ of a ground truth function, such as the signed distance function (SDF).

Later work that used CBNs, exemplified by NeRF \cite{mildenhall2020nerf,barron2021mipnerf}, used the networks to model not SDFs but rather other, vector-valued, volumetric functions, including color, density, normals, BRDF parameters, and specular features \cite{srinivasan2020nerv,hedman2021baking,yu2021plenoctrees,knodt2021neural,zhang2021nerfactor}.  Since these networks were no longer used to represent solutions implicitly, their name started to shift to ``coordinate-based'' networks, e.g., \cite{tancik2021learned}.  An important innovation from this cohort was positional encoding, in which the network's positional input $\xv$ was embedded into a higher dimensional feature space by sinusoidal maps, which greatly improved the spatial resolution of the networks \cite{mildenhall2020nerf,tancik2020fourier,sitzmann2019siren,mehta2021modulated,benbarka2021seeing,zheng2021rethinking}.  Another key innovation was measuring the loss $L(\theta)$ not pointwise between samples of $f_\theta$ and some ground truth volumetric function $f$, but rather between volumetric renderings (to images) of $f_\theta$ and $f$, the latter renderings being ground truth images.

NeRF et al.\ focused on training the CBN $f_\theta(\xv)$ to globally represent a single scene, without benefit of a latent vector $\zv$.  However, subsequent work shifted towards using the CBN with different latent vectors for different objects \cite{stelzner2021decomposing,yu2021unsupervised} or different regions (i.e., blocks or tiles) in the scene \cite{chen2021learning,reiser2021kilonerf,takikawa2021neural,martel2021acorn,mehta2021modulated}.  Partitioning the scene into blocks, and using a CBN with a different latent vector in each block, simultaneously achieves faster rendering \cite{reiser2021kilonerf,takikawa2021neural}, higher resolution \cite{chen2021learning,mehta2021modulated,martel2021acorn}, and scalability to scenes of unbounded size \cite{unconstrained-scene-generation}.  However, this puts much of the burden of the representation on the local latent vectors, rather than on the parameters of the CBN.  This is analogous to conventional block-based image representations, in which the same set of basis functions (e.g., $8\times8$ DCT) is used in each block, and activation of each basis vector is specified by a vector of basis coefficients, different for each block.

Our work borrows heavily from these works.  We partition 3D space into blocks (hierarchically using trees, akin to \cite{takikawa2021neural,martel2021acorn,yu2021plenoctrees}), and represent the color within each block volumetrically using a CBN $f_\theta(\xv;\zv)$, allowing fast, high-resolution, and scalable reconstruction.  Unlike all previous CBN works, however, we train the representation not just for fit but for efficient compression using techniques from learned image compression.

\subsection{Learned Image Compression}

Using neural networks for good compression is non-trivial.  Simply truncating the latent vectors of an existing representation to a certain number of bits is likely to fail, if only because small quantization errors in the latents may easily map into large quantization errors in their reconstructions.  Moreover, the entropy of the quantized latents is a more important determiner of the bit rate than the total number of coefficients in the latent vectors or the number of bits in their binary representation.  Early work on learned image compression could barely exceed the rate-distortion performance of JPEG on low-quality $32\times32$ thumbnails \cite{ToOMHwViMi16}.  However, over the years the rate-distortion performance has consistently improved \cite{BaLaSi16a,ToViJoHwMi17,BaLaSi17,Ba18,BaMiSiHwJo18,MiBaTo18,BalleEtAl:20,hu2021learning} to the point where the best learned image codecs outperform the latest video standard (VVC) in PSNR, albeit at much greater complexity \cite{GuoZFC:21}, and greatly outperform conventional image codecs (by over $2\times$ reduction in bit rate) at the same perceptual distortion \cite{mentzer2020high}.  All current competitive learned image codecs are versions of nonlinear transform coding \cite{BalleEtAl:20}, in which the bottleneck latents in an auto-encoder are uniformly scalar quantized and entropy coded (with a hyperprior), for transmission to a decoder.  The decoder uses a convolutional neural network as a synthesis transform.  The codec is trained end-to-end through a differentiable proxy for the quantizer, often modeled as additive uniform noise.  The loss function is a Lagragian $L(\theta)=D(\theta)+\lambda R(\theta)$, where $D(\theta)$ is an expected distortion and $R(\theta)$ is an expected bit rate (i.e., cross-entropy), and $\lambda>0$ is a Lagrange multiplier.

Our work borrows significantly from this work, in that we use the same uniform scalar quantization and entropy model (though without a hyperprior for now) as used for the best learned image compression.  Moreover, we train our representation using a similar Lagrangian loss function.

\subsection{Point Cloud Compression}

MPEG is standardizing two point cloud codecs: video-based (V-PCC) and geometry-based (G-PCC) \cite{Schwarz:18,JangEtAl:19,GraziosiEtAl:20}.  V-PCC is based on existing video codecs, while G-PCC is based on new, but in many ways classical, geometric approaches.  Like previous works \cite{zhang_icip_2014,ThanouCF16,QuerozC:16,QueirozC:17b,PavezCQO:18,CohenTV16,ChouKK:20,KrivokucaCK:20}, both V-PCC and G-PCC compress geometry first, then compress attributes conditioned on geometry.  Neural networks have been applied with some success to geometry compression \cite{YanSLLLL:19,quach2019learning,GuardaRP:19a,GuardaRP:19b,guarda2020deep,tang2020deep,Quach2020ImprovedDP,Milani:20,Milani:21,LazzarottoAEL21}, but not to attribute compression.  Exceptions may include \cite{QuachVD2020Folding}, which uses learned neural 3D$\rightarrow$2D folding but compresses with conventional image coding, and \cite{ShengLLXLW:21}, which compresses attributes using a PointNet-style architecture, which is not volumetric.  The attribute compression in G-PCC uses linear transforms, which adapt based on the geometry.  A core transform is the region-adaptive hierarchical transform (RAHT) \cite{QuerozC:16,SandriCKQ:19}, which is a linear transform that is orthonormal with respect to a discrete measure whose mass is put on the point cloud geometry \cite{ChouKK:20,SandriFCQL:19}. Thus RAHT compresses attributes conditioned on geometry. Beyond RAHT, G-PCC uses prediction (of the RAHT coefficients) and joint entropy coding to obtain superior performance \cite{LasserreF:19,GPCC,PavezSQO:21}.

Our work borrows heavily from RAHT, as we apply RAHT's orthonormalization formulas to our latent vectors.  It turns out that this is crucial for good rate-distortion performance for point cloud attribute compression.

\section{LVAC Framework}
\label{sec:framework}

\subsection{Approach to Volumetric Representation}

\begin{figure*}[t]
    \centering
    \includegraphics[width=1.0\linewidth, trim=0 0 60 0, clip
    ]{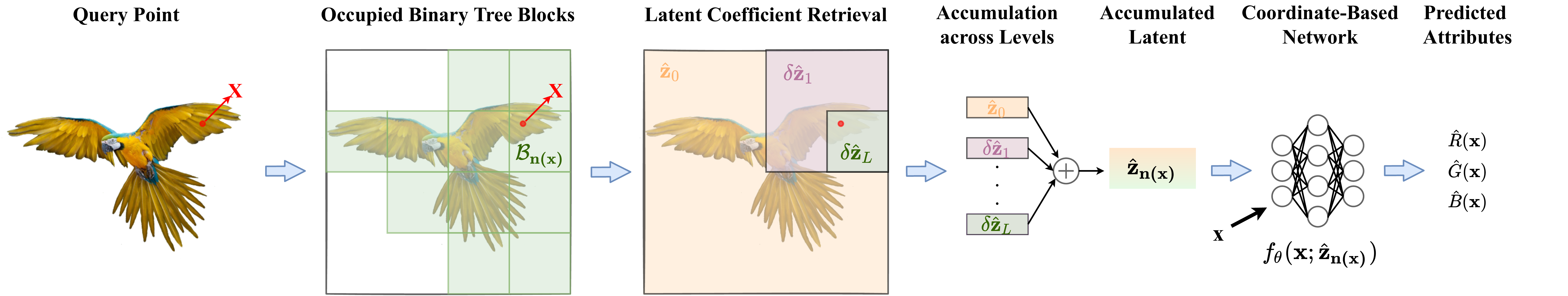}
    \caption{Querying attributes at position $\xv\in\RR^3$. The block $\Bc_{\nv(\xv)}$ at target level $L$ in which $\xv$ is located is identified by traversing a binary space partition tree.  The quantized latent $\hat\zv_0$ at the root and quantized difference latents $\delta\hat\zv_\ell$ at increasing levels of detail $\ell=1,\ldots,L$ are accumulated to form a cumulative latent $\hat\zv_{\nv(\xv)}$, which is input along with $\xv$ to the CBN at level $L$, to produce $\hat\yv=f_\theta(\xv;\hat\zv_{\nv(\xv)})$ (after \cite{takikawa2021neural}).
    }
    \label{fig:framework}
   \vspace{-0.1in}
\end{figure*}

A real-valued (or real vector-valued) function
\begin{equation}
    f:\mathbb{R}^d\rightarrow\mathbb{R}^r
\end{equation}
is said, informally, to be {\em volumetric} if $d=3$ and {\em hyper-volumetric} if $d>3$.  A volumetric (or hyper-volumetric) function $f$ may be fit by another volumetric function $f_\theta$ from a parametric family of volumetric functions $\{f_\theta:\theta\in\Theta\}$ by minimizing an error $d(f,f_\theta)$ over $\theta\in\Theta$.  A simple example is linear regression.  Suppose $\{(\xv_i,\yv_i)\}_{i=1}^N$ is a point cloud with point positions $\xv_i\in\RR^3$ and point attributes $\yv_i=f(\xv_i)\in\RR^r$.  Then an affine function
\begin{equation}
    \yv = f_\theta(\xv) = \Am\xv+\bv,
\end{equation}
with $\theta=(\Am,\bv)$, may be fit to the data by minimizing the squared error $d(f,f_\theta)=||f-f_\theta||^2=\sum_i||f(\xv_i)-f_\theta(\xv_i)||^2$ over~$\theta$.
Although a linear or affine volumetric function may not be able to represent adequately the complex spatial arrangement of colors of point clouds like those in \cref{fig:pointclouds}, two strategies may be used to improve the fit.  The first is to make $f_\theta$ far more complex, e.g., represented by a CBN with millions of parameters.  The second is to partition the scene into blocks, and use a simpler CBN within each block.
LVAC chooses the latter approach.

\subsection{Latent Vectors}

To be precise, in LVAC, the attributes $\yv_i$ in a block $\Bc_\nv$ at offset $\nv$ are fit with a volumetric function $\yv=f_\theta(\xv-\nv;\zv_\nv)$ represented by a simple CBN, shifted to offset $\nv$.  The CBN parameters $\theta$ are learned and fixed for each point cloud.  However, each block $\Bc_\nv$ supplies its own latent vector $\zv_\nv$, which selects the exact volumetric function $f_\theta(\cdot;\zv)$ used in the block.  The role of $\theta$ is to choose the {\em family} of volumetric functions best for each point cloud, or for point clouds in general.
The role of $\zv$ is to choose a member of the family best for each block.  The overall volumetric function may be expressed as
\begin{equation}
    \yv=f_{\theta,\Zm}(\xv)=\sum_\nv f_\theta(\xv-\nv;\zv_\nv)\indicator_{\Bc_\nv}(\xv) ,
    \label{eqn:local_cbn}
\end{equation}
where the sum is over all block offsets $\nv$, $\indicator_{\Bc_\nv}$ is the indicator function for block $\Bc_\nv$ (i.e., $\indicator_{\Bc_\nv}(\xv) =1$ iff the query point $x$ is inside $\Bc_\nv$), and $\Zm=[\zv_\nv]$ is the matrix whose rows $\zv_\nv$ are the blocks' latent vectors.

To compress the point cloud attributes $\{\yv_i\}$ given the geometry $\{\xv_i\}$, LVAC compresses and transmits $\Zm$ and possibly $\theta$ as $\hat{\Zm}$ and $\hat{\theta}$ using $R(\hat{\theta},\hat{\Zm})$ bits. This communicates the volumetric function $f_{\hat{\theta},\hat{\Zm}}$ to the decoder.  The decoder can then use $f_{\hat{\theta},\hat{\Zm}}$ to reconstruct the attributes $\yv_i$ at each point position $\xv_i$ as $\hat{\yv}_i=f_{\hat{\theta},\hat{\Zm}}(\xv_i)$,
incurring distortion
\begin{equation}
    D(\hat{\theta},\hat{\Zm})
    = d(f,f_{\hat{\theta},\hat{\Zm}})
    = \sum_i ||\yv_i - f_{\hat{\theta},\hat{\Zm}}(\xv_i)||^2.
    \label{eqn:distortion}
\end{equation}
The decoder can also use $\yv=f_{\hat{\theta},\hat{\Zm}}(\xv)$ to reconstruct the attributes $\yv$ at an \emph{arbitrary} position $\xv\in\RR^3$.  However,
LVAC minimizes the distortion $D(\hat{\theta},\hat{\Zm})$ subject to a constraint on the bit rate, $R(\hat{\theta},\hat{\Zm})\leq R_0$.  This is done by minimizing the Lagrangian $J(\hat{\theta},\hat{\Zm})=D(\hat{\theta},\hat{\Zm})+\lambda R(\hat{\theta},\hat{\Zm})$ for some Lagrange multiplier $\lambda>0$ matched to $R_0$.

In the regime of interest in our work, $\theta$ has about 250-10K parameters, while $\Zm$ has about 500K-8M parameters.  Hence the focus of this paper is on compression of $\Zm$.  We assume that the simple CBN parameterized by $\theta$ can be compressed using model compression tools, e.g., \cite{BirdBSC:21,isik2021neural}, to a few bits per parameter with little loss in performance.  Alternatively, we assume that the CBN may be trained to generalize across point clouds, obviating the need to transmit $\theta$.  In \cref{sec:experiments}, we explore conservative bounds on the performance of each assumption.  In this section, however, we focus on compression of the latent vectors $\Zm=[\zv_\nv]$.

\subsubsection{Latent Vector Compression in RAHT}

LVAC compresses the latent vectors $[\zv_\nv]$ using, essentially, RAHT.  Hence, we first discuss how RAHT compresses color attributes.  Though there are many ways to view RAHT, one way to view it is as compression of a piecewise constant volumetric function,
\begin{equation}
    \yv=f_{\Zm}(\xv)=\sum_\nv \zv_\nv \indicator_{\Bc_\nv}(\xv) .
    \label{eqn:raht_cbn}
\end{equation}
This is the same as (\ref{eqn:local_cbn}) with an extremely simple CBN: $f_\theta(\xv;\zv)=\zv$.  In this case, each latent $\zv_\nv\in\RR^3$ directly represents a color, which is constant across block $\Bc_\nv$.  It is clear that the squared error $||f-f_{\Zm}||^2$ is minimized by setting every $\zv_\nv$ to the average (DC) value of the colors of the points in $\Bc_\nv$.  RAHT does not quantize and entropy code the colors $\Zm=[\zv_\nv]$ directly, which would be inefficient.  Rather, RAHT first transforms the $N\!\times\! C$ matrix $\Zm$ using a geometry-dependent $N\!\times\! N$ analysis transform $\Tm_a$, to obtain the $N\!\times\! C$ matrix of transform coefficients $\Vm = \Tm_a\Zm$, most of which may be near zero.  (Here, $N$ is the number of blocks $\Bc_\nv$
that are {\em occupied}, i.e., that contain points,
and $C=3$ is the number of color attributes.)  Then $\Vm$ is quantized to $\hat{\Vm}$ and efficiently entropy coded.  Finally $\hat{\Zm}=\Tm_s\hat{\Vm}$ is recovered using the $N\!\times\! N$ synthesis transform $\Tm_s=\Tm_a^{-1}$.

The analysis and synthesis transforms $\Tm_a$ and $\Tm_s$ are defined in terms of a hierarchical space partition represented by a binary tree.  The root of the tree (level $\ell=0$) corresponds to a large block $\Bc_{0,\bf{0}}$ containing the entire point cloud.  The leaves of the tree (level $\ell=L$) correspond to the $N$ blocks $\Bc_{L,\nv}=\Bc_\nv$ in (\ref{eqn:raht_cbn}), which are voxels of a voxelized point cloud.  In between, for each level $\ell=0,1,\ldots,L-1$, each occupied block $\Bc_{\ell,\nv}$ at level $\ell$ is split into left and right child blocks of equal size, say $\Bc_{\ell+1,\nv_L}$ and $\Bc_{\ell+1,\nv_R}$, at level $\ell+1$.  The split is along either the $x$, $y$, or $z$ axis depending on whether $\ell\;\mbox{mod}\;3$ is 0, 1, or 2. Only child blocks that are occupied are retained in the tree.

To perform the linear analysis transform $\Tm_a\Zm$, RAHT starts at level $\ell=L-1$ and works back to level $\ell=0$, computing the average (DC) value of each block $\Bc_{\ell,\nv}$ as
\begin{equation}
    \zv_{\ell,\nv}
    = \frac{w_{\nv_L}}{w_{\nv_L}\!+\!w_{\nv_R}}\zv_{\ell+1,\nv_L} + \frac{w_{\nv_R}}{w_{\nv_L}\!+\!w_{\nv_R}}\zv_{\ell+1,\nv_R} ,
    \label{eqn:parent_from_children}
\end{equation}
where $w_{\nv_L}\!=\!w_{\ell+1,\nv_L}$ and $w_{\nv_R}\!=\!w_{\ell+1,\nv_R}$ are the {\em weights} of, or number of points in, the left and right child blocks of $\Bc_{\ell,\nv}$.  The global DC value of the entire point cloud is $\zv_{0,\bf{0}}$.  Along the way, RAHT computes the difference between the DC values of each child block and its parent as
\begin{eqnarray}
\delta\zv_{\ell+1,\nv_L} & = & \zv_{\ell+1,\nv_L} - \zv_{\ell,\nv} ,
\label{eqn:left_child_delta} \\
\delta\zv_{\ell+1,\nv_R} & = & \zv_{\ell+1,\nv_R} - \zv_{\ell,\nv} .
\label{eqn:right_child_delta}
\end{eqnarray}
These differences are close to zero and are efficient to entropy code.  The $N\times C$ matrix of transform coefficients $\Vm=\Tm_a\Zm$ consist of the global DC value $\zv_{0,\bf{0}}$ in the first row, and the $N-1$ {\em right child} differences $\delta\zv_{\ell+1,\nv_R}$ computed in (\ref{eqn:right_child_delta}) in the remaining rows.

To perform the linear synthesis transform $\Tm_s\Vm$, RAHT starts at level $\ell=0$ and works up to level $L-1$, computing the {\em left child} differences $\delta\zv_{\ell+1,\nv_L}$ (\ref{eqn:left_child_delta}) from the {\em right child} differences $\delta\zv_{\ell+1,\nv_R}$ (\ref{eqn:right_child_delta}) in $\Vm$ using the constraint
\begin{equation}
     {\bf 0}
     =\frac{w_{\nv_L}}{w_{\nv_L}\!+\!w_{\nv_R}}\delta\zv_{\ell+1,\nv_L} + \frac{w_{\nv_R}}{w_{\nv_L}\!+\!w_{\nv_R}}\delta\zv_{\ell+1,\nv_R} ,
\end{equation}
which is obtained from (\ref{eqn:parent_from_children}) using (\ref{eqn:left_child_delta})-(\ref{eqn:right_child_delta}).~Then (\ref{eqn:left_child_delta})-(\ref{eqn:right_child_delta}) are inverted to obtain $\zv_{\ell+1,\nv_L}$ and $\zv_{\ell+1,\nv_R}$ from $\zv_{\ell,\nv}$, ultimately computing the values $\zv_{L,\nv}=\zv_\nv$ for blocks at level $L$.

Expressions for the $N\times N$ matrices $\Tm_a$ and $\Tm_s$ can be worked out from the above linear operations.  In particular, it can be shown that each row of $\Tm_s$ computes the color $\zv_{L,\nv}$ of some leaf voxel $\Bc_{L,\nv}$ by summing the color $\zv_0$ of the root block with the color differences $\delta\zv_{\ell}$ at levels of detail $\ell=1,\ldots,L$ from the root to the leaf.  Moreover, it can be shown that $\Tm_a$ and $\Tm_s$ can be orthonormalized by multiplication by a diagonal matrix $\Sm=\mbox{diag}(s_1,\ldots,s_N)$, where
\begin{eqnarray}
    s_1 & = & (\mbox{\# points in point cloud})^{-1/2}
    \label{eqn:normalization_S0}, \\
    s_m & = & \left(\frac{w_{\ell+1,\nv_L}(w_{\ell+1,\nv_L}\!+\!w_{\ell+1,\nv_R})}{w_{\ell+1,\nv_R}}\right)^{-1/2},
    \label{eqn:normalization_Sn}
\end{eqnarray}
where element $s_1$ of $\Sm$ corresponds to row 1 of $\Vm$ (the global DC value $\zv_{0,\bf 0}$) and element $s_m$ of $\Sm$ corresponds to row $m>1$ of $\Vm$ (a right child difference $\delta\zv_{\ell+1,\nv_R}$).  That is, $\Sm^{-1}\Tm_a$ and $\Tm_s\Sm$ are orthonormal (and transposes of each other).  This implies that the every row of the normalized coefficients $\bar{\Vm}=\Sm^{-1}\Vm$ should be quantized uniformly with the same step size $\Delta$, or equivalently that the rows of the unnormalized coefficients $\Vm=\Tm_a\Zm$ should be quantized with scaled step sizes $s_m\Delta$.  This scaling is crucial for RAHT, as it quantizes with finer precision the coefficients that are more important.  The more important coefficients are generally associated with blocks with more points. This establishes a dependency of the attribute compression on the geometry (see \cref{fig:encoder_decoder}). An alternative way to understand the scaling is that it ensures that the quantization error stays the same, rather than blowing up, after the synthesis transform.

\subsubsection{Latent Vector Compression in LVAC}

LVAC quantizes and entropy codes the latent vectors $\zv_\nv\in\RR^C$ (where now $C\gg3$ typically) for the blocks $\Bc_\nv$ in (\ref{eqn:local_cbn}), adapting RAHT with the following {\em crucial differences}:

First, the blocks $\Bc_\nv=\Bc_{L,\nv}$ are at a {\em target level} of detail $L$, lower (i.e., coarser) than the voxel level.  Thus the blocks $\Bc_{L,\nv}$ contain say $N_x\!\times\! N_y\!\times\! N_z$ voxels, only some of which are occupied.  Then the attributes (typically, colors) of the occupied voxels in $\Bc_{L,\nv}$ are represented by the volumetric function $f_\theta(\xv-\nv;\zv_\nv)$ of a CBN at level $L$, which better models the attributes within the block at certain bit rates.

Second, since the latent vectors $\zv_\nv\in\RR^C$ are not themselves the attributes of the occupied voxels, they are not a direct input to the encoder.  Hence the encoder cannot apply the analysis transform $\Tm_a$ to $\Zm=[\zv_\nv]$ to obtain the transform coefficients $\Vm$.  Instead, LVAC learns $\Vm$ through back-propagation, without an explicit $\Tm_a$, first through the distortion measure and volumetric function (\ref{eqn:distortion}), and then through the synthesis transform $\Tm_s$ and scaling matrix $\Sm$.  The coefficients $\theta$ of the CBN may be optimized at the same time.  In short, LVAC is learned while RAHT is not.

Third, learning gives LVAC the opportunity to optimize $\Vm$ not just to minimize the distortion $D$, 
but to minimize the ultimate rate-distortion objective $D+\lambda R$, which minimizes the distortion subject to a bit rate constraint.

\begin{figure}
    \centering
    \resizebox{\columnwidth}{!}{
    \includegraphics[width=1.0\linewidth, trim=30 2 20 2, clip]{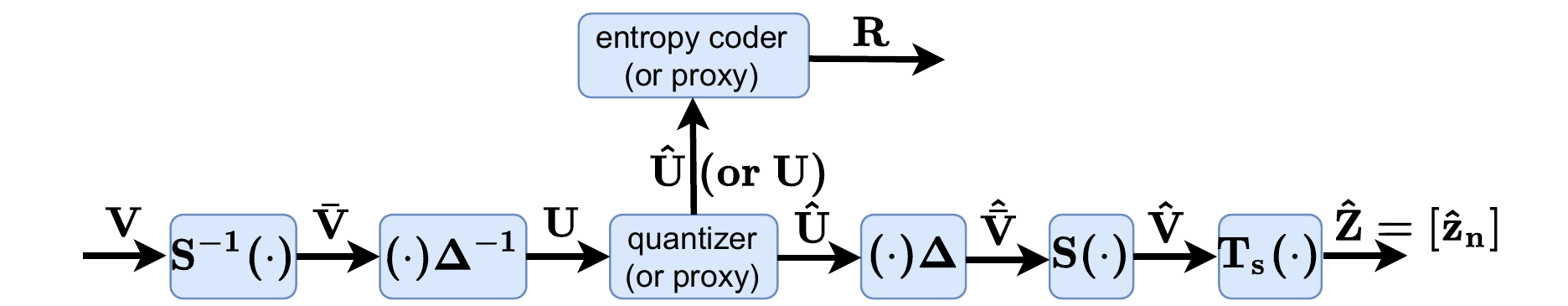}}
    \caption{LVAC pipeline for compressing latents $\Zm=[\zv_\nv]$.  $\Zm$ is represented by difference latents $\Vm$, normalized by $\Sm$ across levels and blocks to obtain $\bar\Vm$, divided by step sizes $\bf\Delta$ across channels to obtain $\Um$, quantized by rounding to obtain $\hat\Um=\lfloor\Um\rceil$, and reconstructed as $\hat\Zm=\Tm_s\Sm\hat\Um\bf\Delta$.  $\Vm$ is optimized by back-propagating through $D(\theta,\Zm)+\lambda R(\theta,\Zm)$ and the pipeline using differentiable proxies for the quantizer and entropy coder.}
    \label{fig:pipeline}
   \vspace{-0.1in}
\end{figure}

\Cref{fig:pipeline} shows the compression pipeline that produces $\hat{\Zm}=[\hat{\zv}_\nv]$ from $\Vm$, through which the back-propagation must be performed.  The diagonal matrix $\Sm$ (defined in (\ref{eqn:normalization_S0})-(\ref{eqn:normalization_Sn})) scales the coefficients in $\Vm$ to produce $\bar{\Vm}=\Sm^{-1}\Vm$, but is constant across channels $c=1,\ldots,C$.  The diagonal matrix ${\bf\Delta}=\mbox{diag}(\Delta_1,\ldots,\Delta_C)$ applies different step sizes $\Delta_c$ to each channel in $\bar{\Vm}$ to produce $\Um=\bar{\Vm}{\bf\Delta}^{-1}$, but is constant across coefficients.  The quantizer rounds the real matrix $\Um$ elementwise to produce the integer matrix $\hat{\Um}=\lfloor\Um\rceil$, which is then entropy coded to produce a bit string of length $R$ in total. The integer matrix $\hat{\Um}$ is also transformed by ${\bf\Delta}$, $\Sm$, and $\Tm_s$ in sequence to produce $\hat{\Zm}=\Tm_s\Sm\hat{\Um}{\bf\Delta}$. Note that learnable parameters in \cref{fig:pipeline} are $\Vm$, $\bf\Delta$, and parameters of the entropy coder.

Since the quantizer and entropy encoder are not differentiable (or more precisely, their derivatives with respect to $\Vm$ are trivially zero almost everywhere), they must be replaced by differentiable {\em proxies} during optimization.  Various differentiable proxies for the quantizer are possible \cite{luo2020ratedistortionaccuracy,AugustssonT:20}, but as in \cite{BaLaSi17} and others, we use the proxy $Q(\Um)=\Um+\Wm$, where $\Wm$ is iid $\mbox{unif}(-0.5,0.5)$.
Various differentiable proxies for the entropy coder are also possible.  As the number of bits in the entropy code for $\Um=[u_{m,c}]$, we use the proxy $R(\Um)=-\sum_{m,c}\log_2 p_{\phi_{\ell,c}}(u_{m,c})$, where
\begin{equation}
    p_{\phi_{\ell,c}}(u)=\mbox{CDF}_{\phi_{\ell,c}}(u+0.5) - \mbox{CDF}_{\phi_{\ell,c}}(u-0.5)
\end{equation}
\cite{BaLaSi17}.  The CDF is modeled by a neural network with parameters $\phi_{\ell,c}$ that depend on the channel $c$ and also the level $\ell$ (but not the offset $\nv$) of the coefficient $u_{m,c}$.
At inference time, the bit rate is $R(\lfloor{\Um}\rceil)$ instead of $R(\Um)$.
These functions are provided by the Continuous Batched Entropy ({\em cbe}) model with the Noisy Deep Factorized prior in \cite{tensorflow_compression}.

Note that the parameters $\Delta_c$ as well as the parameters $\phi_{\ell,c}$, for all $\ell$ and $c$, must be transmitted to the decoder.  However, the overhead for transmitting $\Delta_c$ is negligible, and the overhead for transmitting $\phi_{\ell,c}$ can be circumvented by using a backward-adaptive entropy code, the adaptive Run-Length Golomb-Rice (RLGR) code \cite{Malvar:2006} in its place at inference time.

\subsection{Coordinate Based Network}

Any coordinate based network can be used in the LVAC framework, but in our experiments we use a two-layer MLP,
\begin{equation}
    \yv = f_\theta(\xv;\zv) = \sigma(\bv^3+\Wm^{3\times H}\sigma(\bv^H+\Wm^{H\times(3+C)}[\xv,\zv]))
\end{equation}
where $\theta=(\bv^3,\Wm^{3\times H},\bv^H,\Wm^{H\times(3+C)})$, $H$ is the number of hidden units, and $\sigma(\cdot)$ is pointwise rectification (ReLU).  (Here we take $\xv$, $\yv$, and $\zv$ to be column vectors instead of the row vectors we use elsewhere.)  Note that there is no sinusoidal positional encoding of $\xv$.  But we also define and use a two-layer {\em position-attention} (PA) network,
\begin{equation}
    \yv = f_\theta(\xv;\zv) = \bv^3 + \zv\odot\sin(\bv^C + \Wm^{C\times3}\xv),
\end{equation}
where $\theta=(\bv^3,\bv^C,\Wm^{C\times3})$ and $\odot$ is pointwise multiplication.  The PA network is a simplified version of the modulated periodic activations in \cite{mehta2021modulated}, and has many fewer parameters than the MLPs while being an efficient representation at low bit rates.

Once the latent vectors $\Zm=[\zv_\nv]$ and $\theta$ are transmitted as $\hat{\Zm}=[\hat{\zv}_\nv]$ and $\hat\theta$, the attributes $\hat\yv$ of any point $\xv\in\RR^3$ can be queried at the decoder, as illustrated in \cref{fig:framework}.

\section{Experimental Results}
\label{sec:experiments}

\subsection{Dataset and Platform}

\begin{figure}
  \centering
   \includegraphics[height=2.0in]{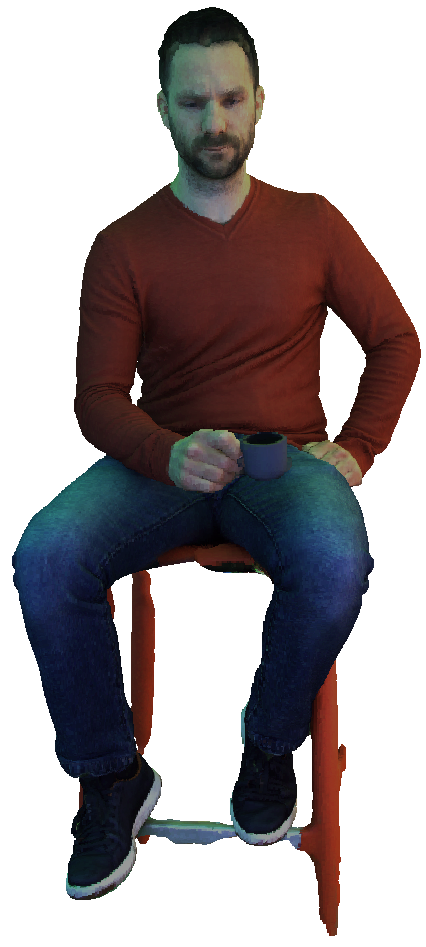}
   \includegraphics[height=2.0in]{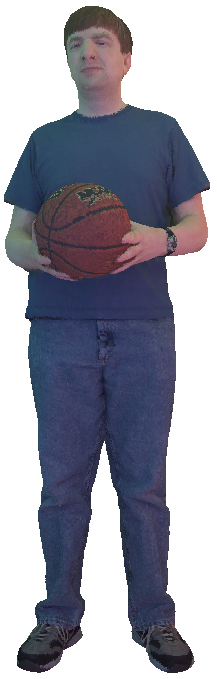}
   \includegraphics[height=2.0in]{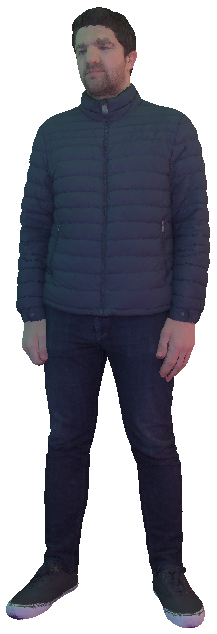}

   \caption{Point clouds {\em chair}, {\em basketball2}, and {\em jacket}.}
   \label{fig:pointclouds2}
   \vspace{-0.1in}
\end{figure}

Our dataset comprises seven full human body voxelized point clouds derived from meshes created in \cite{GuoLDBYWHOPDTTKCDFFRTDI:19, Meka:2020}, shown in \cref{fig:pointclouds,fig:pointclouds2} and summarized in \cref{tab:datasets}.  A voxel is {\em occupied} if any part of the mesh intersects it, and the color $\yv_i$ of that voxel is the average color of the mesh within the voxel.  Integer voxel coordinates are used as the point positions $\xv_i$.  The voxels (and hence the point positions) have 10-bit resolution.  This results in an octree of depth 10, or alternatively a binary tree of depth 30, for every point cloud.
Point clouds are visualized in Meshlab \cite{meshlab}.

\begin{table}
    \centering
    \begin{tabular}{|c|c|} \hline
    Point Cloud & \# points \\ \hline
    {\em rock} & 837434 \\
    {\em chair} & 791416 \\
    {\em scooter} & 959388 \\
    {\em juggling} & 798441 \\
    {\em basketball} & 868224 \\
    {\em basketball2} & 948870 \\
    {\em jacket} & 805882 \\ \hline
    \end{tabular}
    \caption{Voxelized point clouds in our dataset.}
    \label{tab:datasets}
\end{table}

We implement the LVAC framework in Python using Tensorflow.  For most experiments, we train all variables (latents, step sizes, an entropy model per binary level, and a CBN at the target level) on a single point cloud, as the variables are specific to each point cloud.  However, for the generalization experiments, we train only the latents, step sizes, and entropy models on the given point cloud, while using a CBN pre-trained on a different point cloud.  The entire point cloud constitutes one batch.  All configurations are trained in about 25K steps using the Adam optimizer and a learning rate of 0.01, with low bit rate configurations typically taking longer to converge.  Each step takes 0.5-3.0~s on an NVIDIA P100 class GPU in eager mode with various debugging checks in place.  The code will be made available on GitHub.

All results in the main body of this paper are reported for the {\em rock} point cloud, with the {\em basketball} point cloud used for generalization.  Results for the other point clouds in the dataset are provided in the Appendix.

\subsection{Baselines}

\begin{figure}
    \centering
    \includegraphics[width=1.0\linewidth, trim=20 5 35 15, clip]{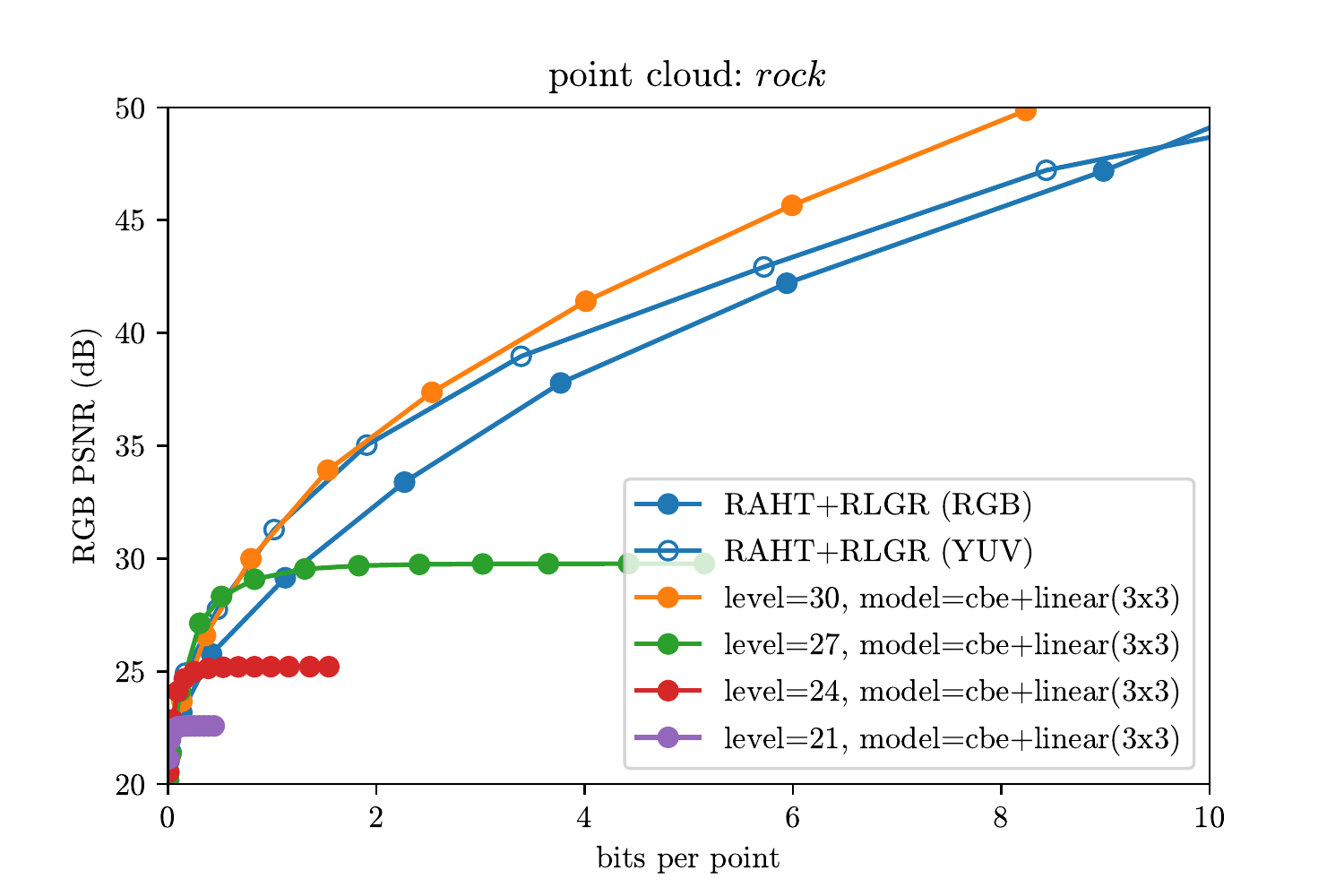}
    \caption{Baselines.  {\em RAHT+RLGR (RGB)} and {\em (YUV)} are shown against $3\!\times\!3$ linear models at levels 30, 27, 24, and 21, which optimize the colorspace by minimizing $D+\lambda R$ using the {\em cbe} entropy model.  Since {\em level=30, model=cbe+linear(3x3)} outperforms {\em RAHT+RLGR (YUV)} we discard the latter and use the others as baselines for more complex CBNs.}
    \label{fig:baselines}
    \vspace{-0.1in}
\end{figure}

Our principal baseline is RAHT, which is the core transform in the MPEG geometry-based point cloud codec (G-PCC), coupled with the adaptive Run-Length Golomb-Rice (RLGR) entropy coder \cite{Malvar:2006}.  \Cref{fig:baselines} shows the rate-distortion (RD) performance of {\em RAHT+RLGR} in RGB PSNR (dB) {\em vs} bit rate (bits per point -- bpp).  As PSNR is a measure of quality, higher is better.  In {\em RAHT+RLGR}, RAHT transforms the point colors conditioned on the geometry.  The resulting coefficients are uniformly scalar quantized with step sizes $2^n$, for $n=0,\ldots,10$.  The quantized coefficients are concatenated by level from the root to the leaves and entropy coded using RLGR, independently for each color component.  The RD performances using RGB and YUV (BT.709) colorspaces are shown in \cref{fig:baselines} in blue with filled and unfilled markers, respectively.  At low bit rates, YUV provides a significant gain in RGB PSNR, but this falls off at high bit rates.

As a secondary baseline, {\em level=30, model=cbe+linear (3x3)} in \cref{fig:baselines} shows the RD performance of our LVAC framework when 3-channel latents ($C=3$) are quantized and entropy coded using the Continuous Batched Entropy ({\em cbe}) model with the Noisy Deep Factorized prior from Tensorflow Compression \cite{tensorflow_compression} followed by a simple $3\!\times\!3$ linear matrix as the CBN, at binary target level 30.
The performance of this simple linear model agrees with that of {\em RAHT-RLGR (YUV)} at low rates, and outperforms it at high rates.  Therefore, it is useful as a pseudo or secondary baseline and we show it in all subsequent plots along with our principal baseline {\em RAHT-RLGR (RGB)}.

\Cref{fig:baselines} also shows that at lower target levels (27, 24, 21), LVAC with the $3\!\times\!3$ matrix saturates at high rates, since the $3\!\times\!3$ matrix has no positional input, and thus represents the volumetric attribute function as a constant across each block.  These constant functions serve as baselines for more complex CBNs at these levels, described next.

\subsection{Coordinate Based Networks}
\label{sec:cbns}

We now compare configurations of the LVAC framework with four different CBNs: {\em linear(3x3)} (as a baseline), {\em mlp(35x256x3)}, {\em mlp(35x64x3)}, and {\em pa(3x32x3)}, at different target levels.  The {\em mlp(35x256x3)} and {\em mlp(35x64x3)} CBNs are two-layer MLPs with 35 inputs (3 for position and 32 for a latent vector) and 3 outputs, having respectively 256 and 64 hidden nodes.  The {\em pa(3x32x3)} CBN is a Position-Attention (PA) network also with 35 inputs (3 for position and 32 for a latent vector) and 3 outputs.  \Cref{tab:cbn_parameters} shows the number of parameters in these networks.  All configurations use the Continuous Batched Entropy ({\em cbe}) model for quantization and entropy coding of the 32-channel latents.

\begin{table}
    \centering
    \begin{tabular}{|c|c|} \hline
    CBN & \# parameters \\ \hline
    {\em linear(3x3)} & 9 \\
    {\em mlp(35x256x3)} & 9987 \\
    {\em mlp(35x64x3)} & 2499 \\
    {\em pa(3x32x3)} & 227 \\ \hline
    \end{tabular}
    \caption{CBNs and number of parameters.}
    \label{tab:cbn_parameters}
    \vspace{-0.1in}
\end{table}

\begin{figure*}
    \centering
    \includegraphics[width=0.33\linewidth, trim=20 5 35 15, clip]{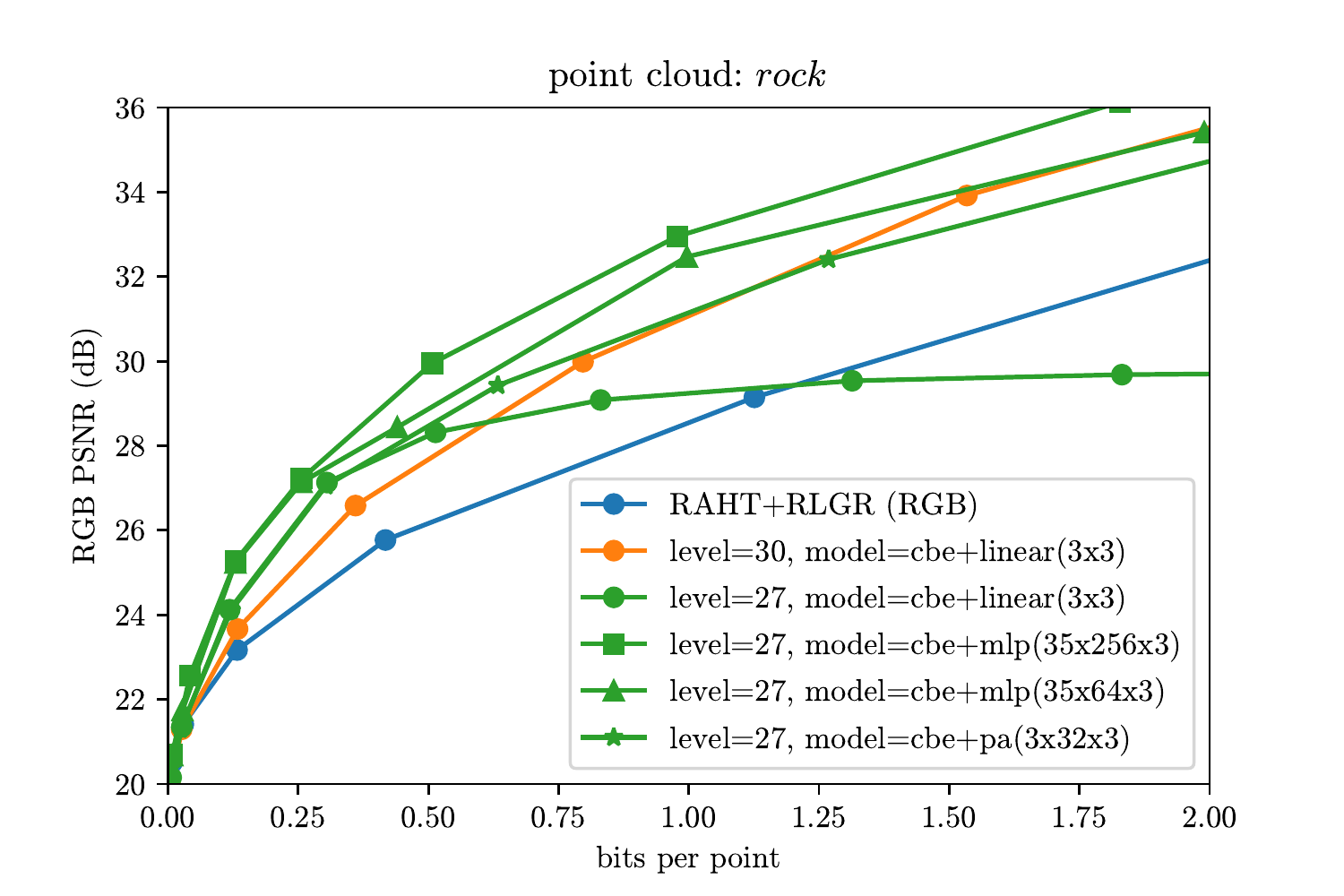}
    \includegraphics[width=0.33\linewidth, trim=20 5 35 15, clip]{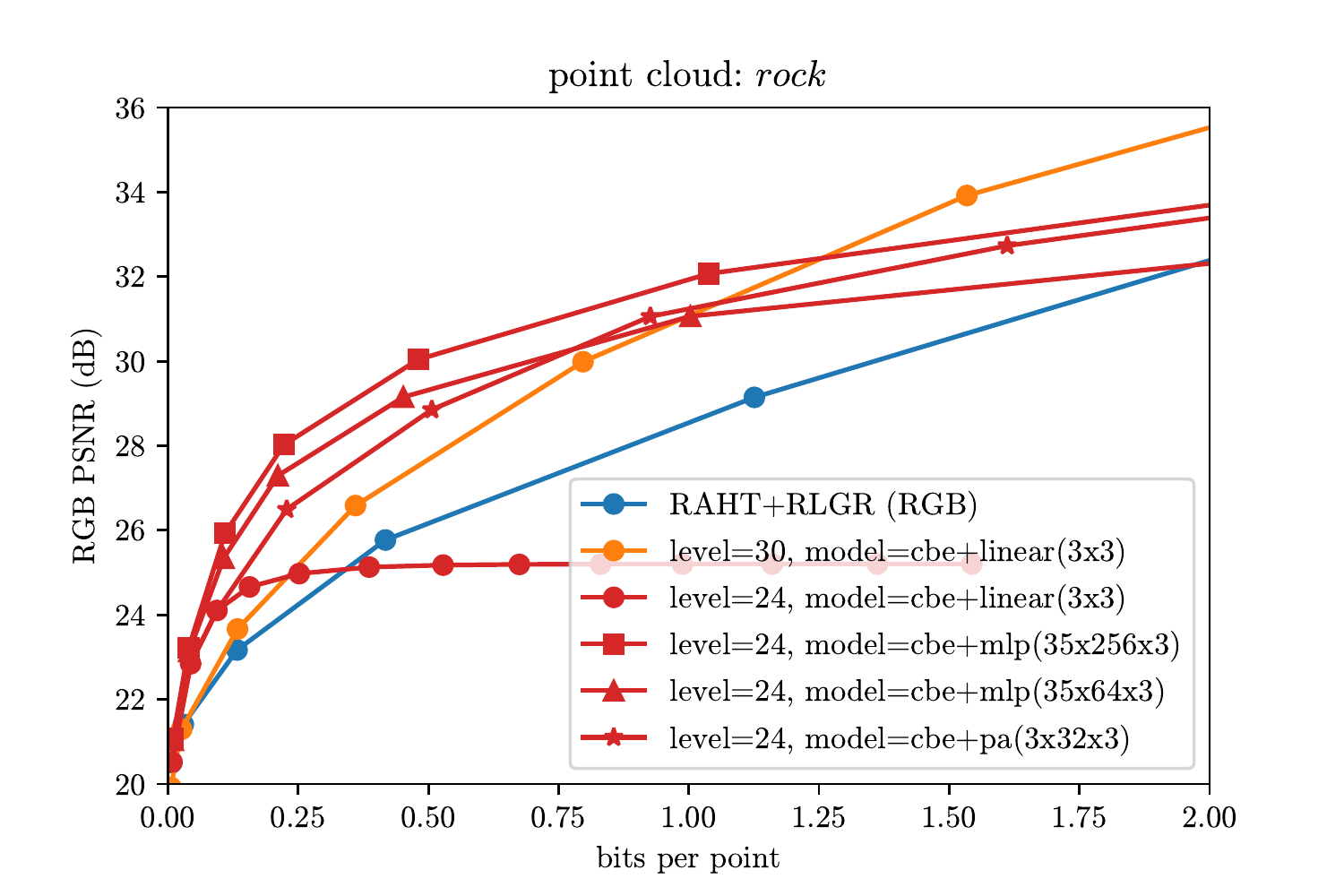}
    \includegraphics[width=0.33\linewidth, trim=20 5 35 15, clip]{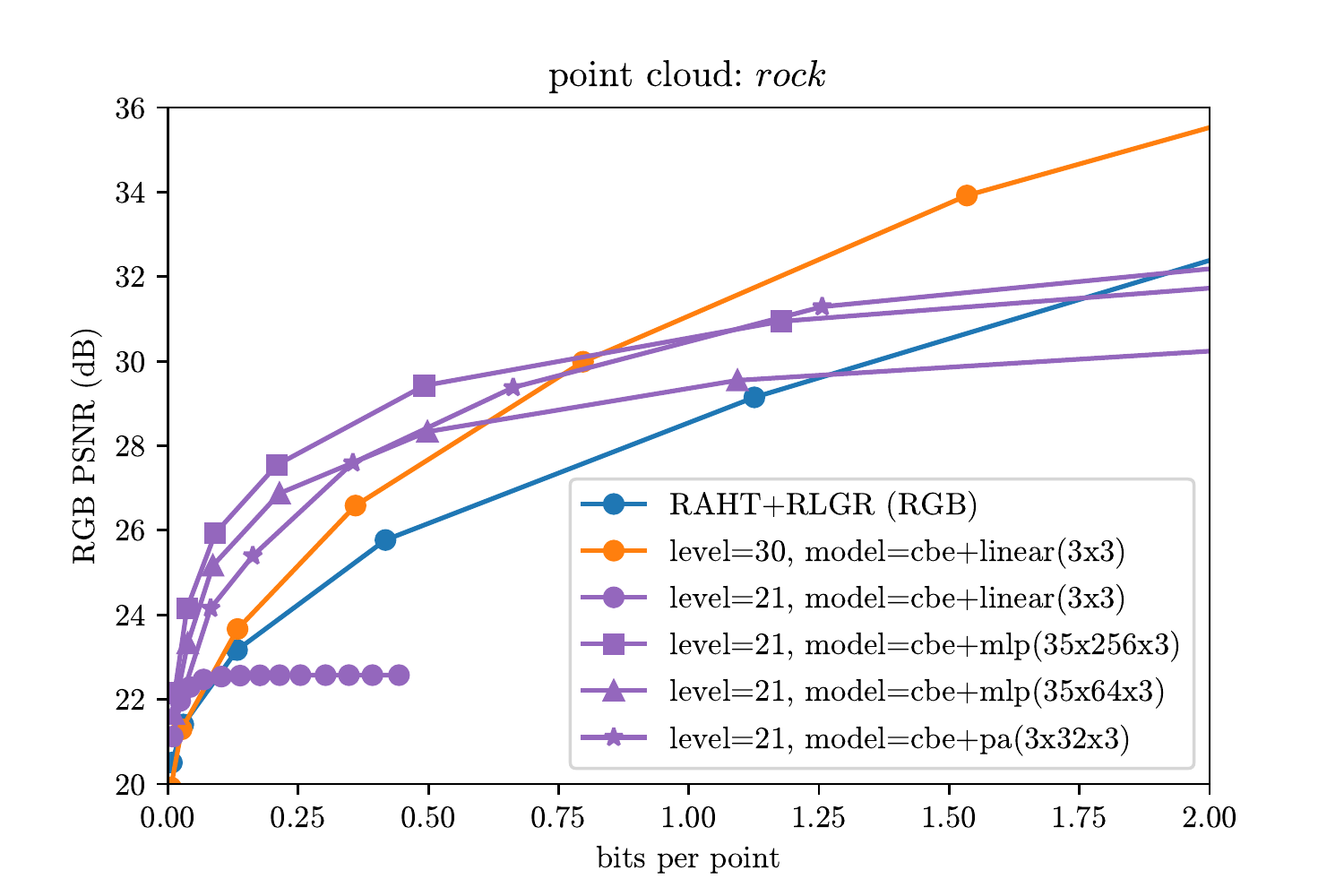}
    \caption{Coordinate Based Networks, by target level.  Left, middle, right each show {\em mlp(35x256x3)}, {\em mlp(35x64x3)}, and {\em pa(3x32x3)} CBNs, along with baselines, at levels 27, 24, 21.  More complex CBNs outperform less complex.  Higher levels are better for higher bit rates.}
    \label{fig:cbns_by_level}
    \vspace{-0.1in}
\end{figure*}

\Cref{fig:cbns_by_level} (left, middle, right) shows (in green, red, purple) the RD performance of these CBNs at different target levels (27, 24, 21), along with the baselines (in blue, orange).
We observe that first, at each target level $L=27,24,21$, the CBNs with more parameters outperform the CBNs with fewer parameters.  In particular, especially at higher bit rates, the MLP and PA networks at level $L$ improve more than 5--10 dB over the linear network at level $L$, whose RD performance saturates as described earlier, for each $L$.  Second, at each target level $L=27,24,21$, there is a range of bit rates over which the MLP and PA networks improve by 2--3 dB over even the {\em level=30, model=cbe+linear(3x3)} baseline, which does not saturate.  The range of bit rates in which this improvement is achieved is higher for level 27, and lower for level 21, reflecting that higher quality requires CBNs with smaller blocksizes.

\Cref{fig:cbns_by_network} in the Appendix shows these same data factored by CBN type instead of by level, to illustrate again that for each CBN type, each level is optimal for a different bit rate range.

\begin{figure}
    \centering
    \includegraphics[width=1.0\linewidth, trim=20 5 35 15, clip]{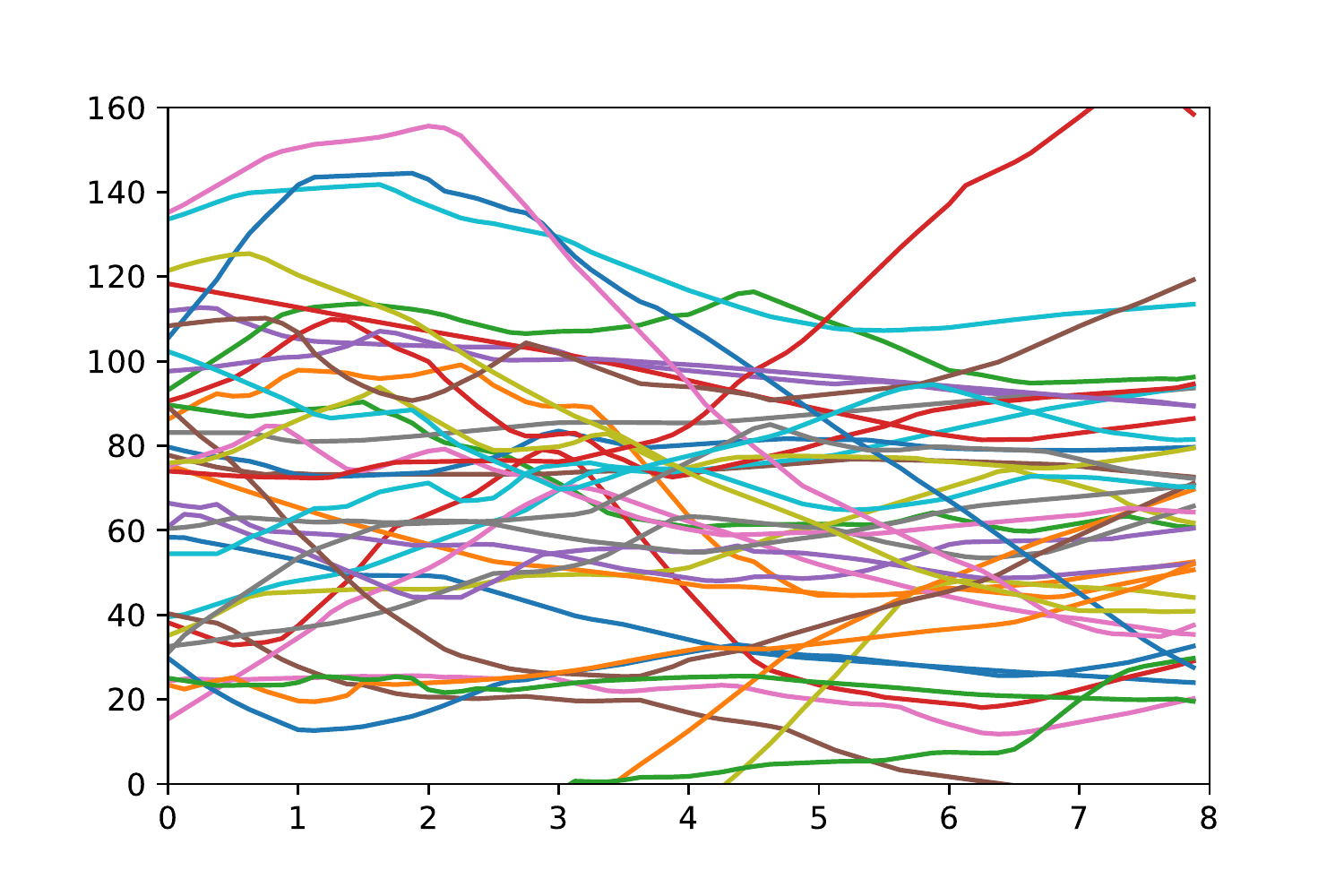}
    \caption{Cuts through the volumetric function $f_{\theta}(\xv;\zv)$ represented by a CBN, for different values of $\zv$, at target level 21.}
    \label{fig:cbn_examples}
    \vspace{-0.1in}
\end{figure}

The nature of a volumetric function $f_{\theta}(\xv;\zv)$ represented by a CBN is illustrated in \cref{fig:cbn_examples}.  To illustrate, we select the CBN {\em mlp(35x256x3)} trained on the {\em rock} point cloud at target level $L=21$, and we plot cuts through the volumetric function $f_{\theta}(\cdot;\zv)$ represented by this CBN.  Specifically, let $n$ be a randomly selected node at the target level $L$, let $\hat{\zv}_n$ be the quantized cumulative latent at that node, and let $\xv_n=(x_n,y_n,z_n)$ be the position of a randomly selected point within the block at that node.  Then we plot the first (red) component of the function $f_{\theta}(\xv;\hat{\zv}_n)$, where $\xv$ varies from $(0,y_n,z_n)$ to $(N_x,y_n,z_n)$, where $N_x$ is the width of a block at level $L$.  We do this for many randomly selected nodes $n$ to get a sense of the distribution of volumetric functions represented at that level.  (The distribution looks similar for green and blue components, and for cuts along $y$ and $z$ axes.)  We observe that for many values of $\hat{\zv}_n$, $f_{\theta}(\cdot;\hat{\zv}_n)$ is a roughly constant function.  Thus, $\hat{\zv}_n$ must encode the colors of the palette used for these functions.  However, we also observe that for some values of $\hat{\zv}_n$, $f_{\theta}(\cdot;\hat{\zv}_n)$ is a ramp or some other nonlinear function across its domain.  Finally, we observe almost no energy at frequencies higher than the Nyquist frequency (half the sampling rate), where the sampling occurs at units of voxels.

\subsection{Generalization}
\label{sec:generalization}

In \cref{sec:cbns}, the CBNs were optimized along with the latents, step sizes, and entropy models for a particular point cloud, {\em rock}.  In this subsection, we take a small step towards exploring the degree to which the CBNs can be generalized across point clouds; that is, whether they can be trained to represent a universal family of volumetric functions.  Towards that end, we pre-train the CBNs on the point cloud {\em basketball}, and fix their parameters while optimizing the other parameters (i.e., latents, step sizes, and entropy model parameters) on the point cloud {\em rock}.  \Cref{fig:cbns_gen} in the Appendix shows the results:  Even with training on only a single other point cloud, the CBNs can indeed generalize across point clouds at low bit rates.  At high bit rates, however, the CBNs trained on just one other point cloud do not perform well, most likely because they have been unable to learn to represent the fine details needed for a different point cloud.  Some of these results will be displayed again in the following subsection.

\subsection{Side Information}

When the latents, step sizes, entropy models, and CBN are all optimized for a specific point cloud, quantizing and entropy coding only the latent vectors $[\zv_\nv]$ is insufficient for reconstructing the point cloud attributes.  The step sizes $[\Delta_c]$, entropy model parameters $[\phi_{\ell,c}]$, and CBN parameters $\theta$ must also be quantized, entropy coded, and sent as {\em side information}.  Sending side information incurs additional bit rate and distortion.  This subsection explores the cost of this side information for both the entropy models and CBN.  The side information for the step sizes is negligible, as there is only one step size for each of $C=32$ channels.

\begin{figure}
    \centering
    \includegraphics[width=1.0\linewidth, trim=20 5 35 15, clip]{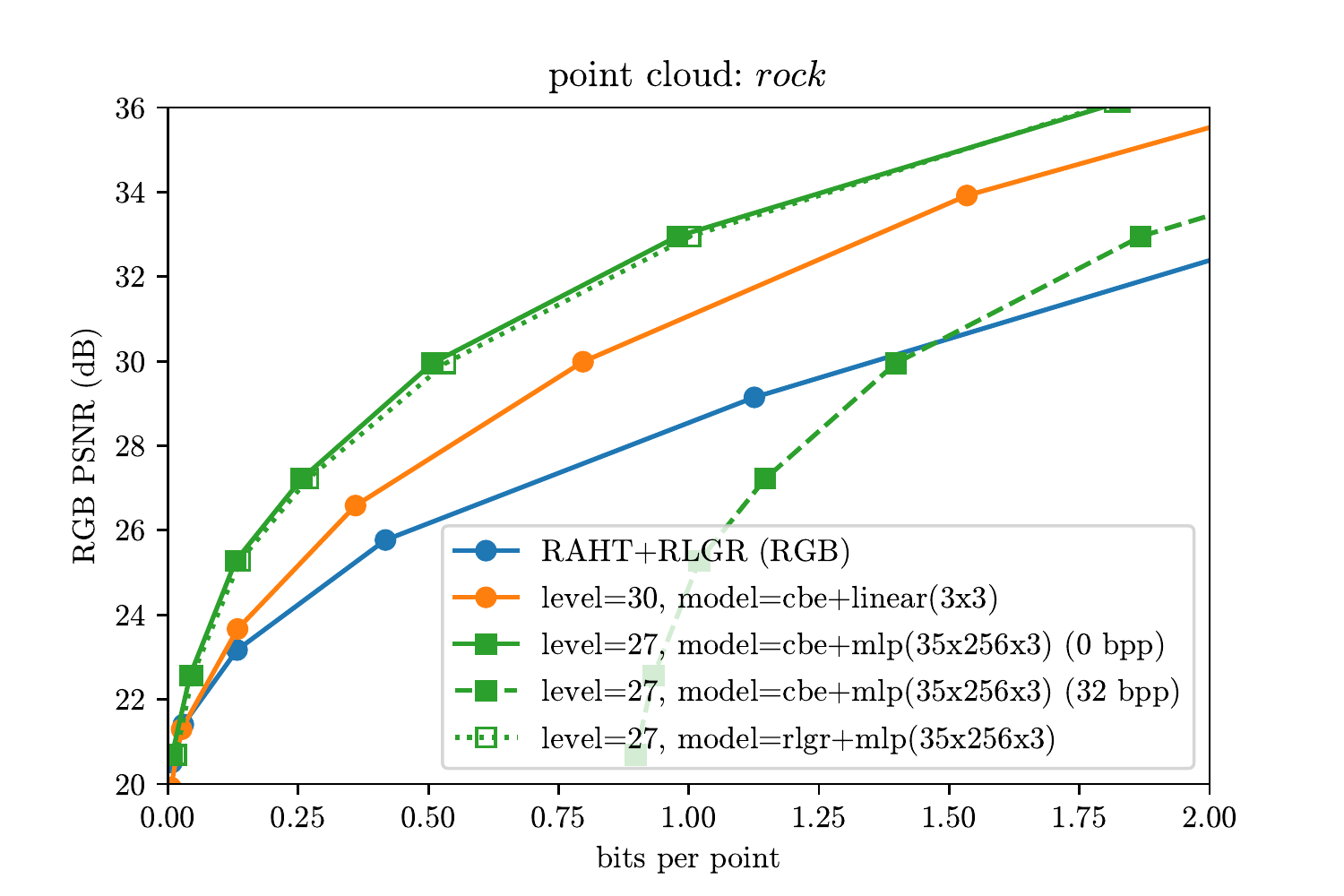}
    \caption{Side information for entropy model.  Sending 32 bits per parameter for the {\em cbe} entropy model would reduce RD performance from solid to dashed green lines.  But the backward-adaptive {\em RLGR} entropy coder (dotted, unfilled) obviates the need to send side information with almost no loss in performance.}
    \label{fig:sideinfo_cbe_vs_rlgr}
    \vspace{-0.1in}
\end{figure}

First, we consider the side information for the entropy models.  For each point cloud, there is one entropy model per binary level\footnote{except binary levels in which each node has only one occupied child}, per channel.  For 26 such binary levels, 32 channels, and the Continuous Batched Entropy ({\em cbe}) model with Noisy Deep Factorized prior, this works out to 23296 floating point parameters.  If we allocate 32 bits per floating point parameter, the bit rate would increase by 0.89 bits per point for the {\em rock} point cloud, which has 837434 points.  Thus the RD performance would move from the solid green line to the dashed green line in \cref{fig:sideinfo_cbe_vs_rlgr}, for {\em level=27, model=cbe+mlp(35x256x3)}.  However, fortunately, this costly side information can be avoided, by using {\em cbe} during training but using RLGR during inference.  Since RLGR is backward adaptive, it can adapt to Laplacian-like distributions without sending any side information.  Of course its coding efficiency may suffer, but our experiments show that this degradation --- to the dotted green line with open markers in \cref{fig:sideinfo_cbe_vs_rlgr} --- is almost negligible.  Henceforth we report RD performance using only RLGR during inference.

\begin{figure*}
    \centering
    \includegraphics[width=0.33\linewidth, trim=20 5 35 15, clip]{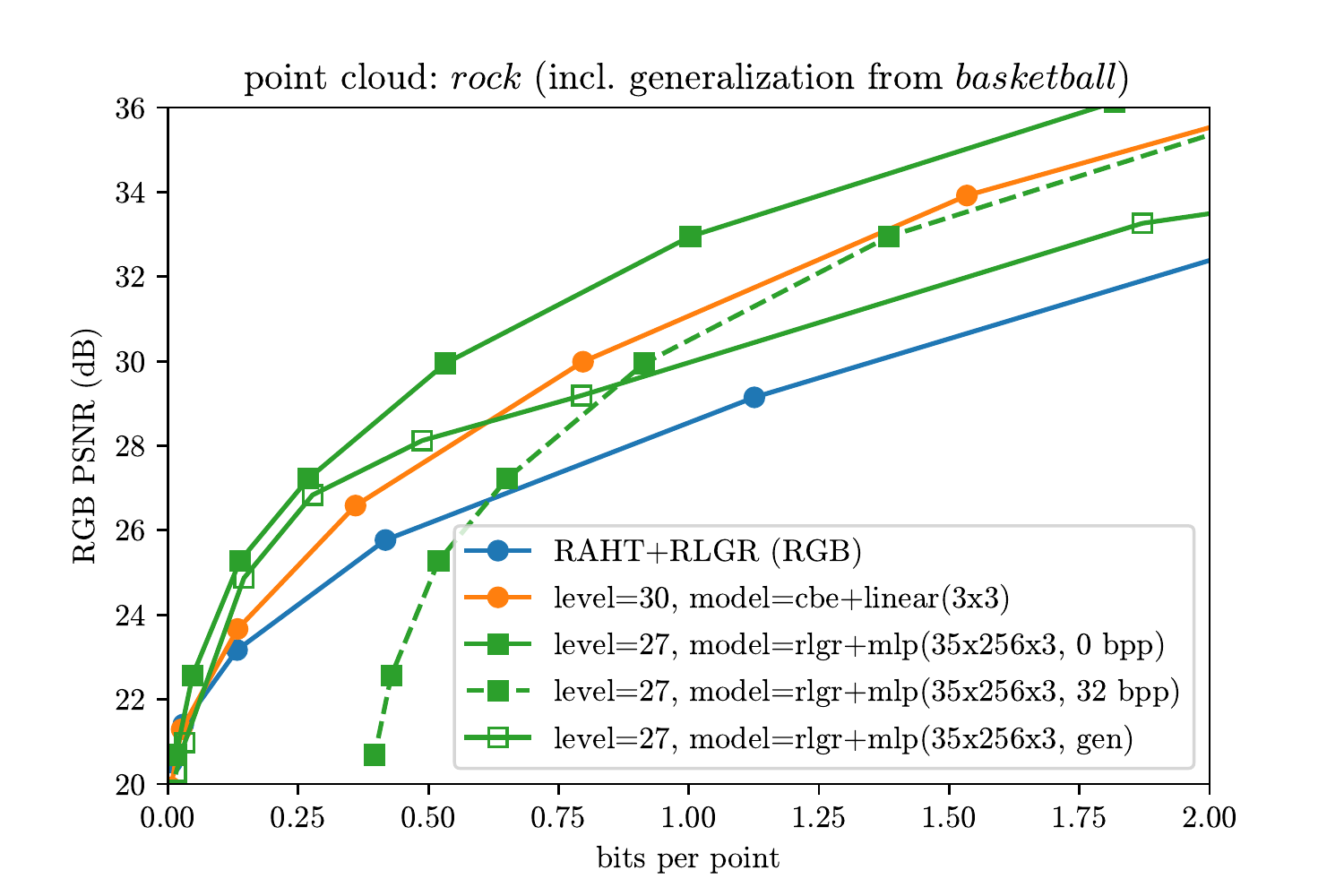}
    \includegraphics[width=0.33\linewidth, trim=20 5 35 15, clip]{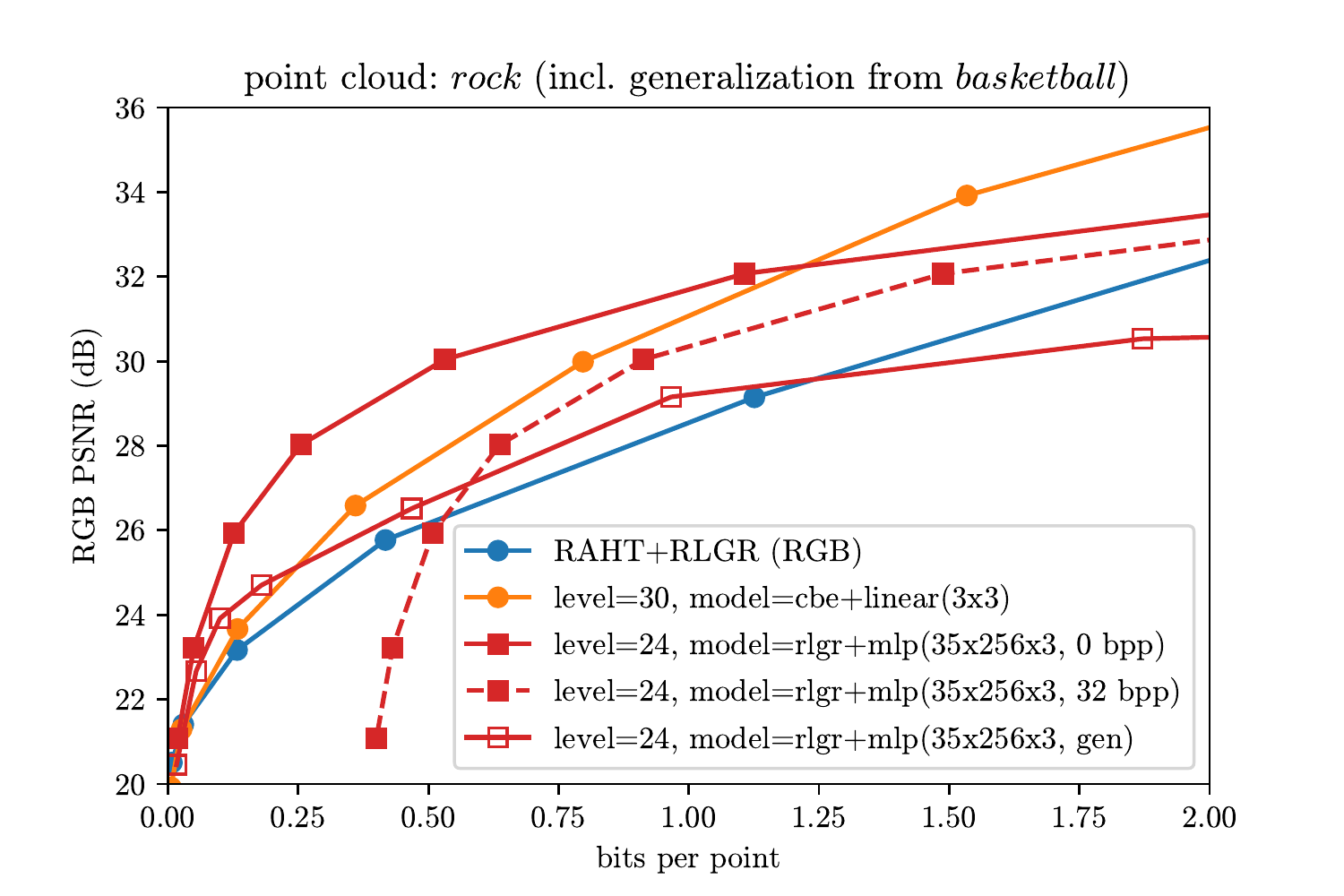}
    \includegraphics[width=0.33\linewidth, trim=20 5 35 15, clip]{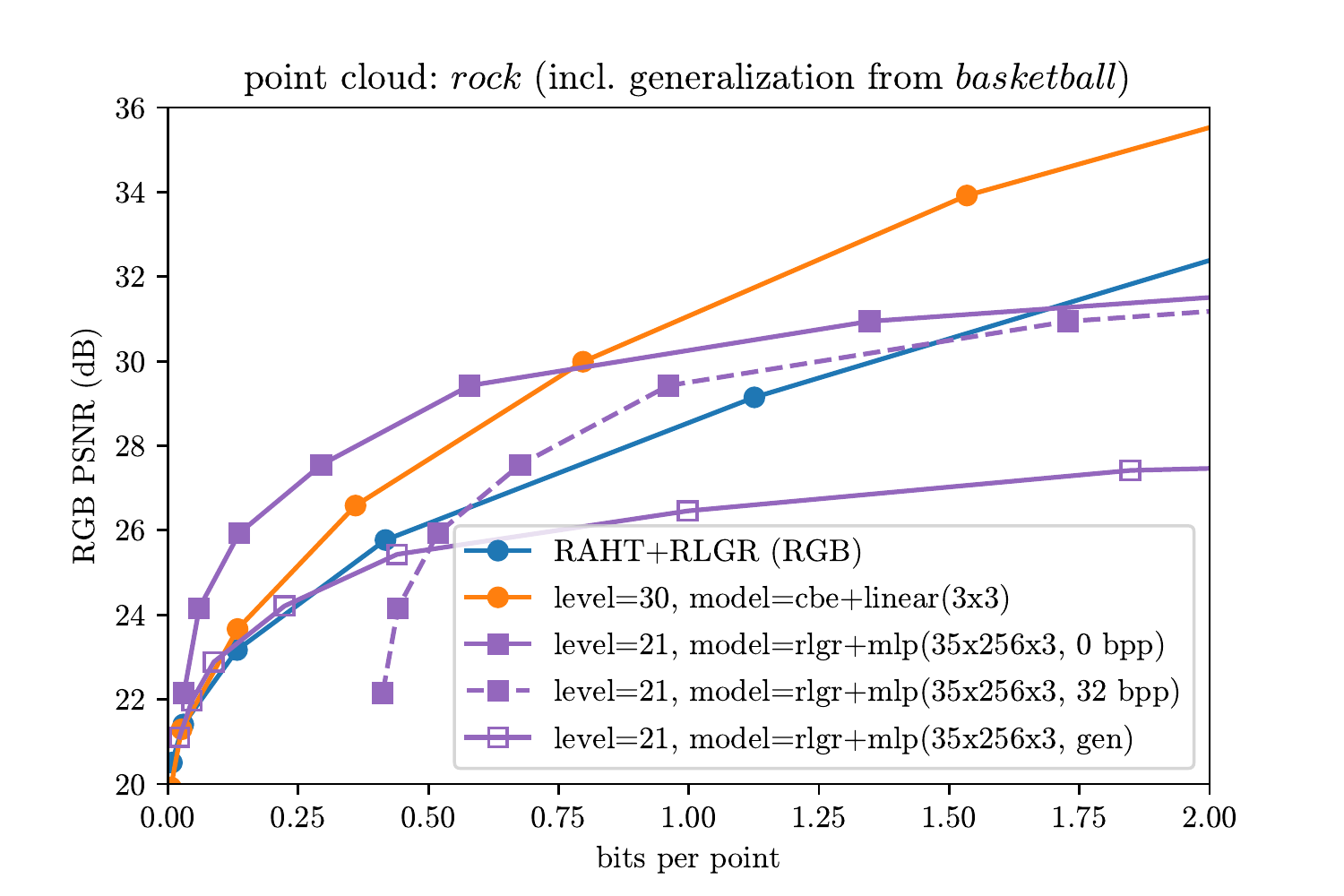}
    \caption{Effect of side information for coordinate based network {\em mlp(35x256x3)} at levels 27 (left), 24 (middle), and 21 (right).  Sending 32 bits per parameter for the CBN would degrade RD performance from solid to dashed lines.  The degradation would be inversely proportional to compression ratio if model compression is used.  Alternatively, generalization (pre-training the CBN on one or more other point clouds), which works well at low bit rates, would obviate the need to transmit any side information.
    Generalization is indicated by ``gen'' in the legend.}
    \label{fig:sideinfo_mlp256}
\end{figure*}

Next, we consider the side information for the CBNs.  For each point cloud, there is one CBN, at the target level.  For the number of parameters in our CBNs (\cref{tab:cbn_parameters}), if we allocate 32 bits per floating point parameter, the side information to transmit the CBN would be 0.38, 0.095, and 0.009 bits per point for {\em mlp(35x256x3)}, {\em mlp(35x64x3)}, and {\em pa(3x32x3)}, respectively.  \Cref{fig:sideinfo_mlp256} shows the resulting RD performance for {\em mlp(35x256x3)}, while \cref{fig:sideinfo_mlp64_and_pa} in the Appendix shows the resulting RD performance for {\em mlp(35x64x3)} and {\em pa(3x32x3)}, at target levels 27, 24, and 21.  It can be seen that 32 bits per parameter of side information to encode {\em mlp(35x256x3)} has a severe effect on RD performance at low bit rates, but less severe at high bit rates.  For {\em mlp(35x64x3)}, the effect is more modest, and for {\em pa(3x32x3)}, it is negligible.  Fortunately for the MLPs, at low bit rates, where the side information penalizes them most, and where generalization works best, they may alternatively be generalized to avoid having to transmit any side information.  The RD performance of the generalized CBNs is also included in the figures.

Also fortunately, it is likely that 32 bits per floating point parameter is an order of magnitude more than necessary.
Prior work has shown that simple model compression can be performed at 8 bits per floating point parameter \cite{Banner8bitQuant, sun2019hybrid8bitQuant, wang2018training8bitQuant} or even more aggressively at 1--4 bits per floating point parameter  \cite{han2015deepQuant, IsikHW:21, oktay2019scalable, stock2019andQuant, wang2019haqQuant, xu2018deepQuant, isik2021noisynn} with very low loss in performance, even with coordinate based networks such as NeRF \cite{BirdBSC:21,isik2021neural}.
However, since model compression is outside the scope of our work, we simply parameterize our results by the number of bits per floating point parameter.  In \cref{sec:convhull}, we will return to the effect of side information under various values for the number of bits per parameter.  First, however, we turn to a key ablation study.

\subsection{Normalization}
\label{sec:normalization}

\begin{figure}
    \centering
    \includegraphics[width=1.0\linewidth, trim=20 15 35 32, clip]{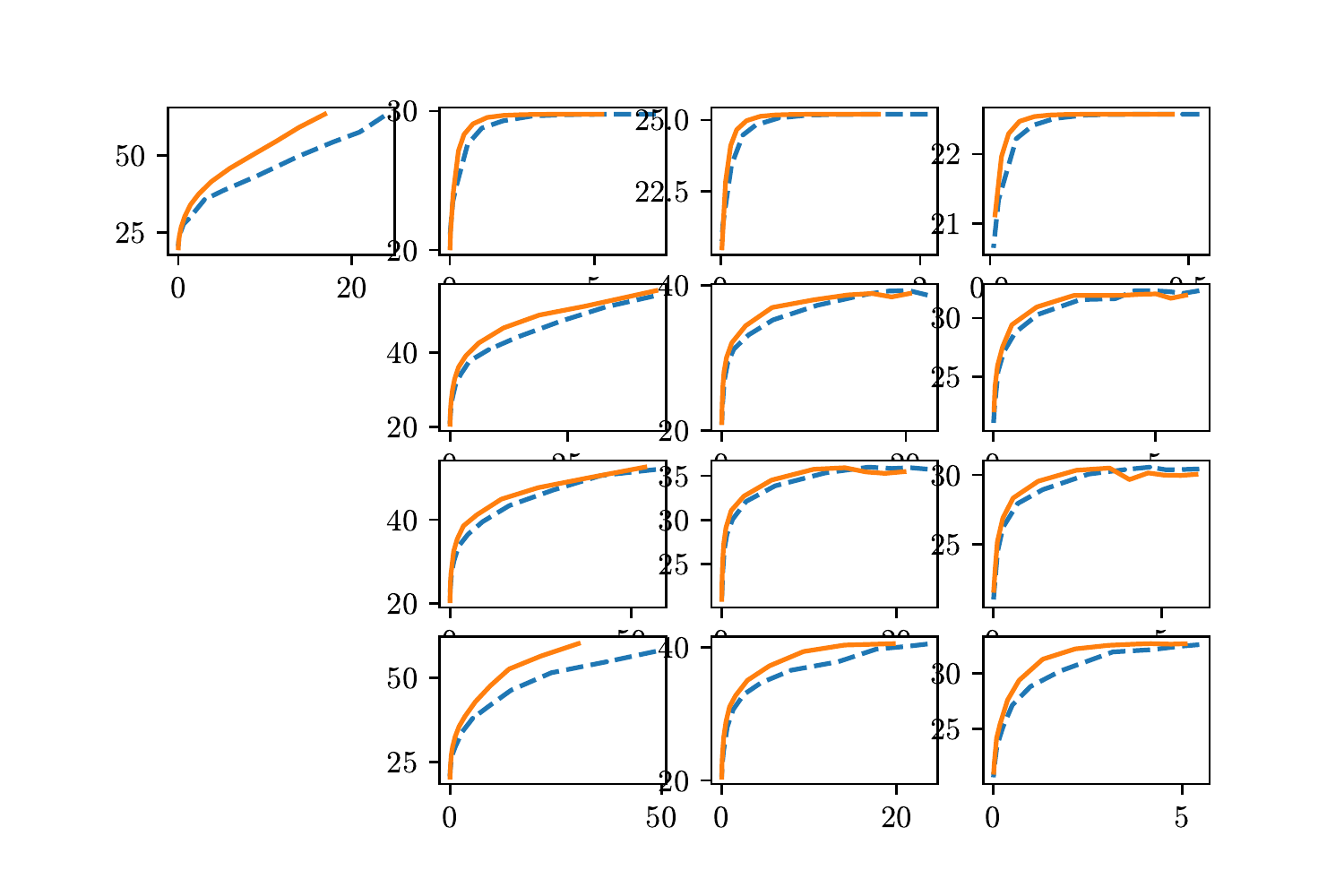}
    \caption{RD performance (RGB PSNR vs.\ bit rate) improvement due to normalization, corresponding to entries in \cref{tab:normalization}.}
    \label{fig:normalization}
    \vspace{-0.1in}
\end{figure}

One of our main contributions is to show that na\"{\i}ve uniform scalar quantization and entropy coding of the latents leads to poor results, and that properly normalizing the coefficients before quantization achieves over a 30\% reduction in bit rate.  In this ablation study, we remove our normalization by setting the scale matrix $\Sm$ in (\ref{eqn:normalization_S0})-(\ref{eqn:normalization_Sn}) and \cref{fig:pipeline} to the identity matrix, thus removing any dependency of the attribute compression on the geometry.  This corresponds to a na\"{\i}ve approach to compression, for example by assuming a fixed number of bits per latent as in \cite{takikawa2021neural}.  \Cref{tab:normalization} shows that compared to this na\"{\i}ve approach, our normalization achieves over 30\% reduction in bit rate on average over all point clouds in the dataset (computed using \cite{Bjntegaard2001CalculationOA,pateux2007excel}).  This quantifies the reduction in bit rate due to conditioning the attribute compression on the geometry.  \Cref{fig:normalization} shows the RD performances of the na\"{\i}ve (dotted blue line) and normalized (solid orange line) approaches, corresponding to the entries in \cref{tab:normalization}.
We observe that,
except in the linear case, normalization tends to help more when the CBN is at a higher target level.  This makes sense, as there are fewer latents to normalize when the CBN is at a lower level.
However, we also observed in \cref{fig:cbns_by_level} that,
at higher bit rates, the CBNs at lower levels are unable to outperform even the normalized linear case at level 30.  Thus, while the normalized linear case is able to condition on detailed geometry all the way to individual voxels, the CBNs despite their much higher complexity perform worse than the linear models at high bit rates because they do not condition on the specific geometry within their blocks.
This points to a continued need to research how to fully condition on geometry with learned attribute compression.

\begin{table}
    \centering
    \begin{tabular}{|c|cccc|} \hline
    & \multicolumn{4}{c|}{level} \\
    CBN & 30 & 27 & 24 & 21 \\ \hline
    linear(3x3) & -31.6\% & -18.6\% & -28.3\% & -37.7\% \\
    mlp(35x256x3) & N/A & -29.9\% & -34.3\% & -27.4\% \\
    mlp(35x64x3) & N/A & -23.8\% & -32.1\% & -31.1\% \\
    pa(3x32x3) & N/A & -41.1\% & -40.3\% & -38.7\% \\ \hline
    \end{tabular}
    \caption{BD-Rate reductions due to normalization, averaged over point clouds.  Normalization is crucial for good performance.  Without normalization, there is no dependence on geometry.}
    \label{tab:normalization}
\end{table}

\subsection{Convex Hull}
\label{sec:convhull}

\begin{figure*}
    \centering
    \includegraphics[width=0.33\linewidth, trim=20 5 35 15, clip]{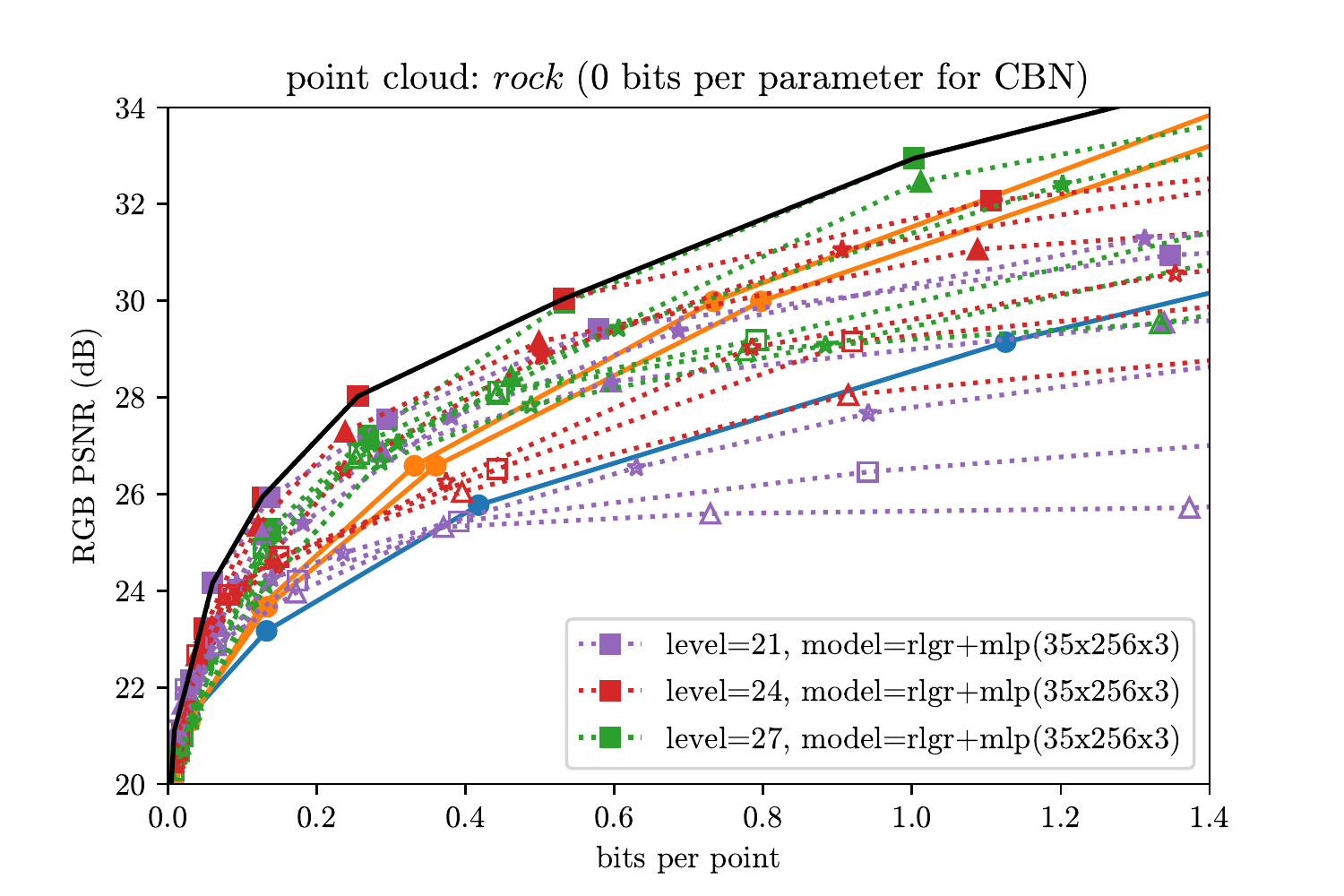}
    \includegraphics[width=0.33\linewidth, trim=20 5 35 15, clip]{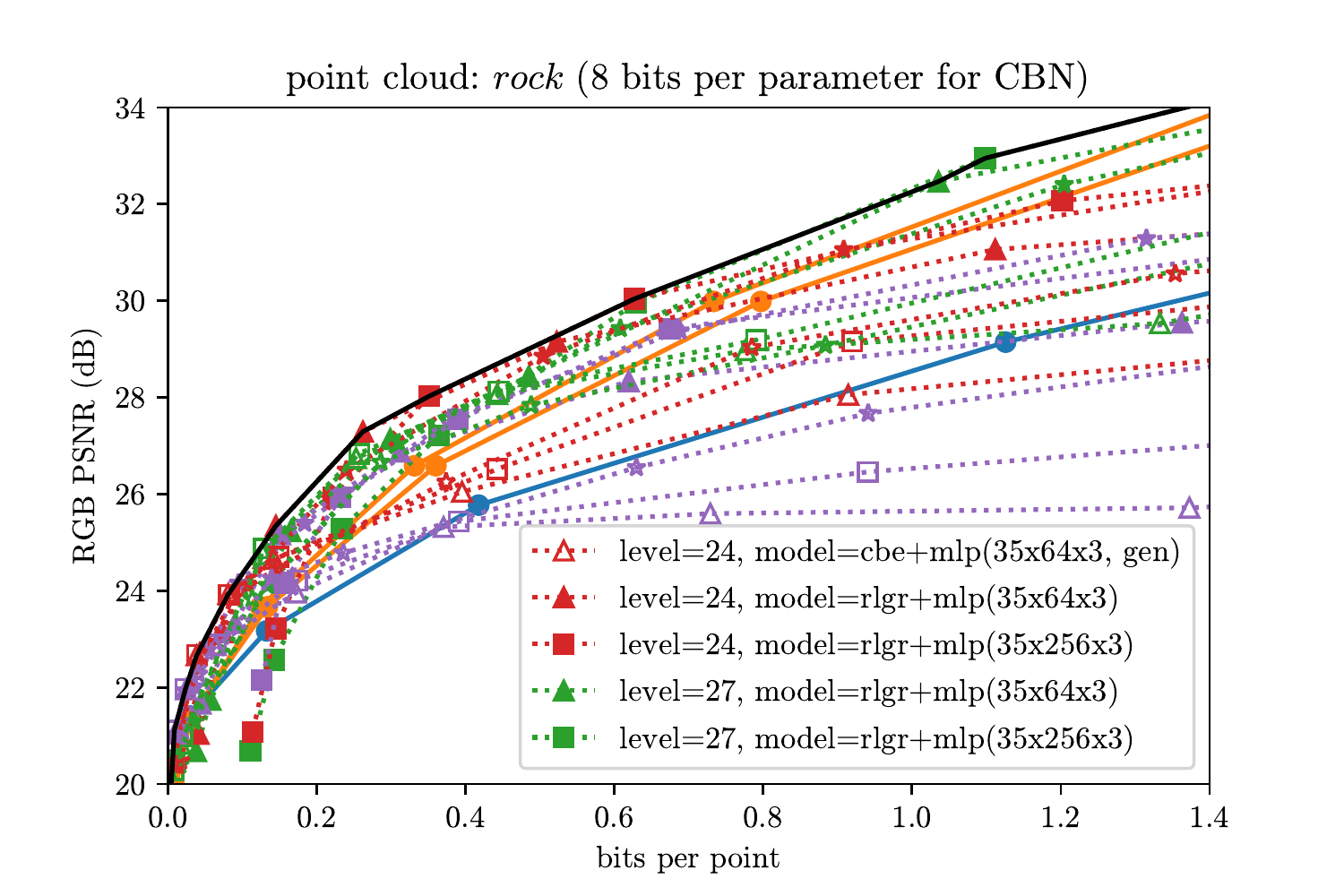}
    \includegraphics[width=0.33\linewidth, trim=20 5 35 15, clip]{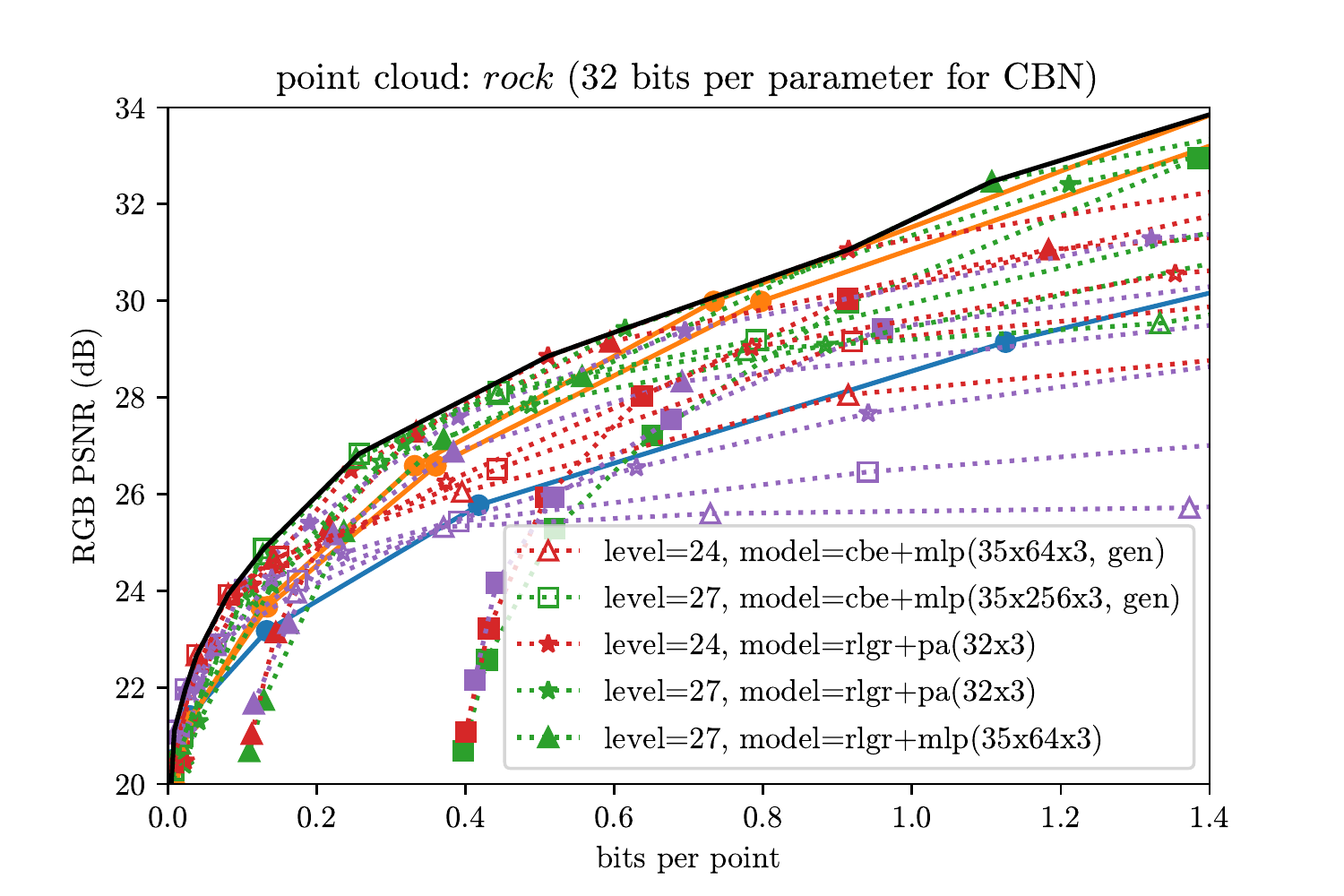}
    \caption{Convex hull (solid black line) of RD performances of all CBN configurations across all levels, including side information using 0 (left), 8 (middle), and 32 (right) bits per CBN parameter.  Configurations that participate in the convex hull are listed in the legend.  At 0 bits per parameter, the more complex CBNs dominate.  At higher bits per parameter, the generalized and less complex CBNs begin to participate, especially at lower bit rates.}
    \label{fig:convhull}
    \vspace{-0.1in}
\end{figure*}

For different bit rate ranges, and for different assumptions on the cost of side information, different configurations of the LVAC framework may be optimal.  \Cref{fig:convhull} shows the convex hull, or Pareto fontier, of all configurations under the assumptions of 0 (left), 8 (middle), and 32 (right) bits per floating point parameter.  All configurations that we have examined in this paper appear in each plot.  However, only those that participate in the convex hull are listed in the legend.  We observe:
first, when the side information costs nothing (0 bits per parameter), the convex hull contains exclusively the largest CBN ({\em mlp(35x256x3)}), at higher target levels for higher bit rates.
Second, as the cost of the side information increases, the smaller CBNs ({\em mlp(35x64x3)} and {\em pa(3x32x3)}), and those that are generalized from another point cloud ({\em mlp(35x256x3, gen)} and {\em mlp(35x64x3 gen)}), begin to participate in the convex hull, especially at lower bit rates.  Eventually, at 32 bits per parameters, the largest CBN is excluded entirely.
Third, the generalizations participate in the convex hull only at the lowest bit rates, despite not incurring any penalty due to side information.  This could be because they are trained only on a single other point cloud in these experiments.  Training the CBNs on more representative data would probably improve their generalization performance
across a wider range of bit rates
but is left for future work.

\subsection{Subjective Quality}

\begin{figure*}
    \centering
    \begin{minipage}{0.245\textwidth}
    \centering\small
    \includegraphics[width=1.0\linewidth]{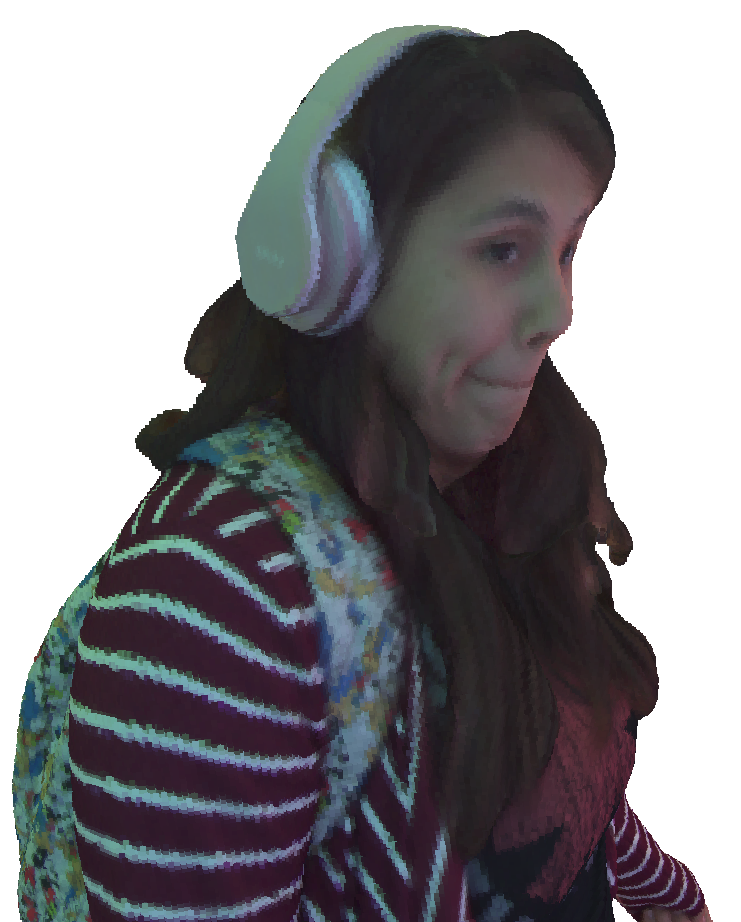} \\
    (a) Original
    \end{minipage}
    \begin{minipage}{0.245\textwidth}
    \centering\small
    \includegraphics[width=1.0\linewidth]{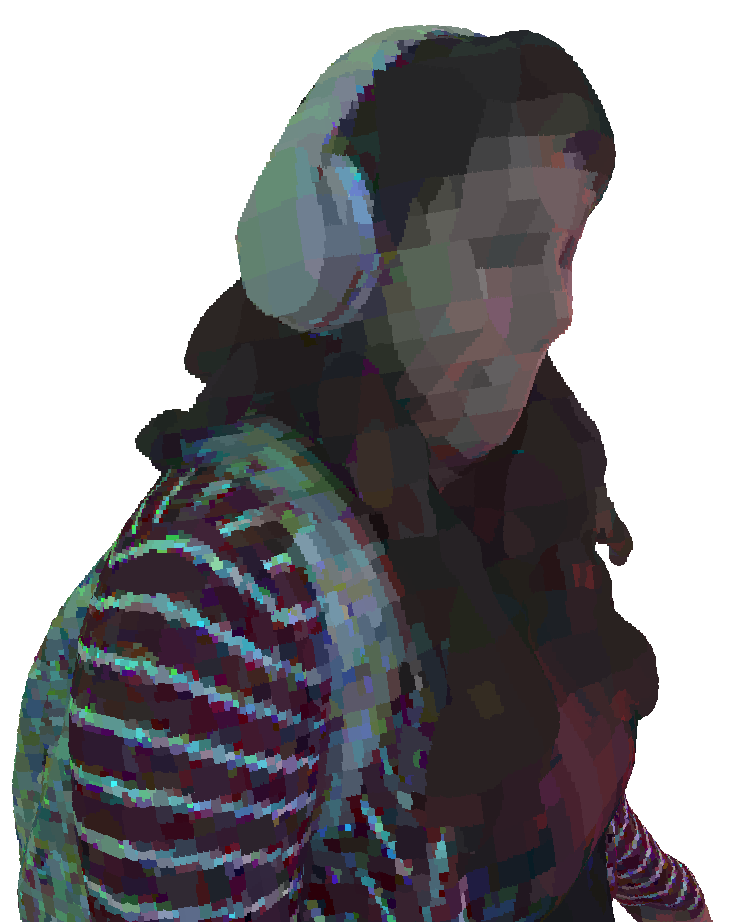} \\
    (b) {\em RAHT-RLGR (RGB)}
    \end{minipage}
    \begin{minipage}{0.245\textwidth}
    \centering\small
    \includegraphics[width=1.0\linewidth]{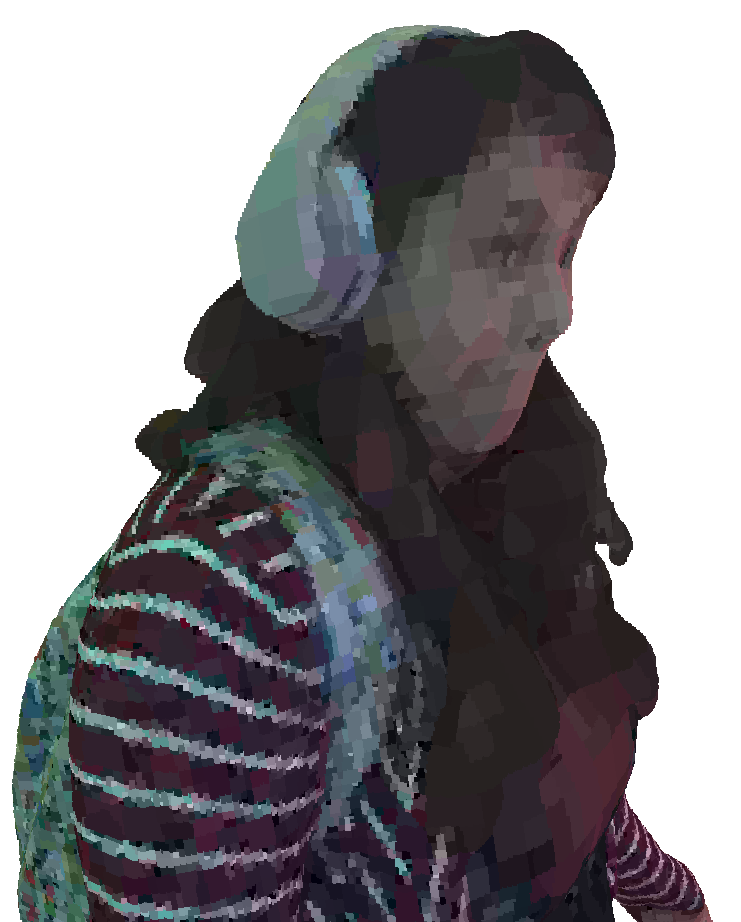} \\
    (c) {\em RAHT-RLGR (YUV)}
    \end{minipage}
    \begin{minipage}{0.245\textwidth}
    \centering\small
    \includegraphics[width=1.0\linewidth]{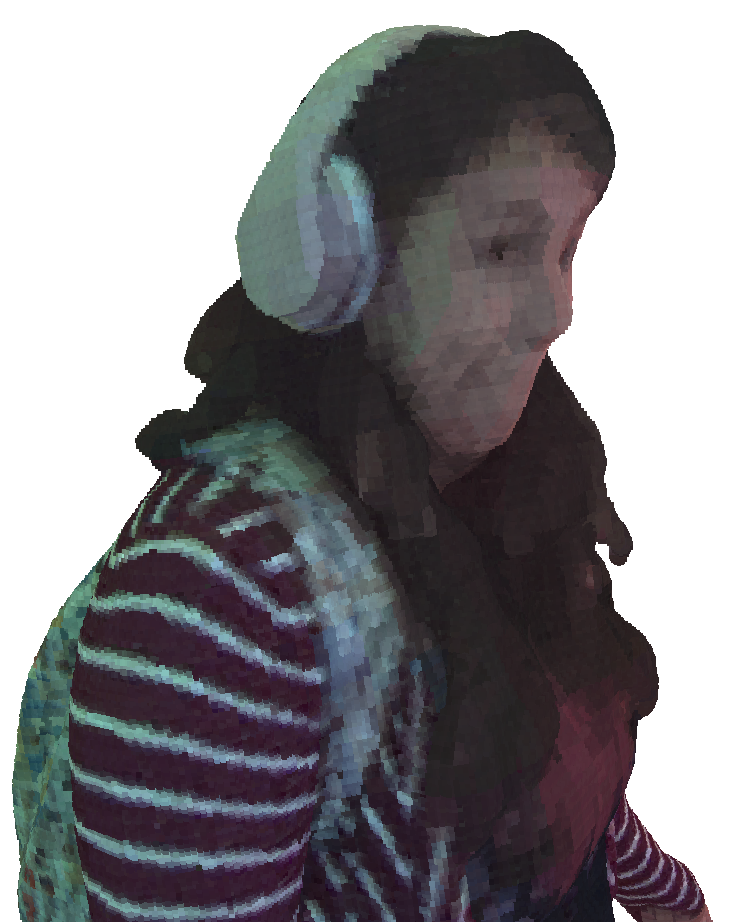} \\
    (d) {\em LVAC mlp(35x256x3)}
    \end{minipage}
    
    \caption{Subjective quality around 0.25 bpp.
    (a) Original. (b) 0.258 bpp, 24.6 dB. (c) 0.255 bpp, 25.9 dB. (d) 0.255 bpp, 28.0 dB.
    }
    \label{fig:subjective}
    \vspace{-0.1in}
\end{figure*}

\Cref{fig:subjective} shows the subjective compression quality around 0.25 bpp, under the assumption of 0 bits per floating point parameter.  Additional bit rates are shown in the Appendix.

\subsection{Baselines, Revisited}

\begin{figure}
    \centering
    \includegraphics[width=1.0\linewidth, trim=20 5 35 15, clip]{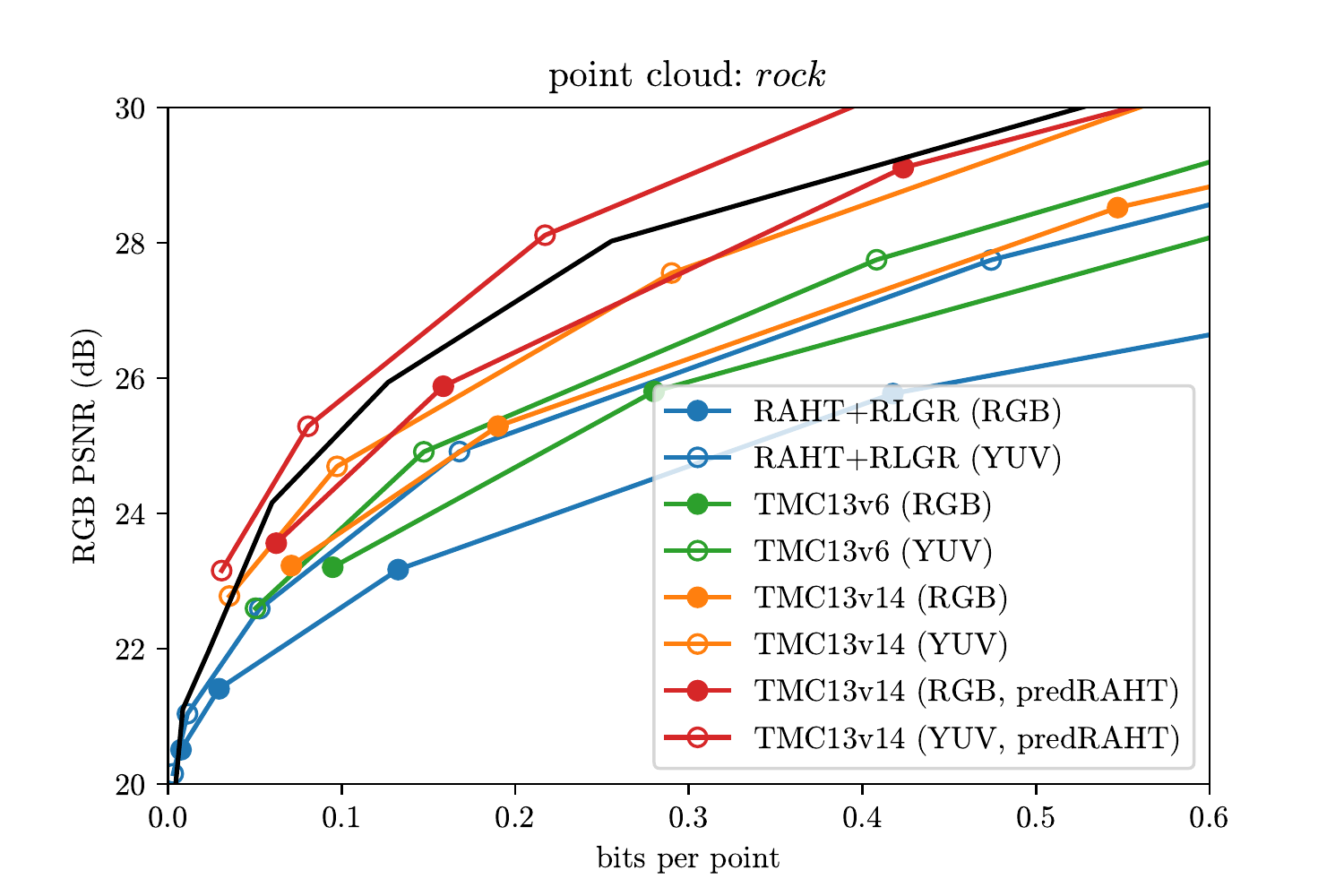}
    \caption{Baselines, revisited. In both RGB and YUV color spaces, MPEG G-PCC reference software TMC13 v6.0 improves over {\em RAHT+RLGR}, primarily due to context-adaptive (i.e., dependent) entropy coding.  TMC13 v14.0 improves still further, primarily due to predictive RAHT.  Better entropy coding (e.g., hyperprior) and predictive RAHT can also be applied to LVAC.}
    \label{fig:baselines_revisted}
    \vspace{-0.1in}
\end{figure}

We now return to the matter of baselines.  \Cref{fig:baselines_revisted} shows our previous baseline, {\em RAHT+RLGR}, for both RGB and YUV colorspaces (blue lines).  Although RAHT is the transform used in MPEG G-PCC, the reference software TMC13 v6.0 (July 2019) offers improved RD performance (green lines) compared to {\em RAHT+RLGR}, due principally to better entropy coding.  In particular, TMC13 uses context-adaptive binary arithmetic coding with various coding modes, while {\em RAHT+RLGR} uses RLGR.  We use {\em RAHT+RLGR} as our baseline because our experiments use RLGR as our entropy coder; the specific entropy coder used in TMC13 is difficult to extract from the standard.  The latest version, TMC13 v14.0 (October 2021), offers even better RD performance, by introducing for example joint coding modes for color channels that are all zero (orange lines).  It also introduces predictive RAHT, in which the RAHT coefficients at each level are predicted from the decoded RAHT coefficients at the previous level \cite{LasserreF:19,GPCC,PavezSQO:21}.  The prediction residuals, instead of the RAHT coefficients themselves, are quantized and entropy coded.  Predictive RAHT alone improves RD performance by 2--3 dB (red lines).
Nevertheless, LVAC with RLGR and no RAHT prediction (solid black line, from \cref{fig:convhull} (left)) is within 1 dB of TMC13 v14.0
with joint entropy coding and RAHT prediction, and it outperforms all other versions.
Moreover, we believe that the RD performance of LVAC can be further improved significantly.  In particular, the principal advances of TMC13 over {\em RAHT+RLGR} ---  joint entropy coding and predictive RAHT --- are equally applicable to the LVAC framework.  For example, joint entropy coding could be done with a hyperprior \cite{BaMiSiHwJo18}, and predictive RAHT could be applied to the latent vectors.  These explorations are left for future work.

\section{Discussion and Conclusion}
\label{sec:conclusion}

This work is the first to compress volumetric functions $\yv = f_\theta(\xv)$ modeled by local coordinate-based networks.  Though we focused on RGB attributes $\yv$, the extension to other attributes (such as signed distance, reflectance, normals, transparency, density, spherical harmonics, etc.) is straightforward.  Also, though we focused on $\xv\in\RR^3$, extensions to hyper-volumetric functions (such as $\yv = f_\theta(\xv,\dv)$ where $\dv$ is a view direction) is also straightforward.  Thus LVAC should be applicable to plenoptic point clouds \cite{Krivokuca:2018,Sandri:2018,Sandri:2019,Zhang:2018,Zhang:2019} as well as radiance fields (e.g.,\cite{mildenhall2020nerf,yu2021plenoctrees,zhang2021nerfactor,takikawa2021neural,martel2021acorn}).
We believe that the main difference between plenoptic point clouds and radiance fields is the distortion measure $d(f,f_\theta)$.  For point clouds, $d(f,f_\theta)$ is measured in the domain of $f$, such as the MSE between colors on points in 3D.  For radiance fields, $d(f,f_\theta)$ is measured in the domain of {\em projections} or renderings of $f$ onto 2D images, such as the MSE between colors of pixels that are renderings of $f$ and $f_\theta$ onto 2D images.  In \cite{QueirozC:17}, the former distortion measures are called {\em matching distortions}, while the latter are called {\em projection distortions}.  The change in distortion measure from matching to projection distortions may be all that is required to apply LVAC properly to radiance field compression.

This work is also among the first to compress point cloud attributes using neural networks,
outperforming RAHT, used in MPEG G-PCC, by 2--4 dB.
Although MPEG G-PCC uses additional coding tools to further improve compression, such as context adaptive arithmetic coding, joint entropy coding of color, and predictive RAHT, these tools are also at our disposal, and may be the subject of further work.
It should be recalled that learned image compression evolved over dozens of papers and a half dozen years, being competitive at first with only JPEG on thumbnails, and then successively with JPEG-2000, WebP, and BGP.  Only recently has learned image compression been able to outperform VVC in PSNR \cite{GuoZFC:21}.  Learned volumetric attribute compression (LVAC), like learned image compression, is a work in progress.  Now is a particularly good time to publicise progress in this area ahead of the JPEG Pleno Call for Proposals on learned point cloud compression, with submissions scheduled in 2022 \cite{JPEG_Pleno_PC_CFE:20}.

We believe that LVAC still has two big weaknesses that need to be addressed.  The first is lack of conditioning on \emph{detailed} geometry.  Currently LVAC conditions on geometry by using RAHT's mechanism to normalize the latents up to the target level.  Unfortunately this conditioning does not make its way to the finest levels of geometry within the CBN.  We believe this is the reason that at high bit rates, CBNs at low target levels fall short of a simple linear network at the voxel level, in RD performance.  In short, the CBNs are leaving big RD improvements on the table.

Another big weakness of LVAC, so far, is its auto-decoder approach, in which the encoder must perform optimization through back-propagation, in order to produce a bit stream for a point cloud.  This is extremely slow relative to a feed-forward encoder.  We recognize the need to develop a feed-forward encoder, and with it, the need for conditioning on the geometry to the finest levels of detail.

These provide plenty of room for future work.

{\small
\bibliographystyle{ieee_fullname}
\bibliography{refs}
}

\clearpage

\section*{Appendix}
\renewcommand{\thesubsection}{\Alph{subsection}}
\label{sec:appendix}

\subsection{Coordinate Based Networks}

\begin{figure*}
    \centering
    \includegraphics[width=0.33\linewidth, trim=20 5 35 15, clip]{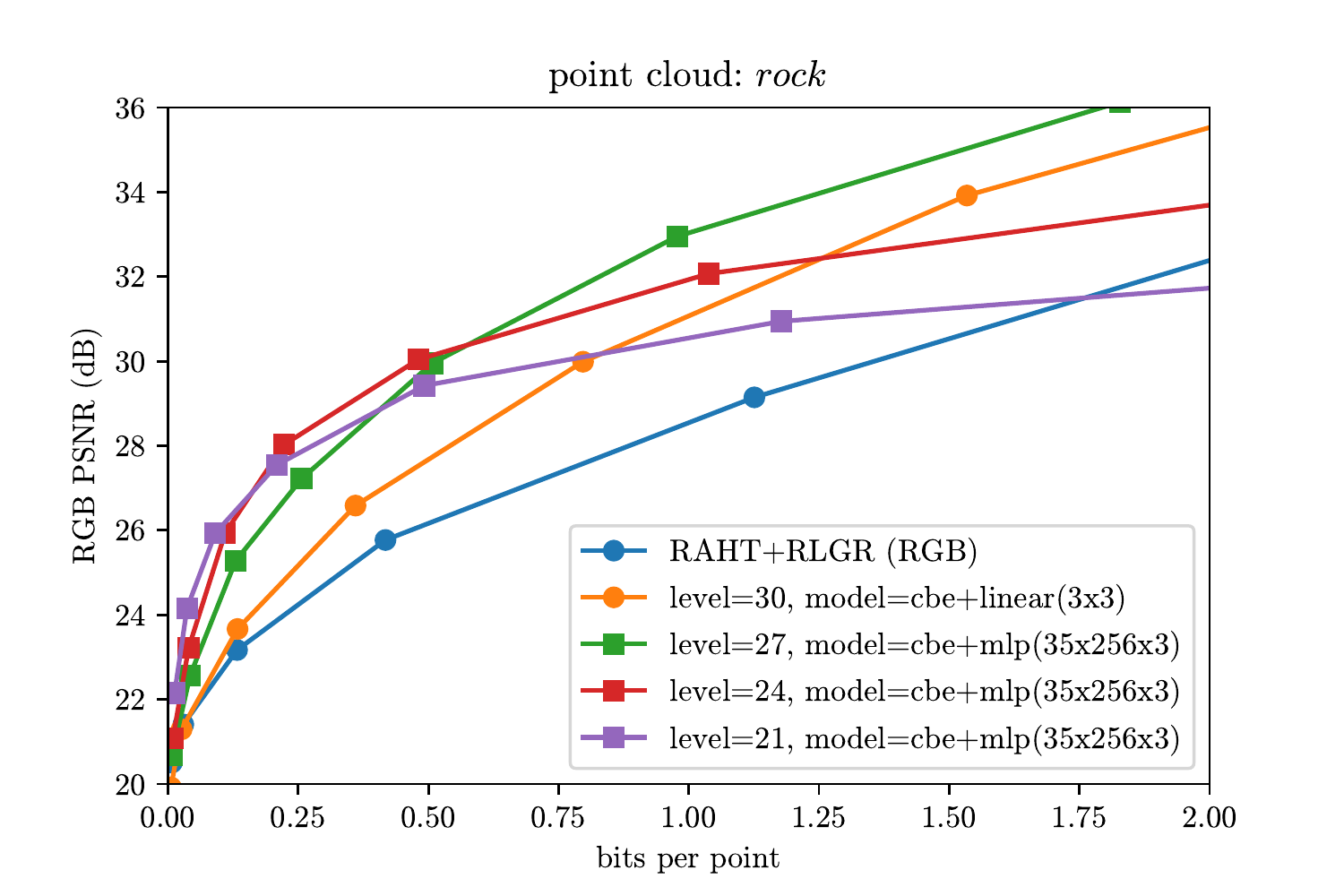}
    \includegraphics[width=0.33\linewidth, trim=20 5 35 15, clip]{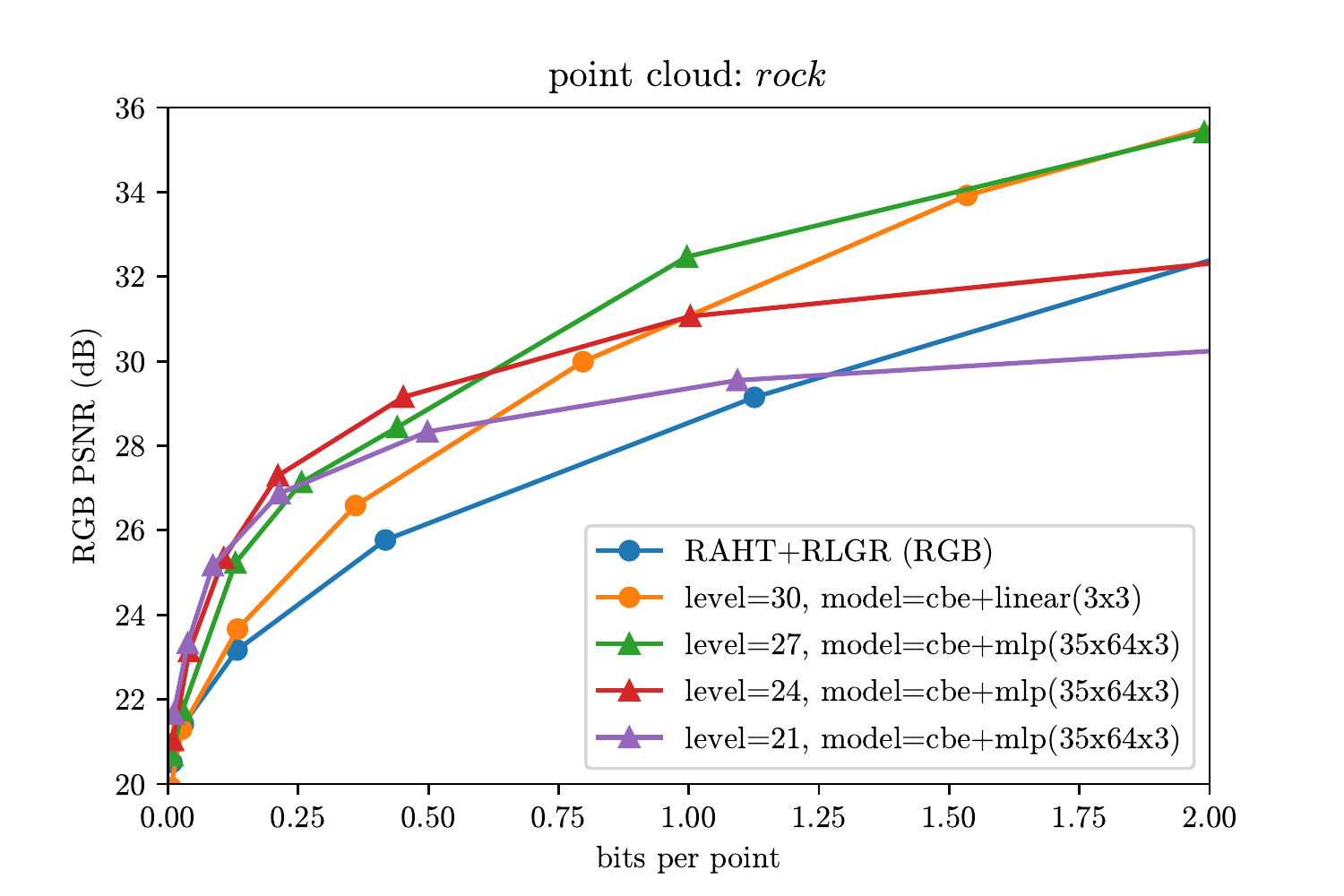}
    \includegraphics[width=0.33\linewidth, trim=20 5 35 15, clip]{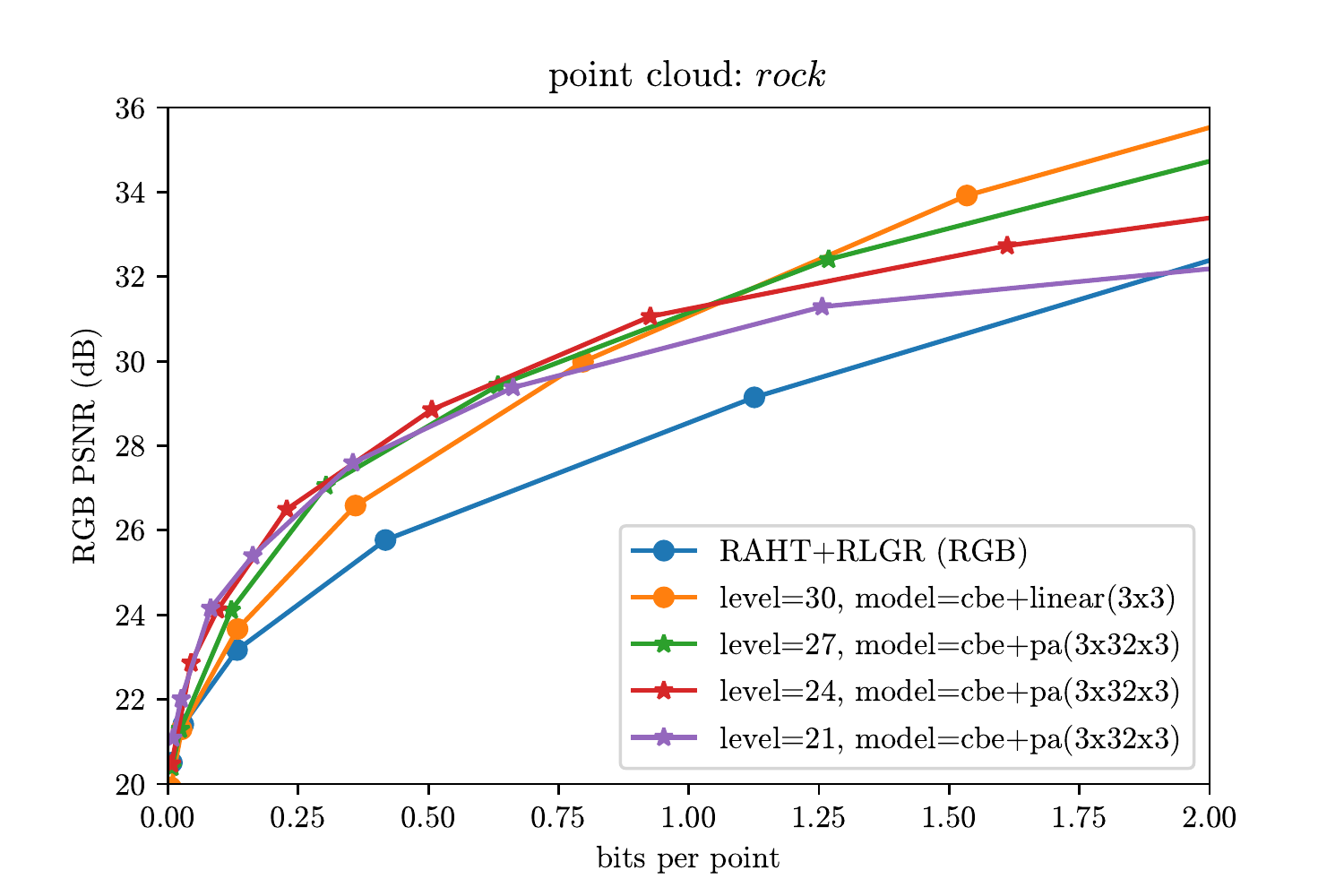}
    \caption{Coordinate Based Networks, by network.}
    \label{fig:cbns_by_network}
\end{figure*}

\Cref{fig:cbns_by_network} shows the RD performance of different networks: (left) {\em mlp(35x256x3)}, (middle) {\em mlp(35x64x3)}, and (right) {\em pa(3x32x3)}, along with baselines.  At higher bit rates, higher target levels perform better.

\subsection{Generalization}

\begin{figure*}
    \centering
    \includegraphics[width=0.33\linewidth, trim=20 5 35 15, clip]{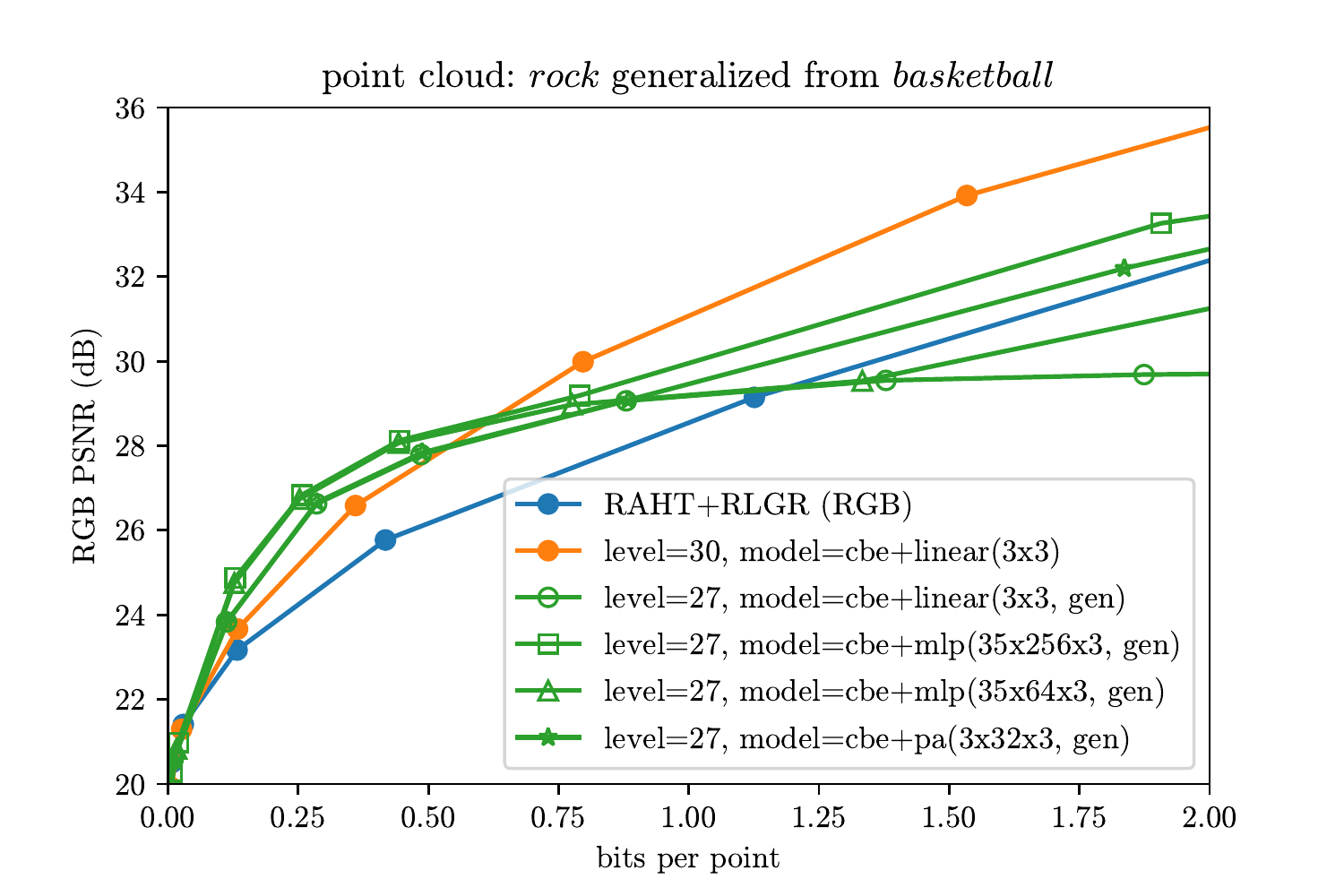}
    \includegraphics[width=0.33\linewidth, trim=20 5 35 15, clip]{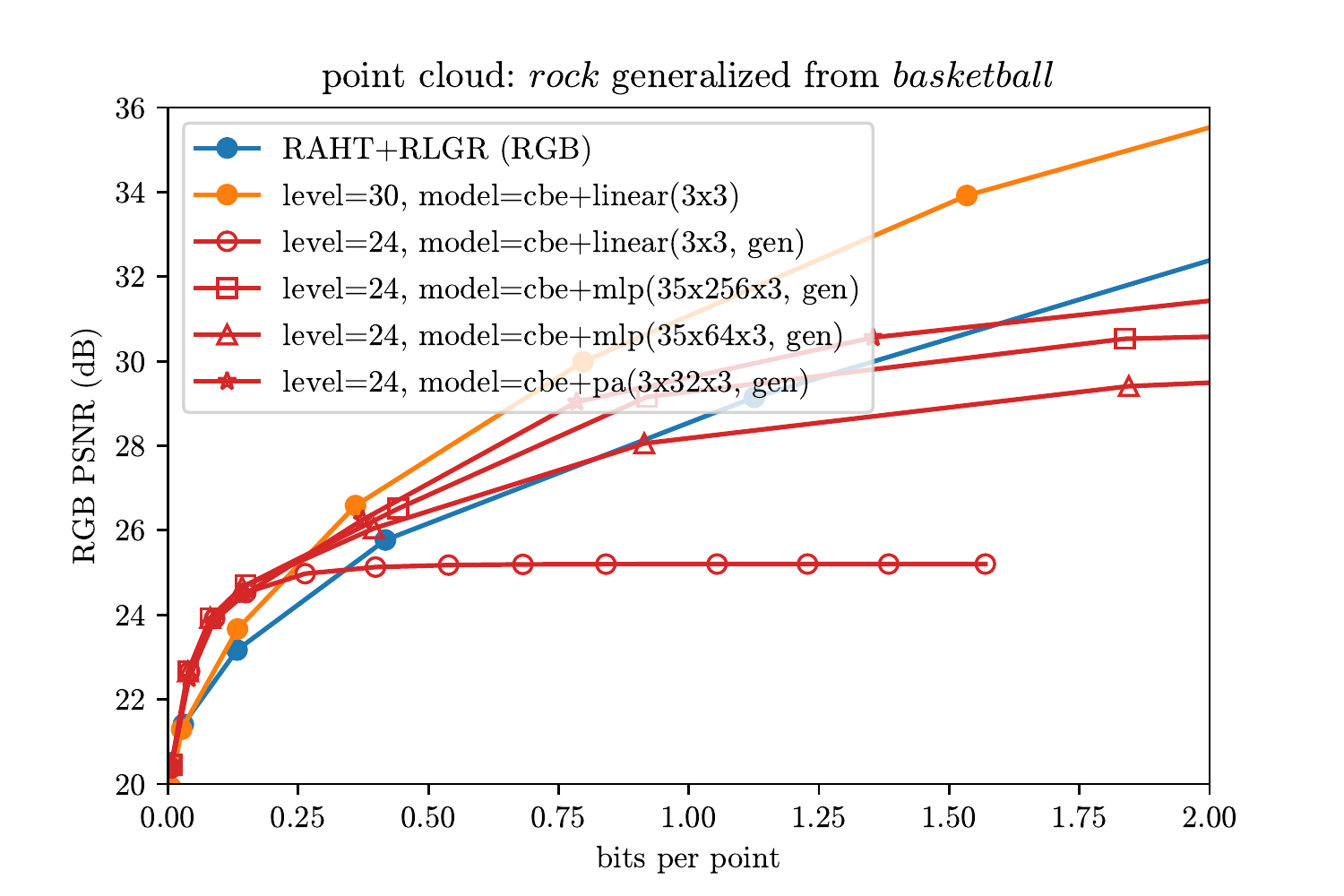}
    \includegraphics[width=0.33\linewidth, trim=20 5 35 15, clip]{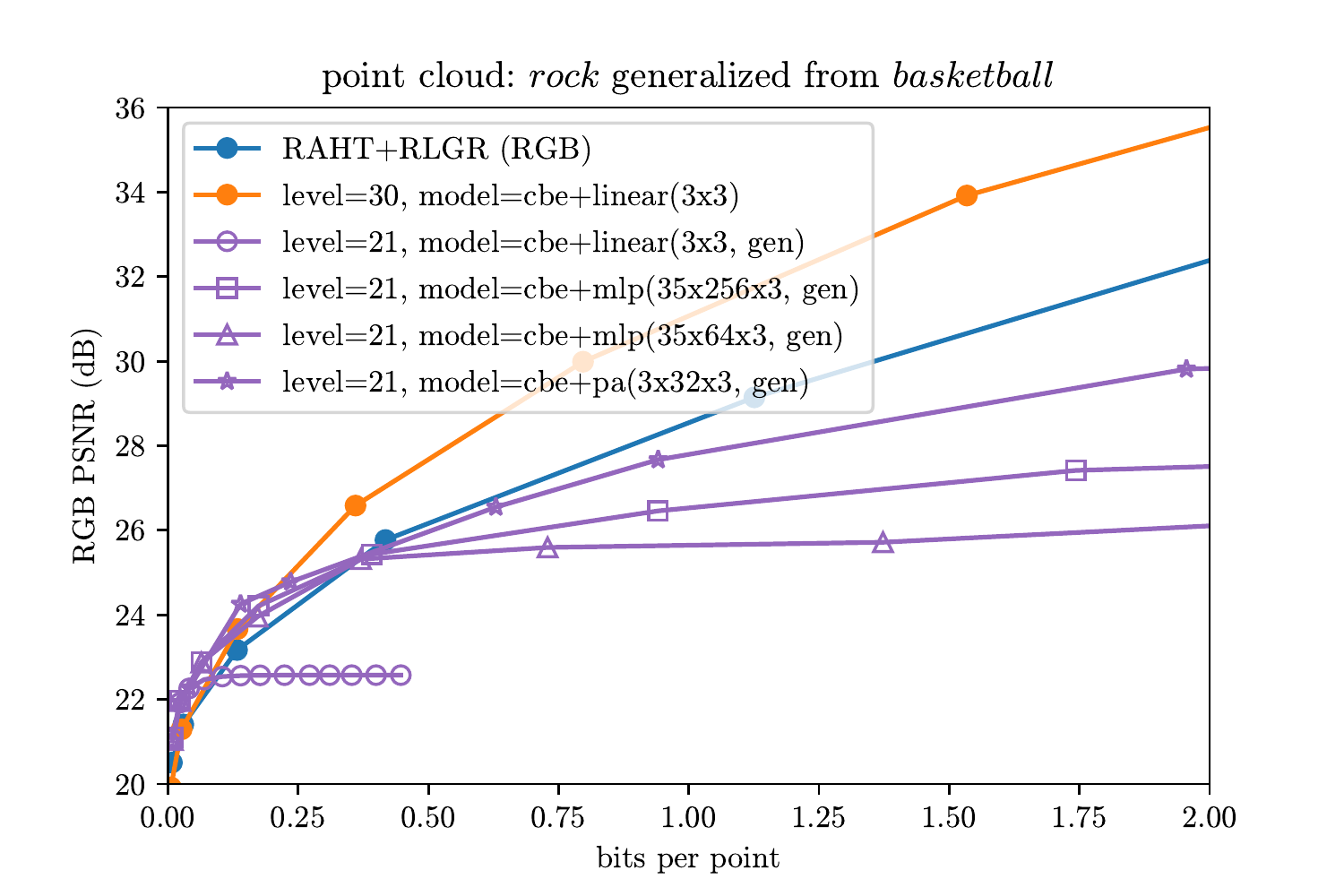}
    \\
    \includegraphics[width=0.33\linewidth, trim=20 5 35 15, clip]{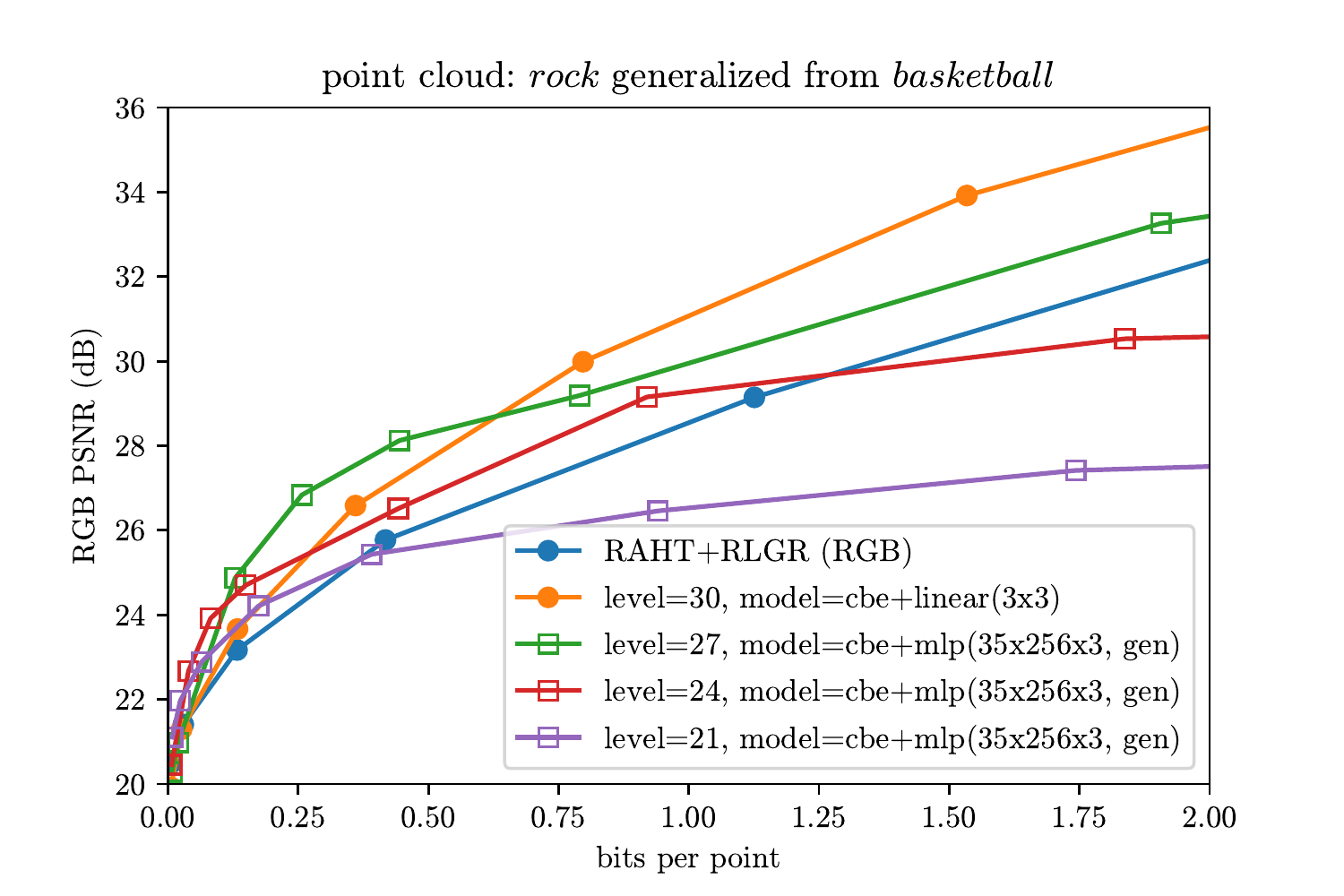}
    \includegraphics[width=0.33\linewidth, trim=20 5 35 15, clip]{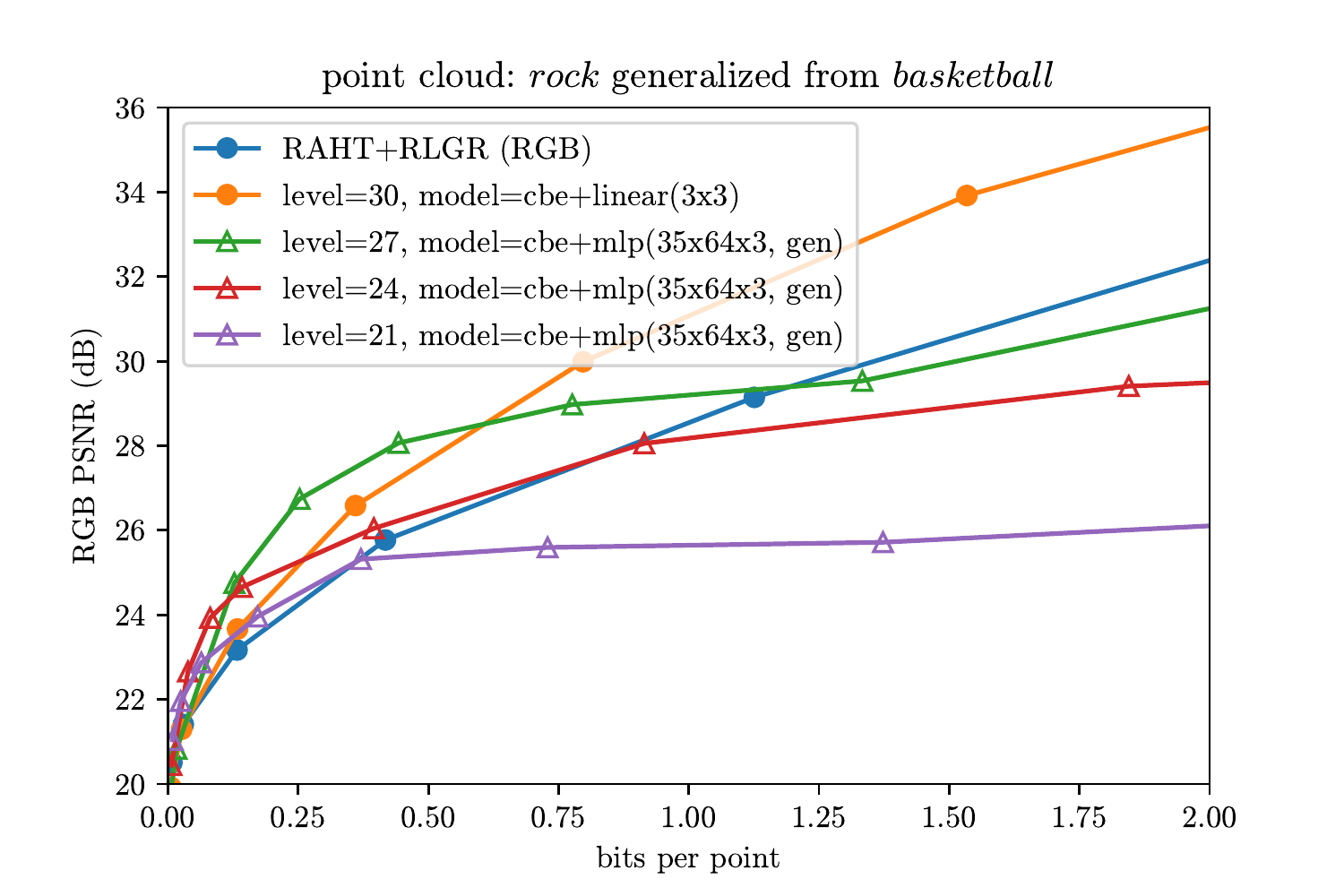}
    \includegraphics[width=0.33\linewidth, trim=20 5 35 15, clip]{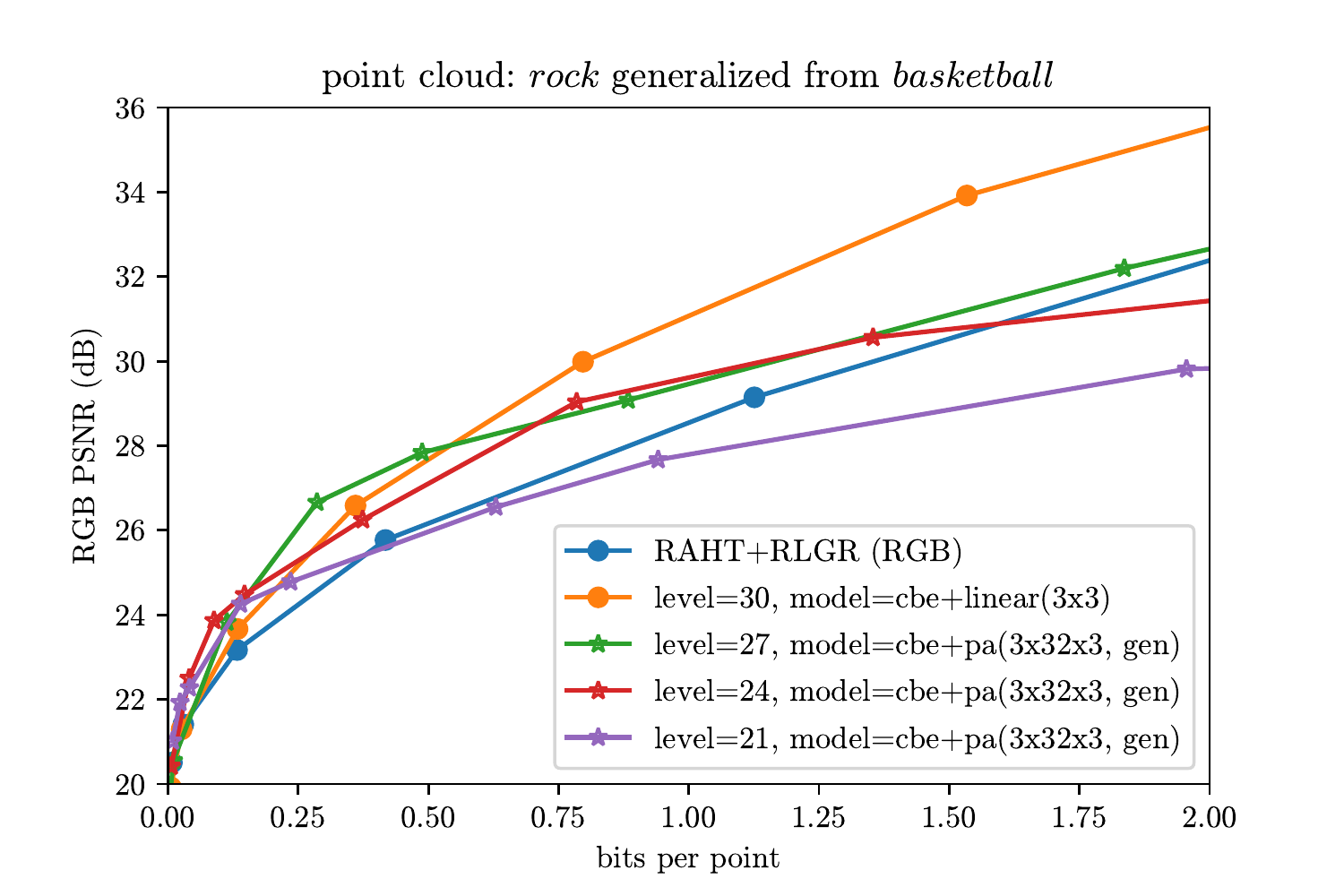}
    \caption{Coordinate Based Networks with generalization, by level (top row) and by network (bottom row).  CBNs that are generalized (i.e., pre-trained on another point cloud) are able to outperform the baselines at low bit rates.}
    \label{fig:cbns_gen}
    \vspace{-0.2in}
\end{figure*}

\Cref{fig:cbns_gen} shows the RD performance of CBNs generalized from, or pre-trained on, another point cloud.  Even with training on only a single other point cloud, the CBNs can indeed generalize across point clouds at low bit rates.  At high bit rates, however, the CBNs trained on just one other point cloud do not perform well, most likely because they have been unable to learn to represent the fine details needed for a different point cloud.

\subsection{Side Information}

\begin{figure*}
    \centering
    \includegraphics[width=0.33\linewidth, trim=20 5 35 15, clip]{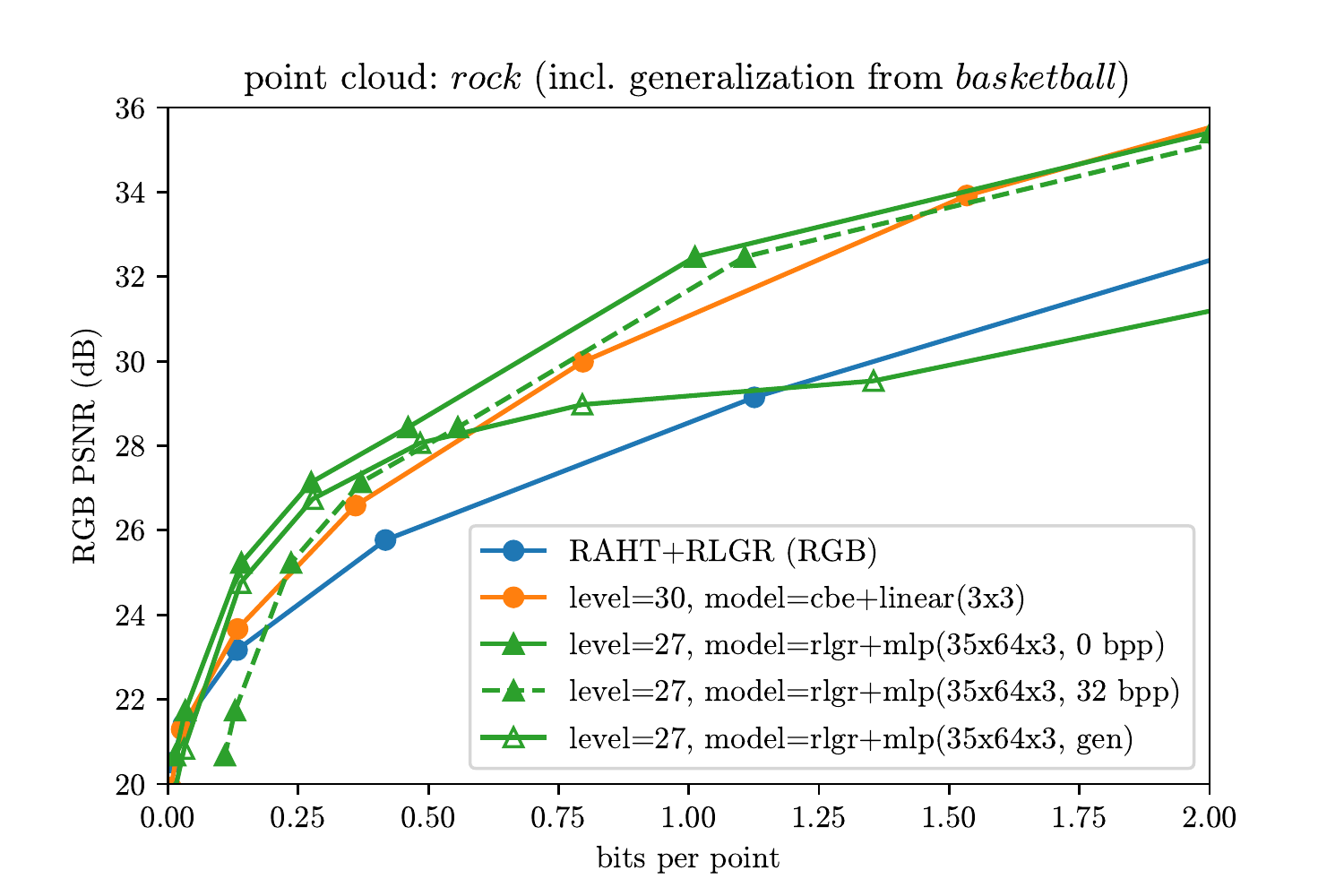}
    \includegraphics[width=0.33\linewidth, trim=20 5 35 15, clip]{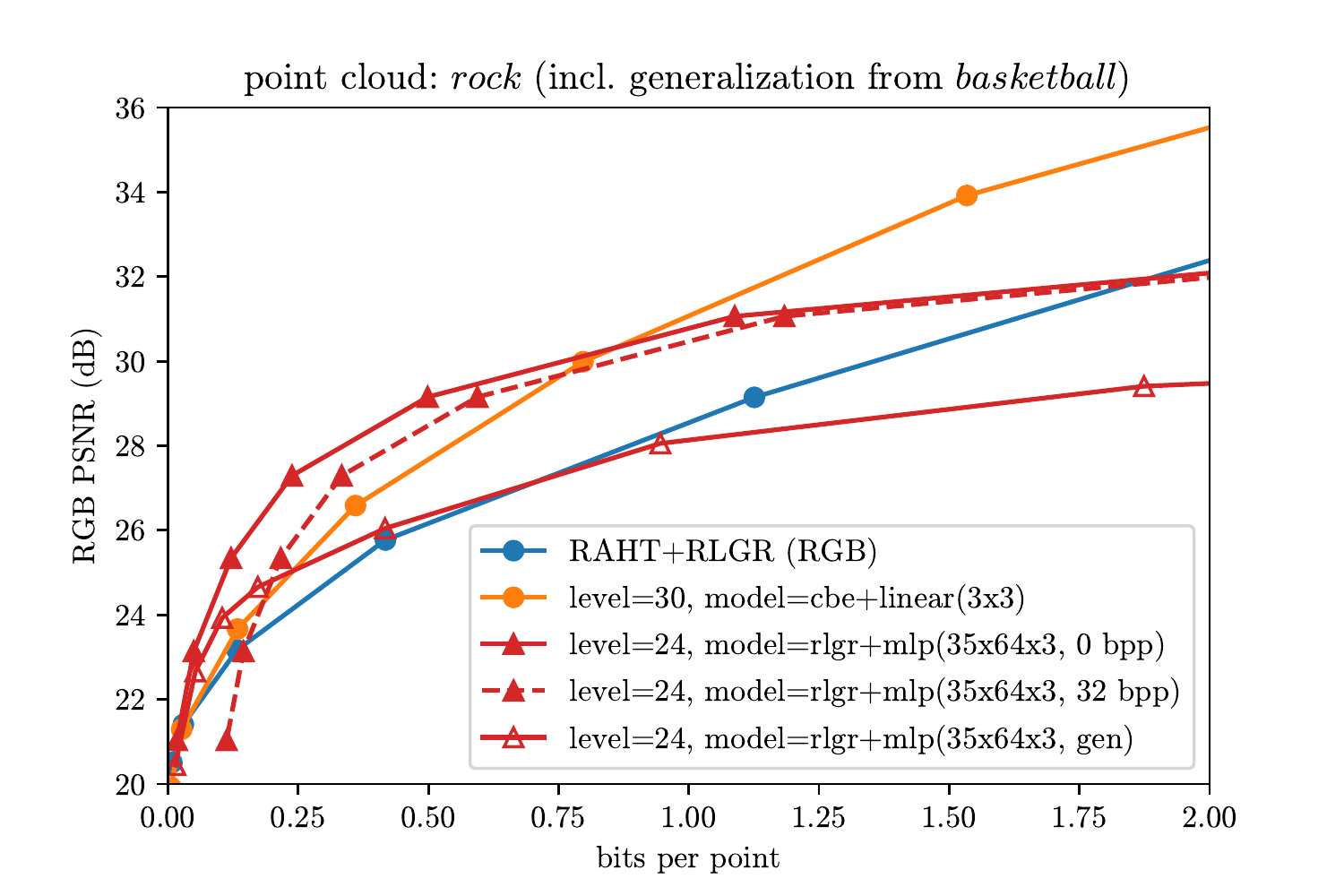}
    \includegraphics[width=0.33\linewidth, trim=20 5 35 15, clip]{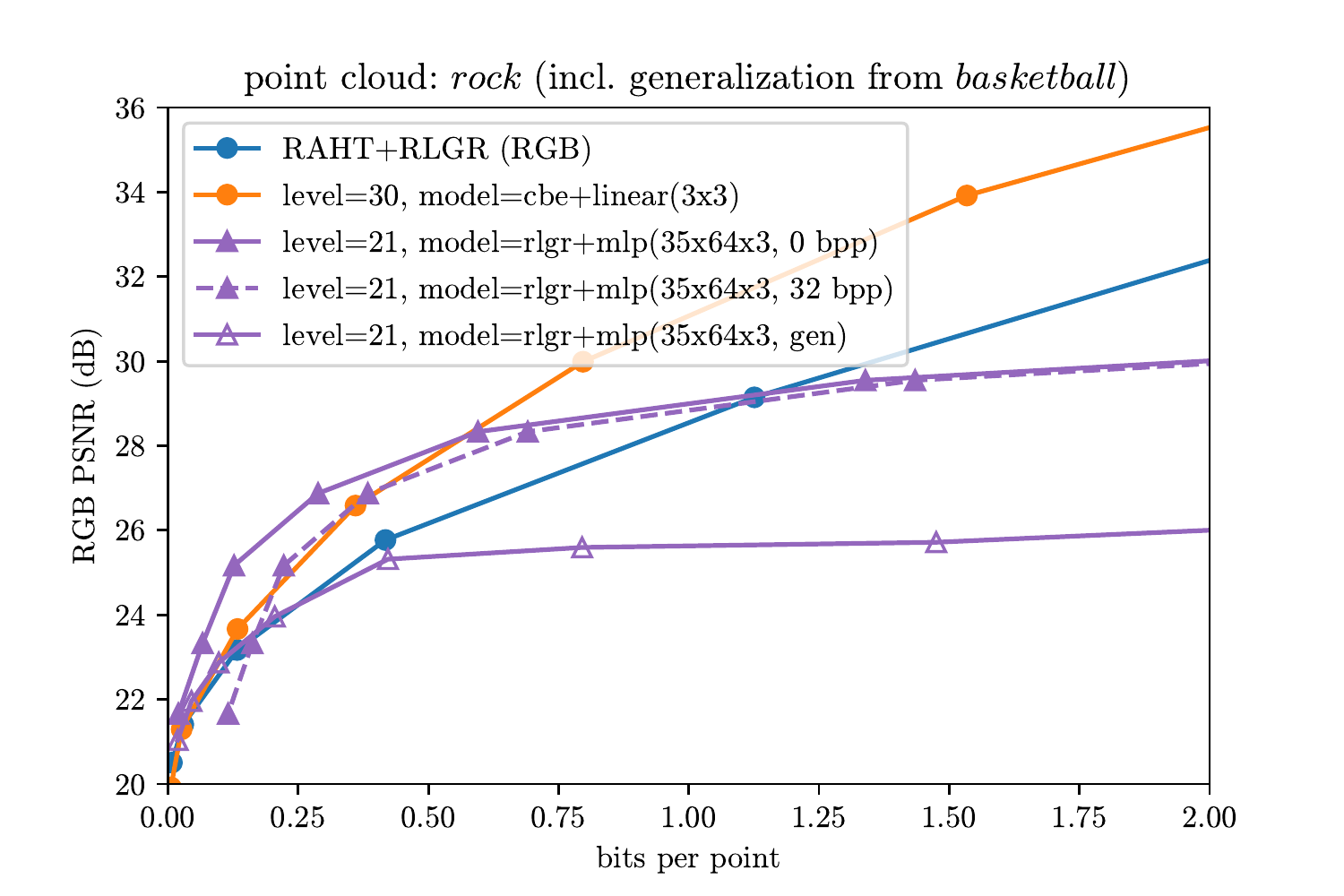}
    \\
    \includegraphics[width=0.33\linewidth, trim=20 5 35 15, clip]{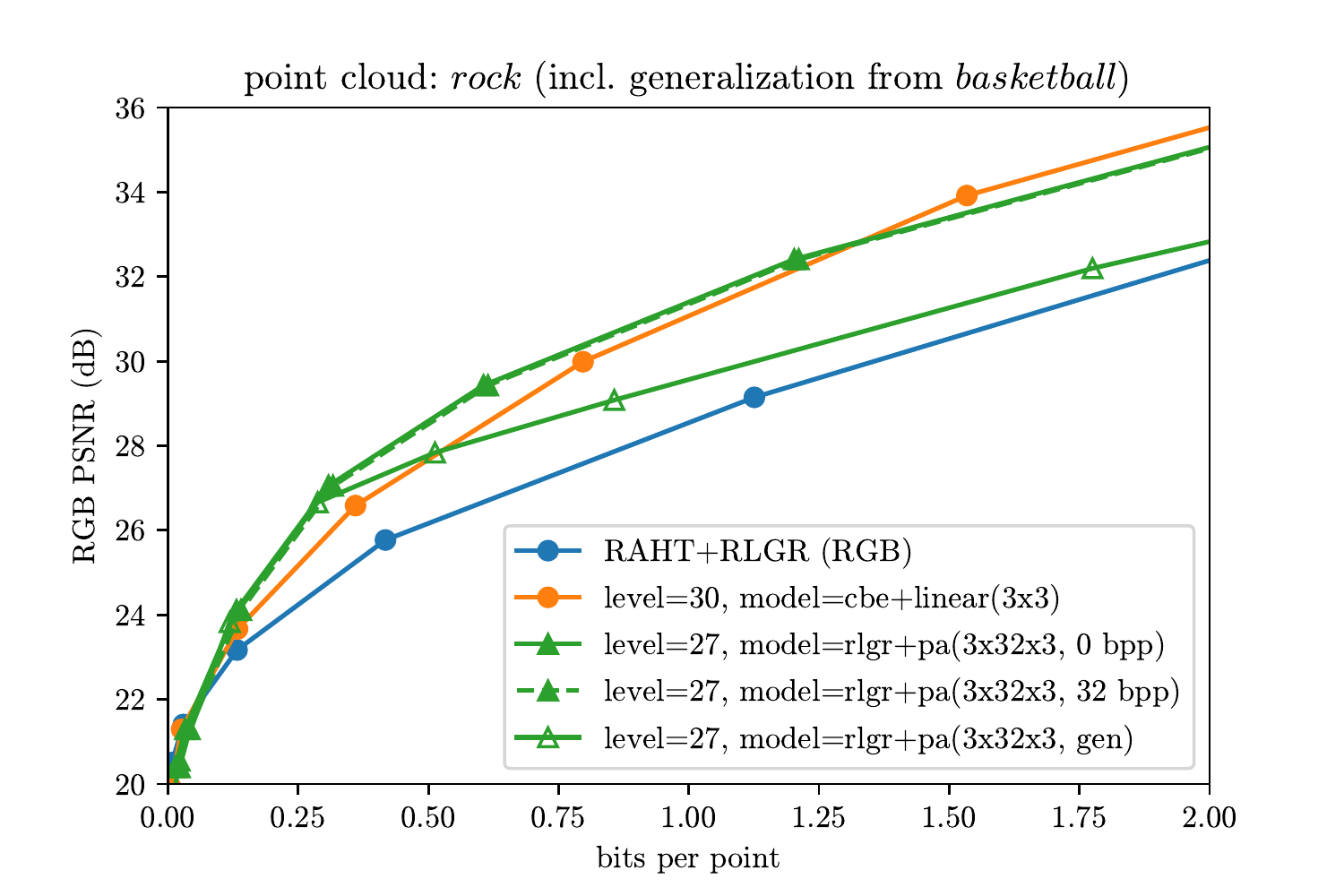}
    \includegraphics[width=0.33\linewidth, trim=20 5 35 15, clip]{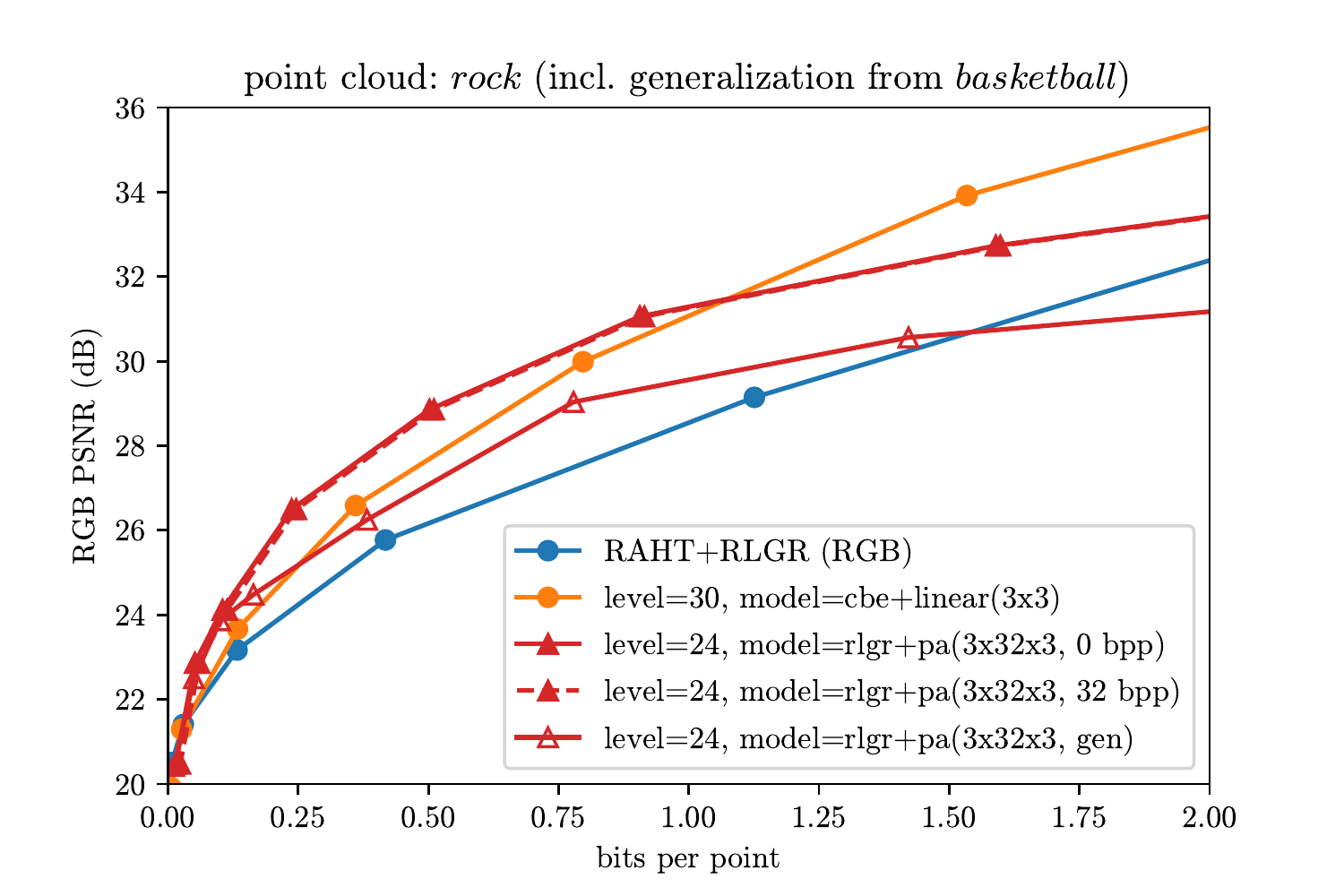}
    \includegraphics[width=0.33\linewidth, trim=20 5 35 15, clip]{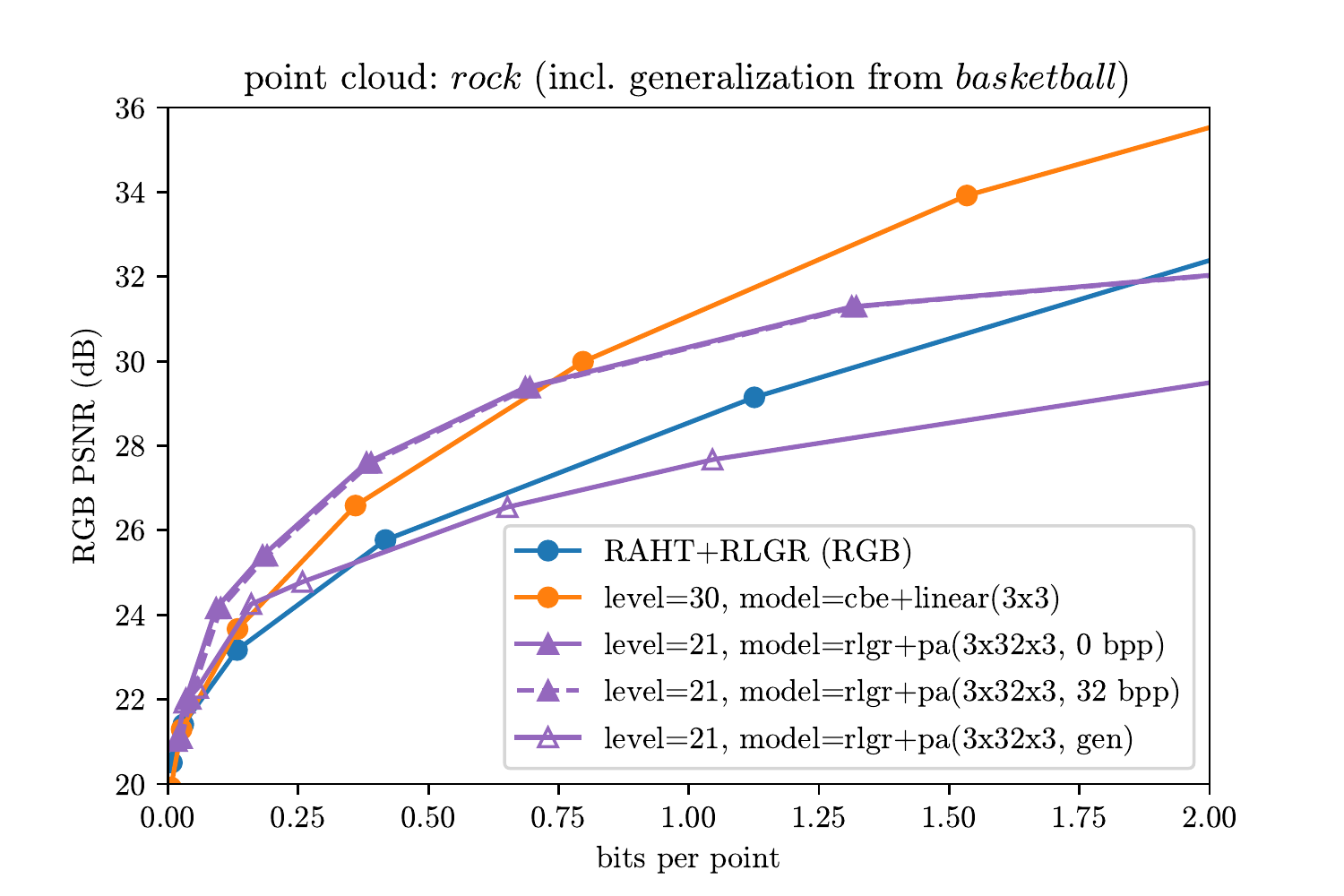}
    \caption{Effect of side information for coordinate based networks {\em mlp(35x64x3)} (top) and {\em pa(3x32x3)} (bottom) at levels 27 (left), 24 (middle), and 21 (right).  See \cref{fig:sideinfo_mlp256} for {\em mlp(32x256x3)}.}
    \label{fig:sideinfo_mlp64_and_pa}
\end{figure*}

\Cref{fig:sideinfo_mlp64_and_pa} shows results for {\em mlp(35x64x3)} and {\em pa(3x32x3)} (top and bottom, respectively) corresponding to the results in \cref{fig:sideinfo_mlp256} for {\em mlp(35x256x3)}.

\subsection{Subjective Quality}

\begin{figure*}
    \centering
    \begin{minipage}{0.30\textwidth}
    \centering\small
    \includegraphics[width=1.0\linewidth]{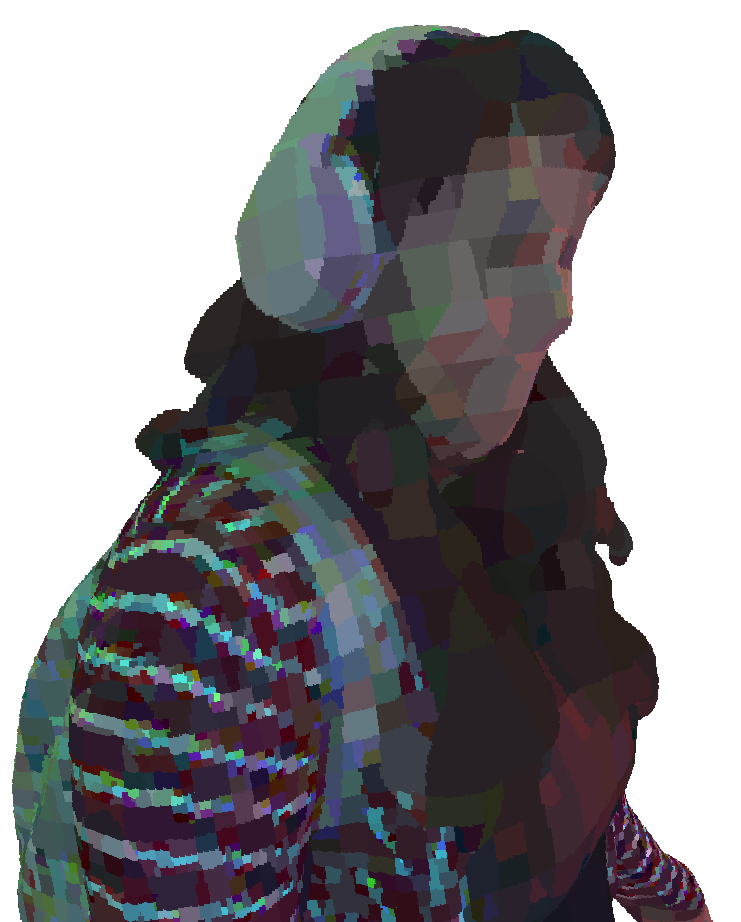} \\
    0.129 bpp, 23.2 dB
    \end{minipage}
    \begin{minipage}{0.30\textwidth}
    \centering\small
    \includegraphics[width=1.0\linewidth]{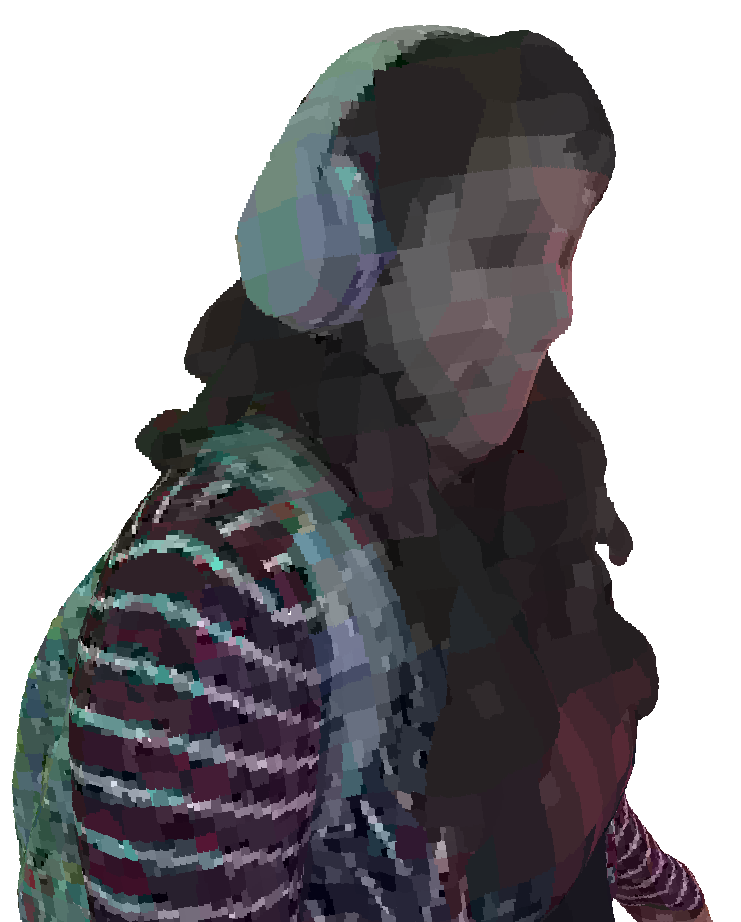} \\
    0.127 bpp, 24.3 dB
    \end{minipage}
    \begin{minipage}{0.30\textwidth}
    \centering\small
    \includegraphics[width=1.0\linewidth]{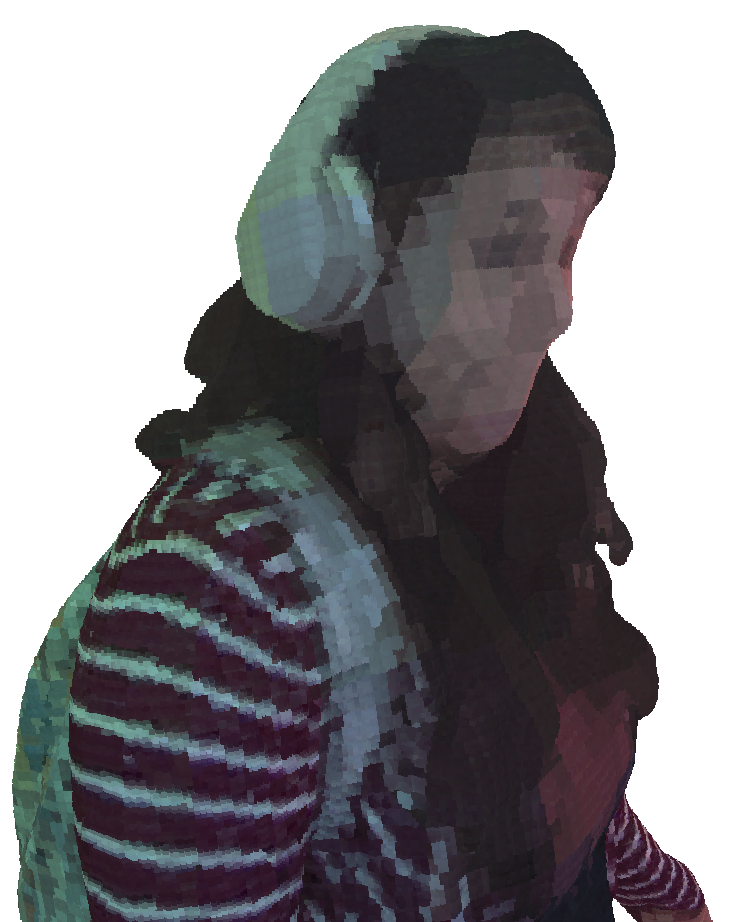} \\
    0.127 bpp, 25.9 dB
    \end{minipage}


    \begin{minipage}{0.30\textwidth}
    \centering\small
    \includegraphics[width=1.0\linewidth]{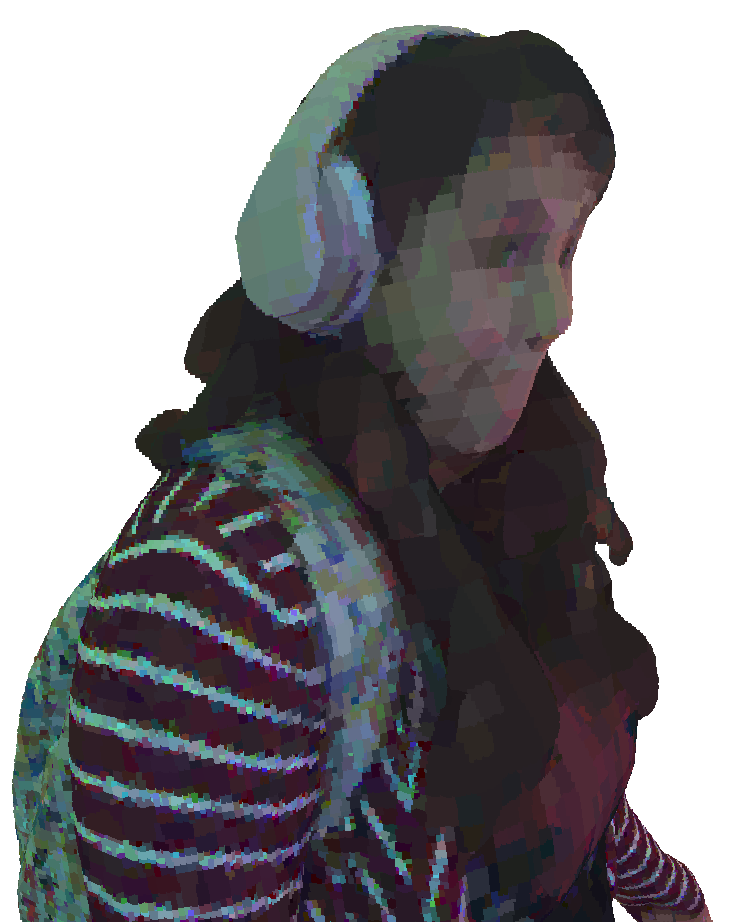} \\
    0.540 bpp, 26.5 dB
    \end{minipage}
    \begin{minipage}{0.30\textwidth}
    \centering\small
    \includegraphics[width=1.0\linewidth]{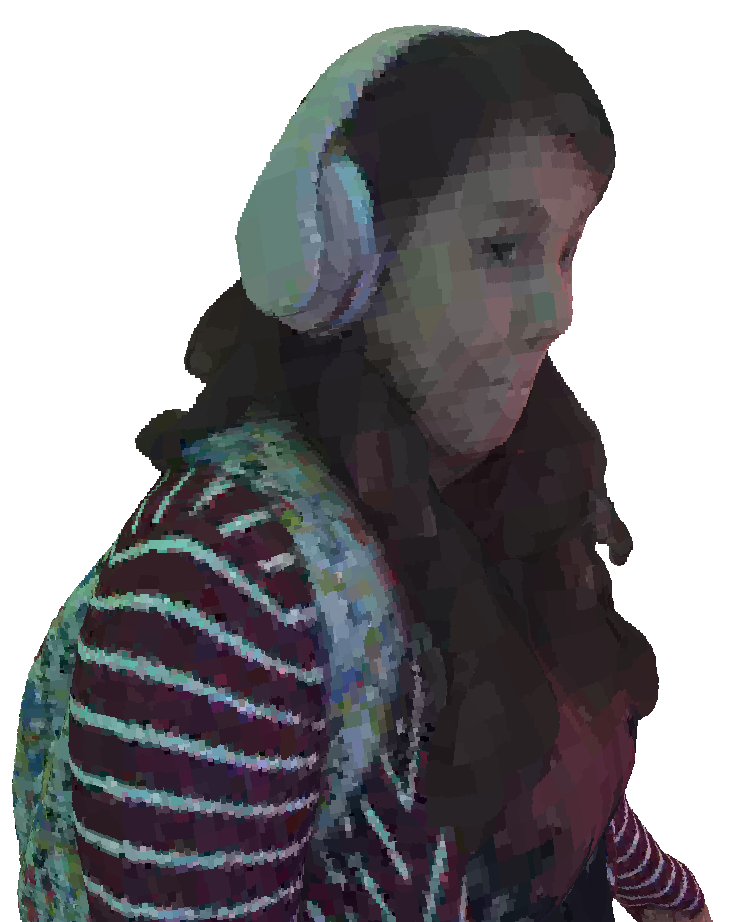} \\
    0.536 bpp, 28.2 dB
    \end{minipage}
    \begin{minipage}{0.30\textwidth}
    \centering\small
    \includegraphics[width=1.0\linewidth]{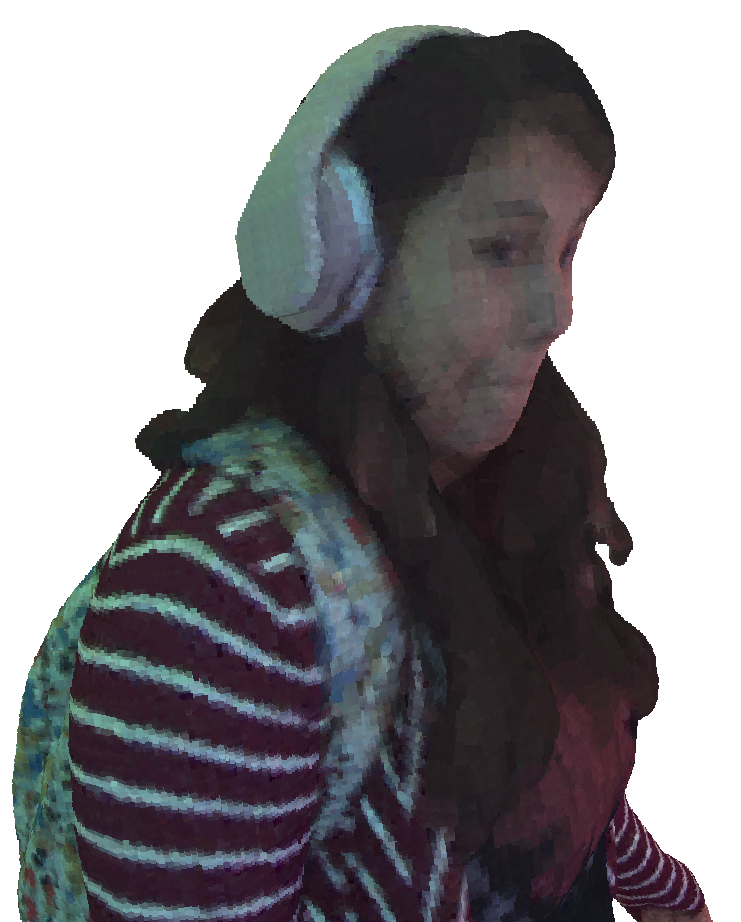} \\
    0.531 bpp, 30.0 dB
    \end{minipage}

    \begin{minipage}{0.30\textwidth}
    \centering\small
    \includegraphics[width=1.0\linewidth]{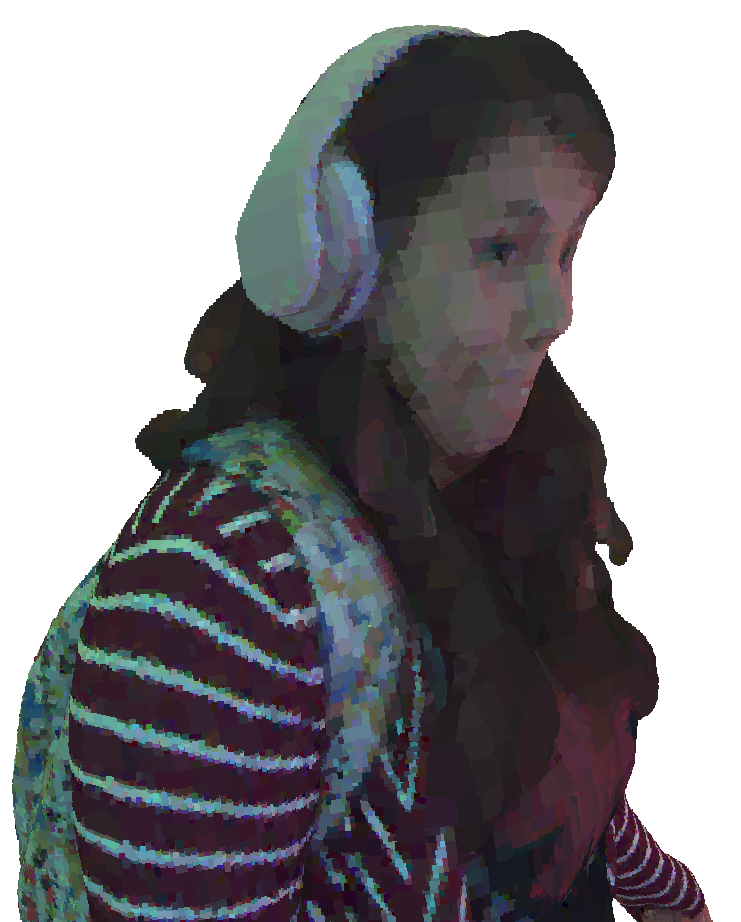} \\
    1.005 bpp, 28.7 dB \\
    {\em RAHT-RLGR (RGB)}
    \end{minipage}
    \begin{minipage}{0.30\textwidth}
    \centering\small
    \includegraphics[width=1.0\linewidth]{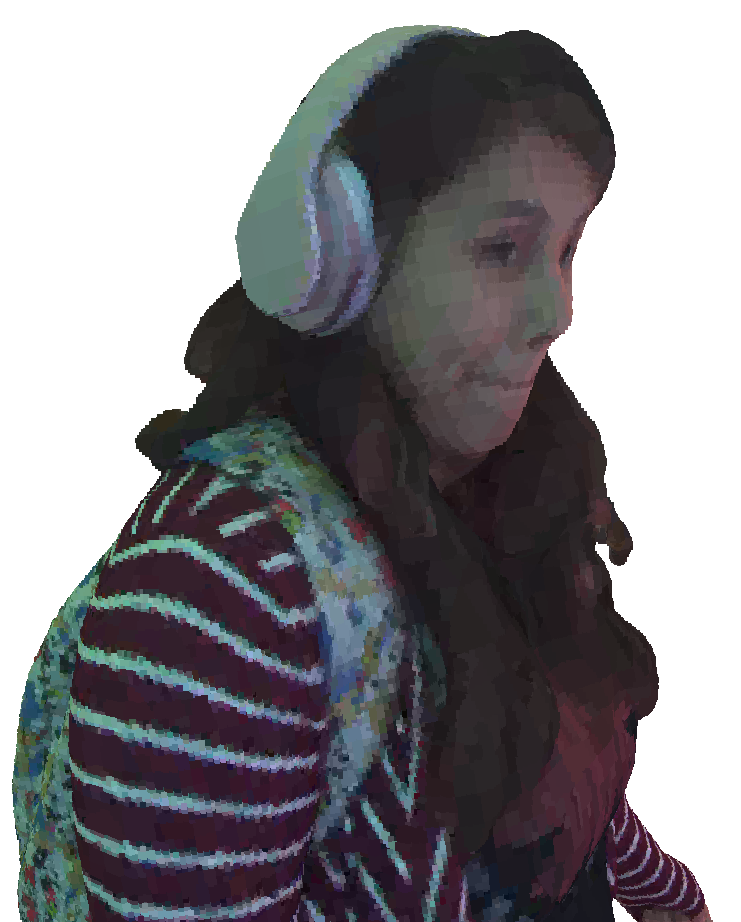} \\
    1.019 bpp, 31.3 dB \\
    {\em RAHT-RLGR (YUV)}
    \end{minipage}
    \begin{minipage}{0.30\textwidth}
    \centering\small
    \includegraphics[width=1.0\linewidth]{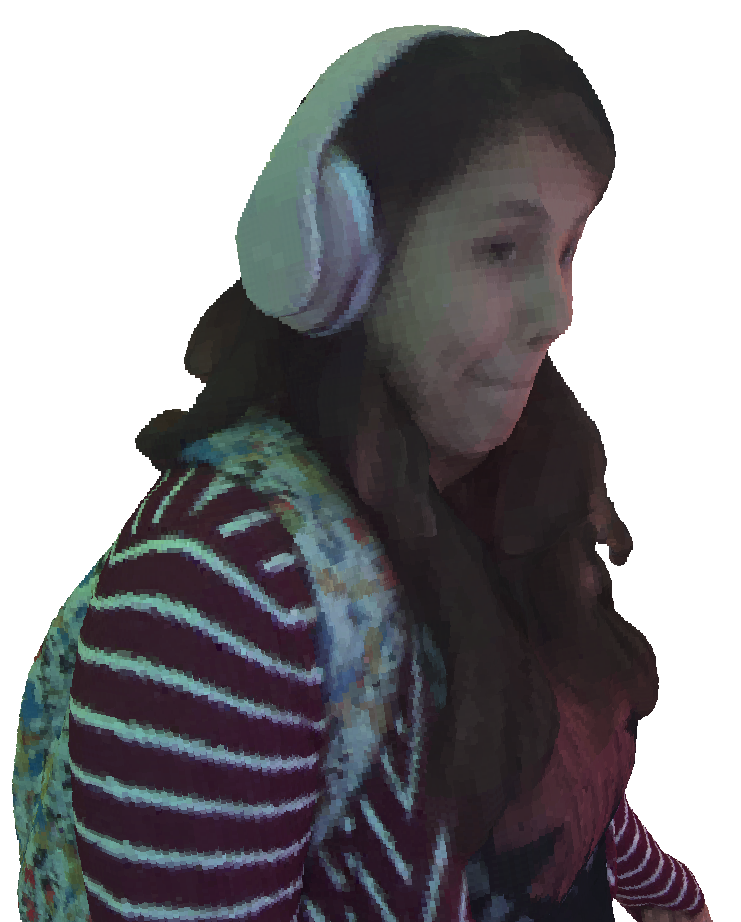} \\
    1.003 bpp, 33.0 dB \\
    {\em LVAC mlp(35x256x3)}
    \end{minipage}

    \caption{Subjective quality of point cloud {\em rock}.  Each row is a different bit rate.  Original is shown in \cref{fig:subjective}. Zoom in to see differences.  See \cref{fig:subjective} for 0.25 bpp.}
    \label{fig:subjective2}
\end{figure*}

\Cref{fig:subjective2} shows subjective results for 0.125, 0.5, and 1.0 bpp corresponding to the subjective results in \cref{fig:subjective} for 0.25 bpp.

\subsection{Additional Point Clouds}

\begin{figure*}
    \centering
    \includegraphics[width=0.33\textwidth, trim=20 5 35 15, clip]{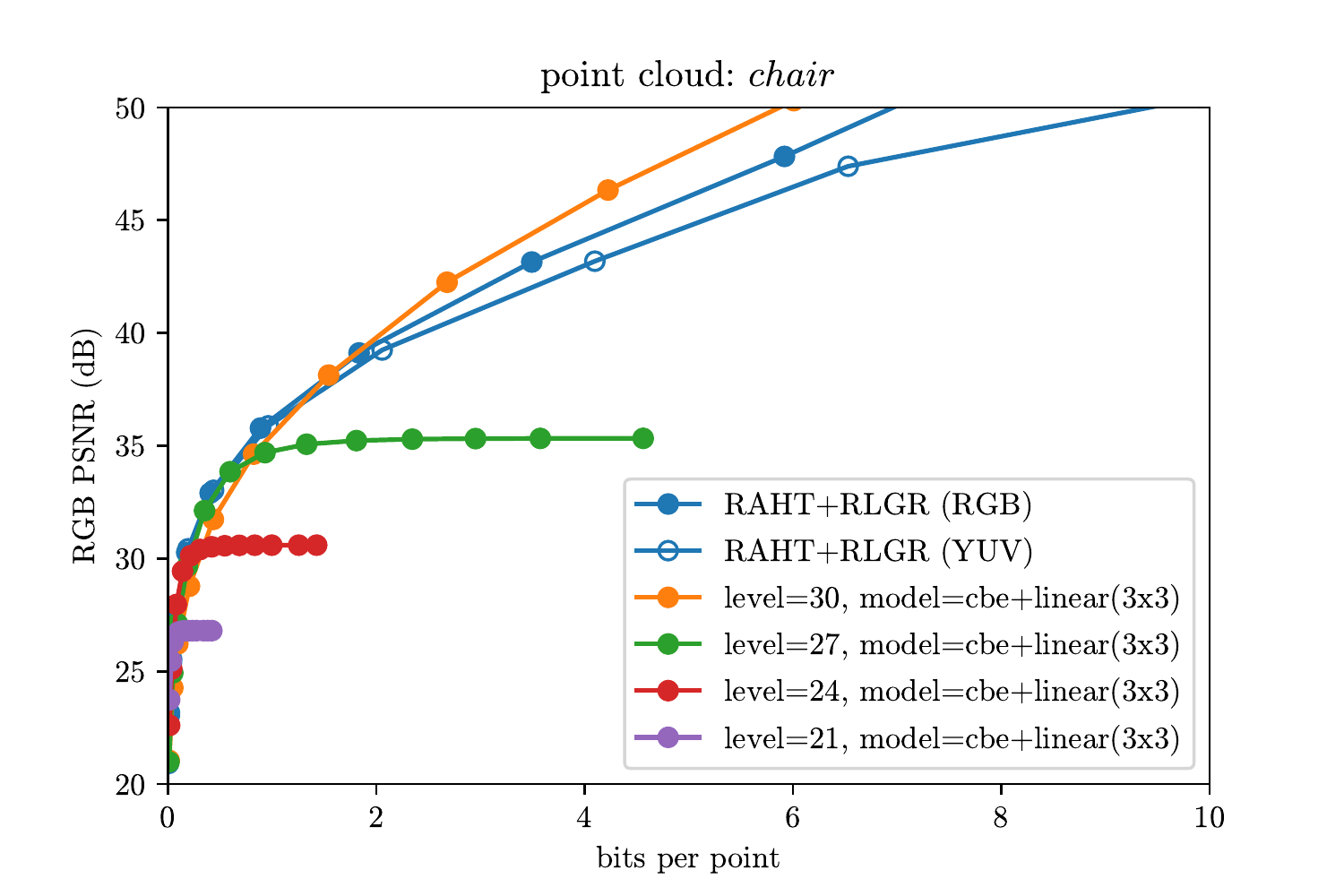}
    \includegraphics[width=0.33\textwidth, trim=20 5 35 15, clip]{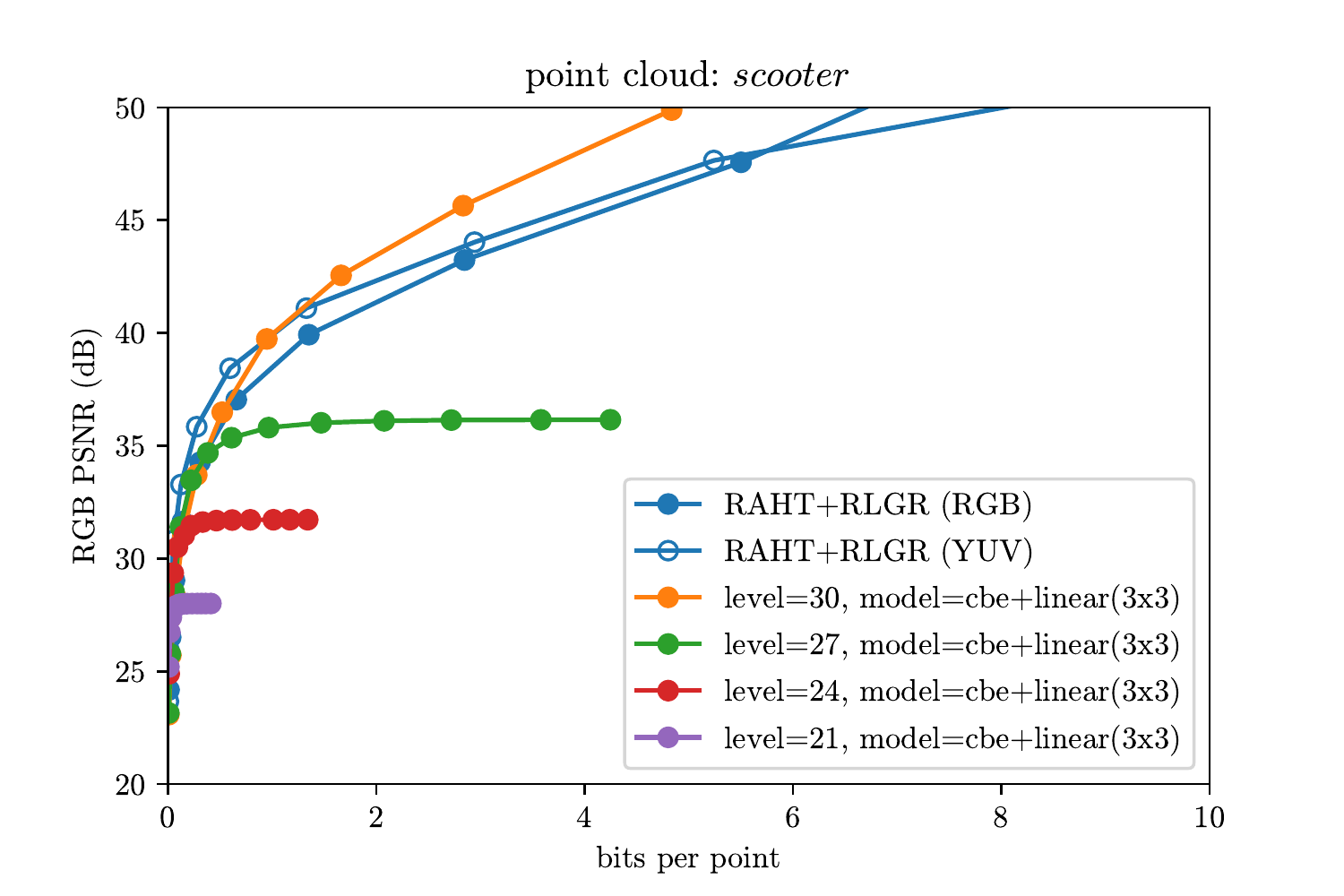}
    \includegraphics[width=0.33\textwidth, trim=20 5 35 15, clip]{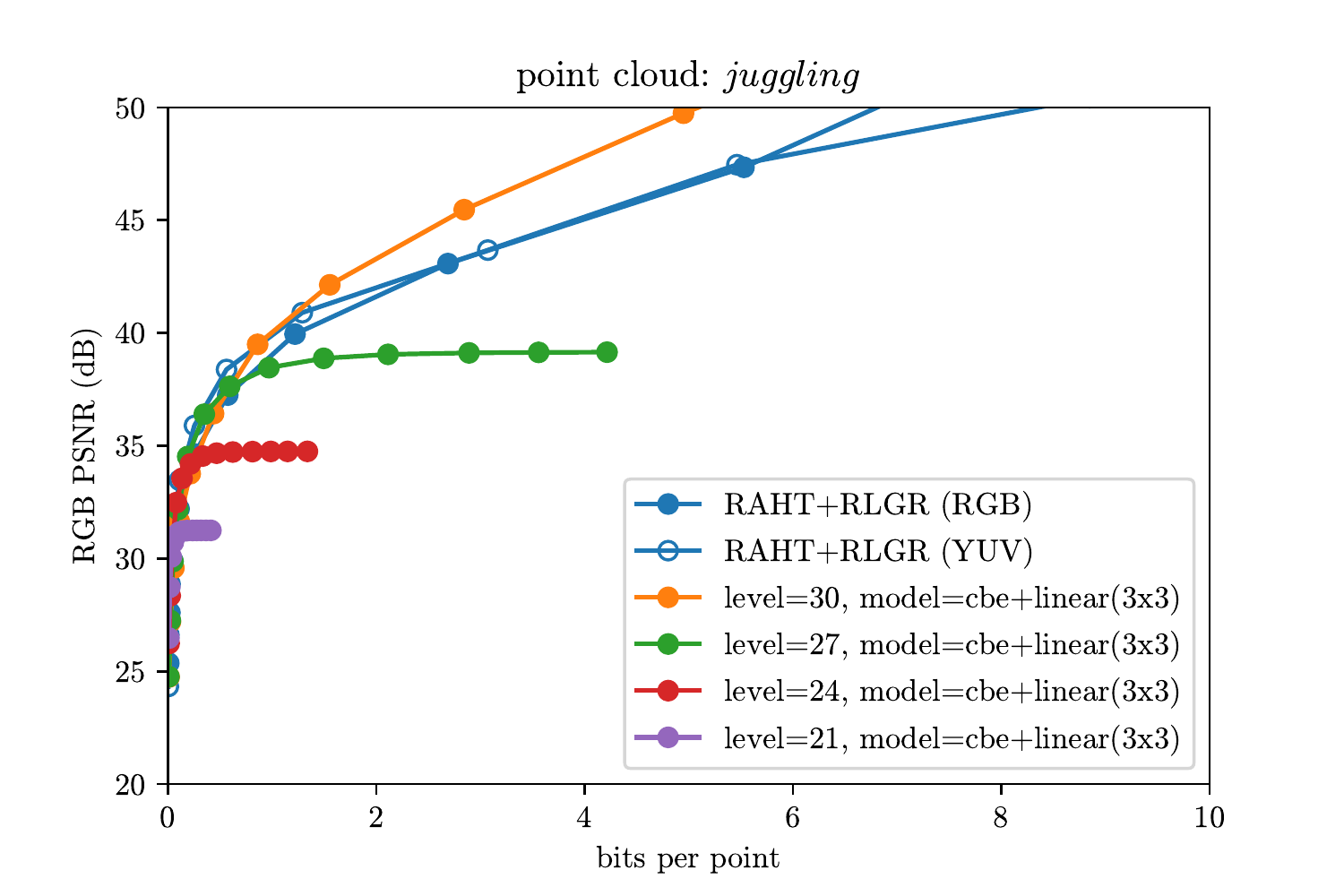}
    \includegraphics[width=0.33\textwidth, trim=20 5 35 15, clip]{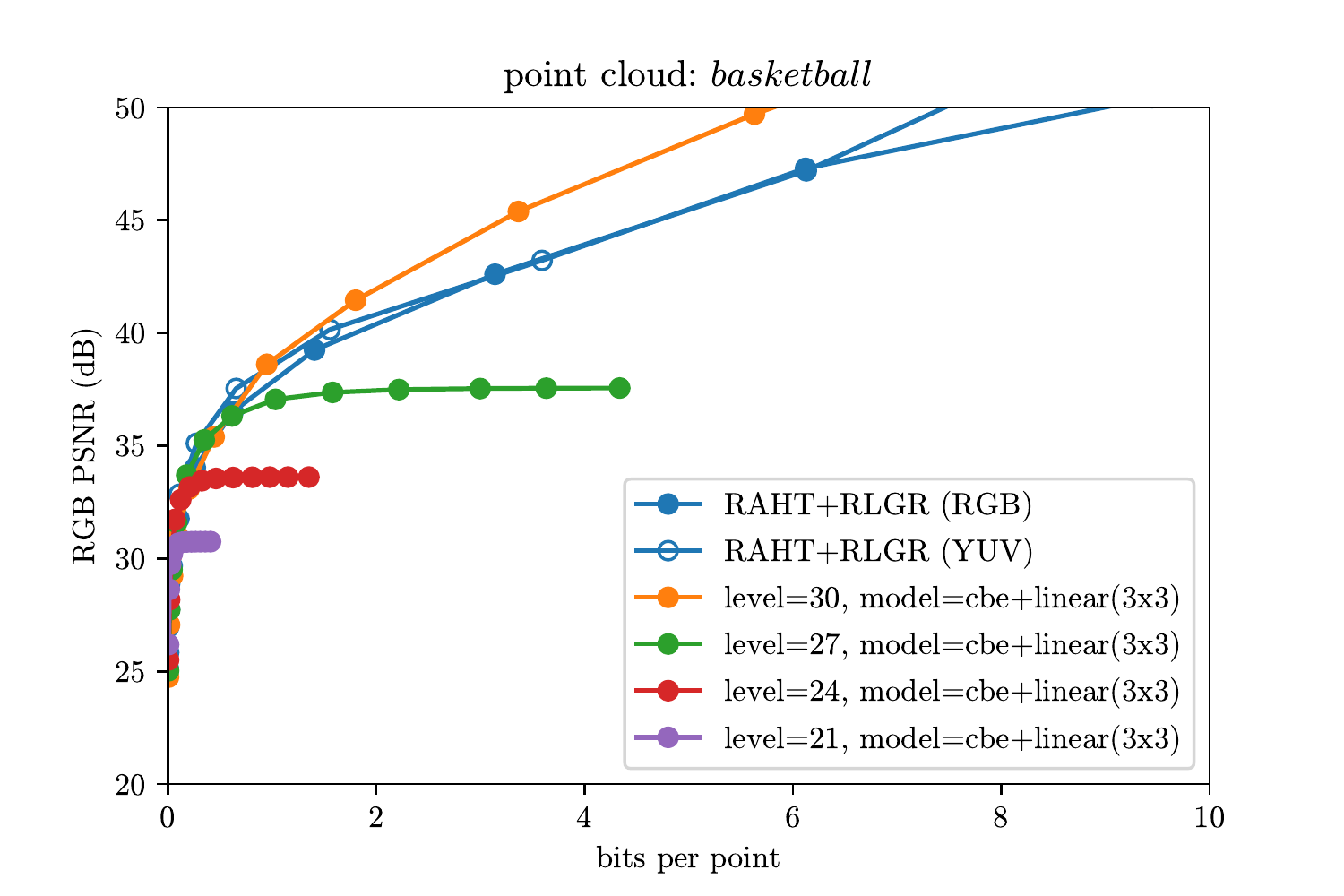}
    \includegraphics[width=0.33\textwidth, trim=20 5 35 15, clip]{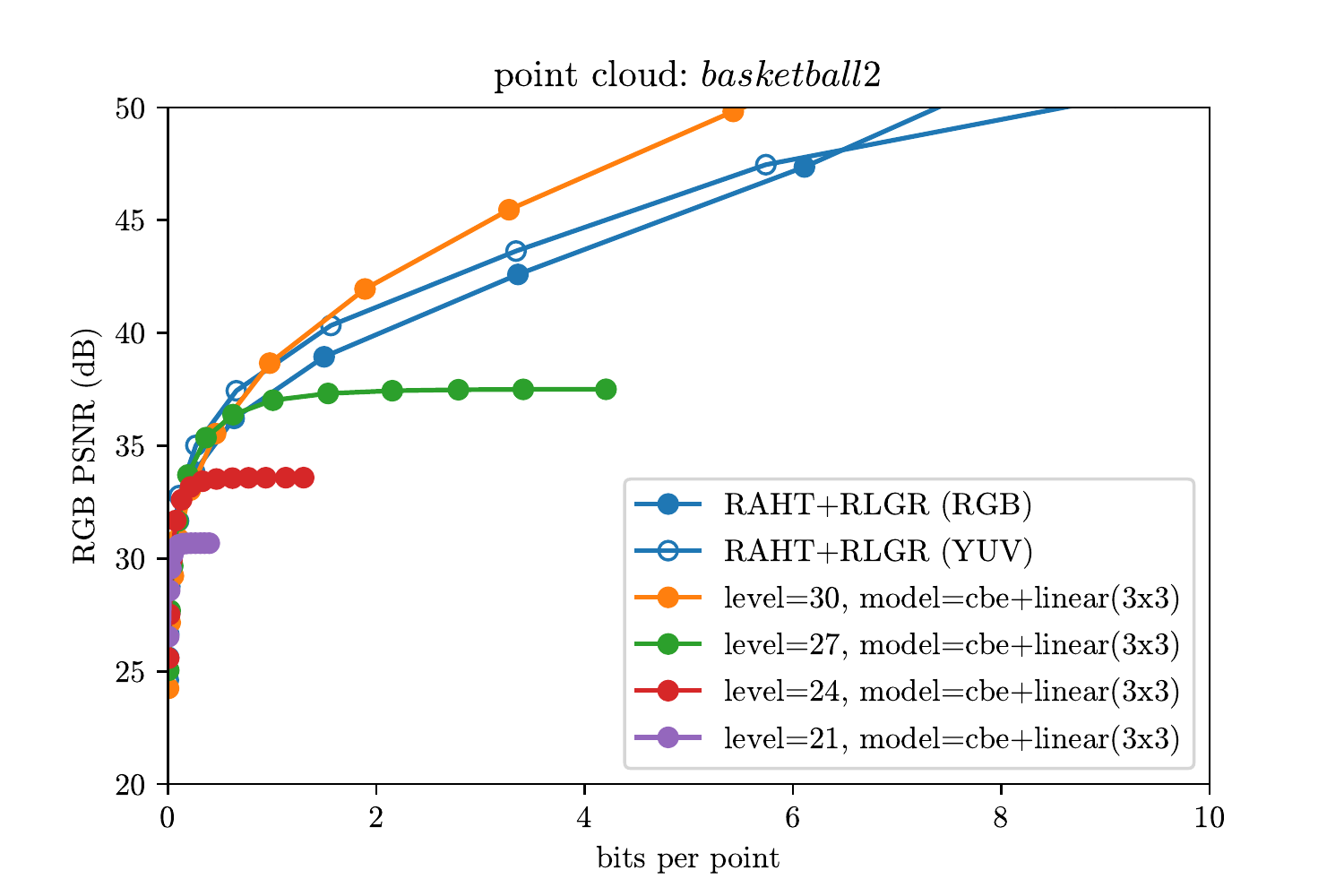}
    \includegraphics[width=0.33\textwidth, trim=20 5 35 15, clip]{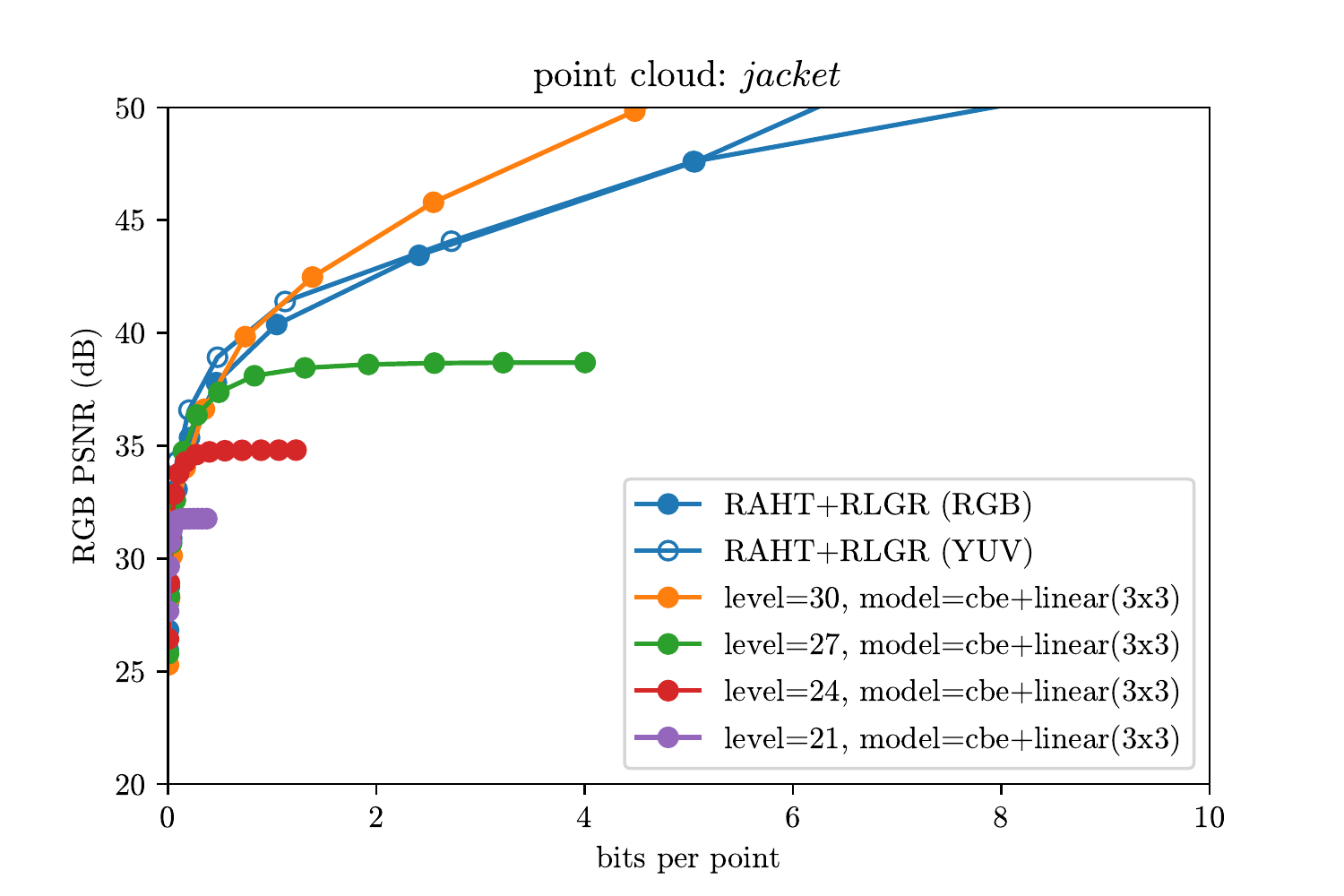}
    \caption{Baselines for six point clouds.  {\em RAHT+RLGR (RGB)} and {\em (YUV)} are shown against $3\!\times\!3$ linear models at levels 30, 27, 24, and 21, which optimize the colorspace by minimizing $D+\lambda R$ using the {\em cbe} entropy model.  See \cref{fig:baselines} for point cloud {\em rock}.}
    \label{fig:baselines_supp}
    \vspace{-0.1in}
\end{figure*}

\begin{figure*}
    \centering
    \includegraphics[width=0.29\linewidth, trim=20 5 35 15, clip]{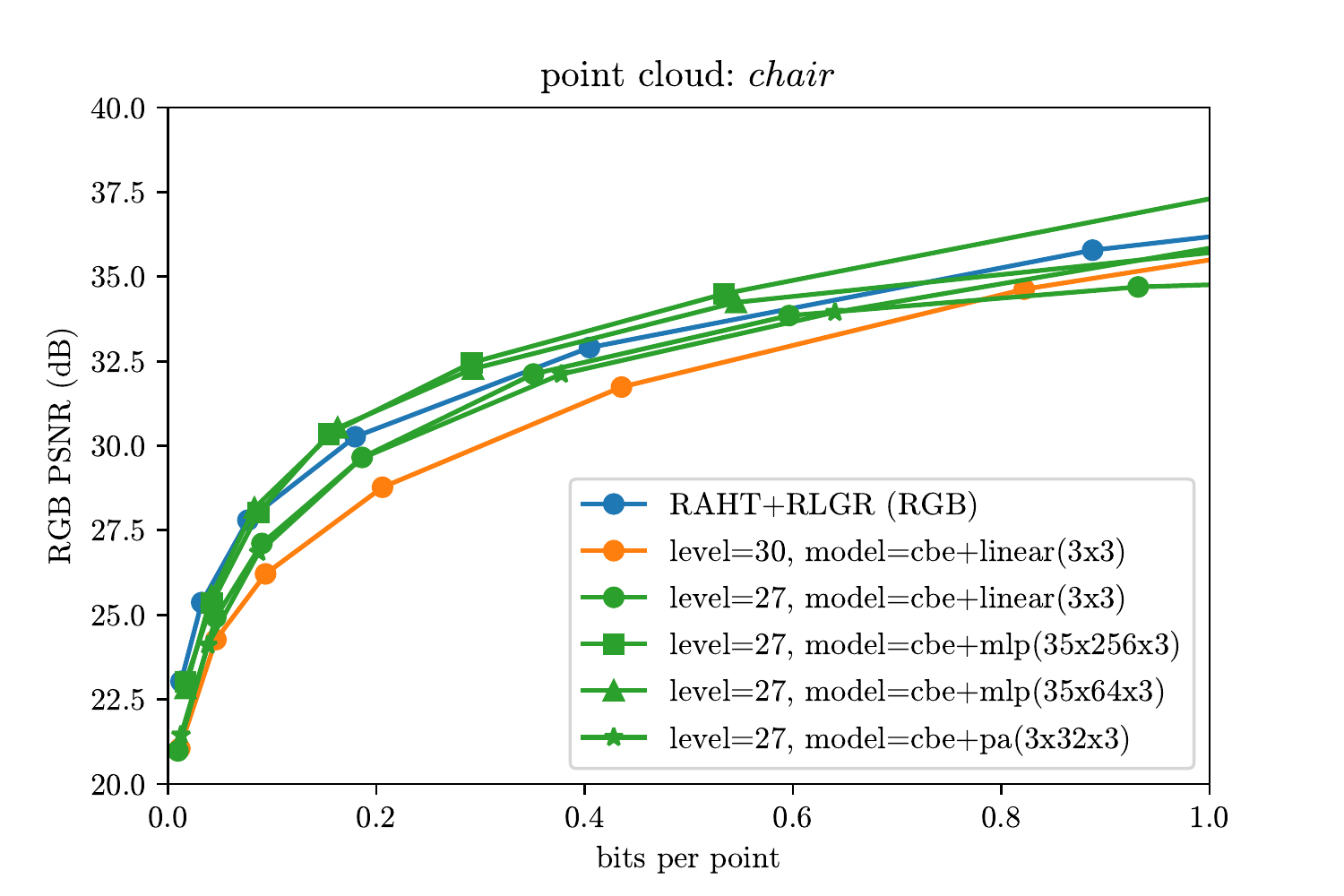}
    \includegraphics[width=0.29\linewidth, trim=20 5 35 15, clip]{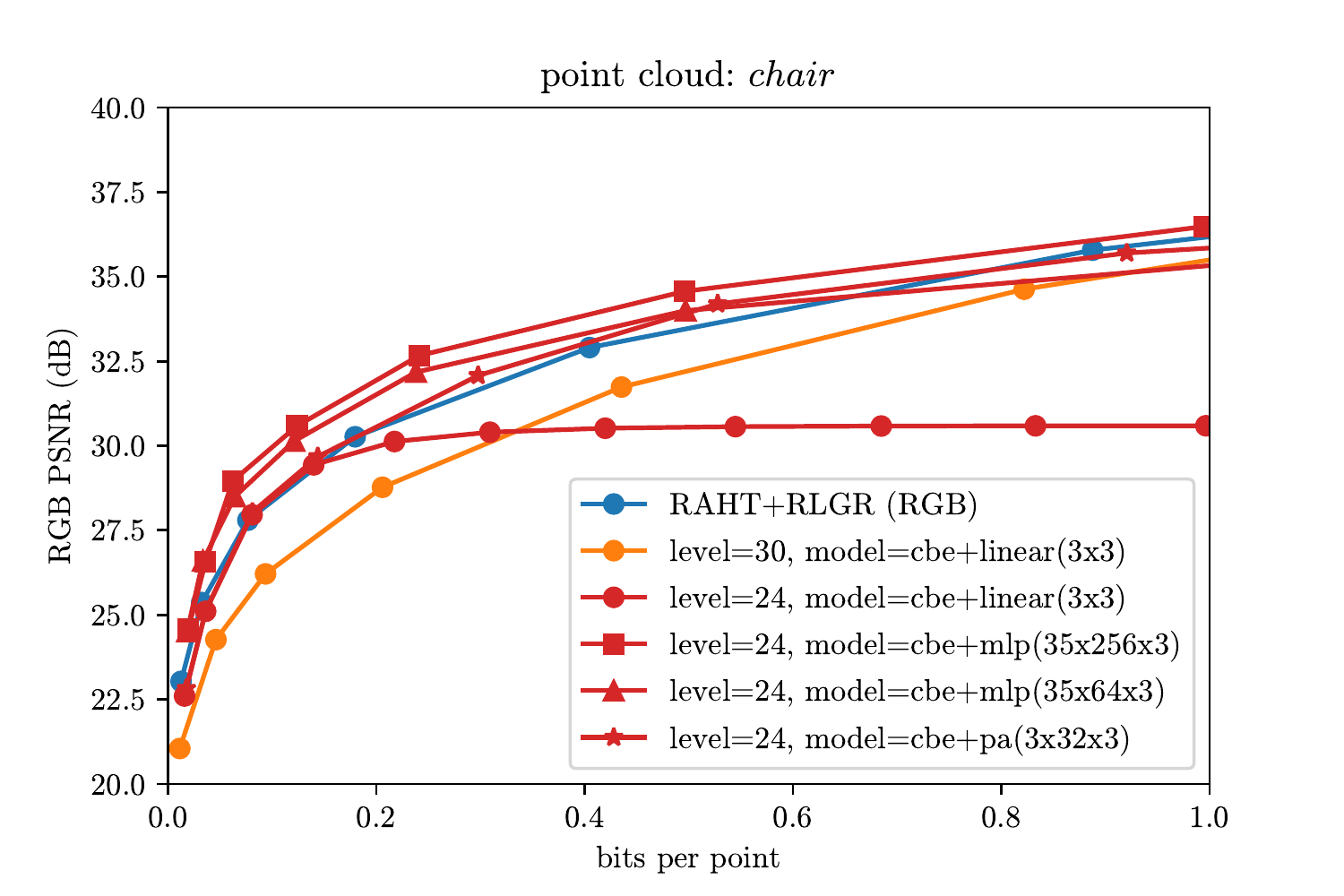}
    \includegraphics[width=0.29\linewidth, trim=20 5 35 15, clip]{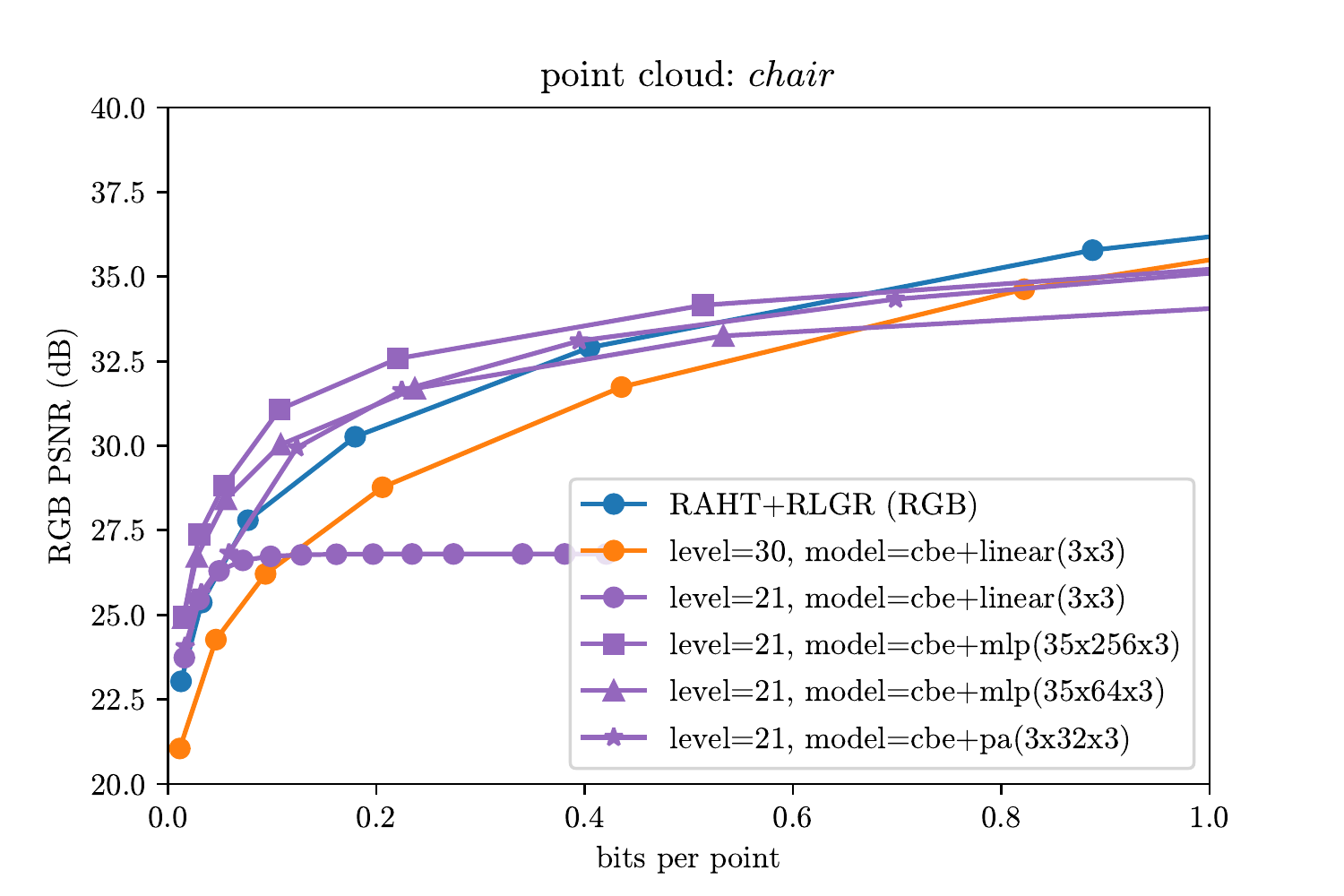}
    
    \includegraphics[width=0.29\linewidth, trim=20 5 35 15, clip]{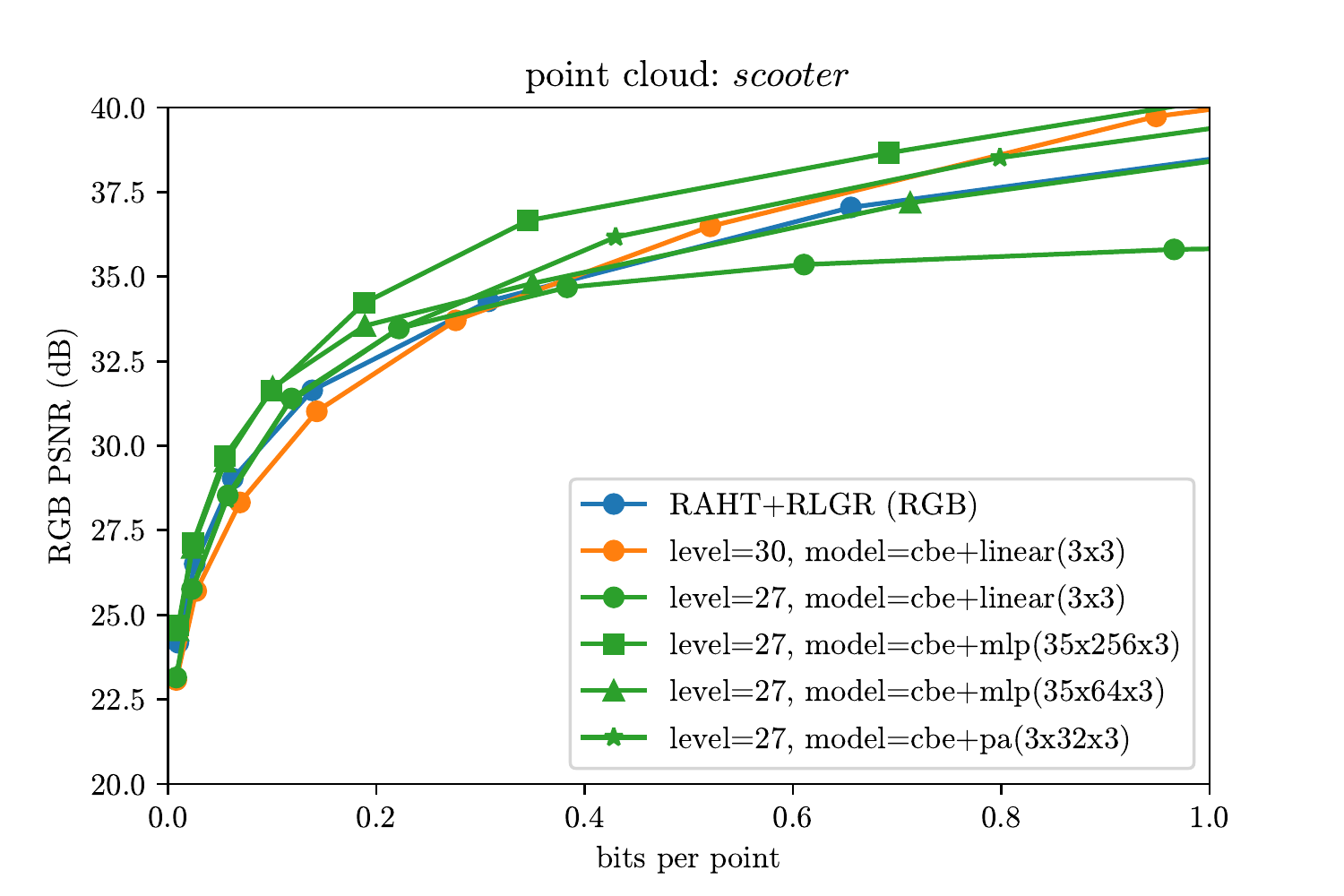}
    \includegraphics[width=0.29\linewidth, trim=20 5 35 15, clip]{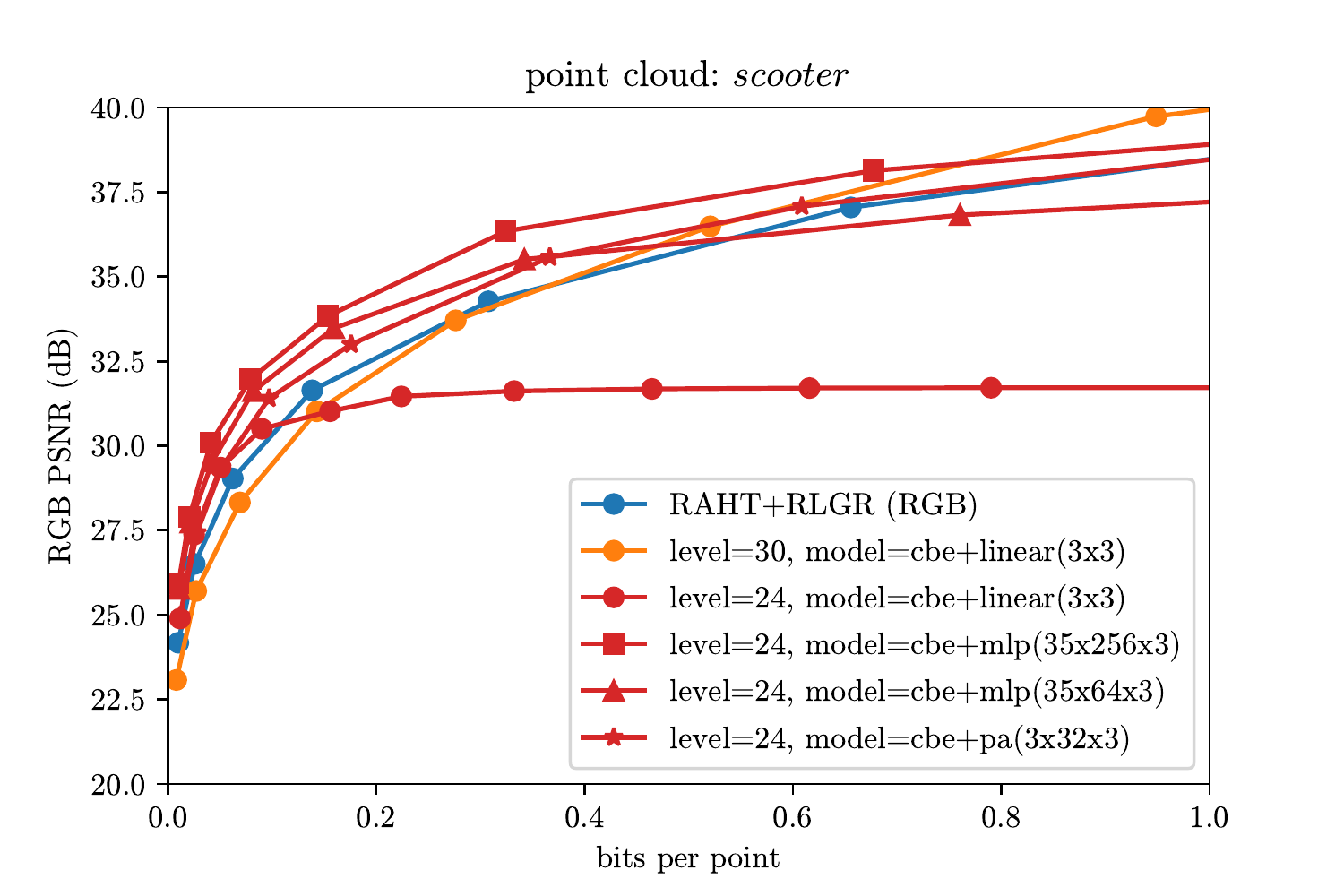}
    \includegraphics[width=0.29\linewidth, trim=20 5 35 15, clip]{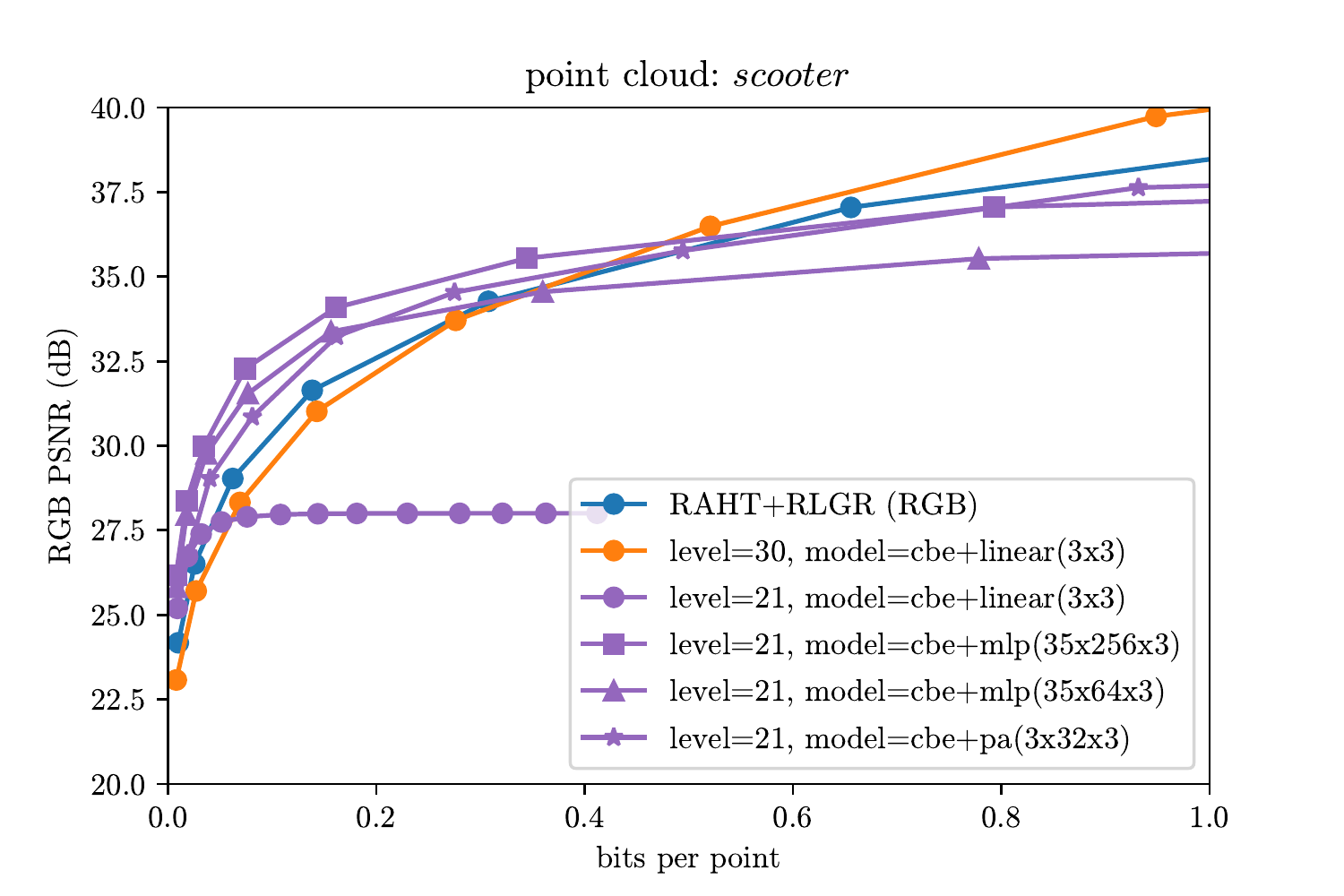}
    
    \includegraphics[width=0.29\linewidth, trim=20 5 35 15, clip]{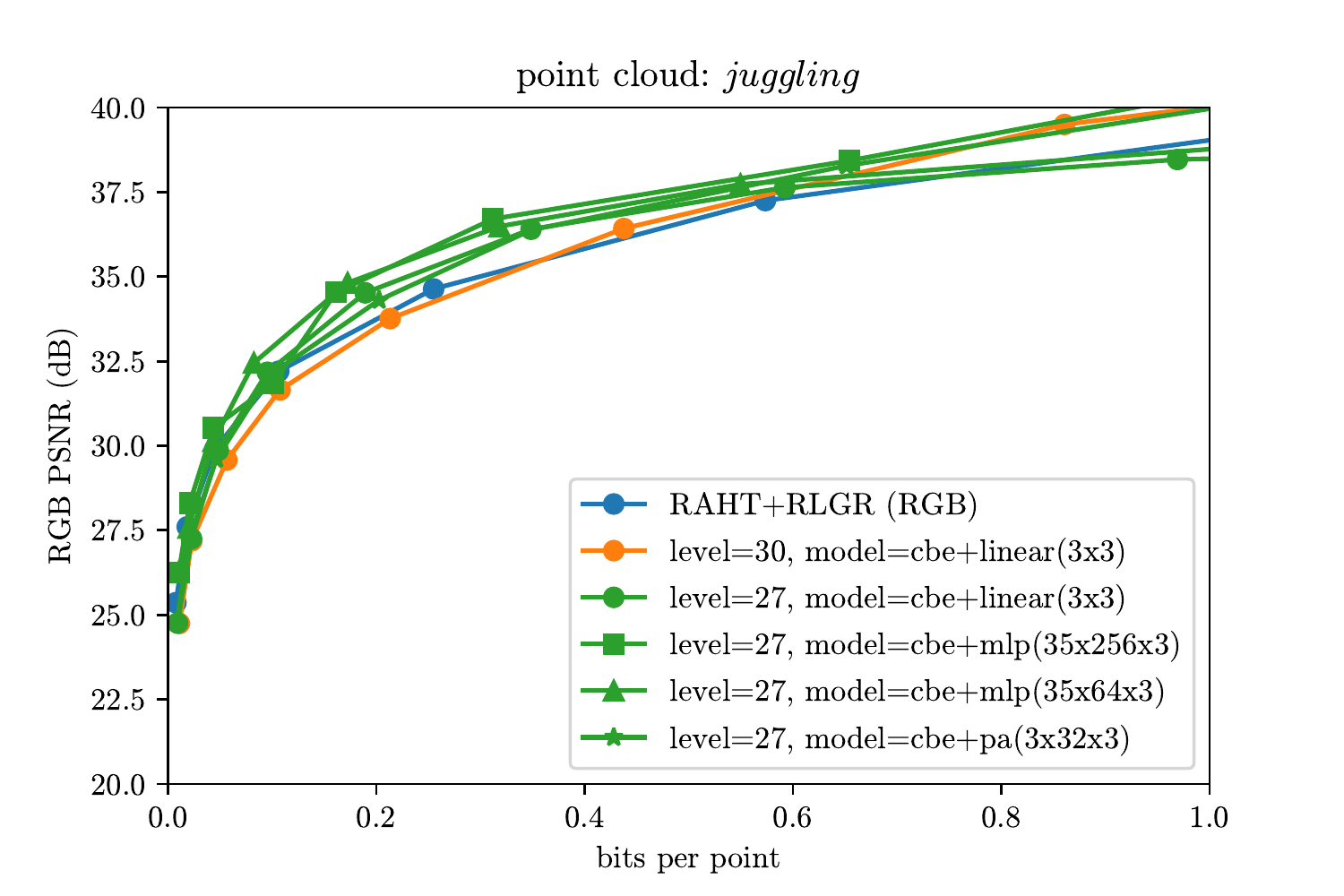}
    \includegraphics[width=0.29\linewidth, trim=20 5 35 15, clip]{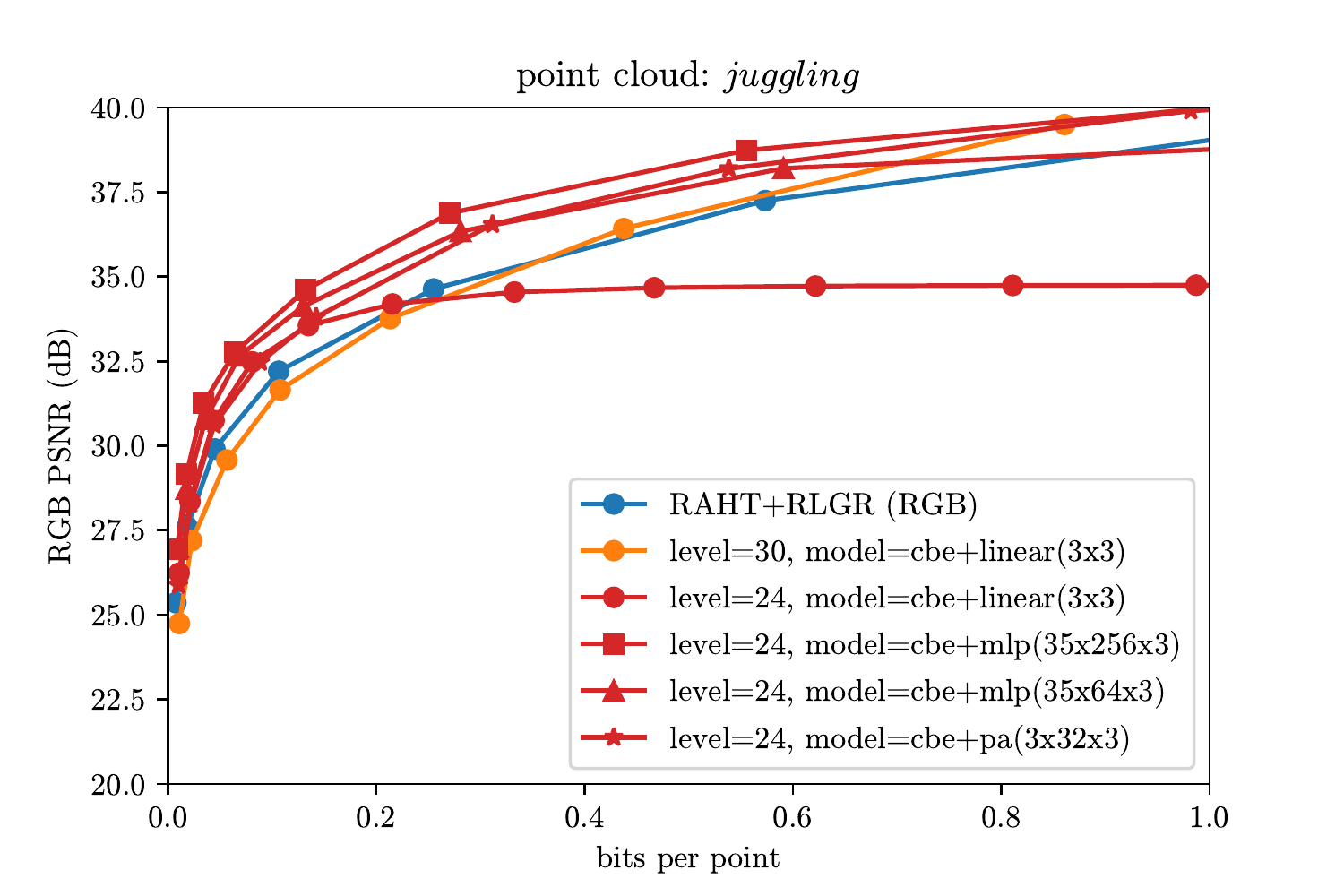}
    \includegraphics[width=0.29\linewidth, trim=20 5 35 15, clip]{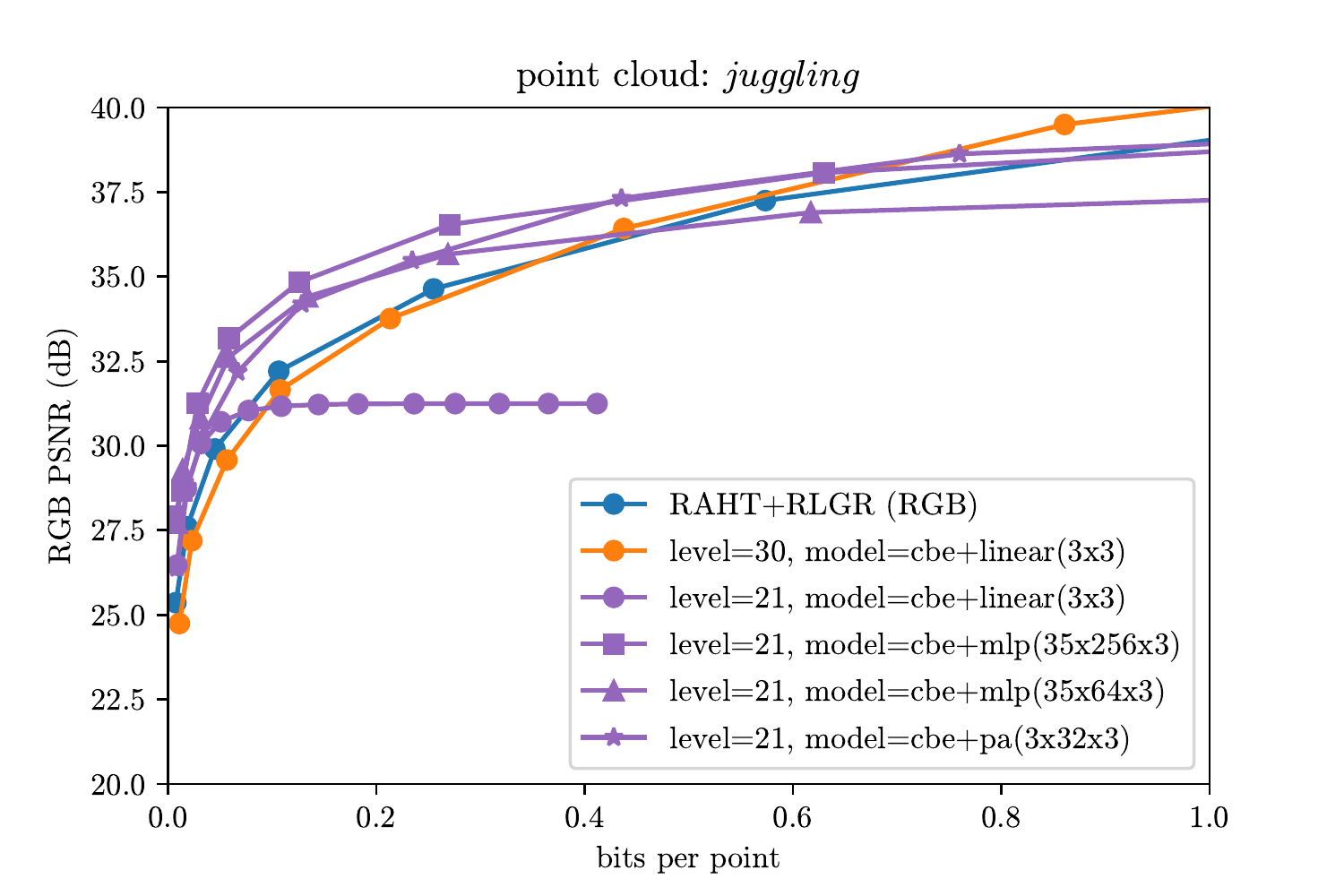}
    
    \includegraphics[width=0.29\linewidth, trim=20 5 35 15, clip]{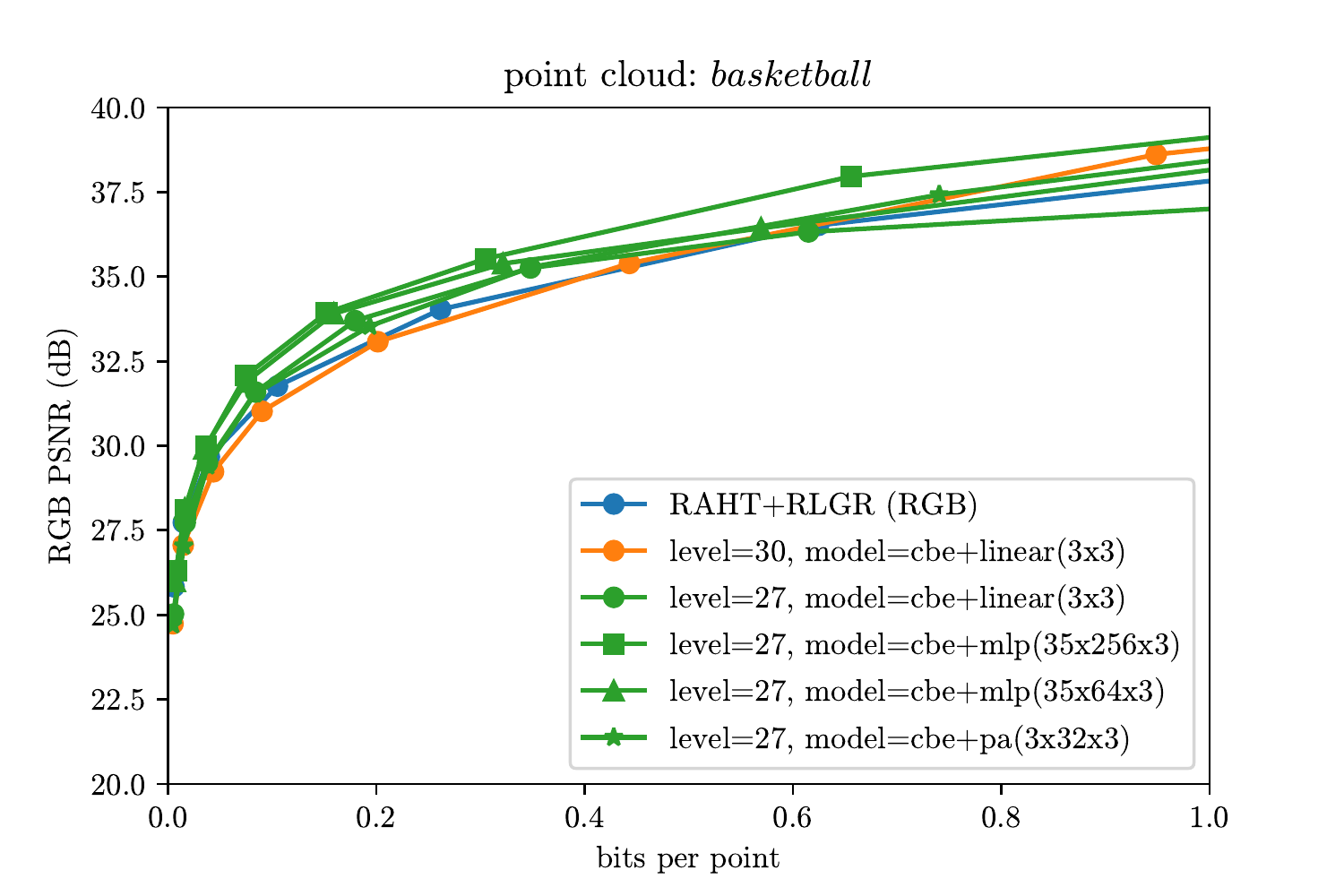}
    \includegraphics[width=0.29\linewidth, trim=20 5 35 15, clip]{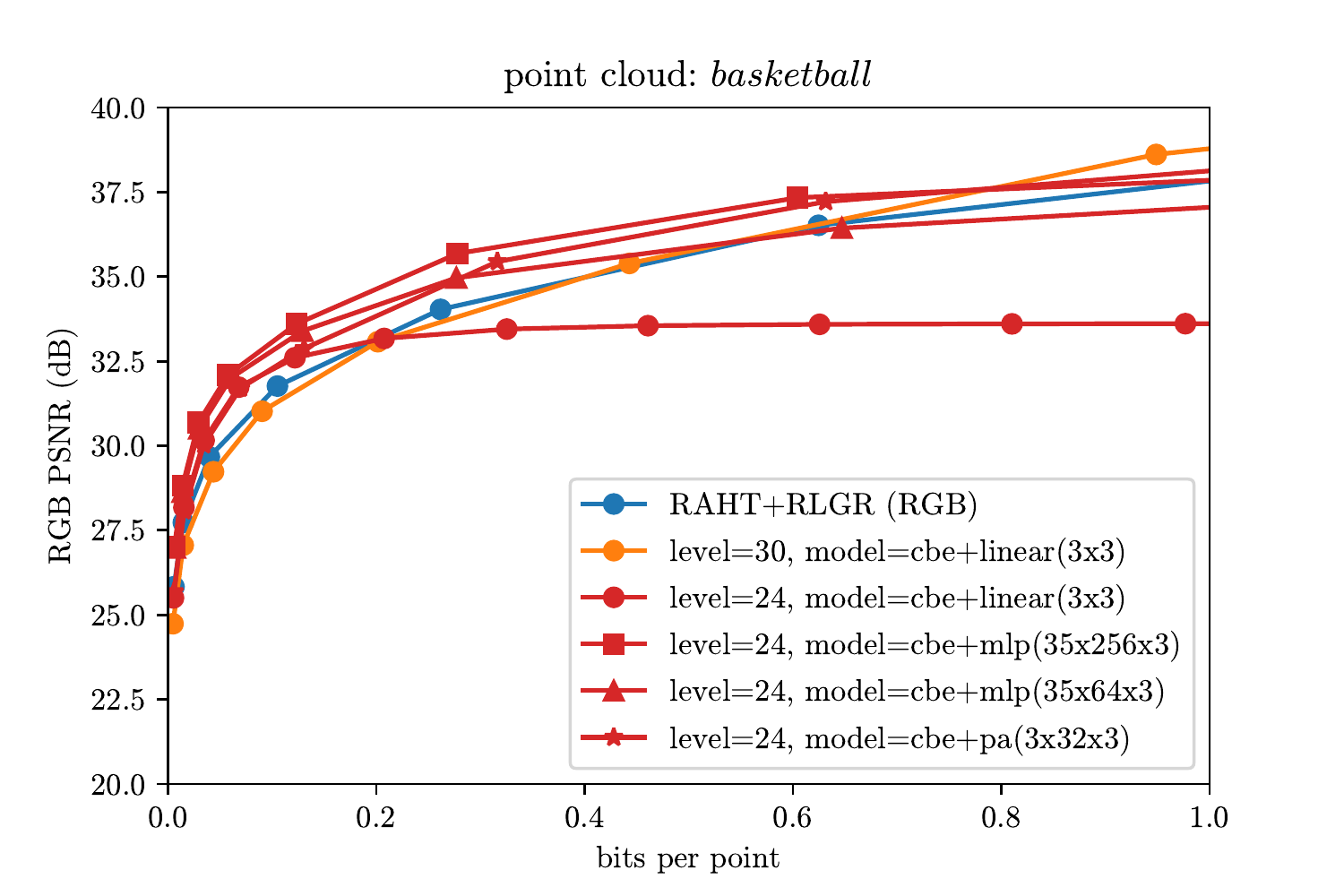}
    \includegraphics[width=0.29\linewidth, trim=20 5 35 15, clip]{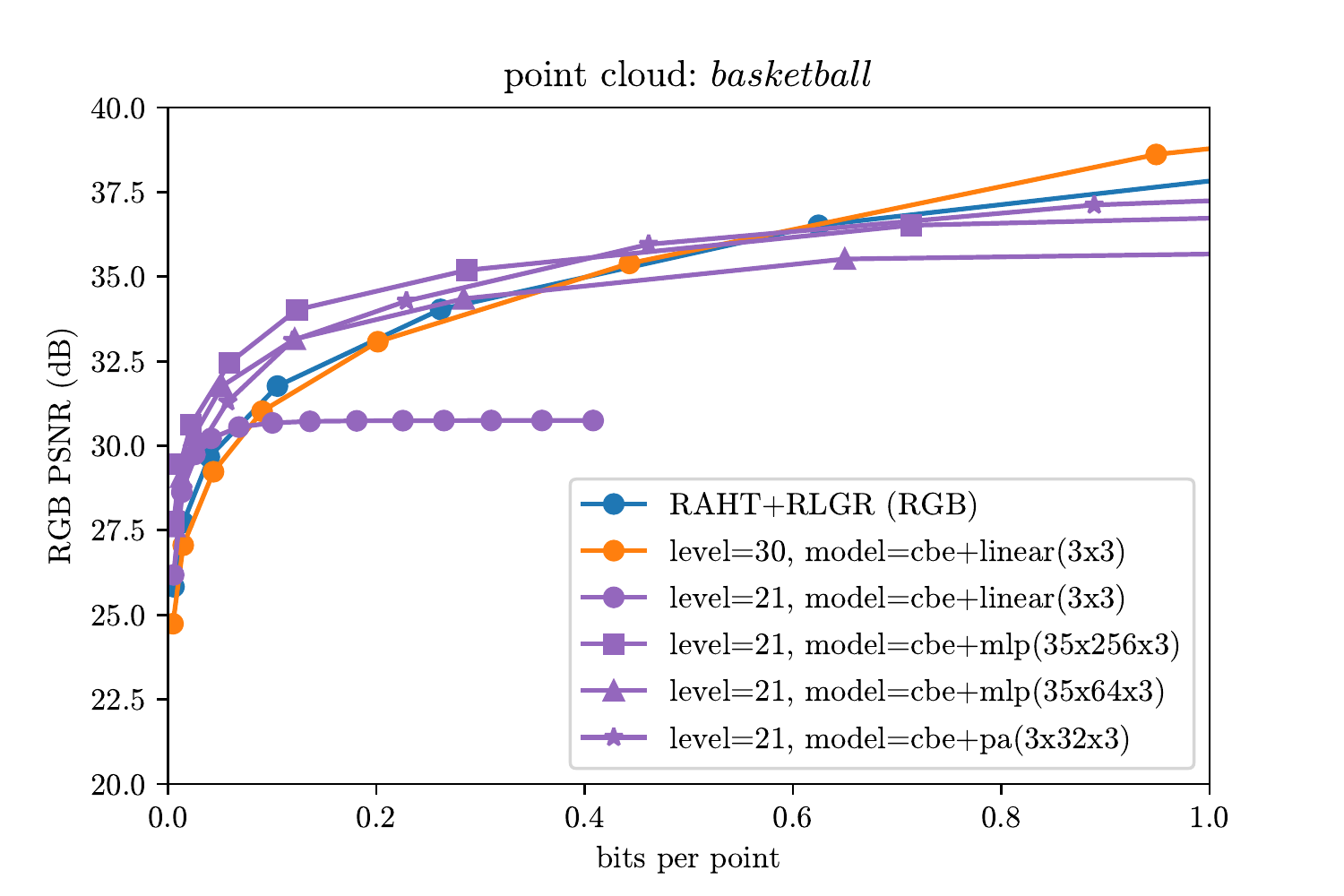}
    
    \includegraphics[width=0.29\linewidth, trim=20 5 35 15, clip]{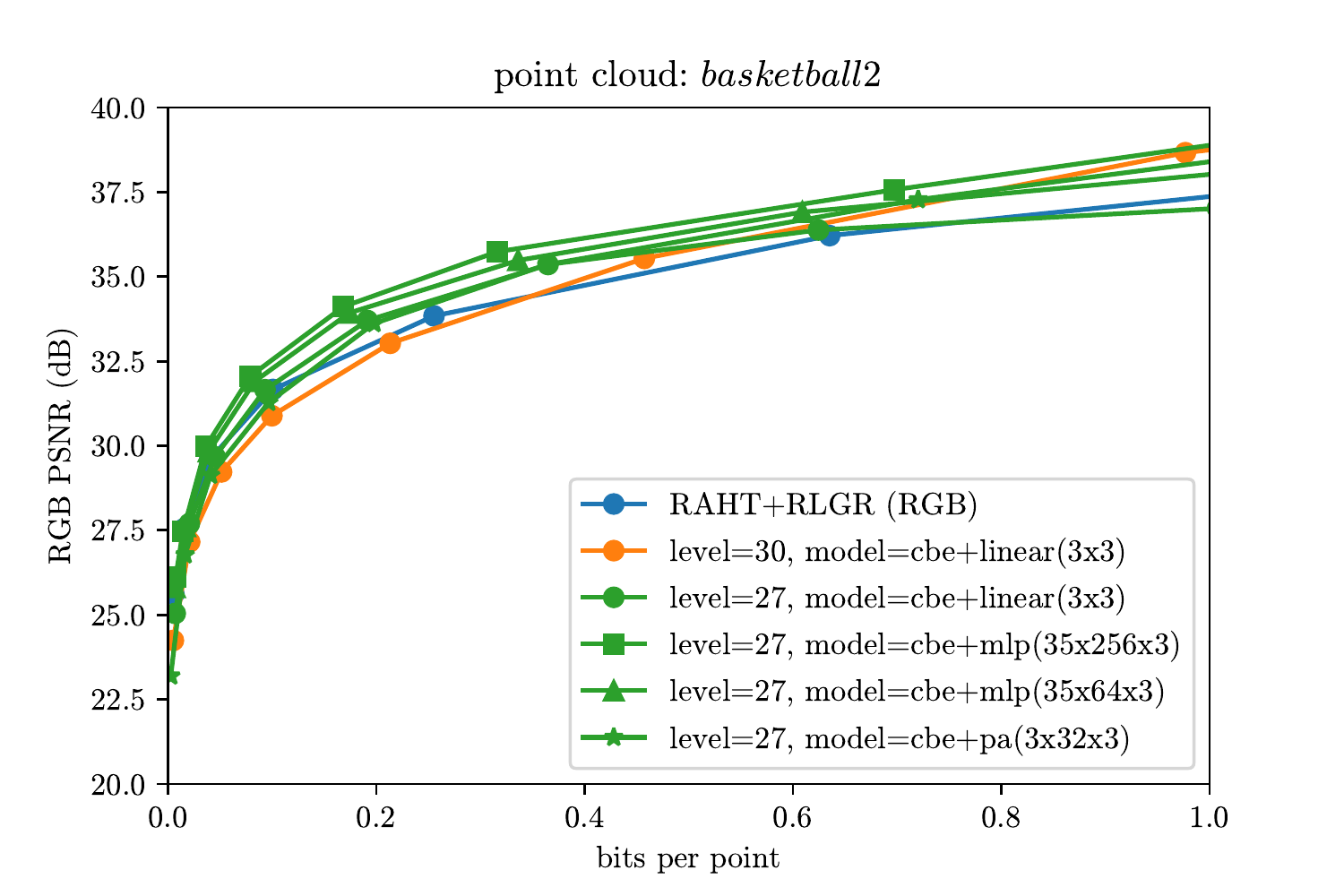}
    \includegraphics[width=0.29\linewidth, trim=20 5 35 15, clip]{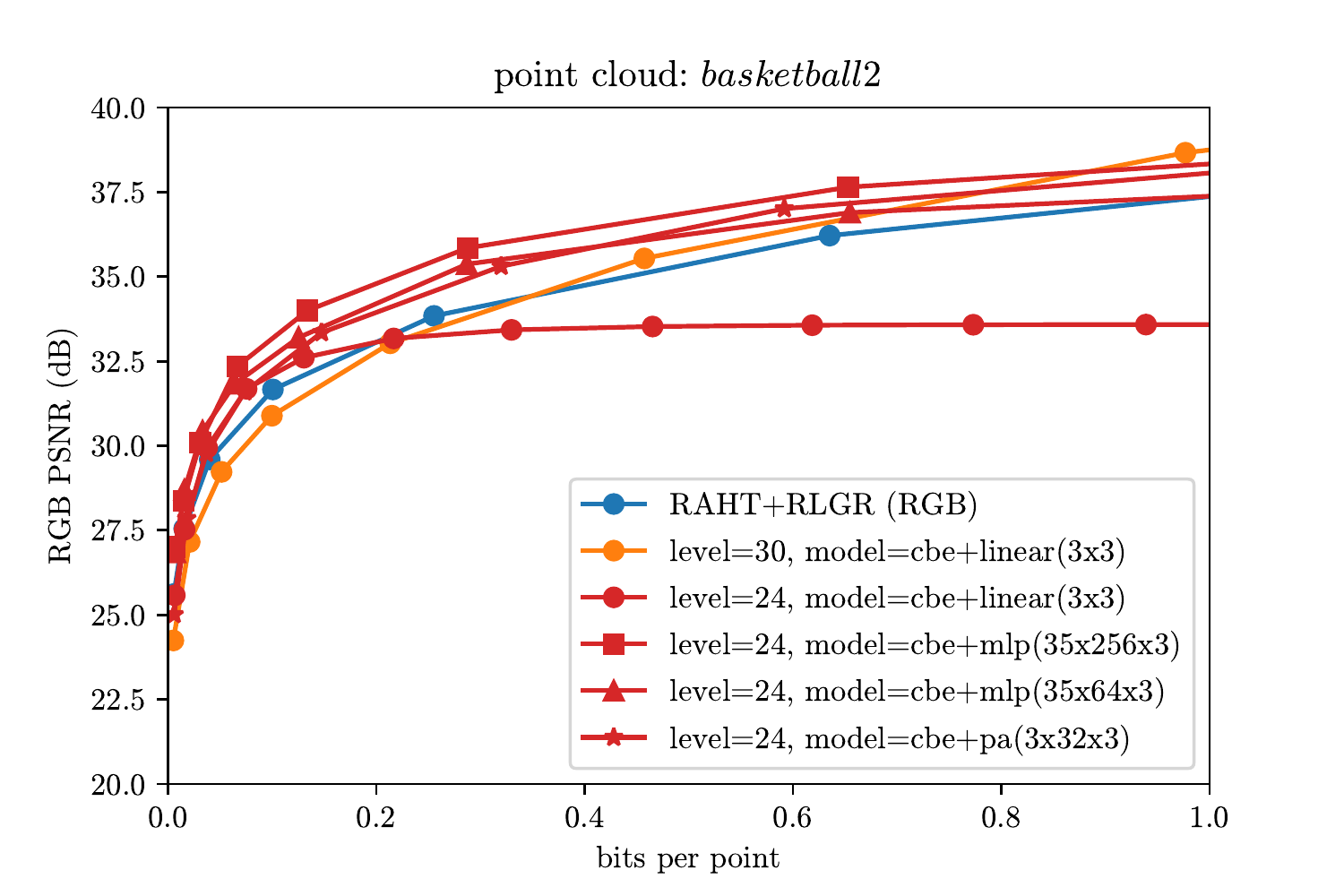}
    \includegraphics[width=0.29\linewidth, trim=20 5 35 15, clip]{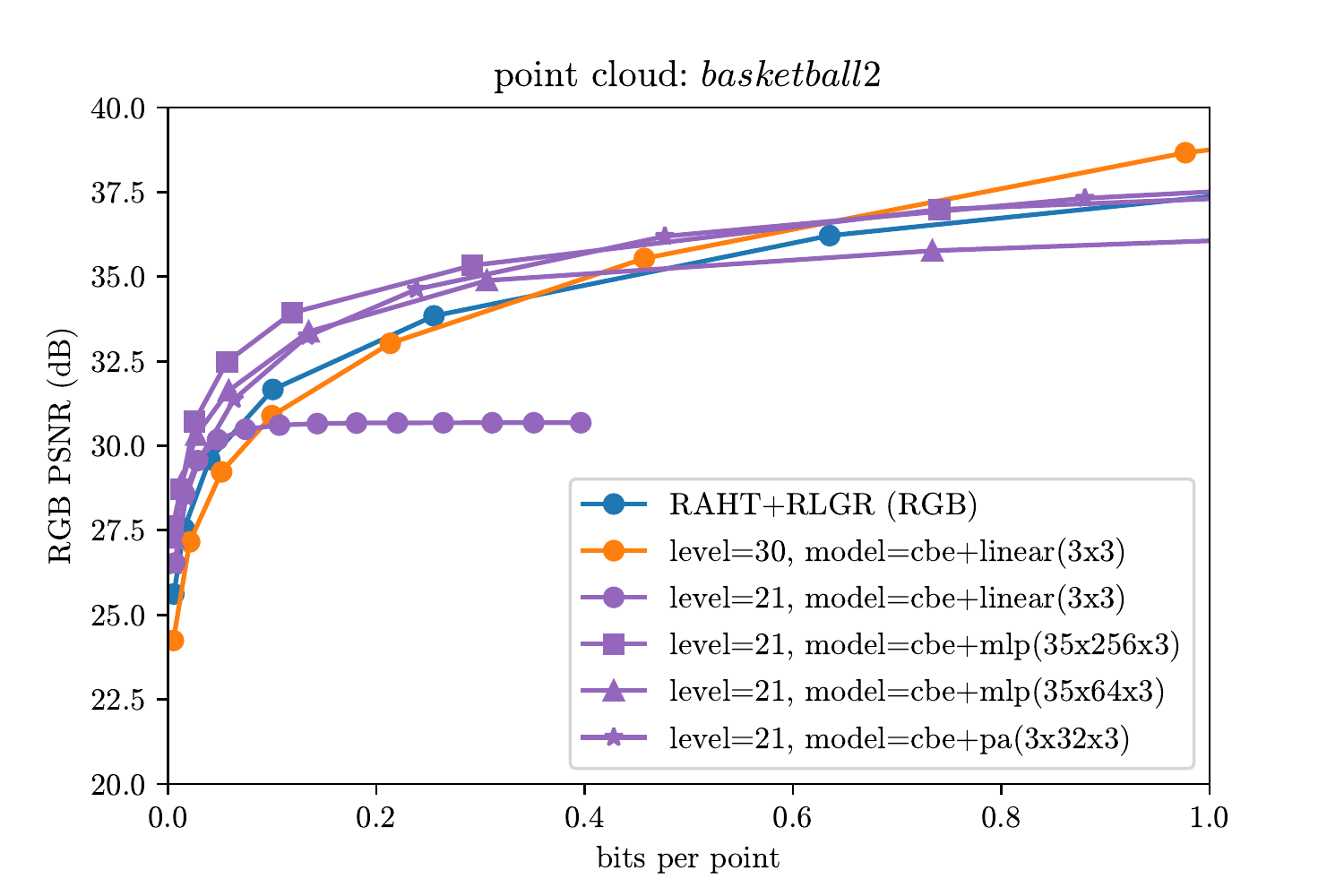}
    
    \includegraphics[width=0.29\linewidth, trim=20 5 35 15, clip]{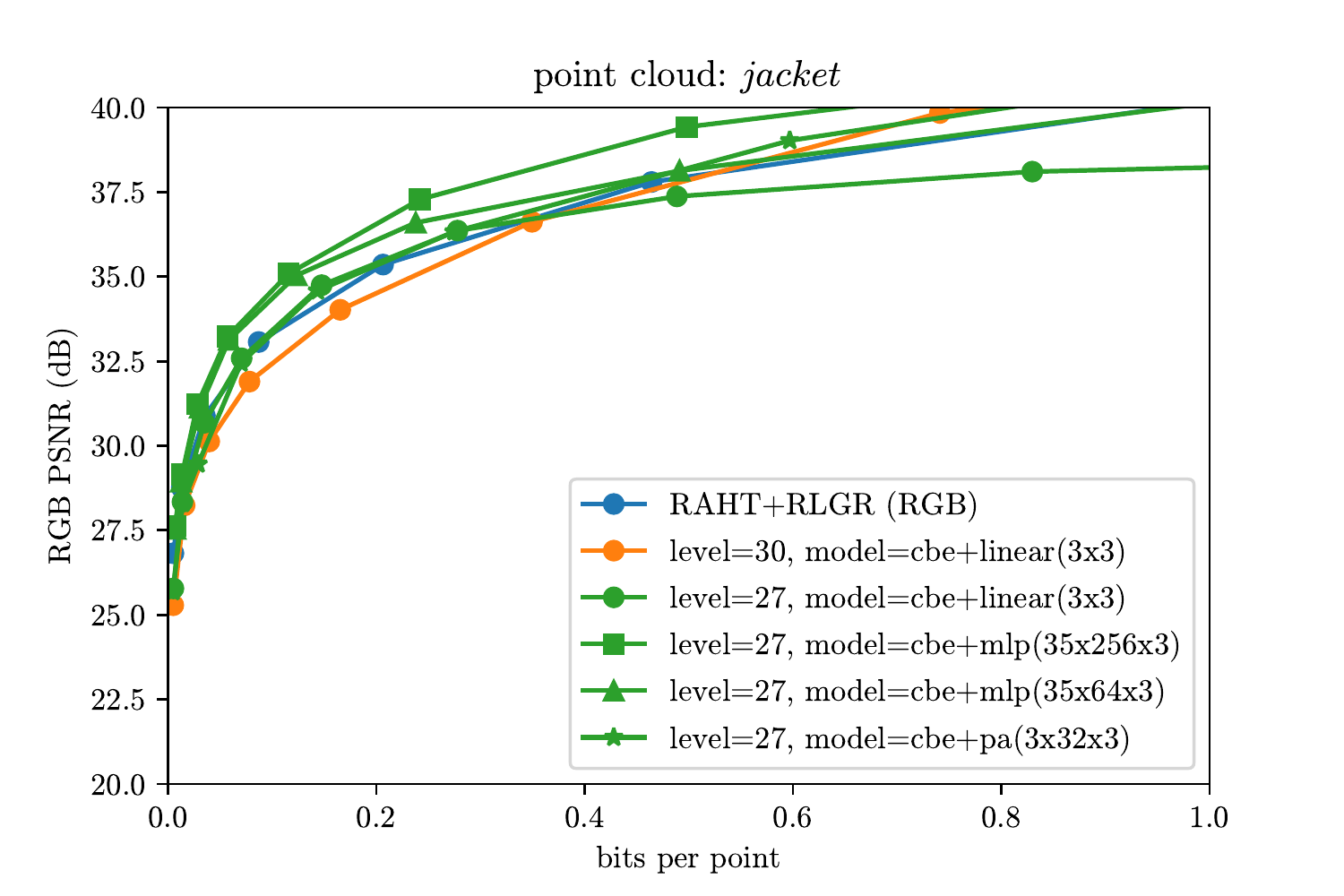}
    \includegraphics[width=0.29\linewidth, trim=20 5 35 15, clip]{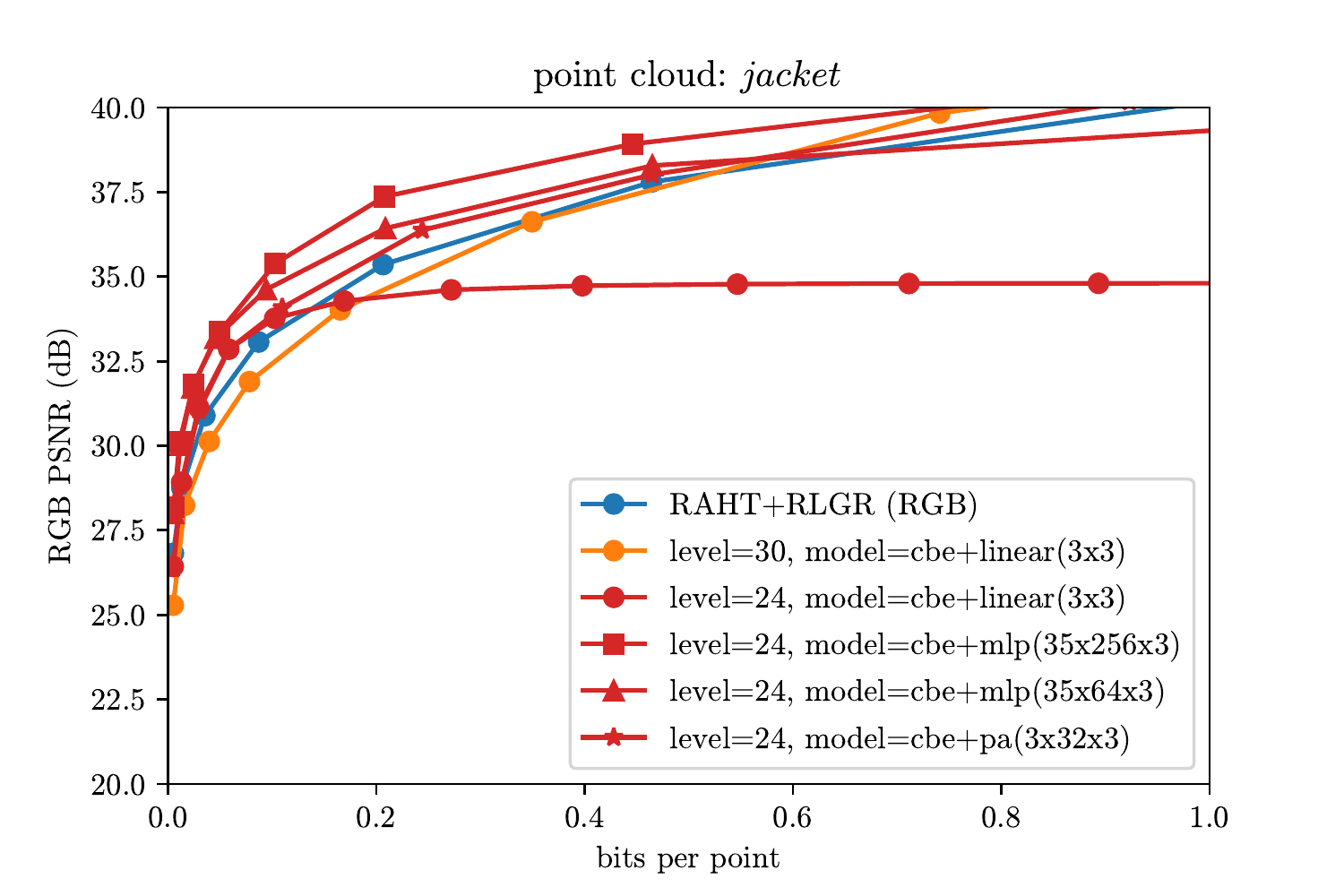}
    \includegraphics[width=0.29\linewidth, trim=20 5 35 15, clip]{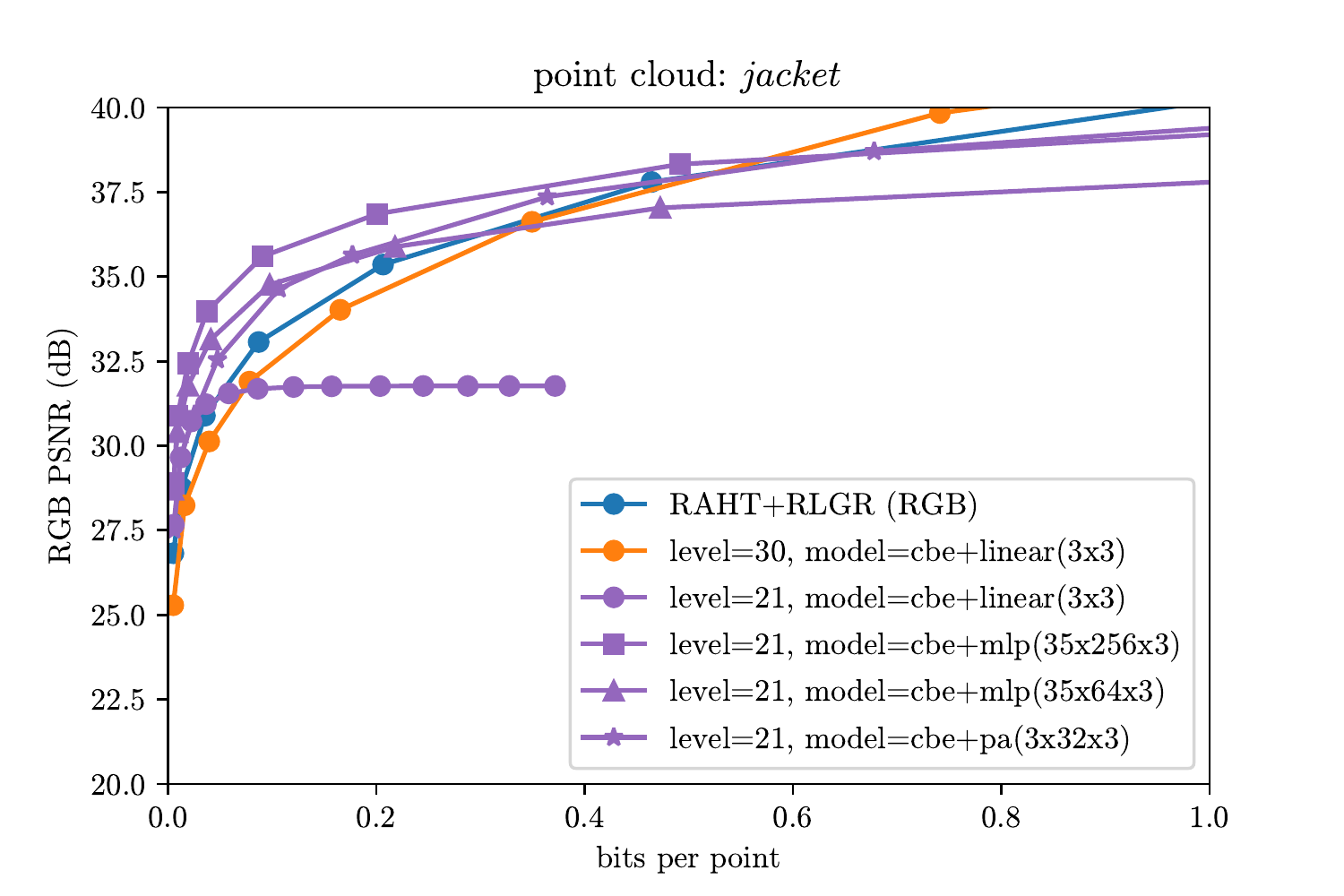}
    \caption{Coordinate Based Networks, by target level.  Each row is a different point cloud.  Left, middle, right columns each show {\em mlp(35x256x3)}, {\em mlp(35x64x3)}, and {\em pa(3x32x3)} CBNs, along with baselines, at levels 27, 24, 21.  See \cref{fig:cbns_by_level} for point cloud {\em rock}.}
    \label{fig:cbns_by_level_supp}
\end{figure*}

\begin{figure*}
    \centering
    \includegraphics[width=0.29\linewidth, trim=20 5 35 15, clip]{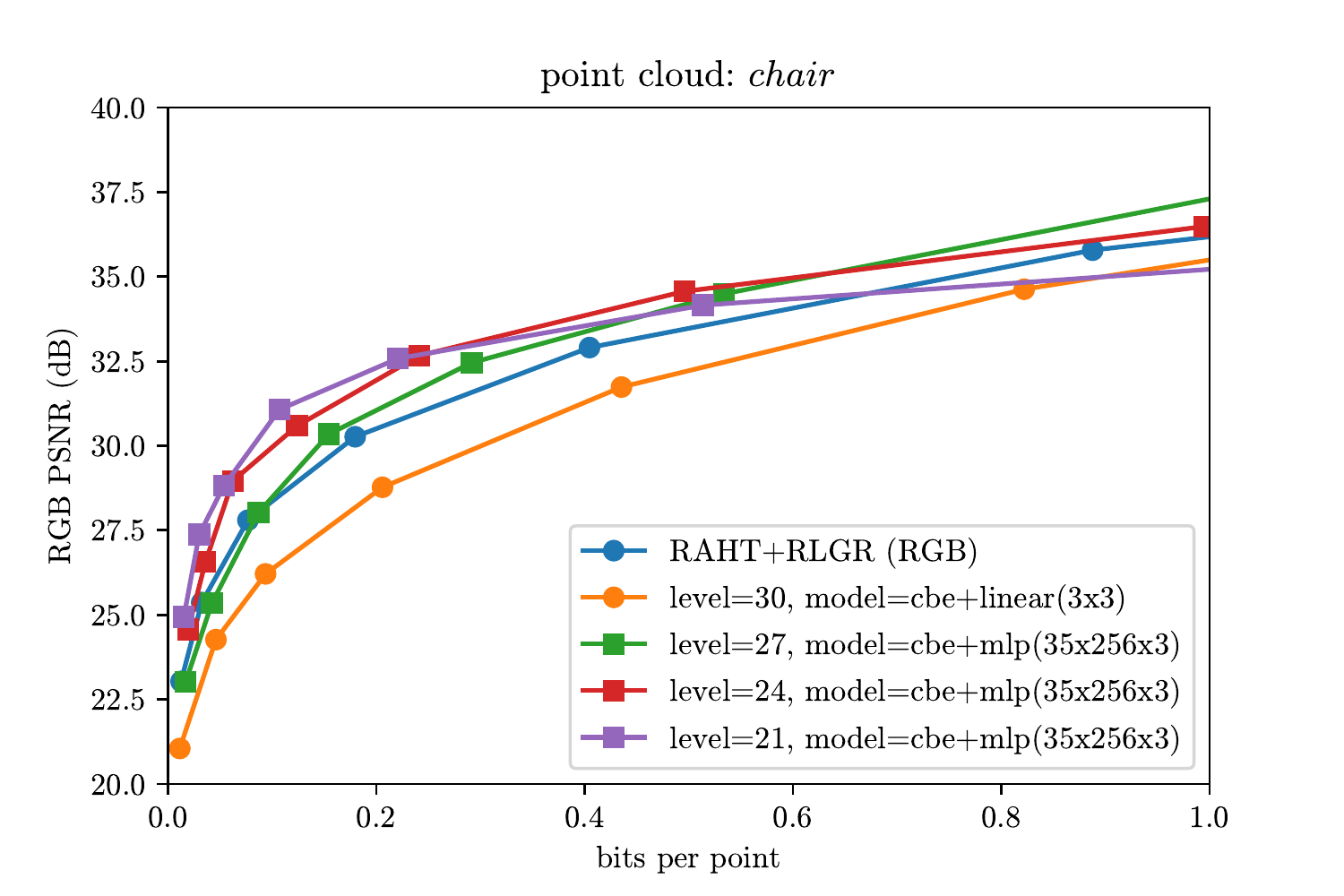}
    \includegraphics[width=0.29\linewidth, trim=20 5 35 15, clip]{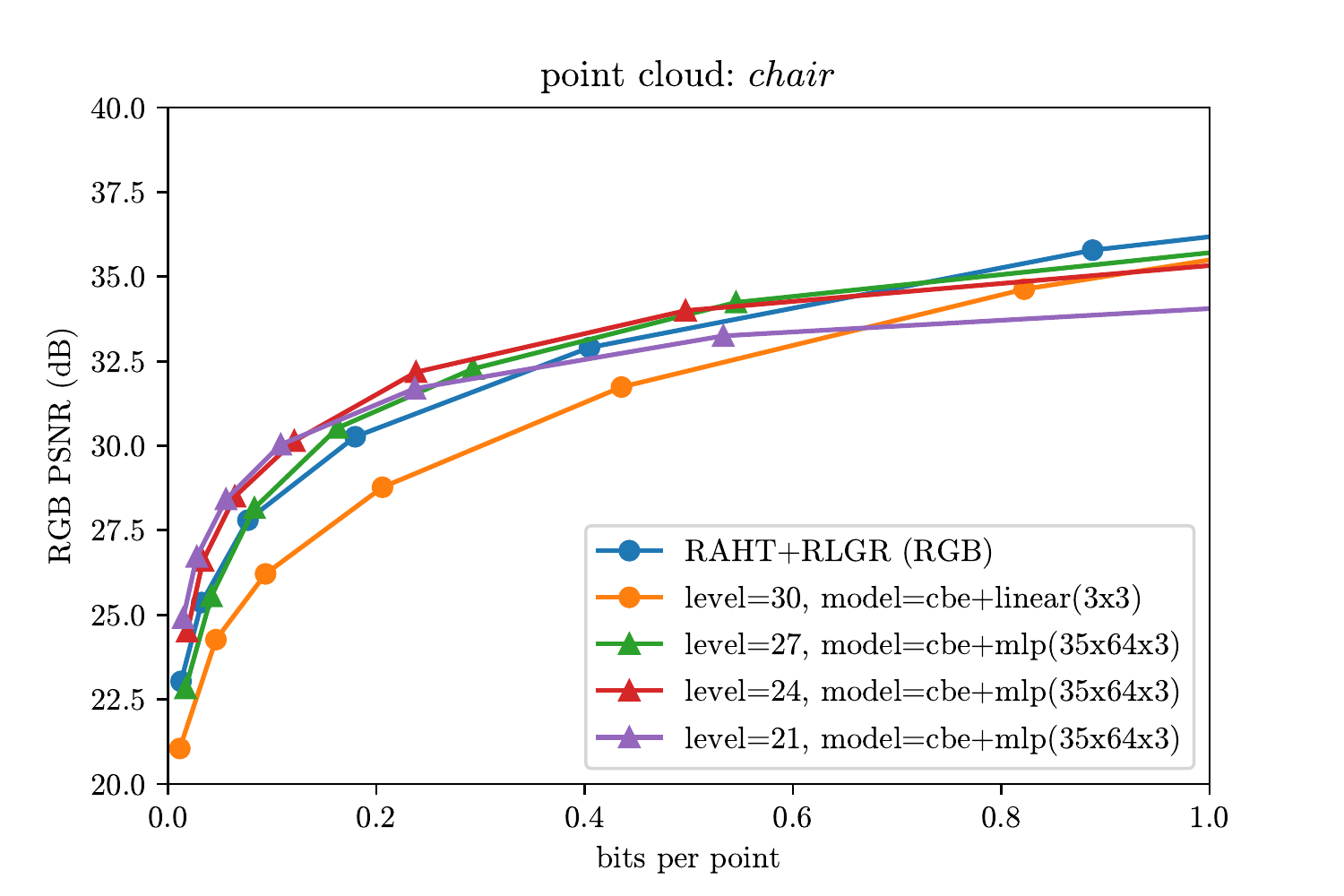}
    \includegraphics[width=0.29\linewidth, trim=20 5 35 15, clip]{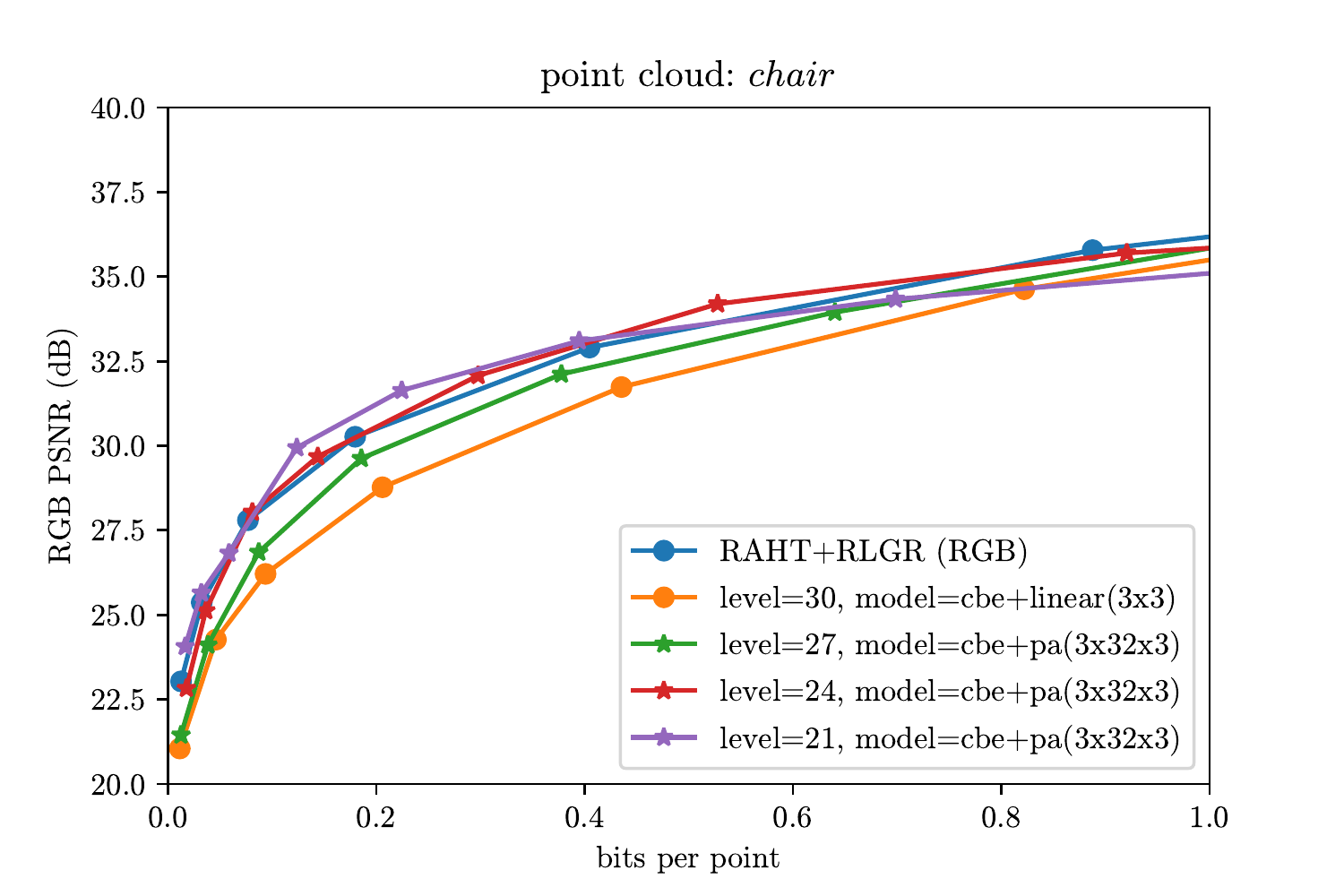}
    
    \includegraphics[width=0.29\linewidth, trim=20 5 35 15, clip]{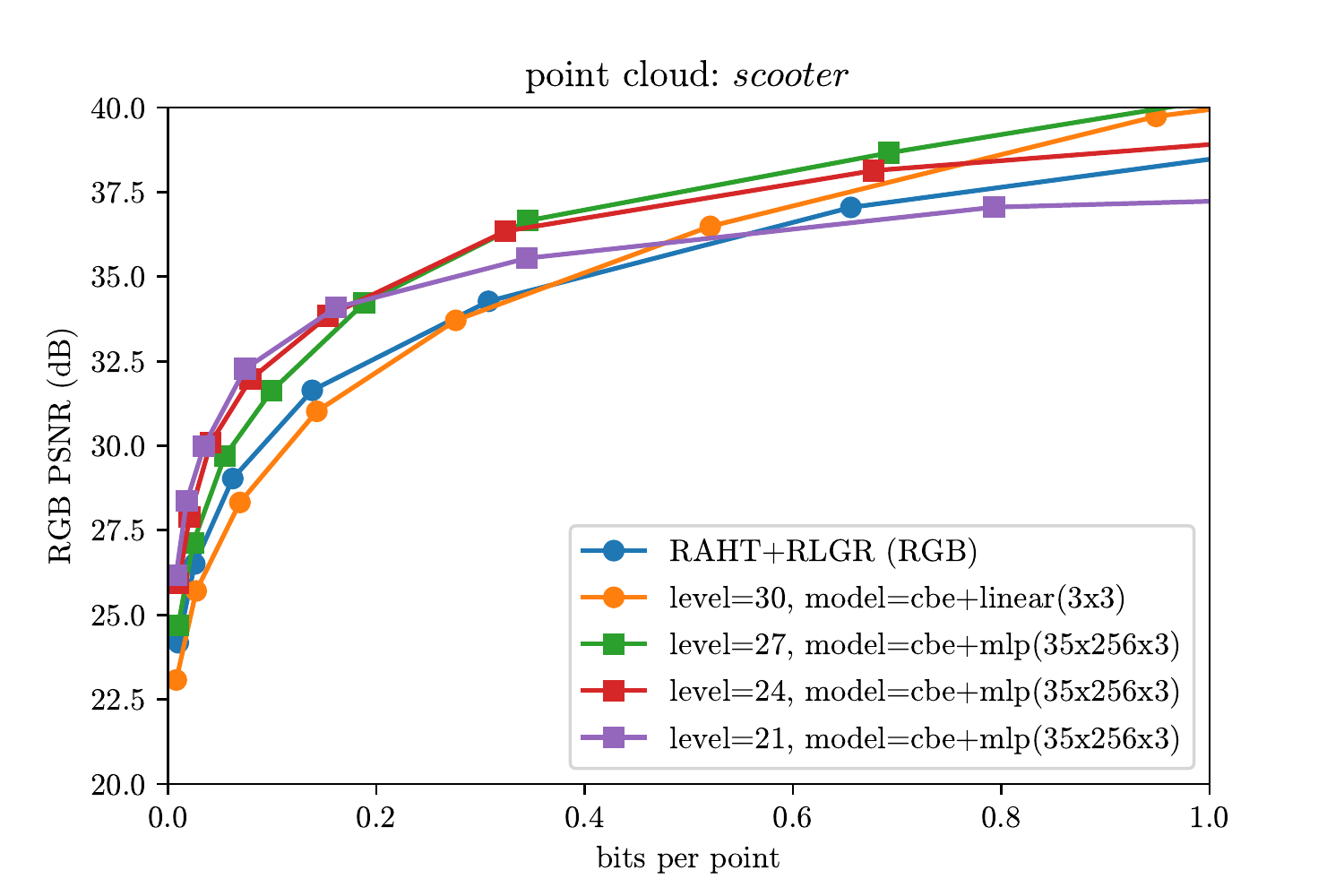}
    \includegraphics[width=0.29\linewidth, trim=20 5 35 15, clip]{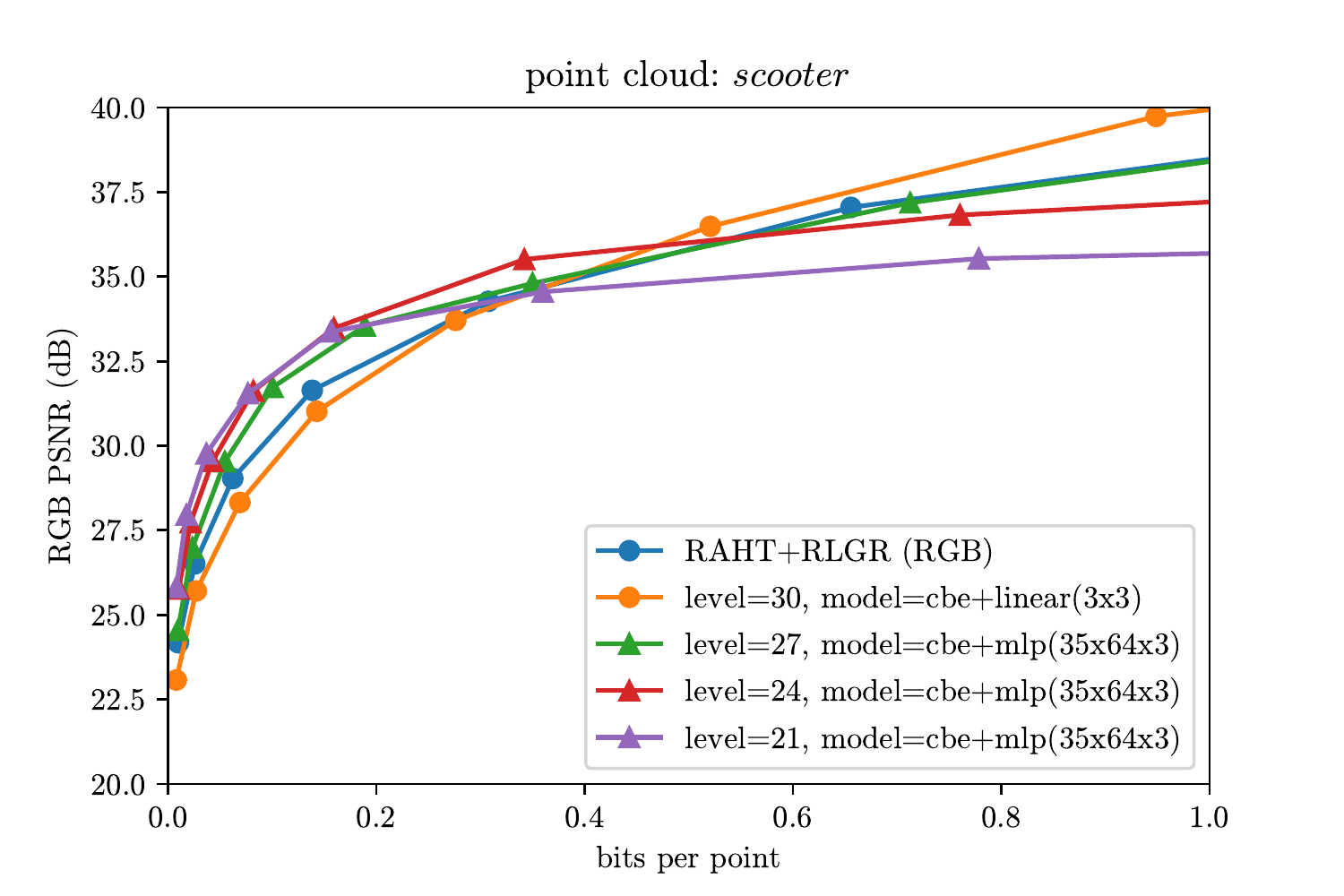}
    \includegraphics[width=0.29\linewidth, trim=20 5 35 15, clip]{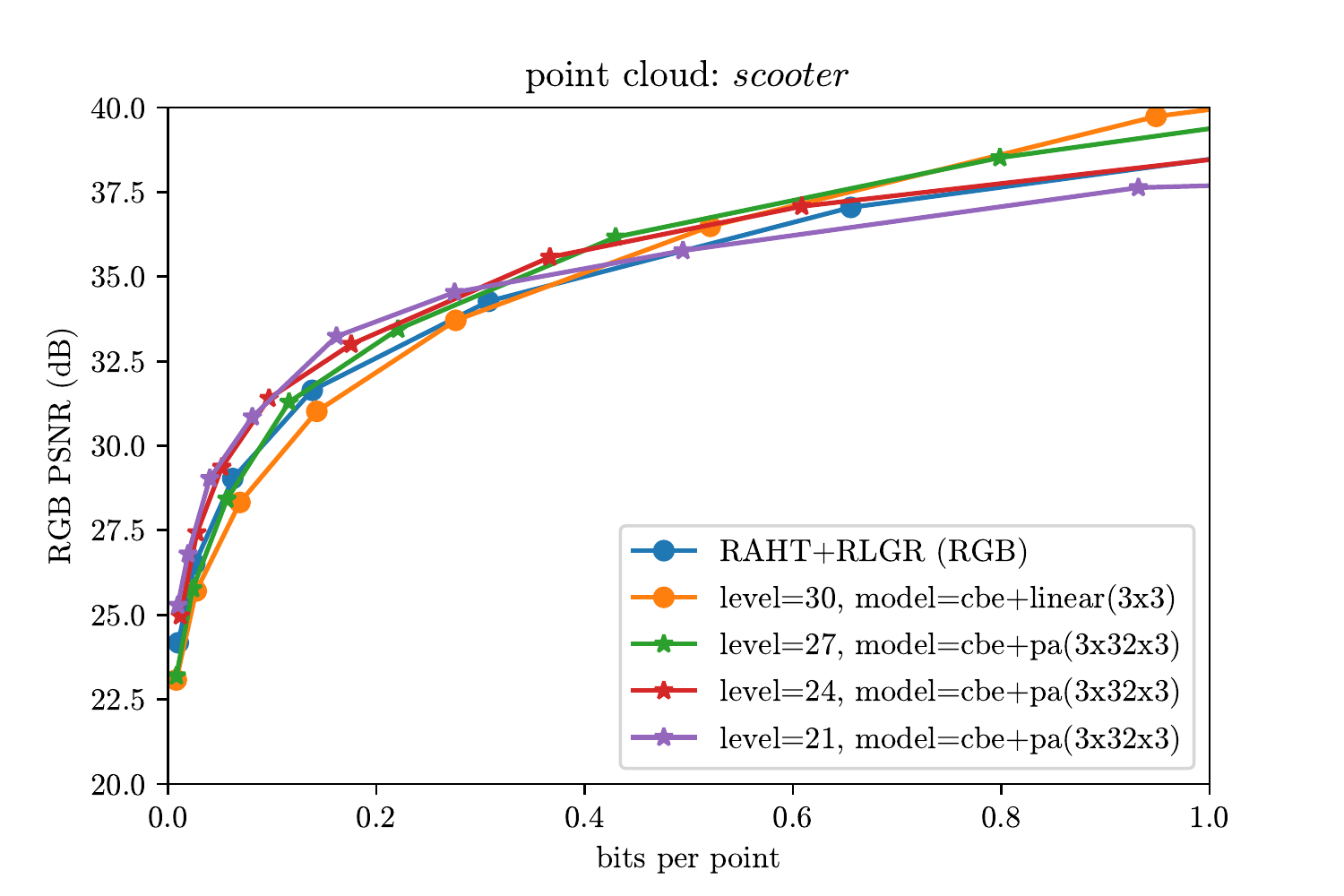}
    
    \includegraphics[width=0.29\linewidth, trim=20 5 35 15, clip]{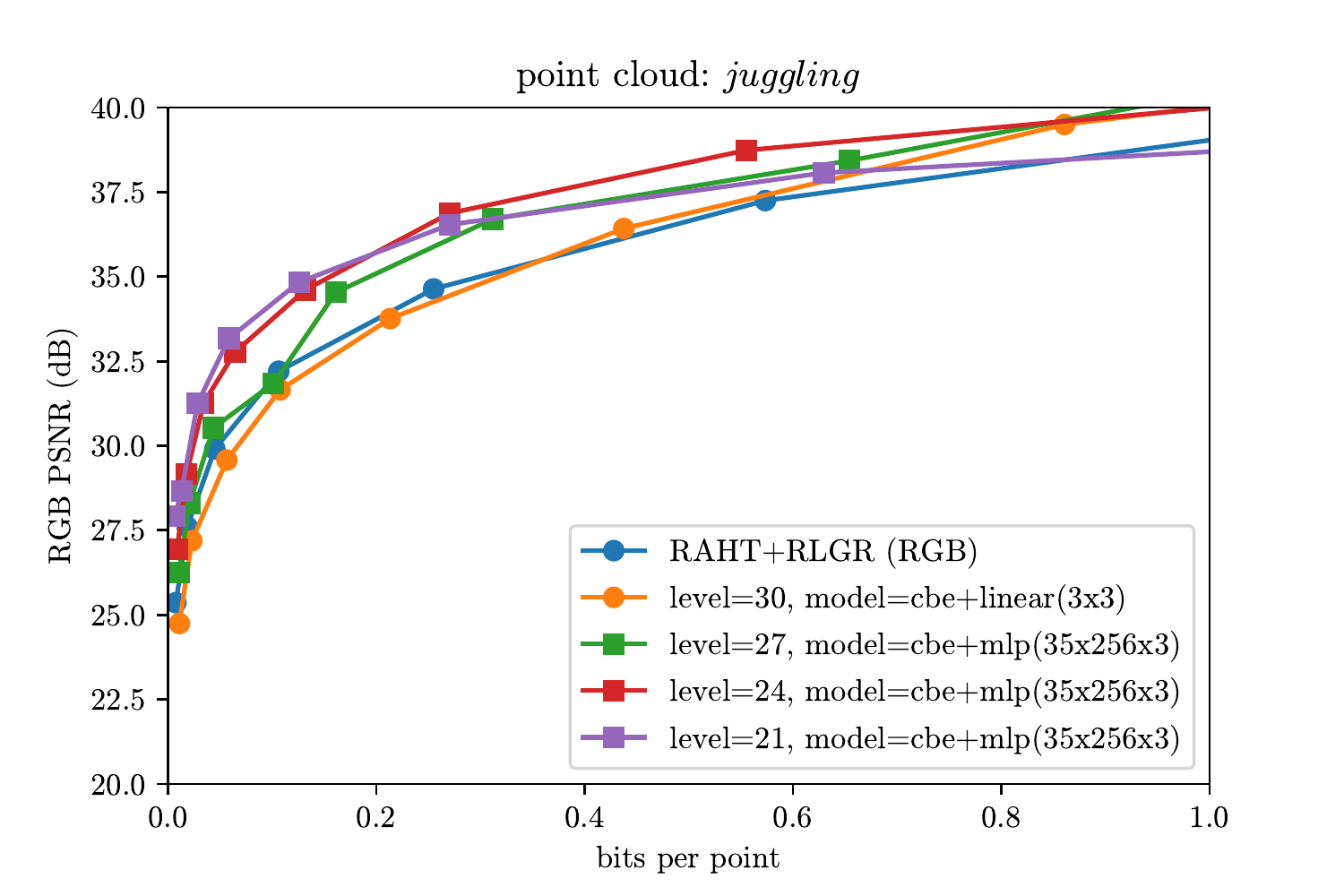}
    \includegraphics[width=0.29\linewidth, trim=20 5 35 15, clip]{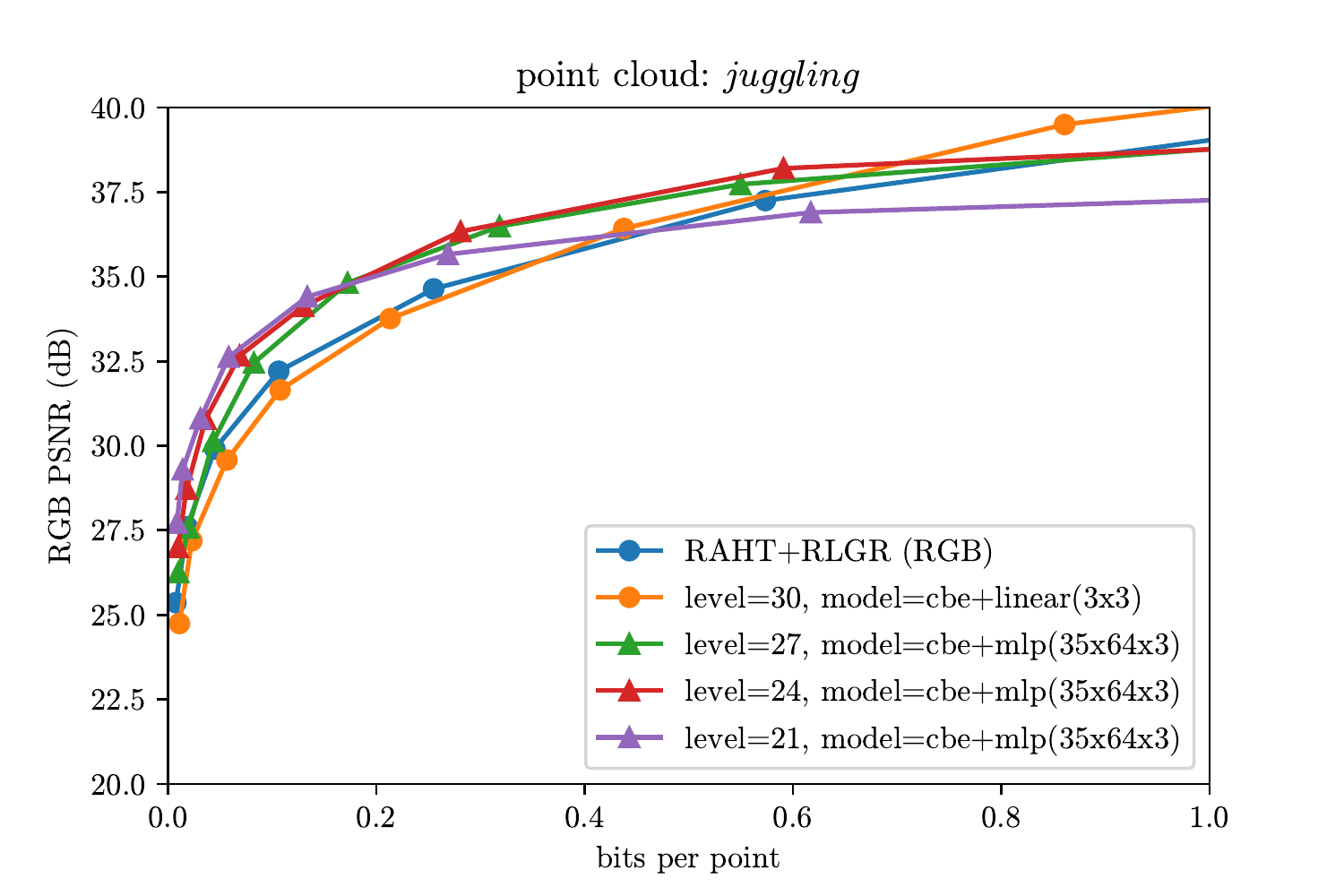}
    \includegraphics[width=0.29\linewidth, trim=20 5 35 15, clip]{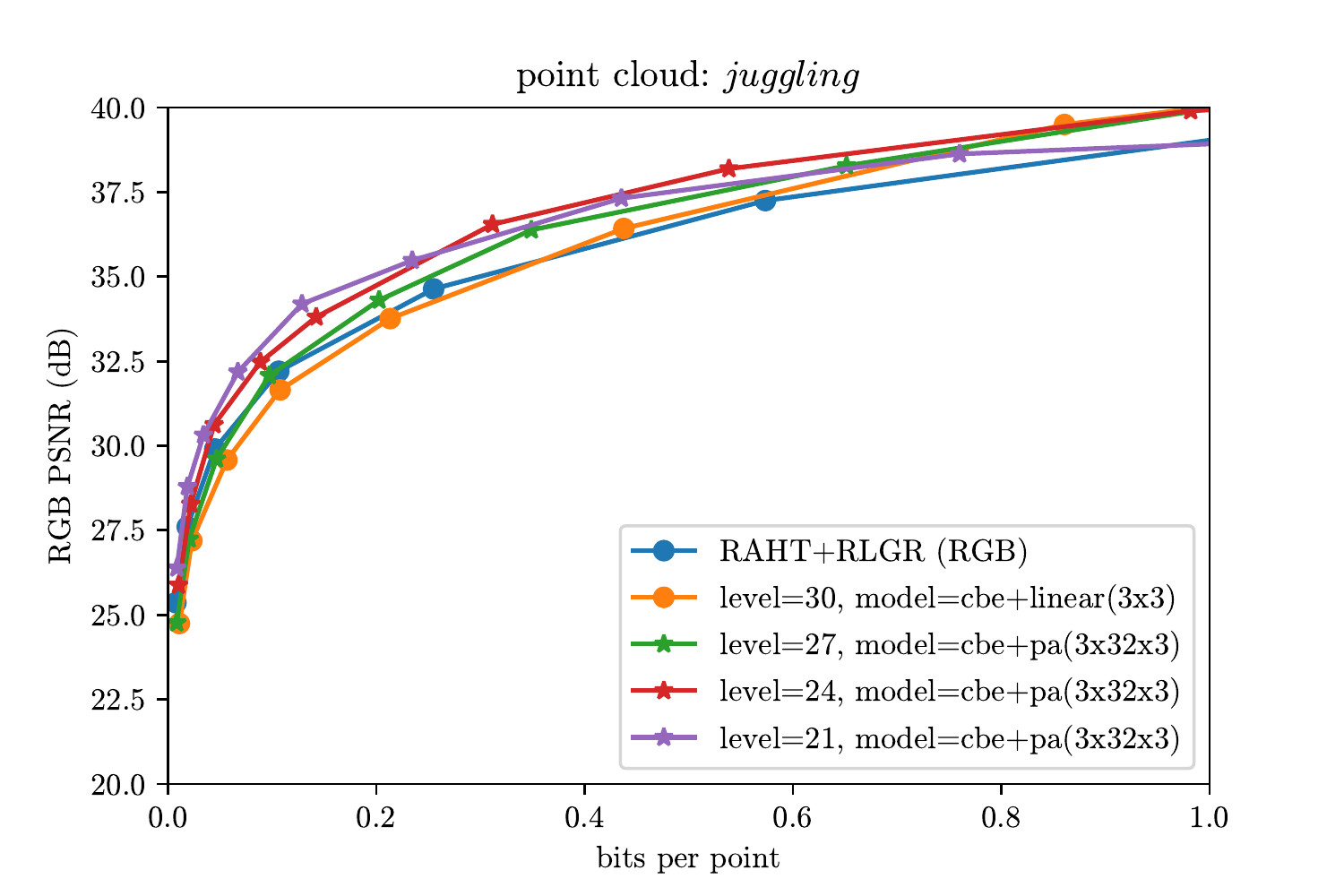}
    
    \includegraphics[width=0.29\linewidth, trim=20 5 35 15, clip]{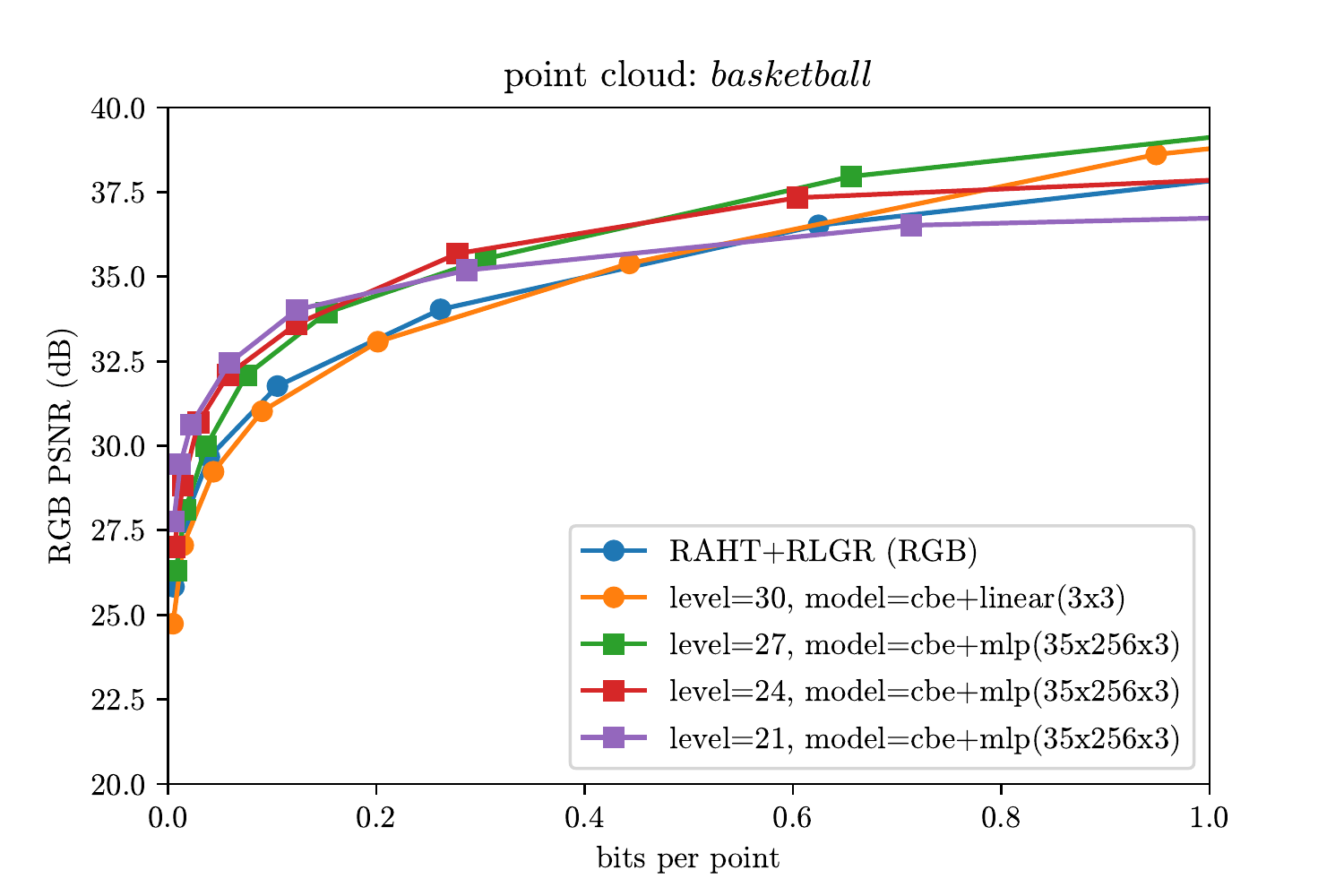}
    \includegraphics[width=0.29\linewidth, trim=20 5 35 15, clip]{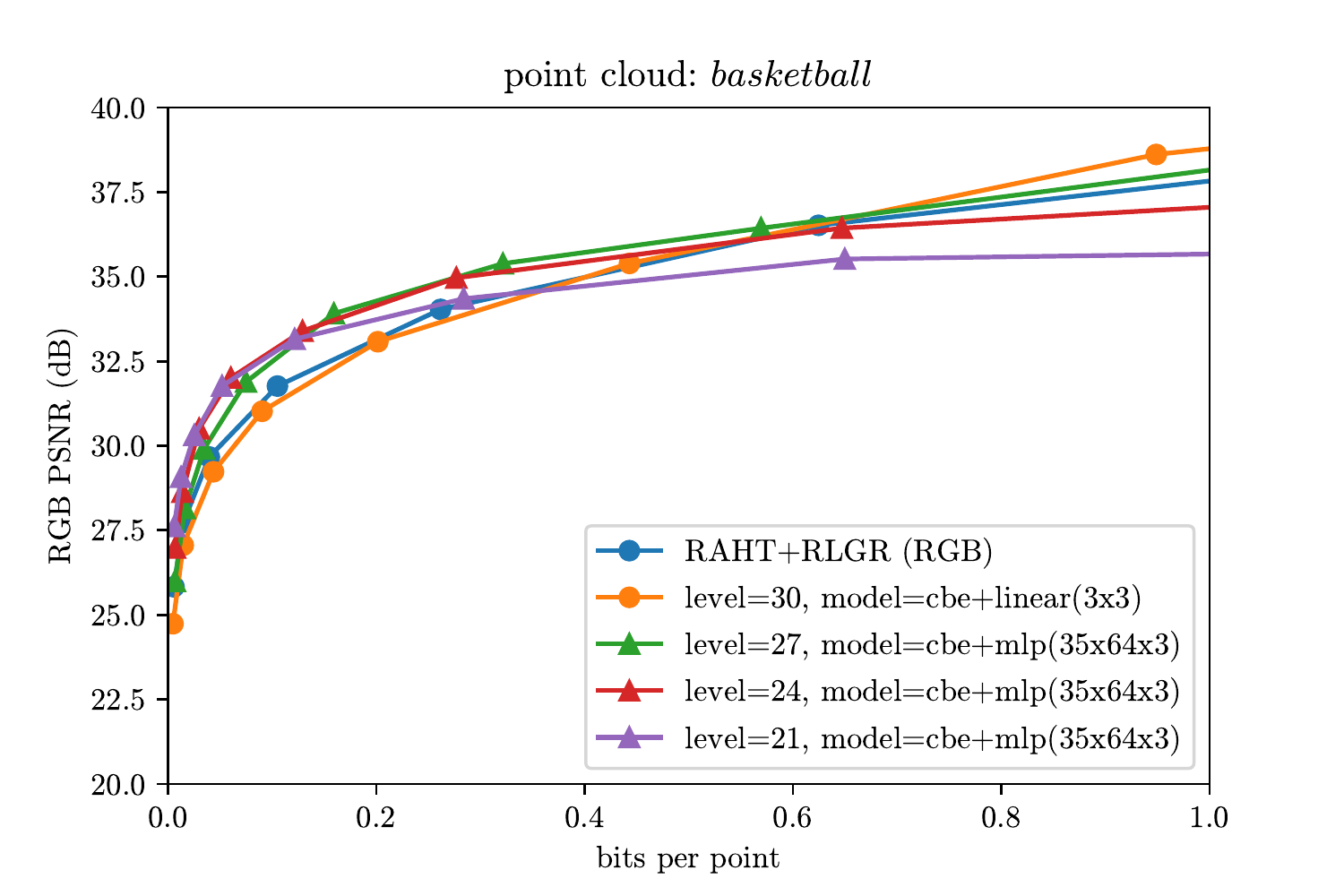}
    \includegraphics[width=0.29\linewidth, trim=20 5 35 15, clip]{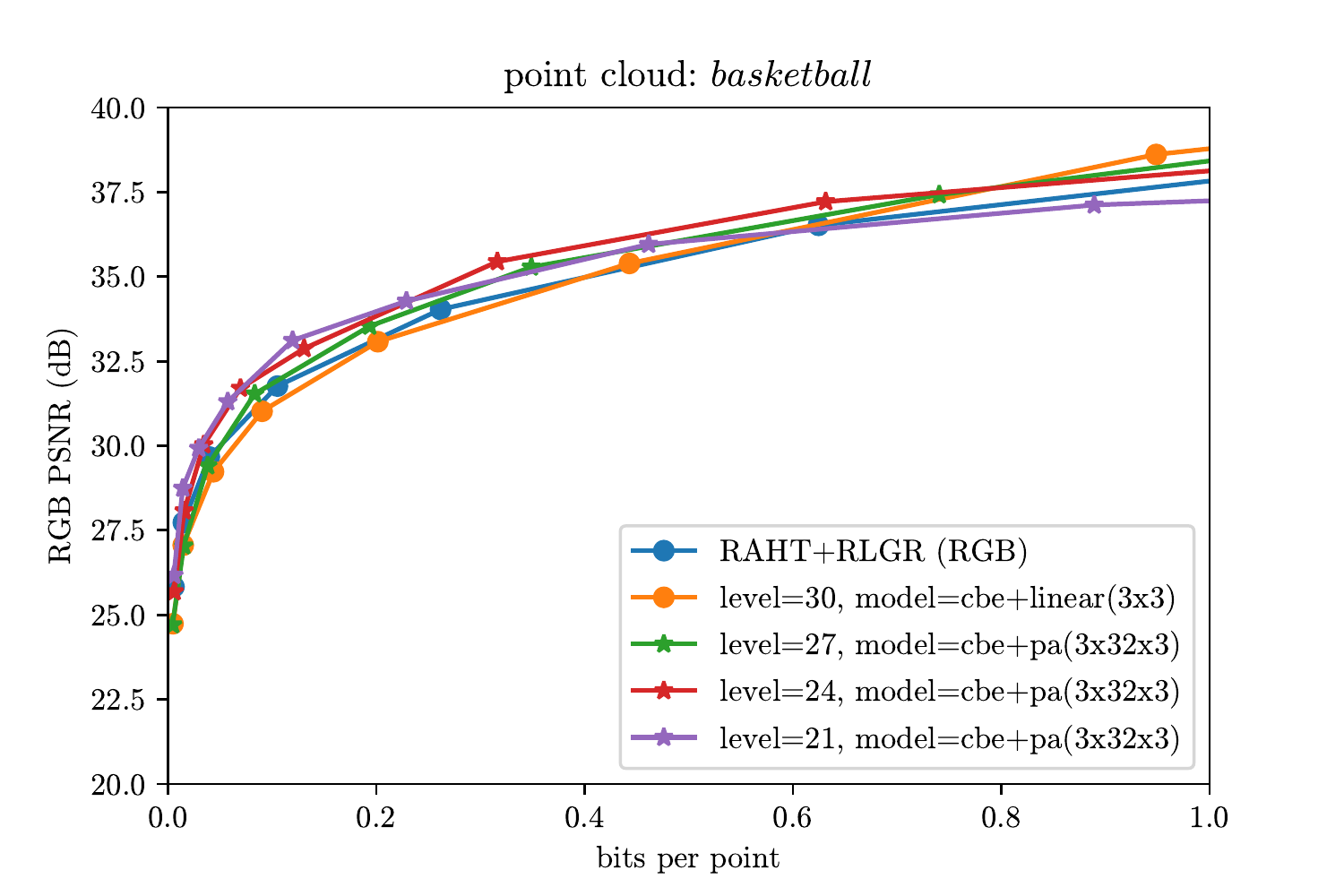}
    
    \includegraphics[width=0.29\linewidth, trim=20 5 35 15, clip]{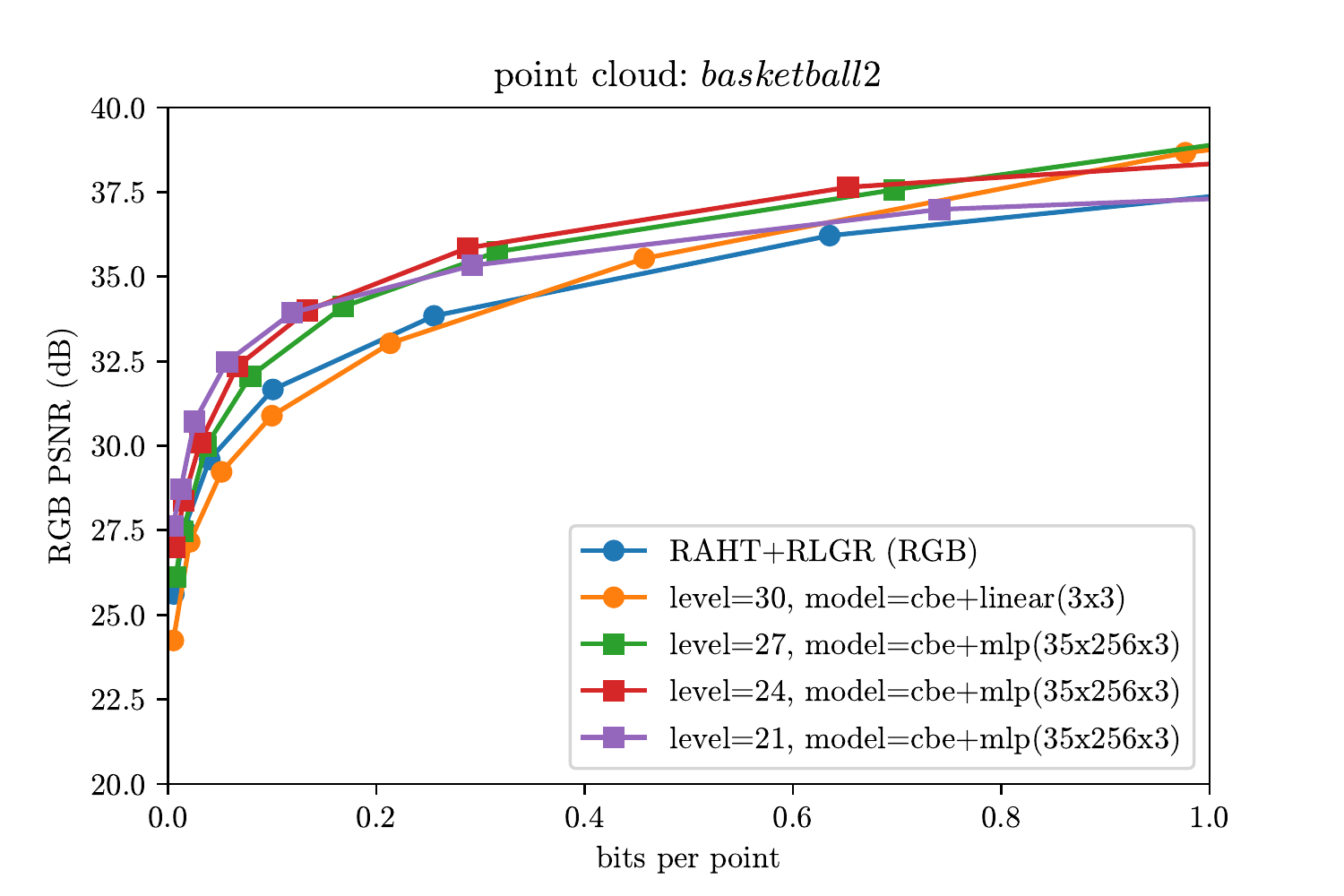}
    \includegraphics[width=0.29\linewidth, trim=20 5 35 15, clip]{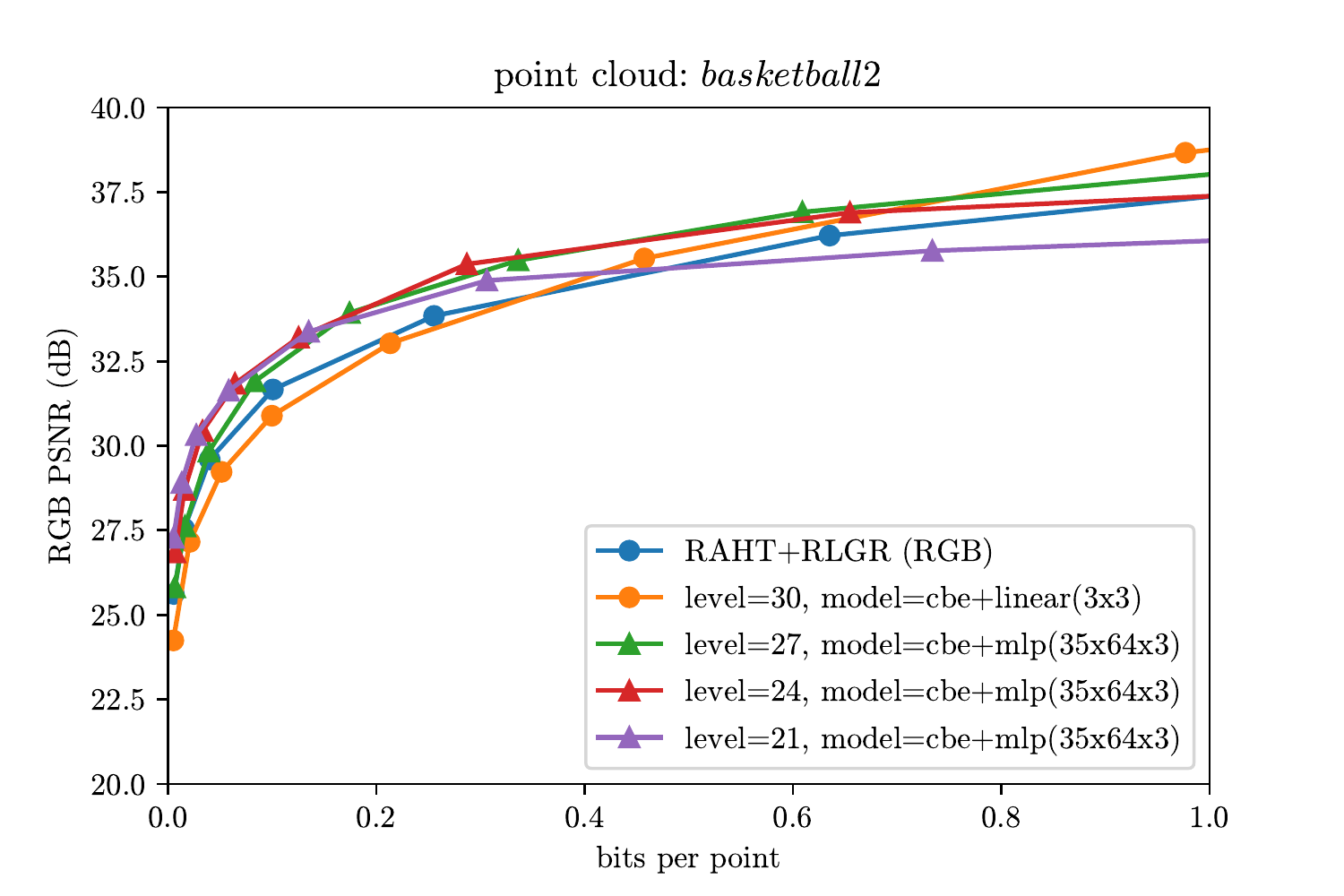}
    \includegraphics[width=0.29\linewidth, trim=20 5 35 15, clip]{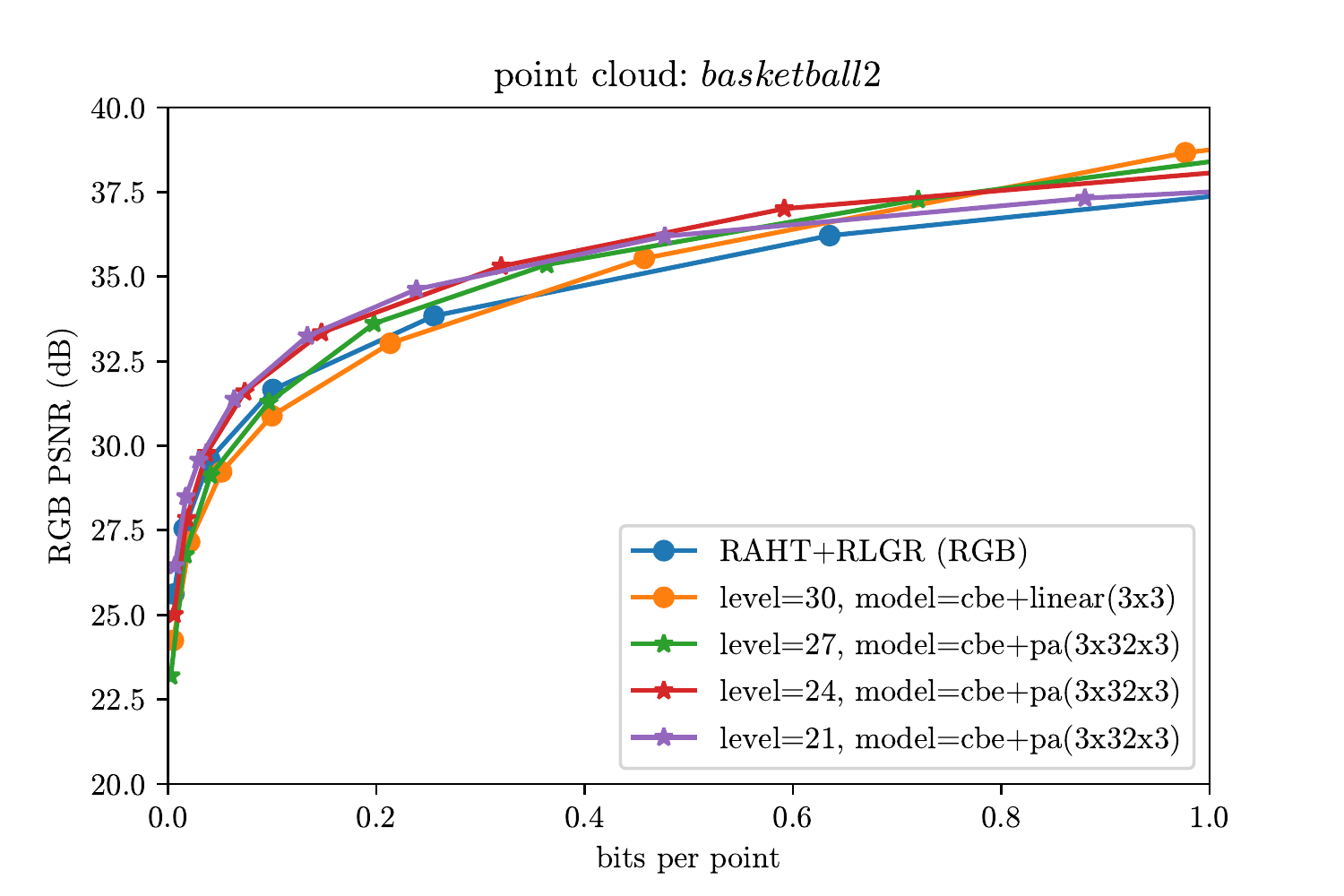}
    
    \includegraphics[width=0.29\linewidth, trim=20 5 35 15, clip]{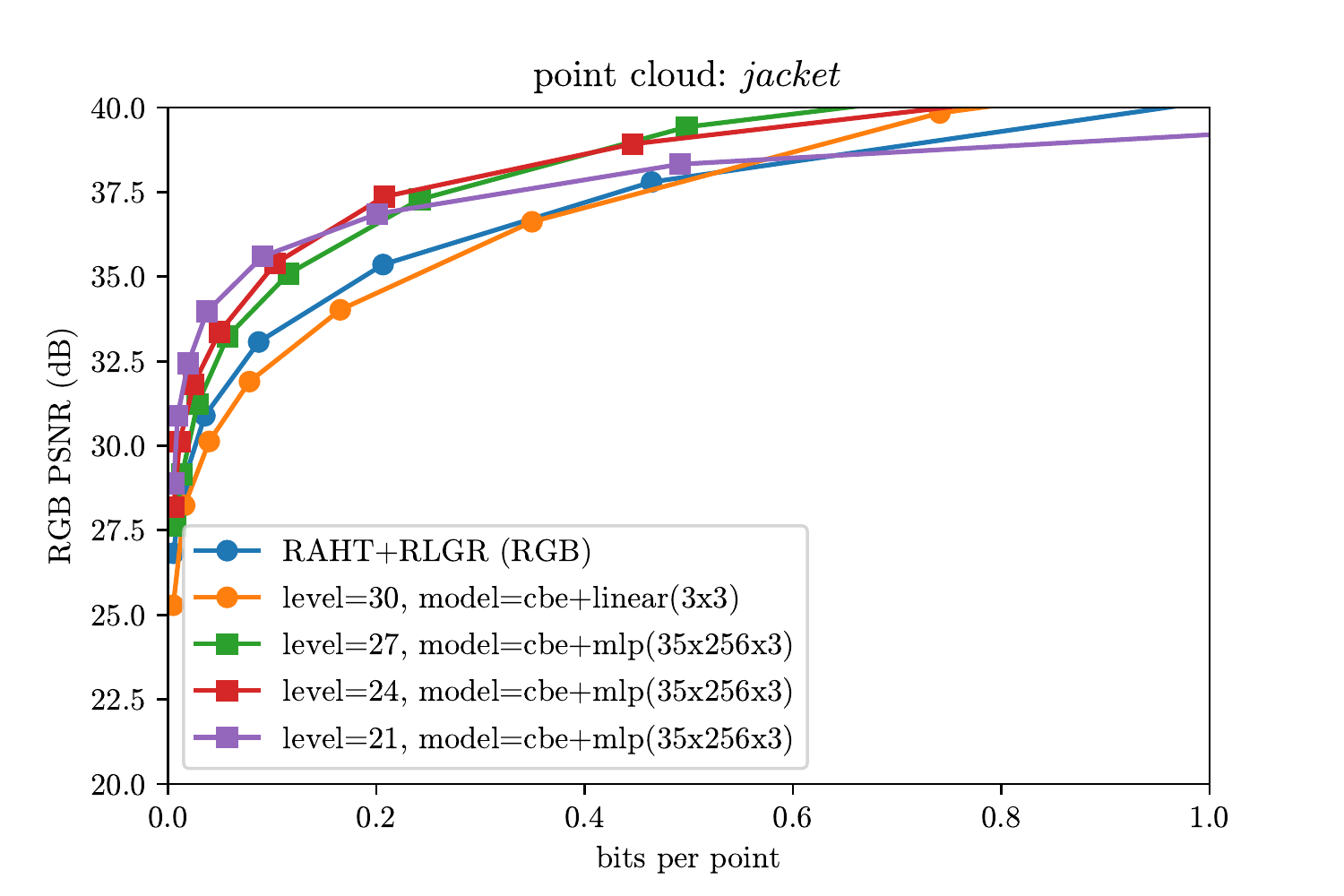}
    \includegraphics[width=0.29\linewidth, trim=20 5 35 15, clip]{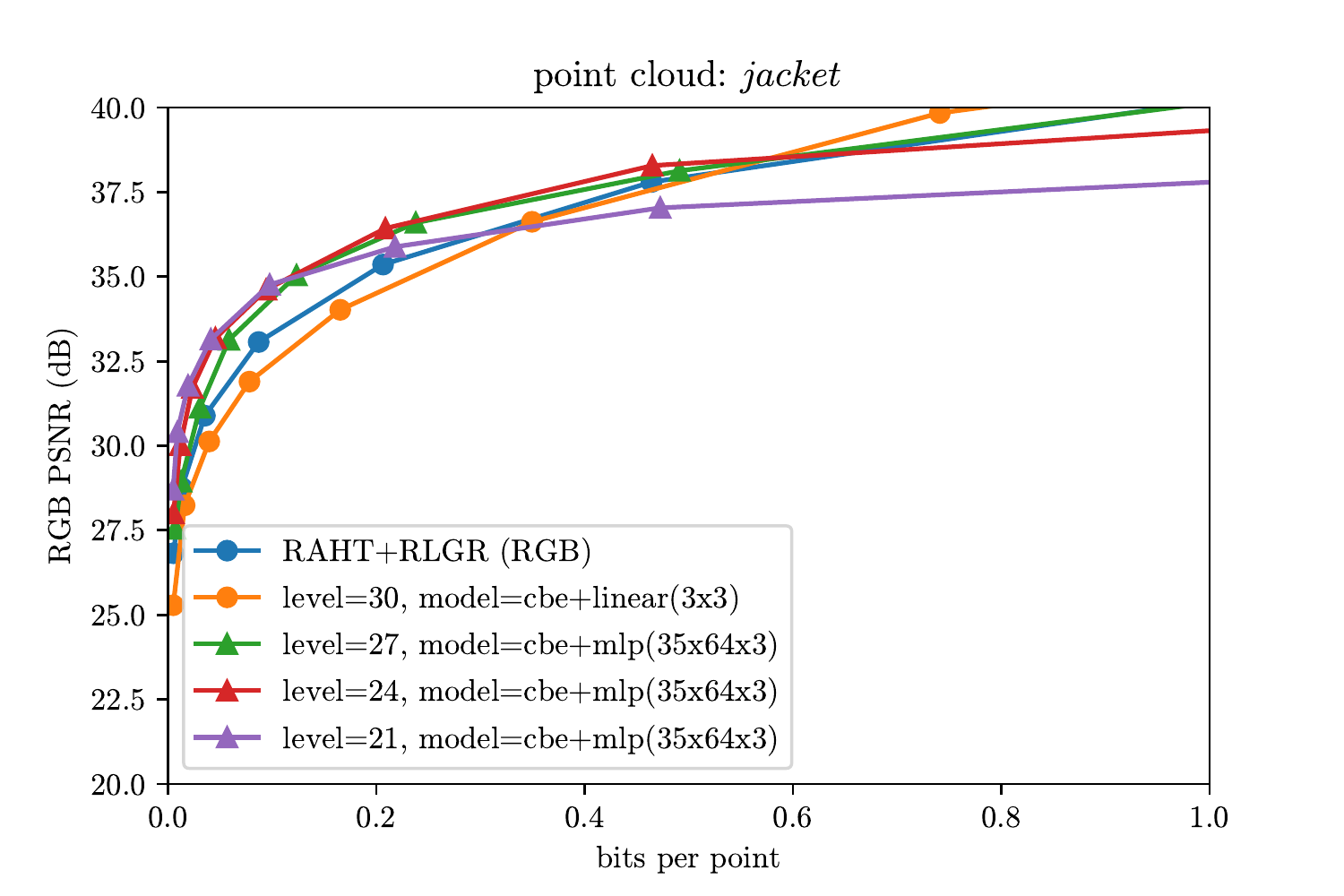}
    \includegraphics[width=0.29\linewidth, trim=20 5 35 15, clip]{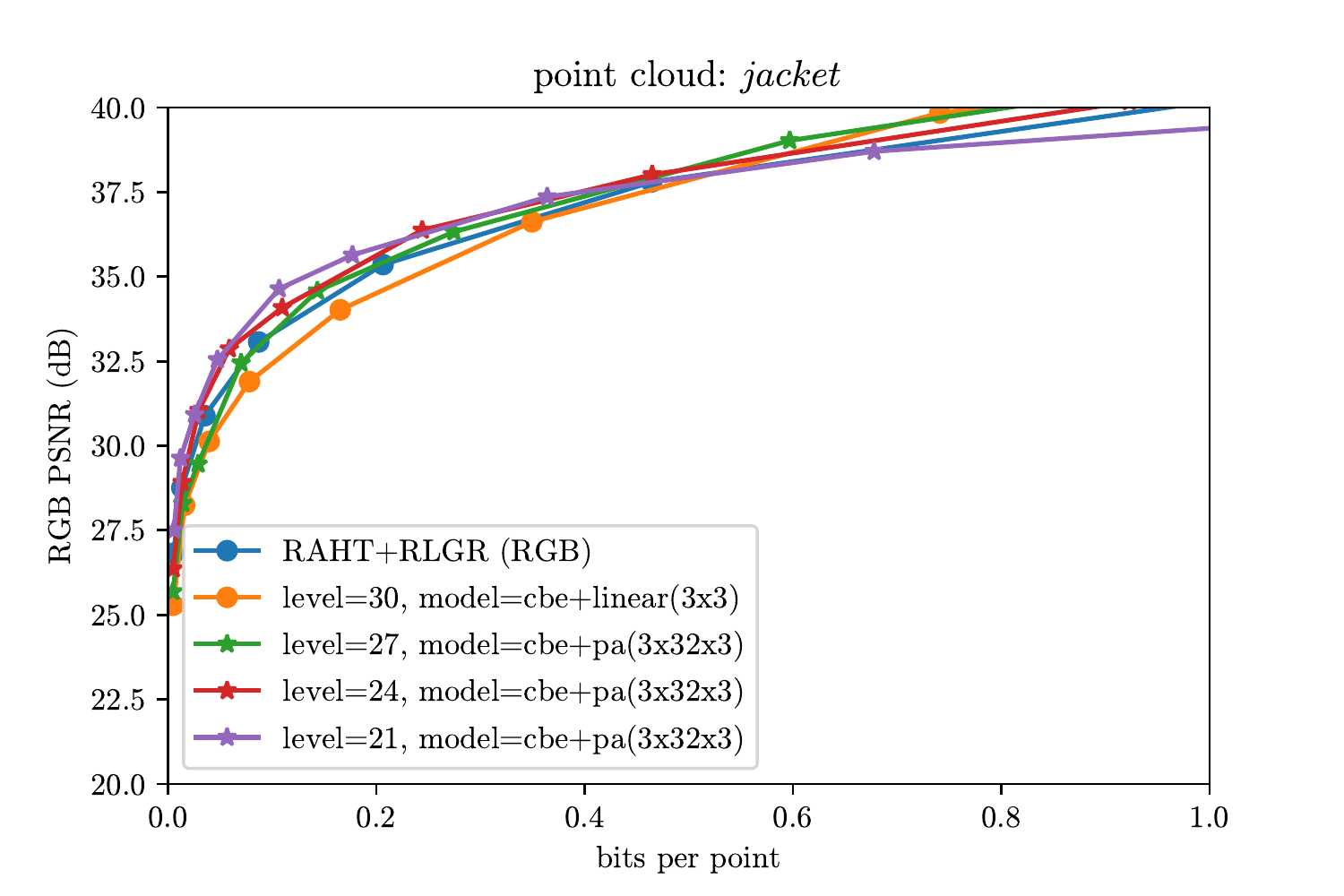}
    
    \caption{Coordinate Based Networks, by network.  Each row is a different point cloud.  Left, middle, right columns each show levels 27, 24, and 21, along with baselines, for CBNs {\em mlp(35x256x3)}, {\em mlp(35x64x3)}, and {\em pa(3x32x3)}.  See \cref{fig:cbns_by_network} for point cloud {\em rock}.}
    \label{fig:cbns_by_network_supp}
\end{figure*}

\begin{figure*}
    \centering
    \includegraphics[width=0.29\linewidth, trim=20 5 35 15, clip]{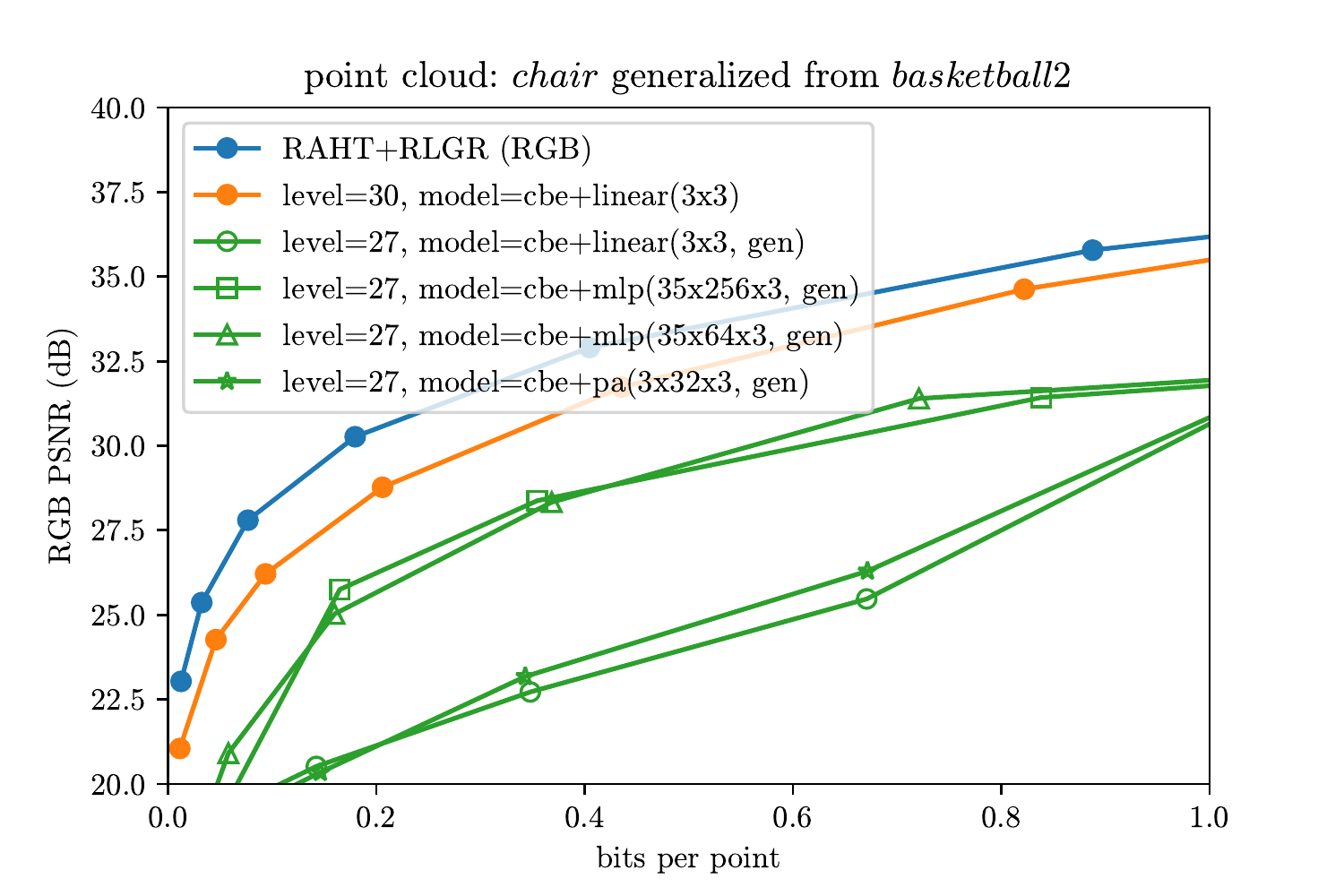}
    \includegraphics[width=0.29\linewidth, trim=20 5 35 15, clip]{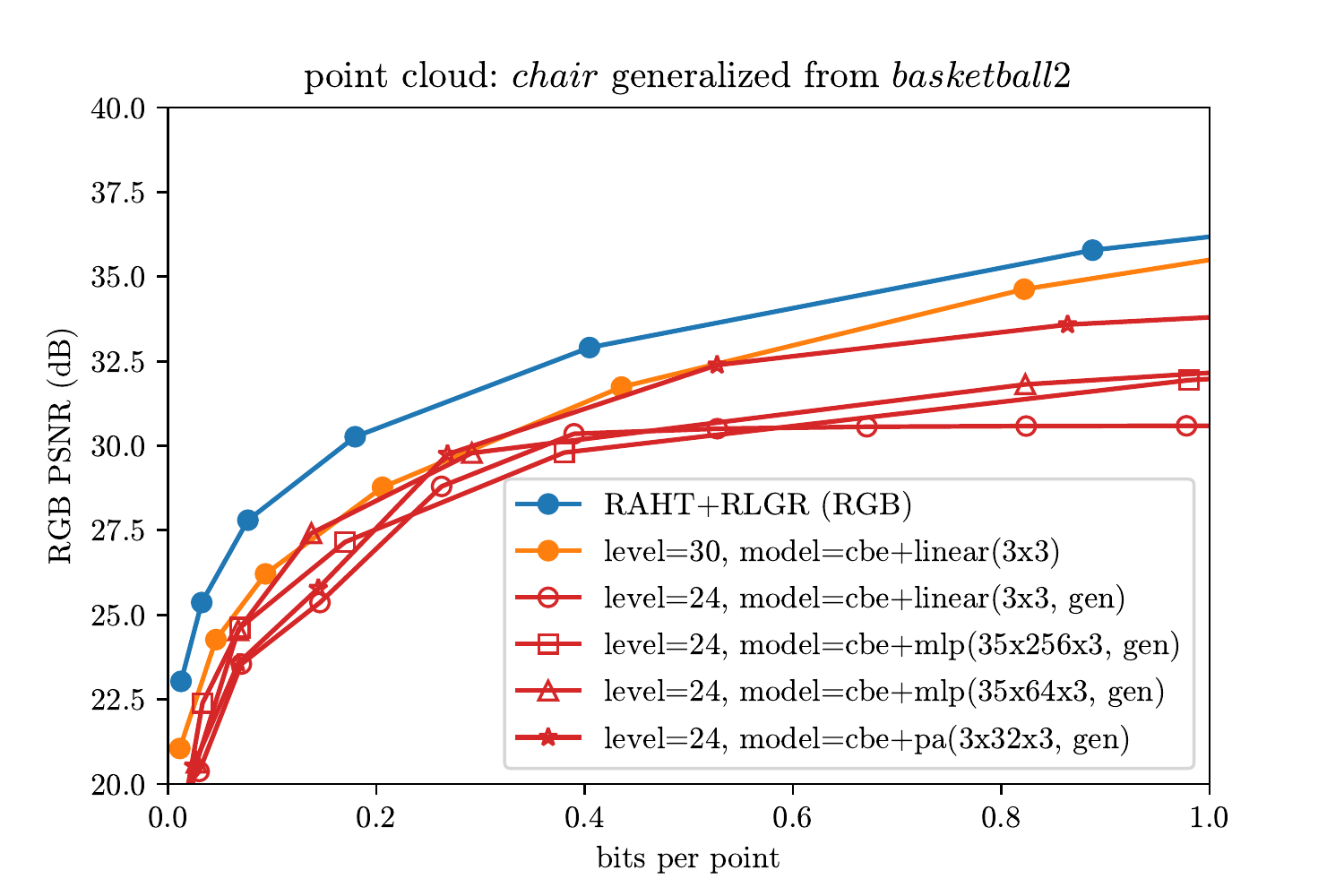}
    \includegraphics[width=0.29\linewidth, trim=20 5 35 15, clip]{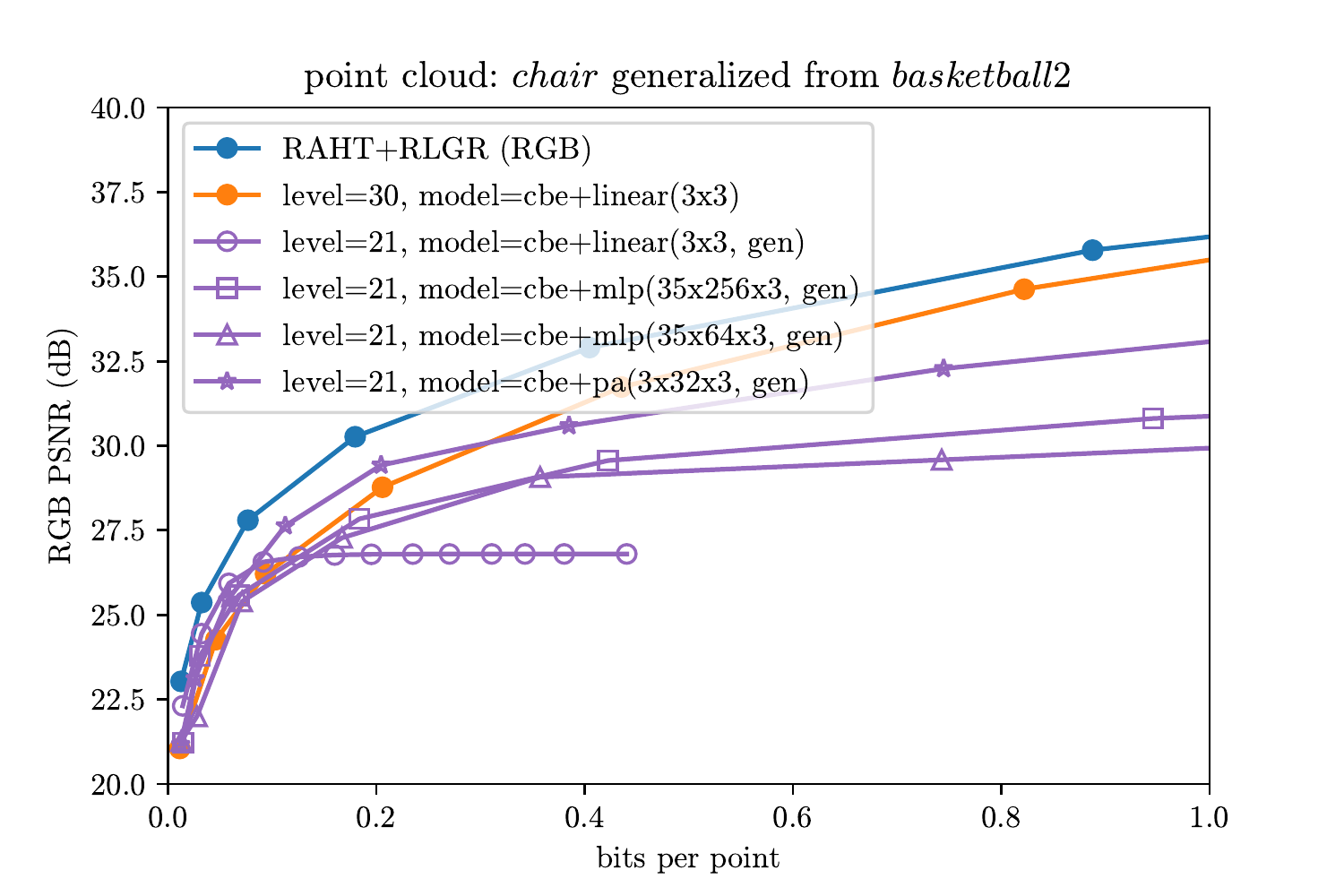}
    
    \includegraphics[width=0.29\linewidth, trim=20 5 35 15, clip]{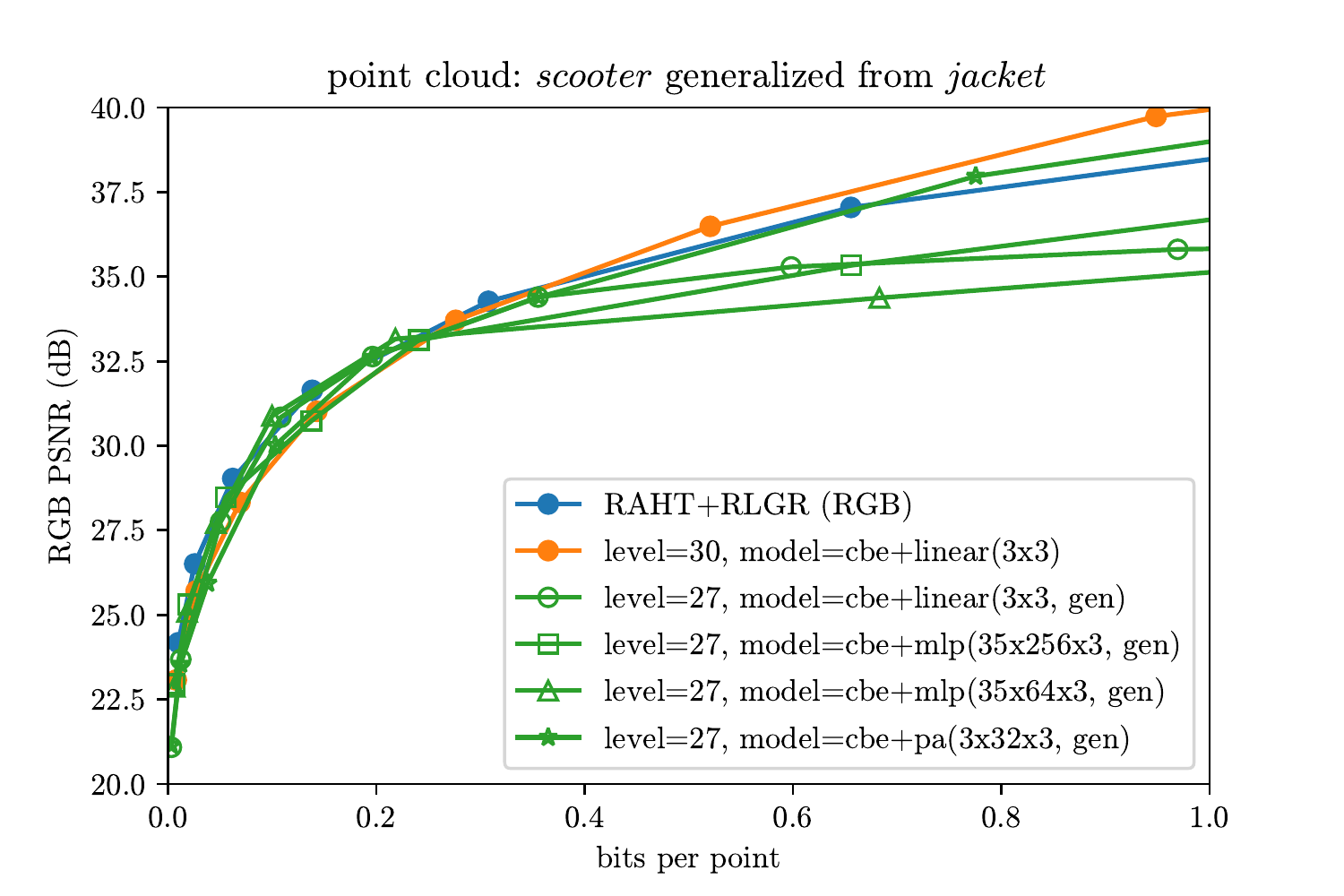}
    \includegraphics[width=0.29\linewidth, trim=20 5 35 15, clip]{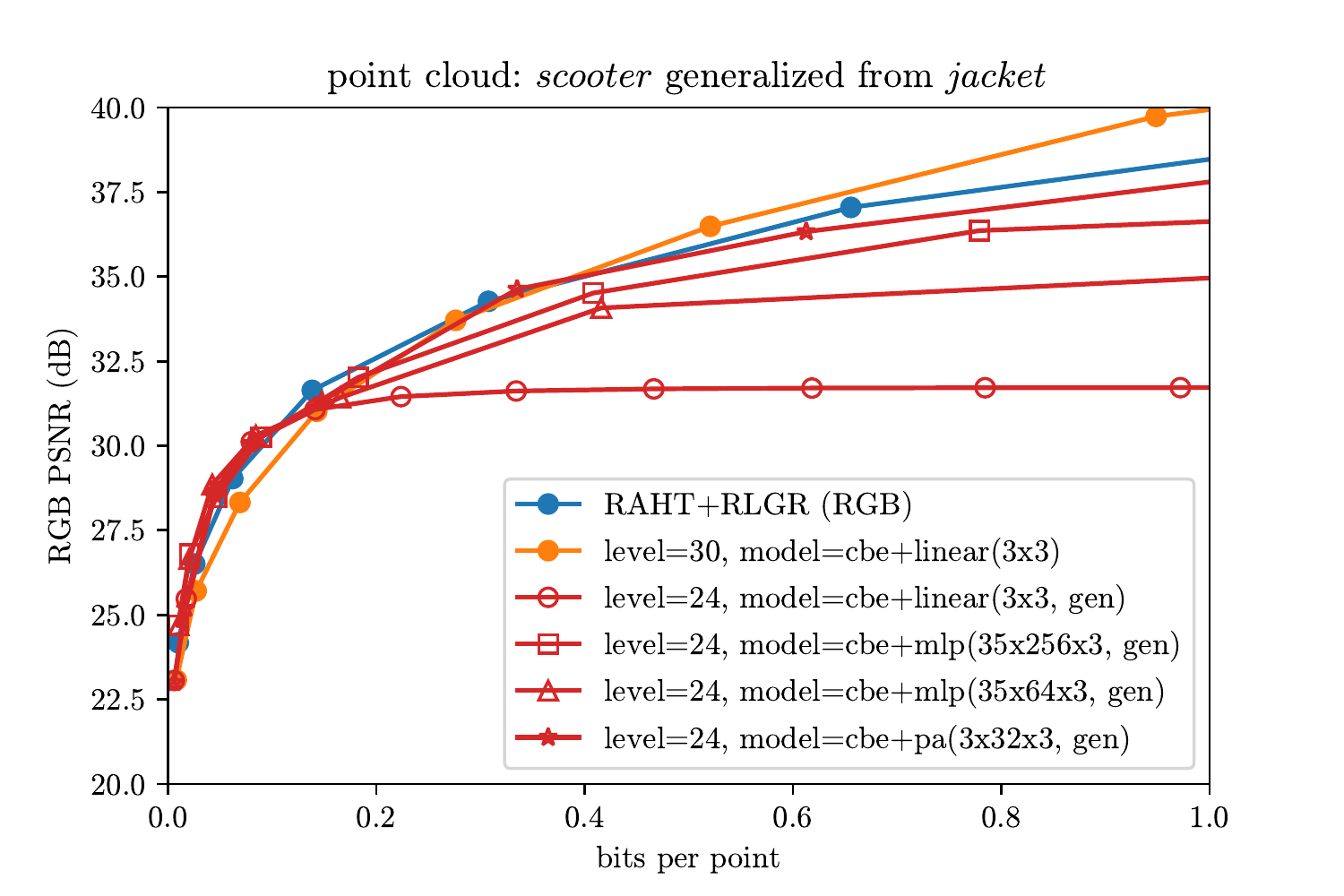}
    \includegraphics[width=0.29\linewidth, trim=20 5 35 15, clip]{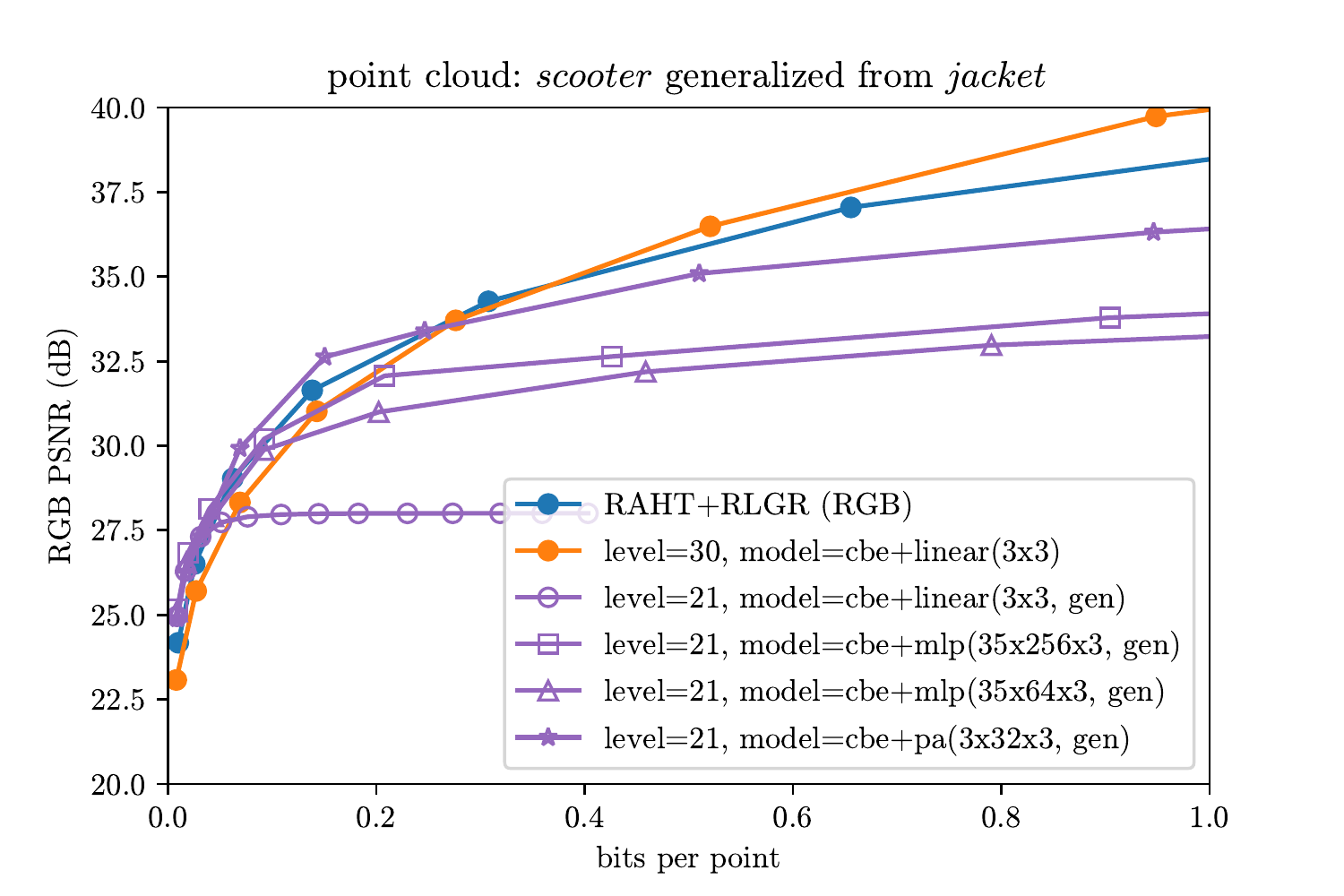}
    
    \includegraphics[width=0.29\linewidth, trim=20 5 35 15, clip]{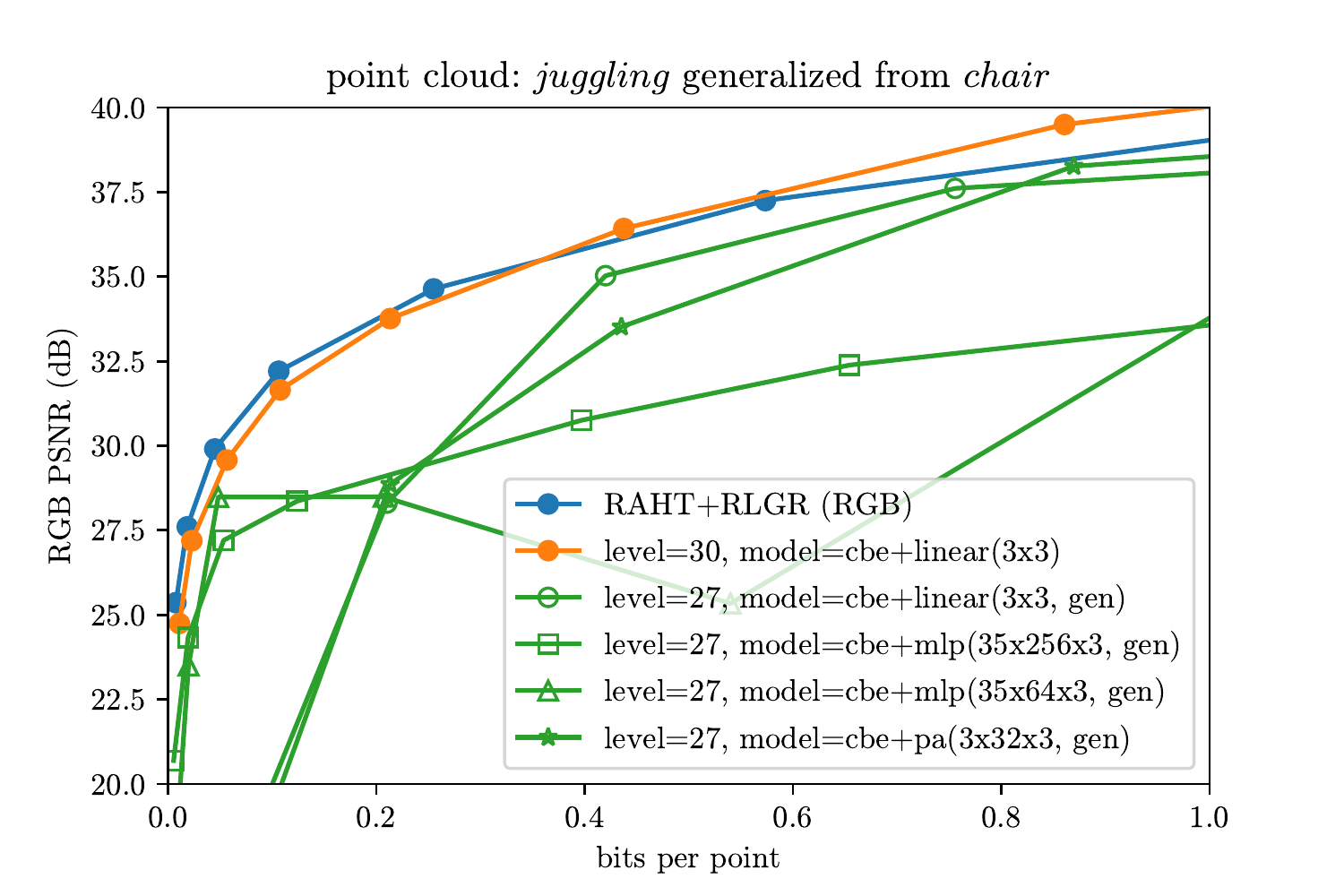}
    \includegraphics[width=0.29\linewidth, trim=20 5 35 15, clip]{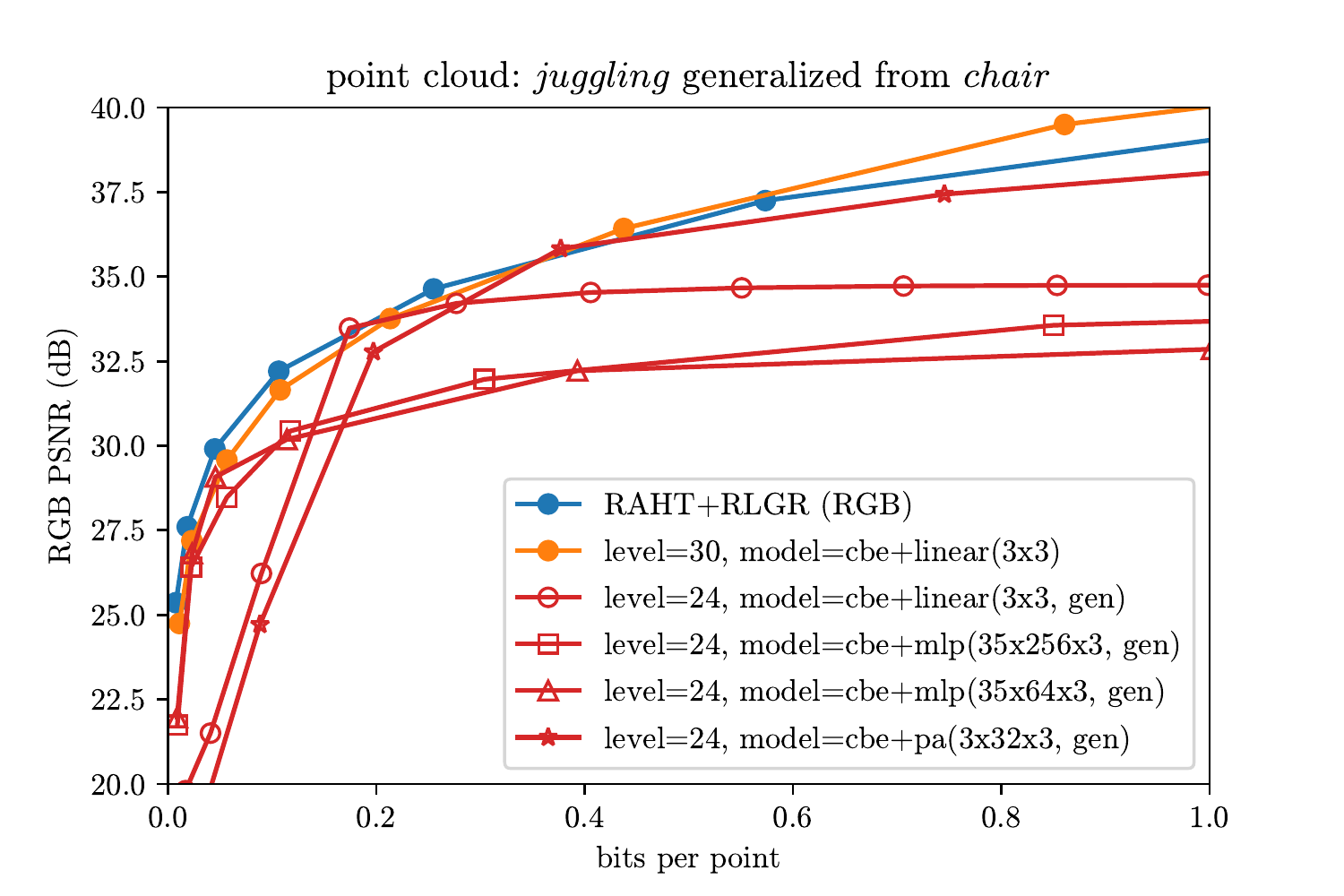}
    \includegraphics[width=0.29\linewidth, trim=20 5 35 15, clip]{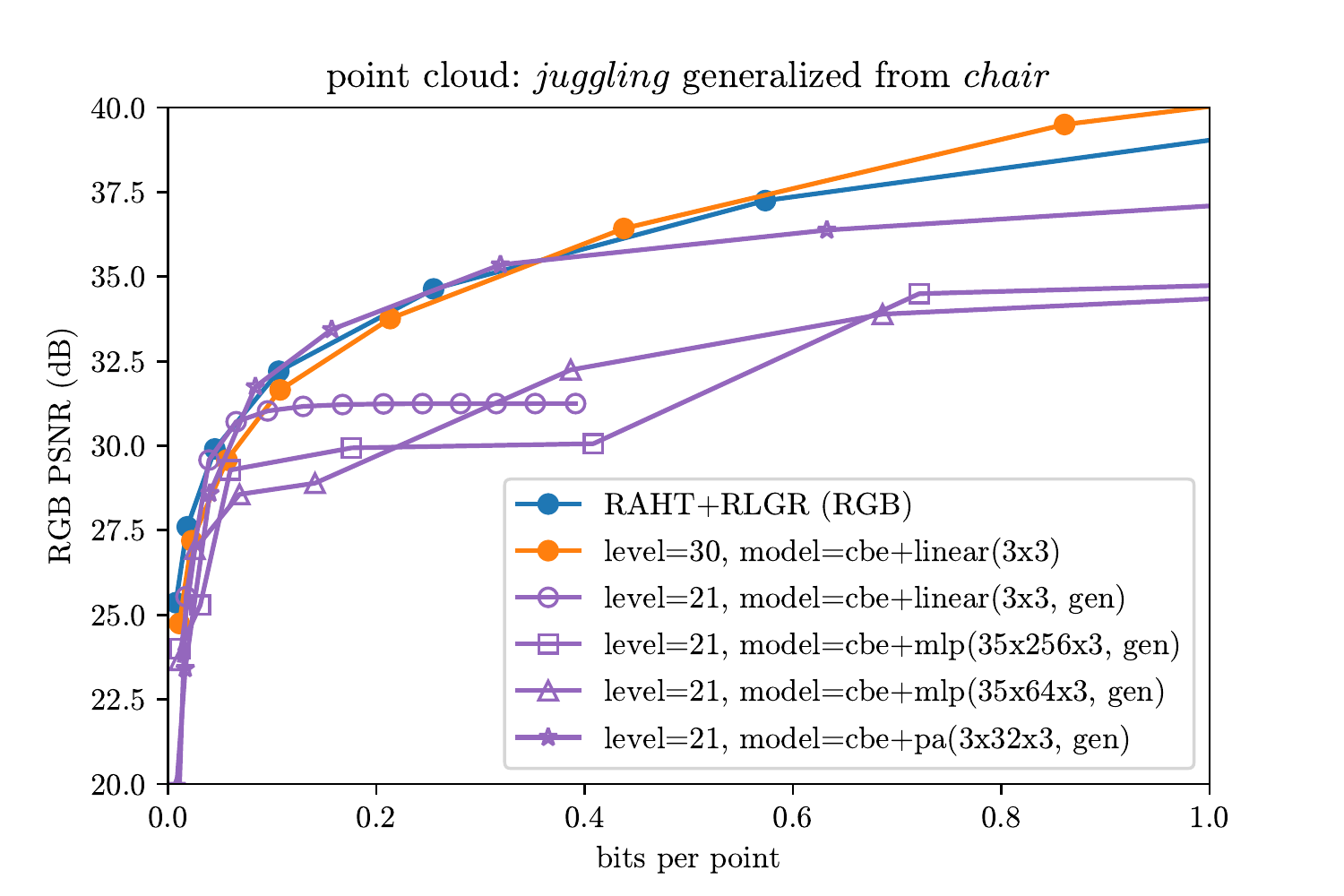}
    
    \includegraphics[width=0.29\linewidth, trim=20 5 35 15, clip]{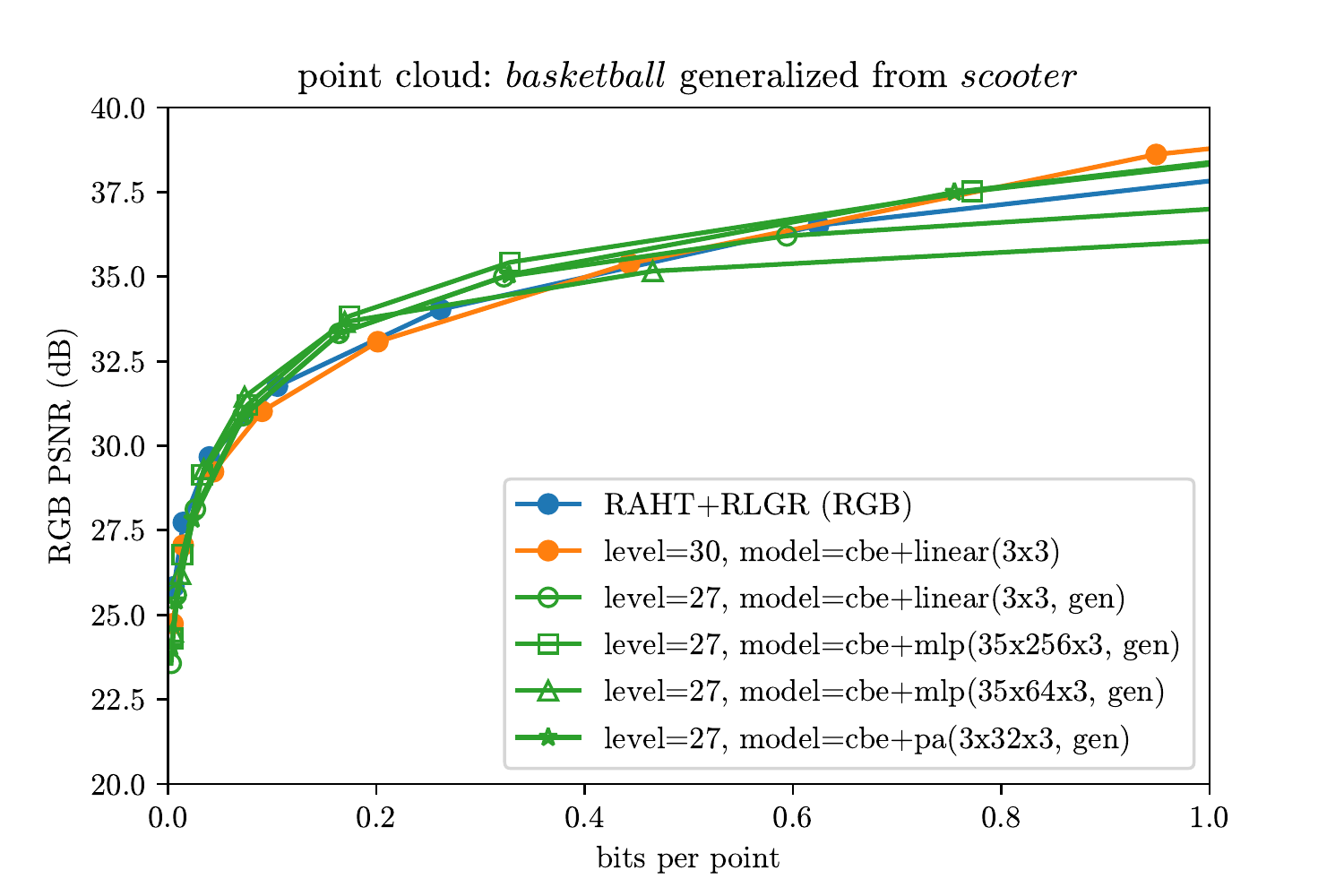}
    \includegraphics[width=0.29\linewidth, trim=20 5 35 15, clip]{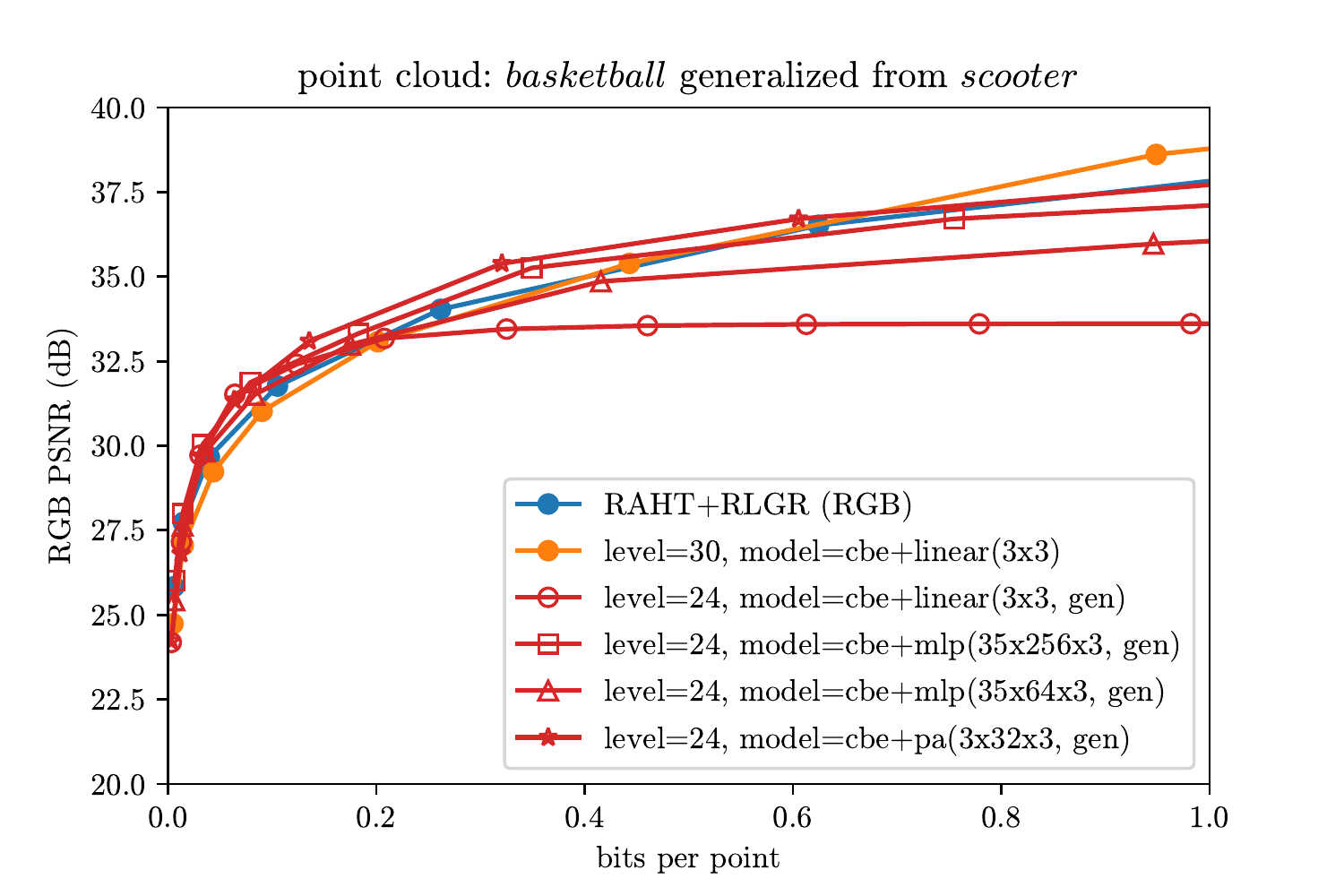}
    \includegraphics[width=0.29\linewidth, trim=20 5 35 15, clip]{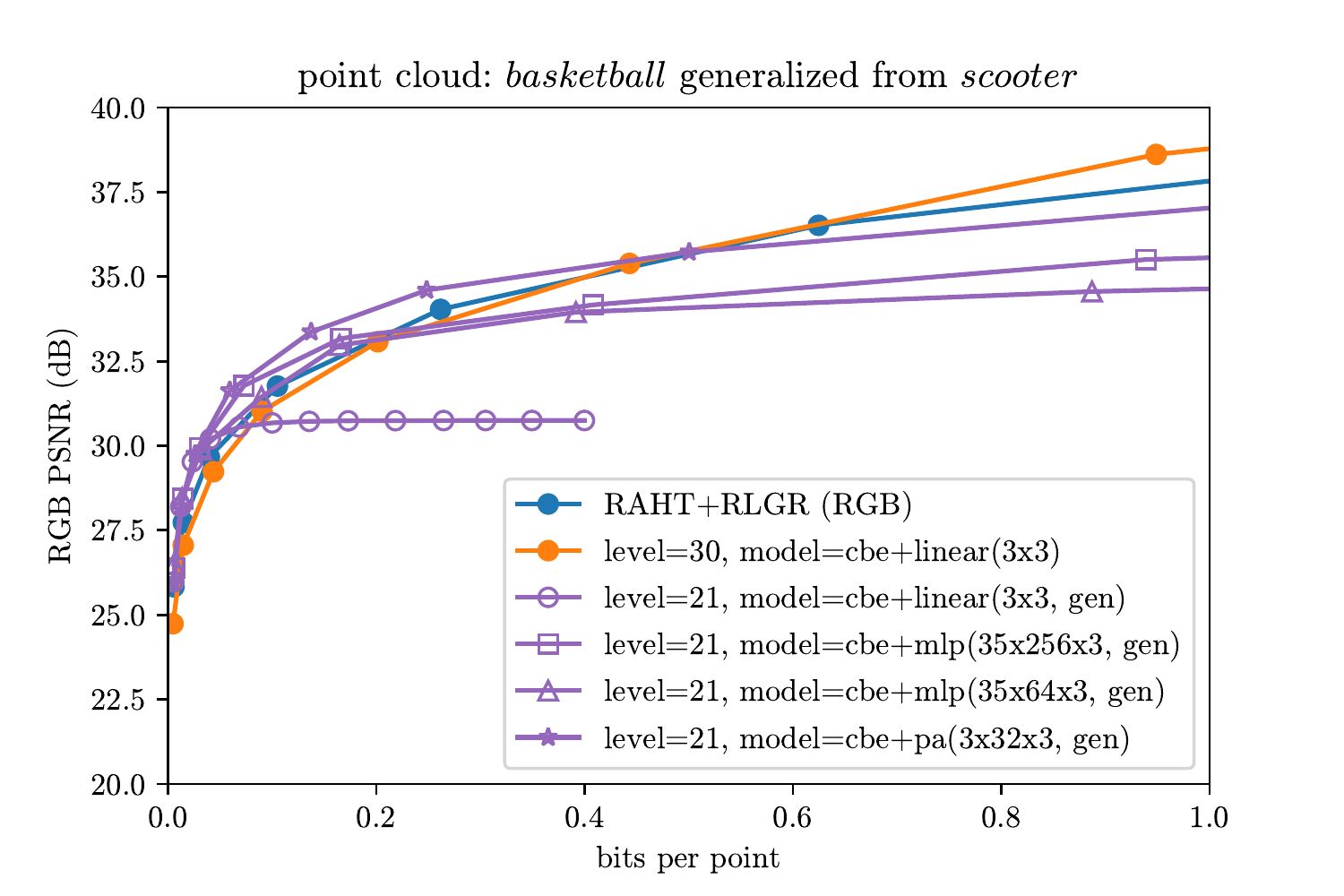}
    
    \includegraphics[width=0.29\linewidth, trim=20 5 35 15, clip]{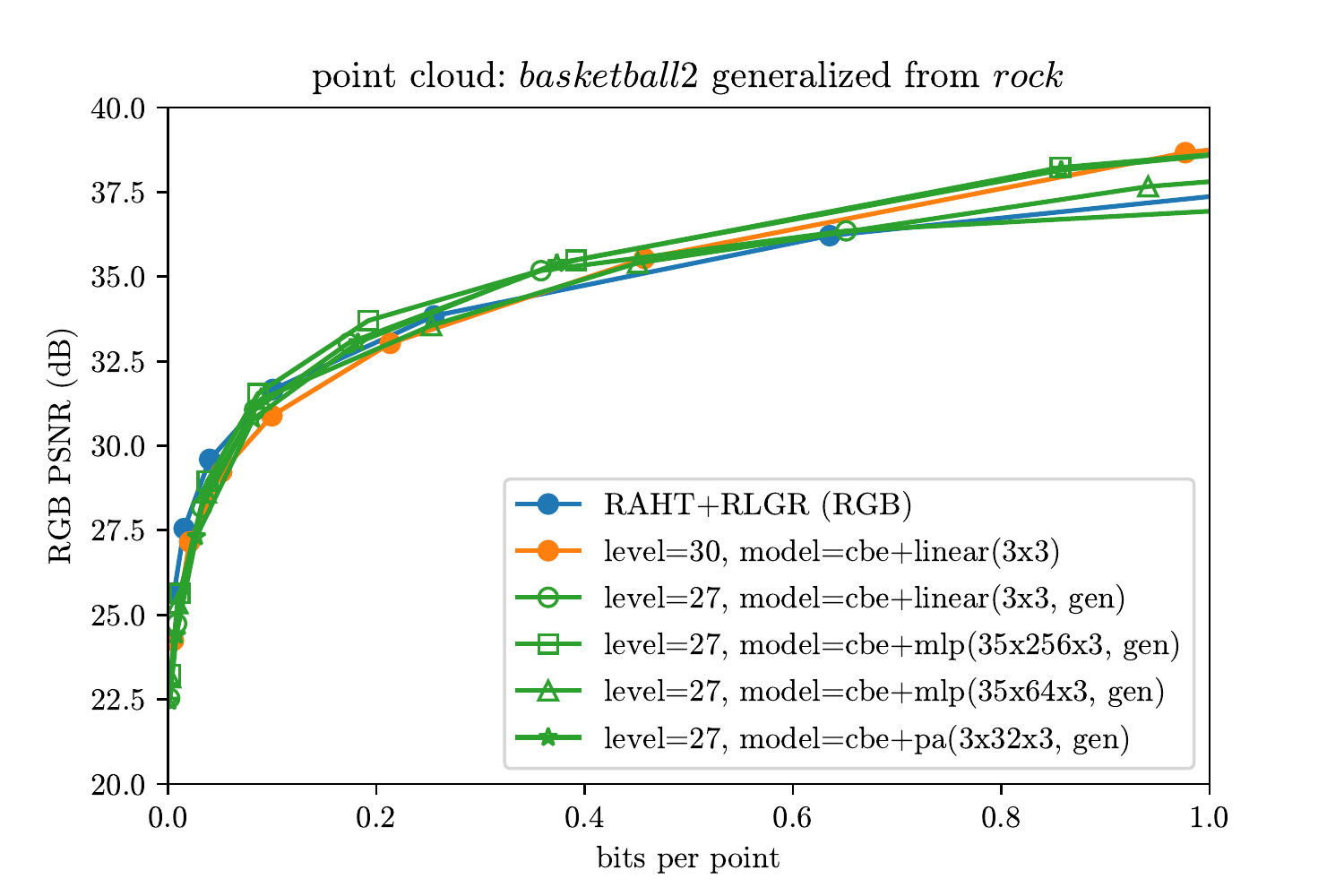}
    \includegraphics[width=0.29\linewidth, trim=20 5 35 15, clip]{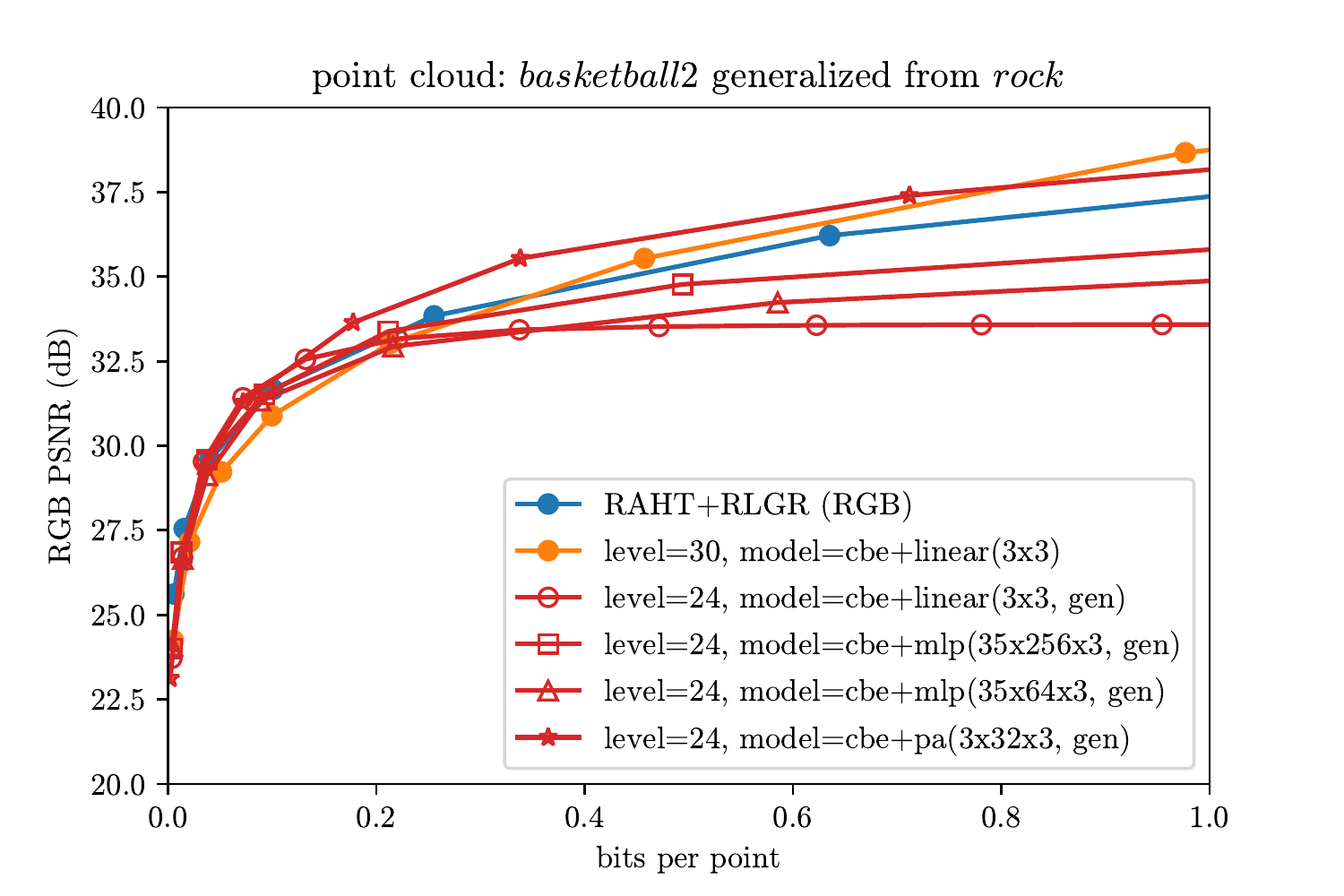}
    \includegraphics[width=0.29\linewidth, trim=20 5 35 15, clip]{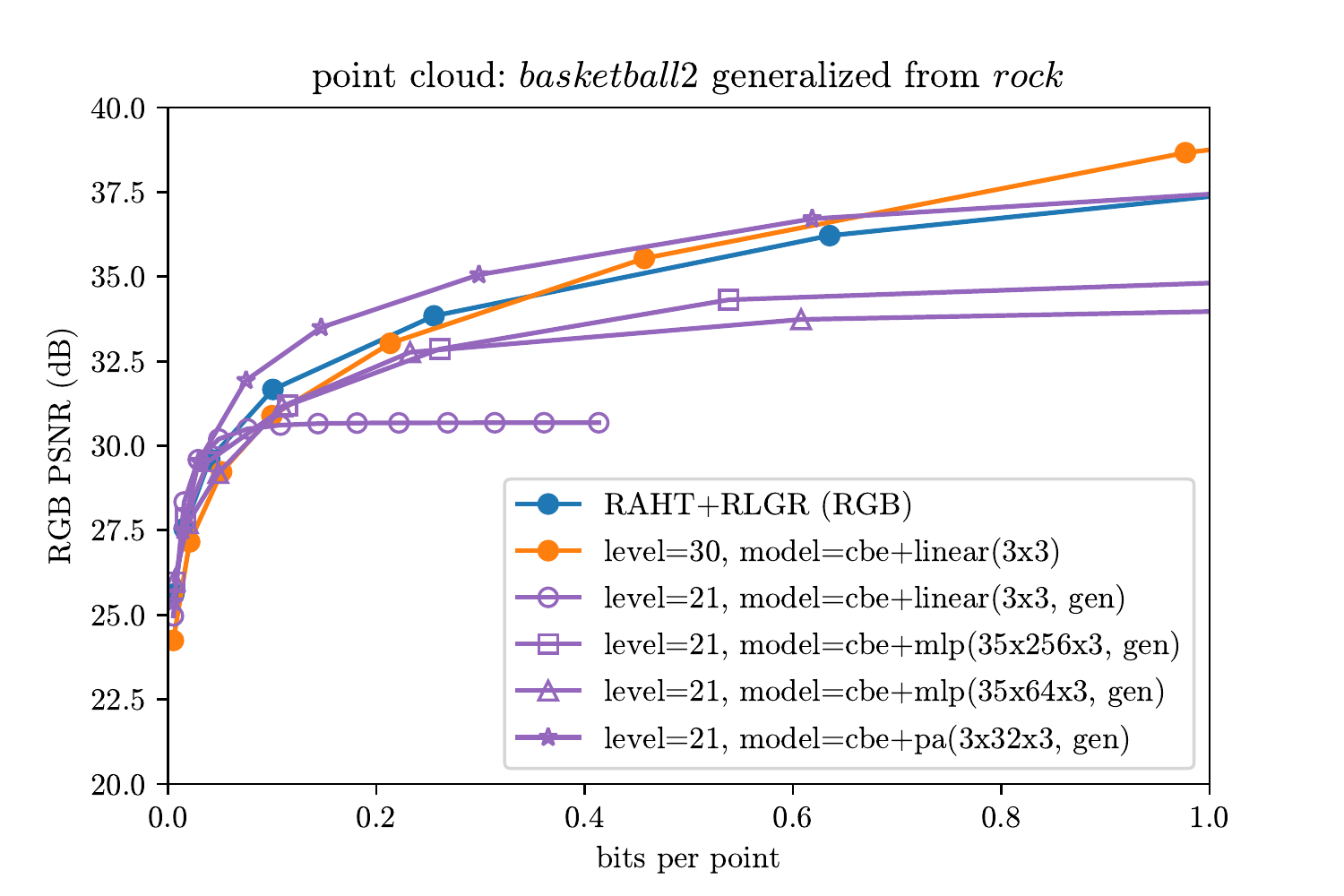}
    
    \includegraphics[width=0.29\linewidth, trim=20 5 35 15, clip]{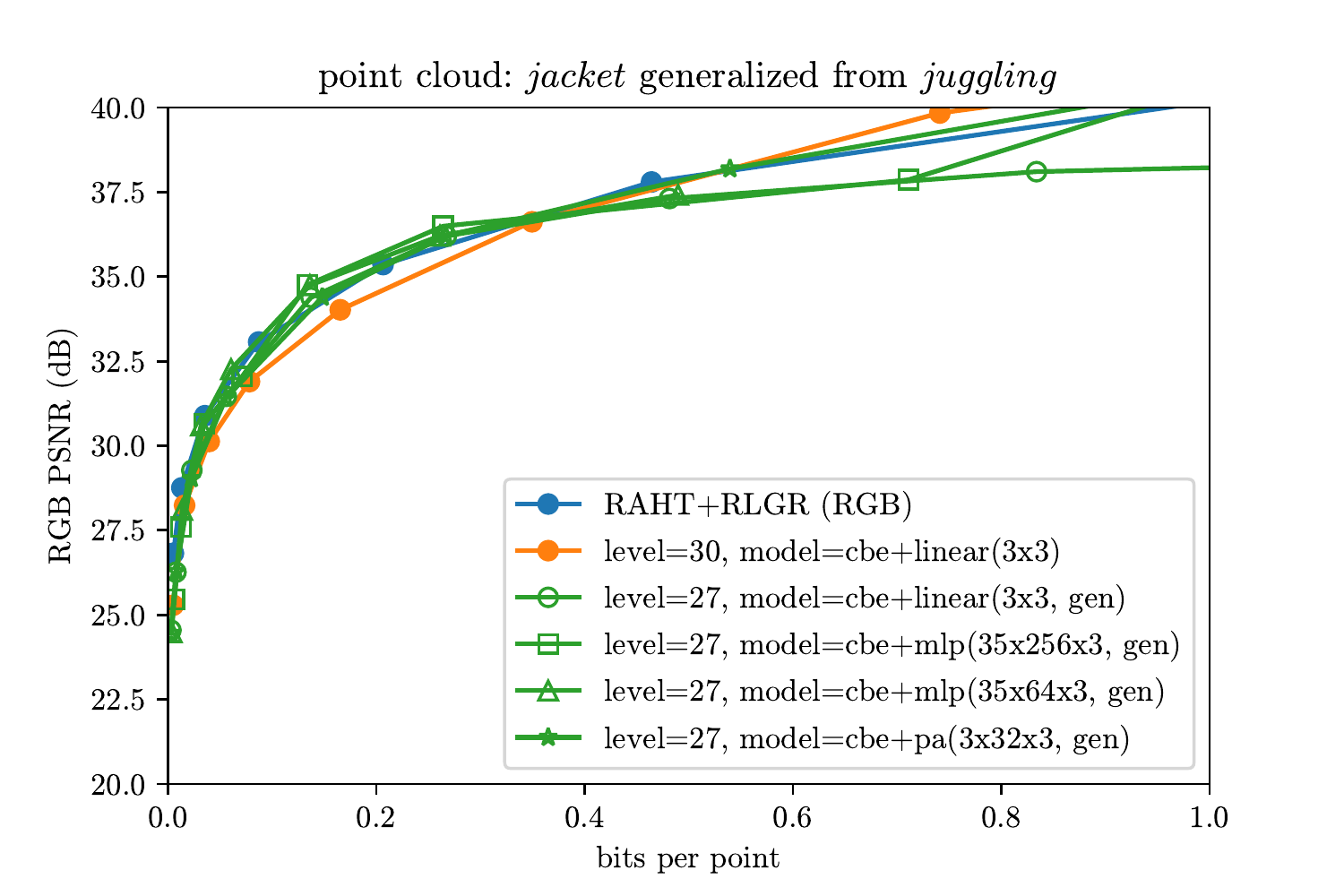}
    \includegraphics[width=0.29\linewidth, trim=20 5 35 15, clip]{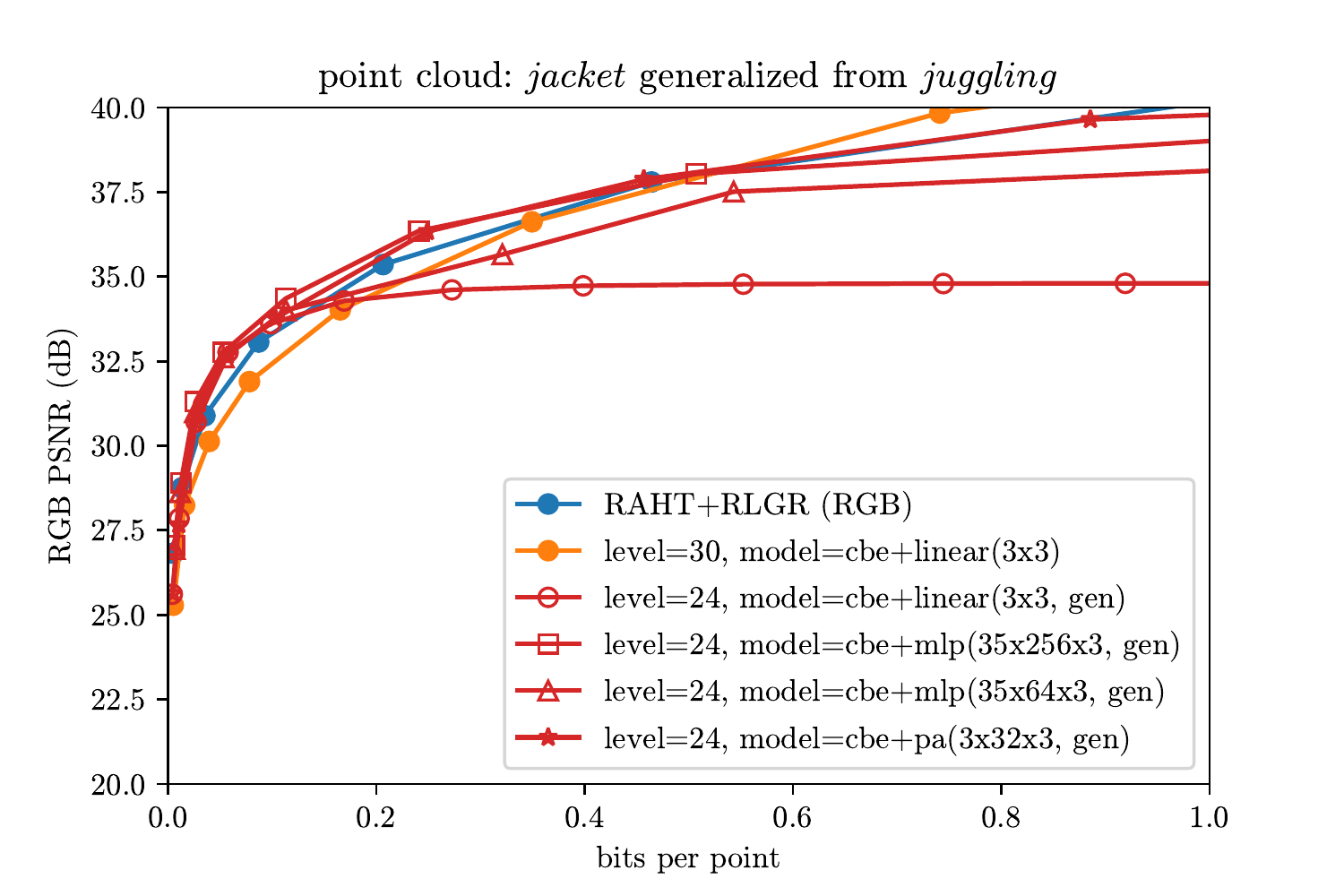}
    \includegraphics[width=0.29\linewidth, trim=20 5 35 15, clip]{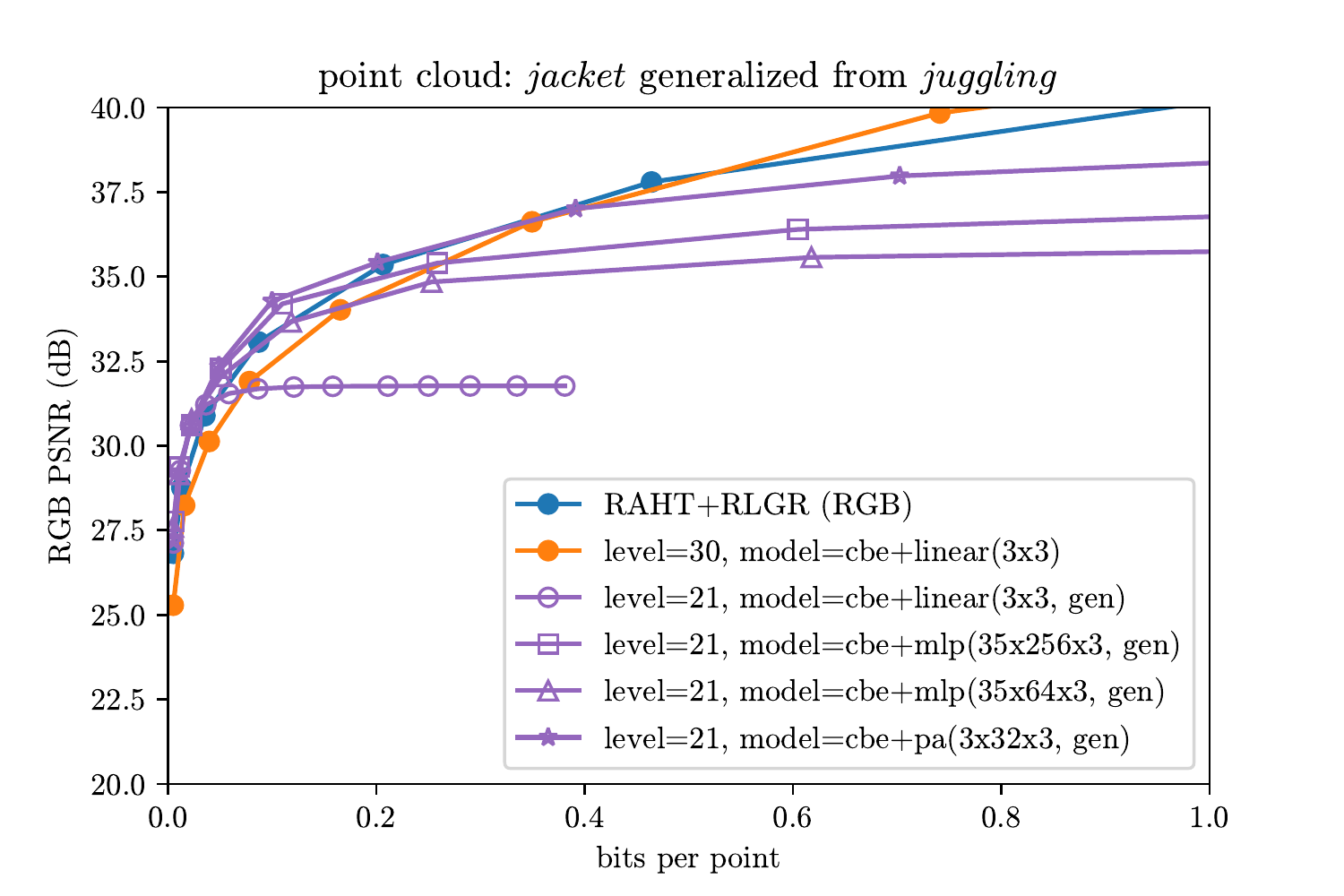}
    \caption{Coordinate Based Networks with generalization, by target level.  Each row is a different point cloud.  Left, middle, right columns each show {\em mlp(35x256x3)}, {\em mlp(35x64x3)}, and {\em pa(3x32x3)} CBNs, along with baselines, at levels 27, 24, 21.  See \cref{fig:cbns_gen} (top) for point cloud {\em rock}.}
    \label{fig:cbns_by_level_gen_supp}
\end{figure*}

\begin{figure*}
    \centering
    \includegraphics[width=0.29\linewidth, trim=20 5 35 15, clip]{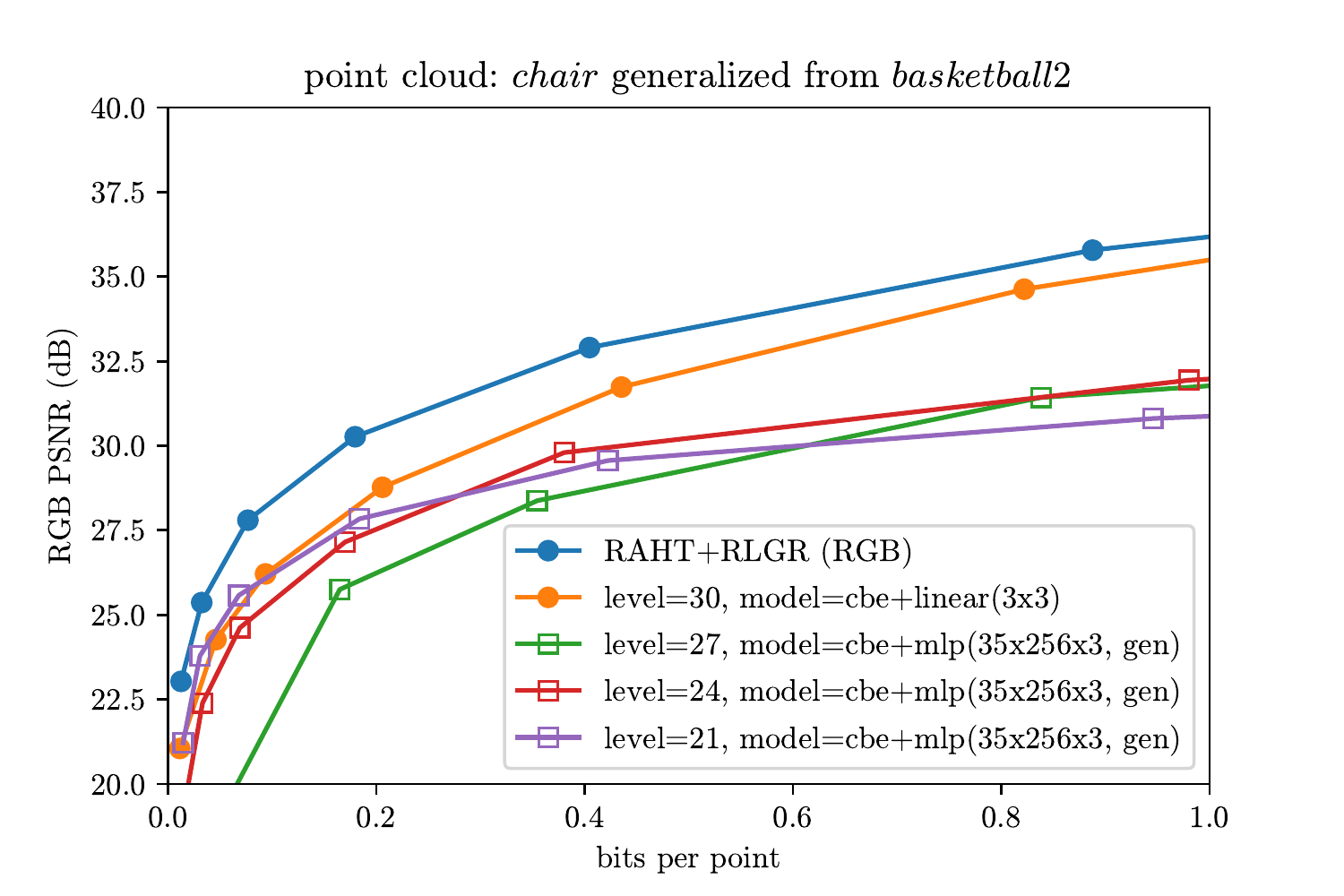}
    \includegraphics[width=0.29\linewidth, trim=20 5 35 15, clip]{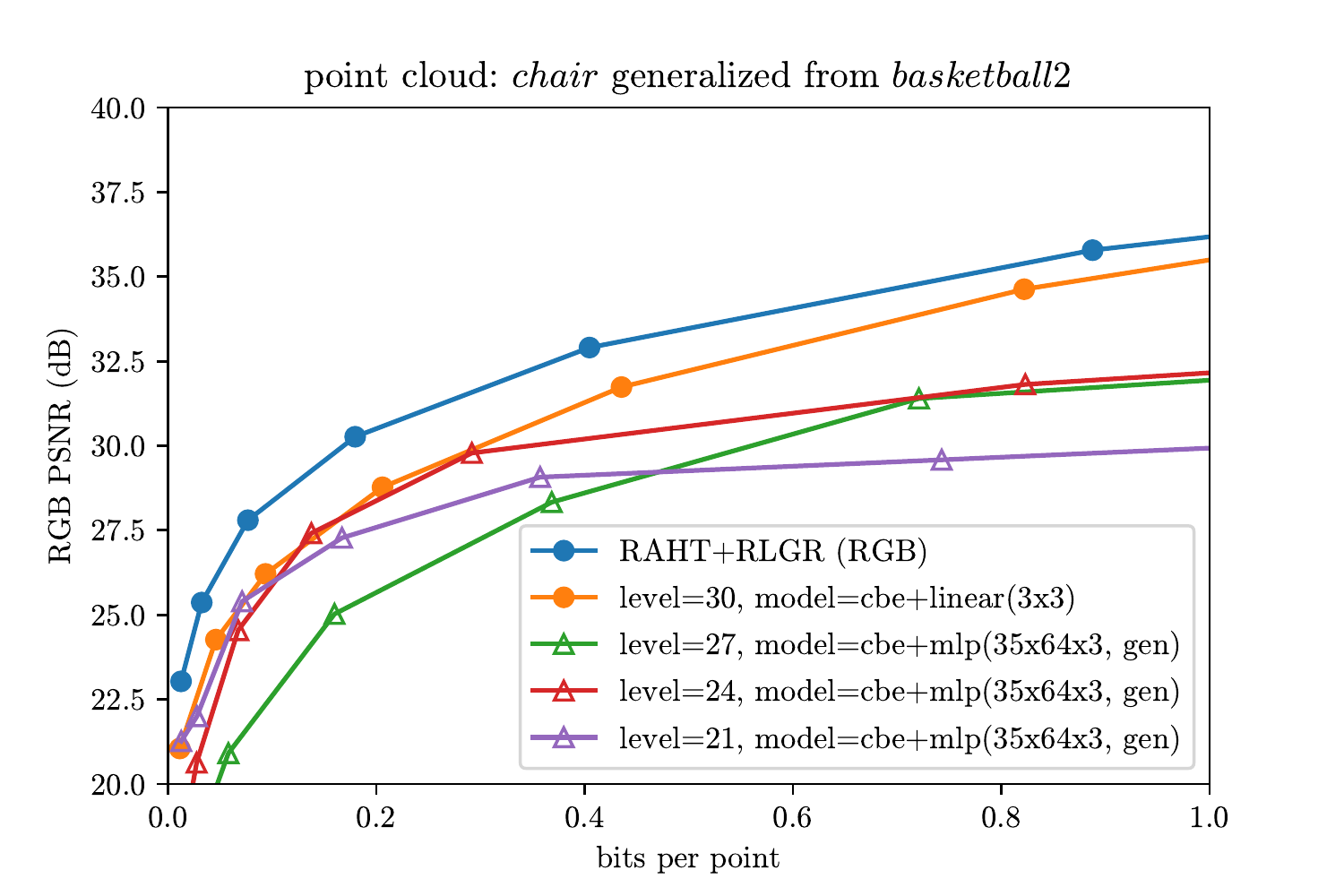}
    \includegraphics[width=0.29\linewidth, trim=20 5 35 15, clip]{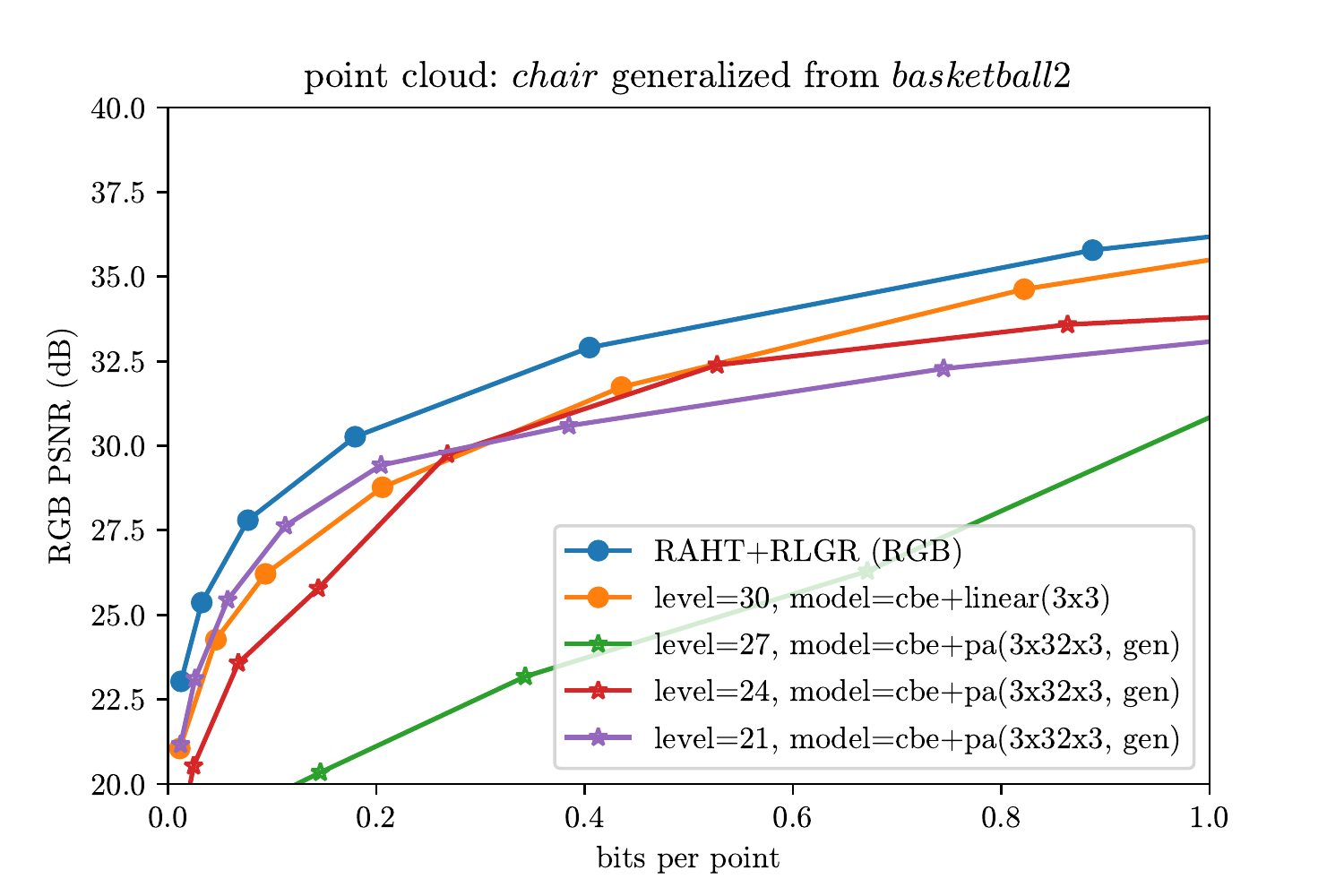}
    
    \includegraphics[width=0.29\linewidth, trim=20 5 35 15, clip]{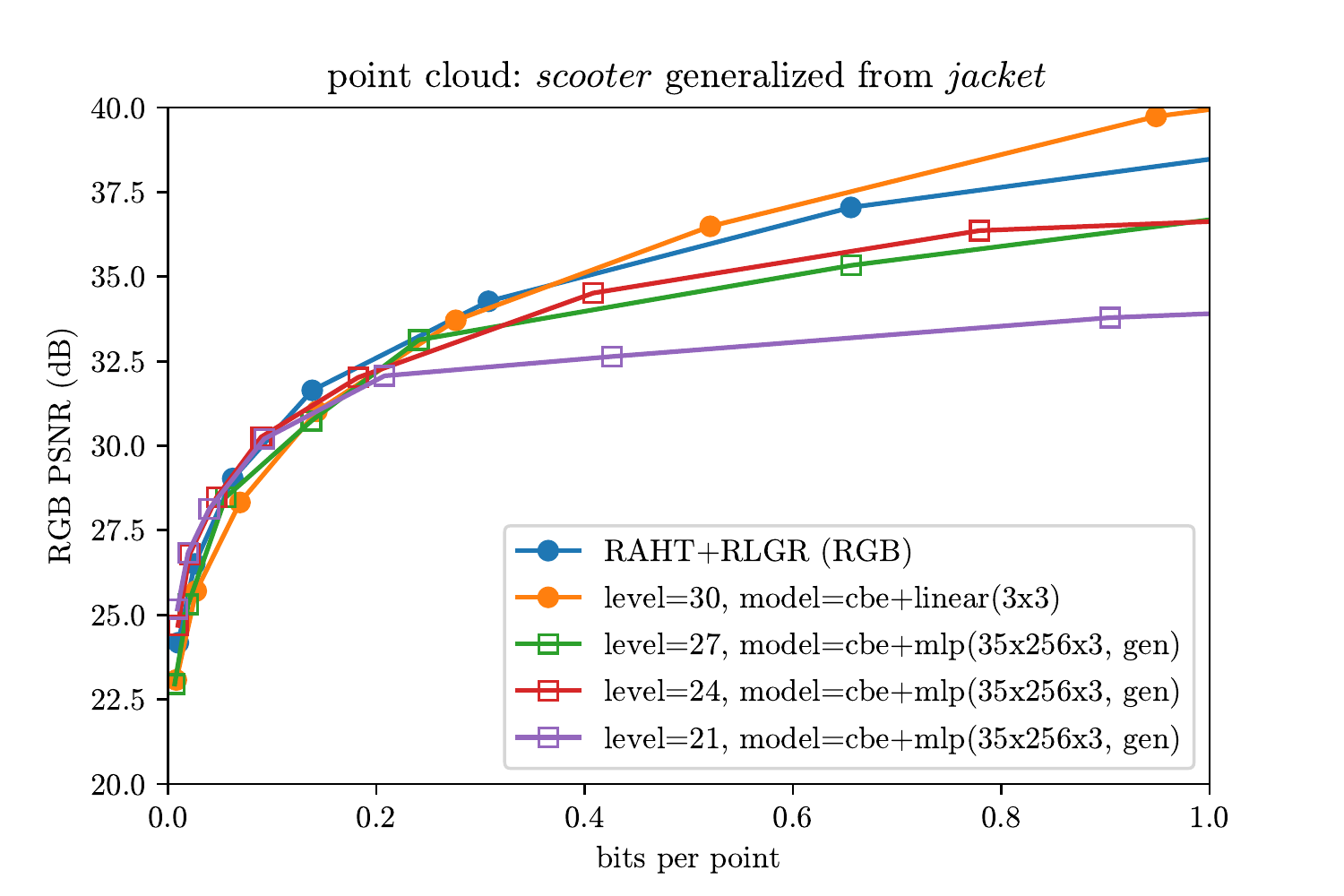}
    \includegraphics[width=0.29\linewidth, trim=20 5 35 15, clip]{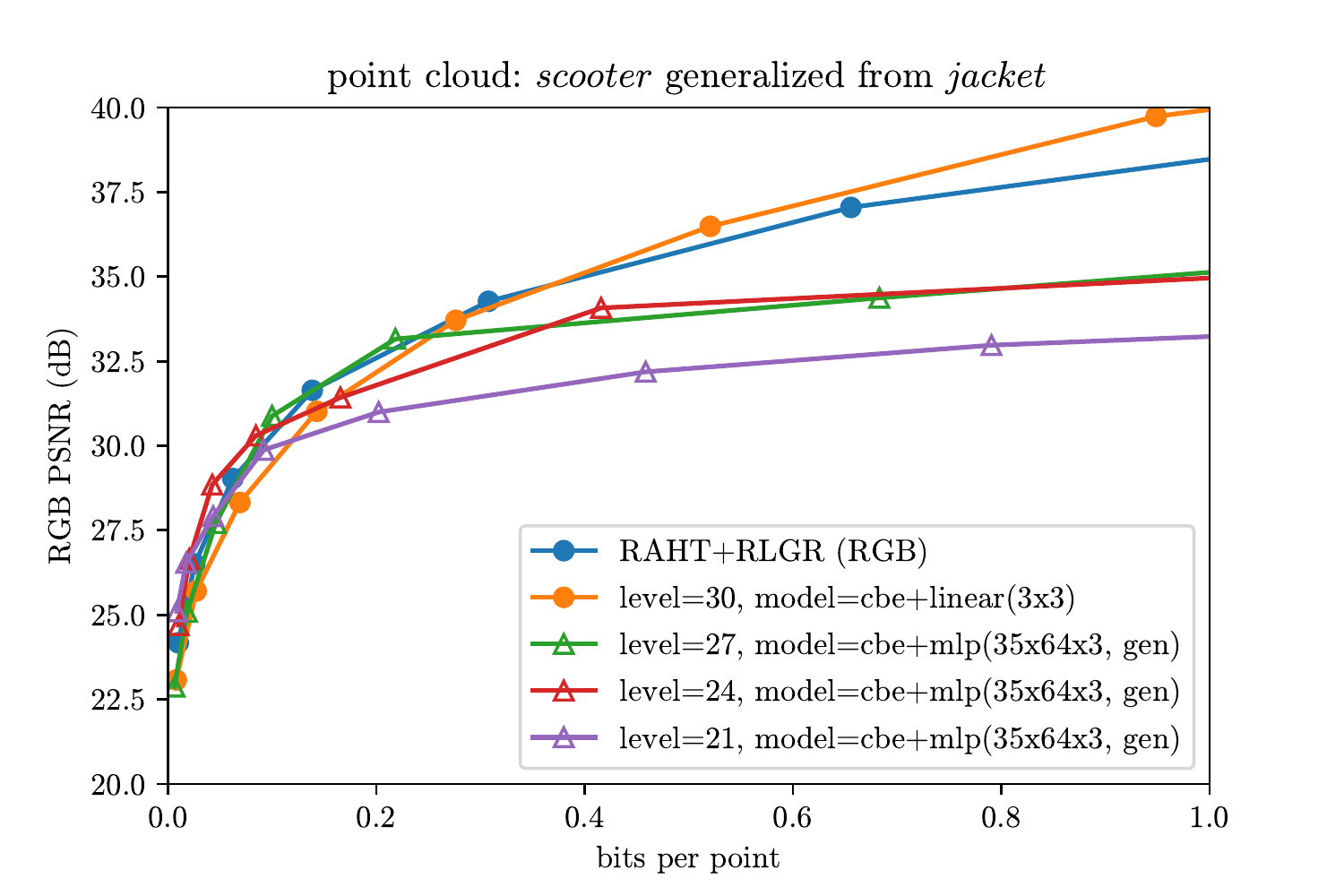}
    \includegraphics[width=0.29\linewidth, trim=20 5 35 15, clip]{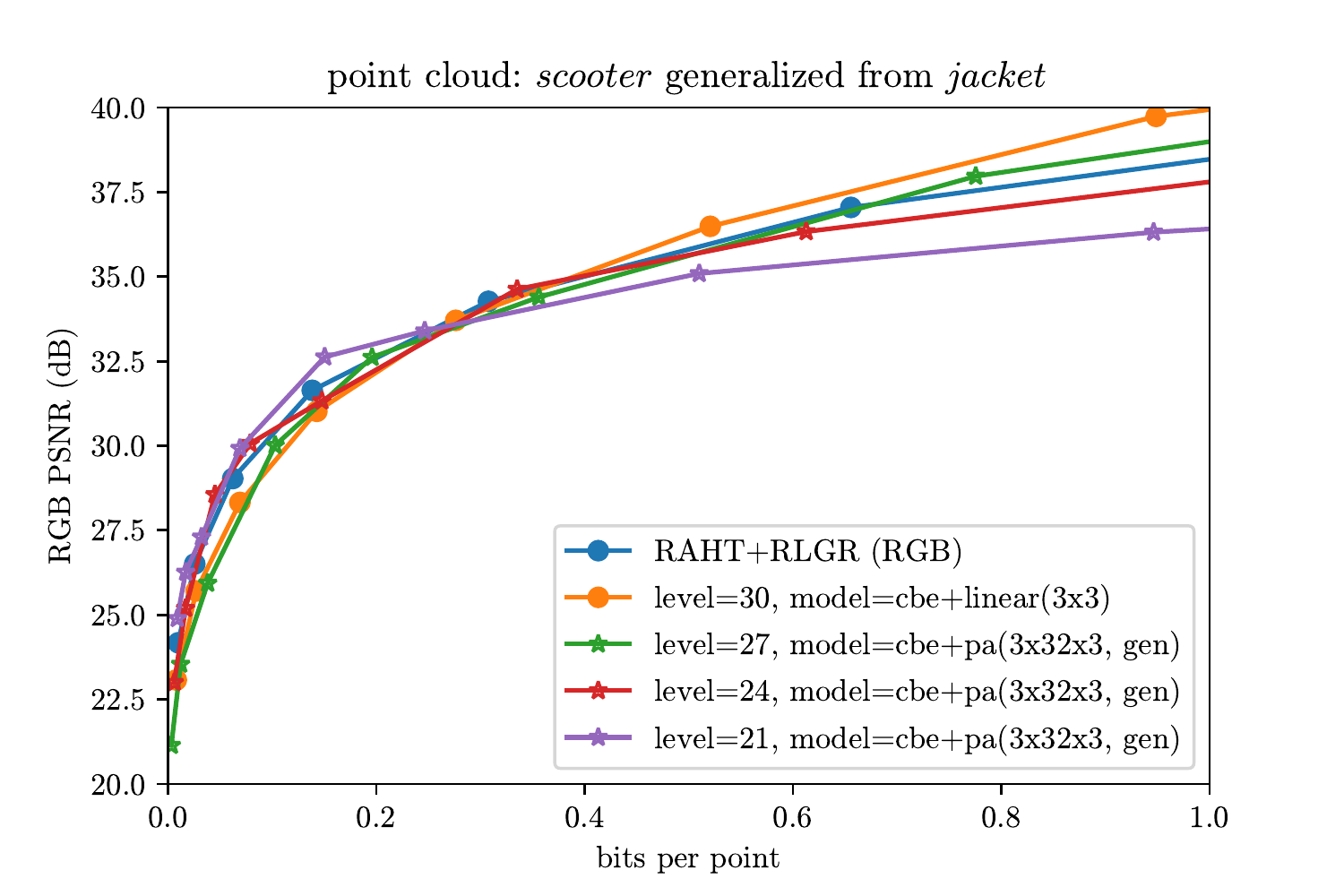}
    
    \includegraphics[width=0.29\linewidth, trim=20 5 35 15, clip]{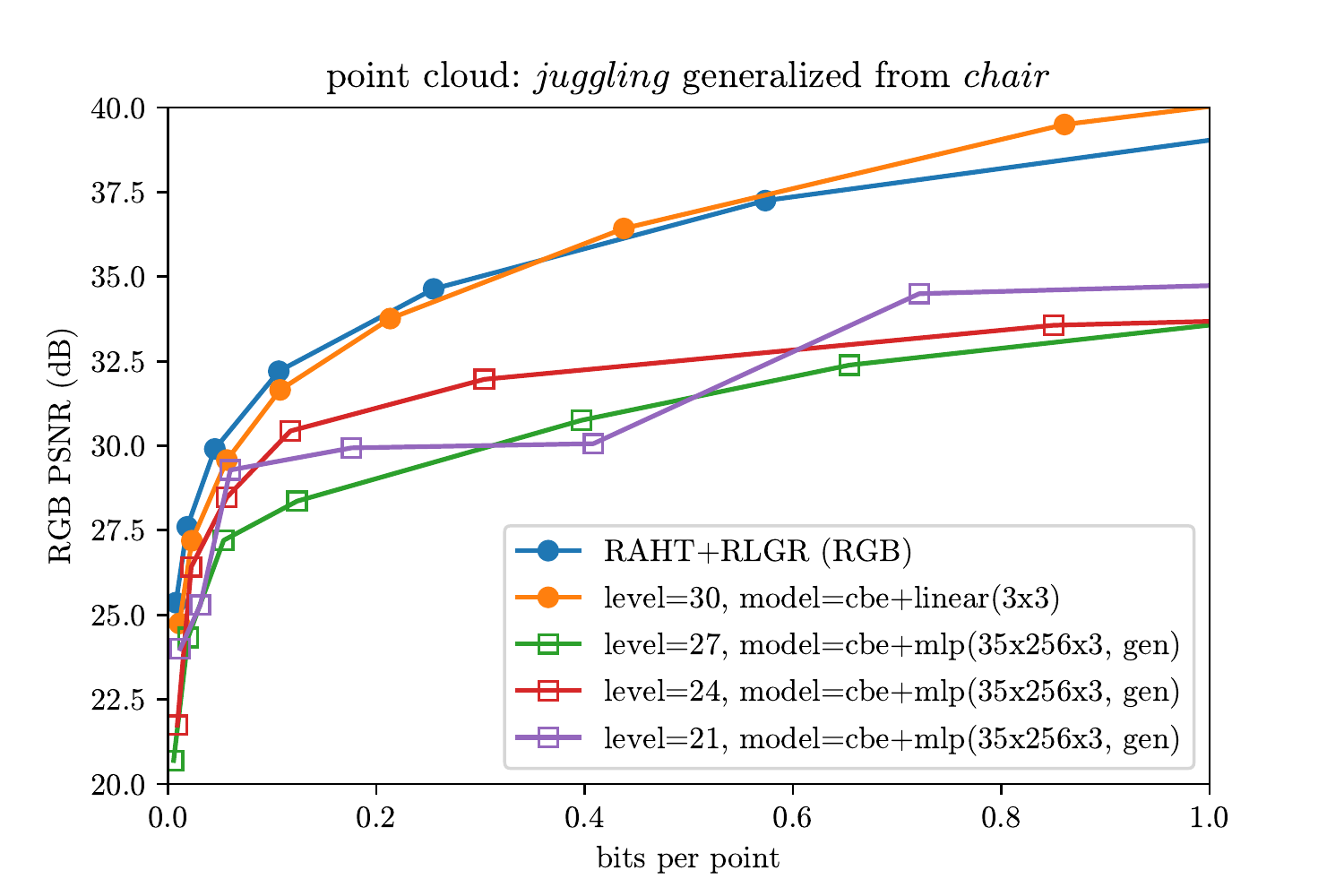}
    \includegraphics[width=0.29\linewidth, trim=20 5 35 15, clip]{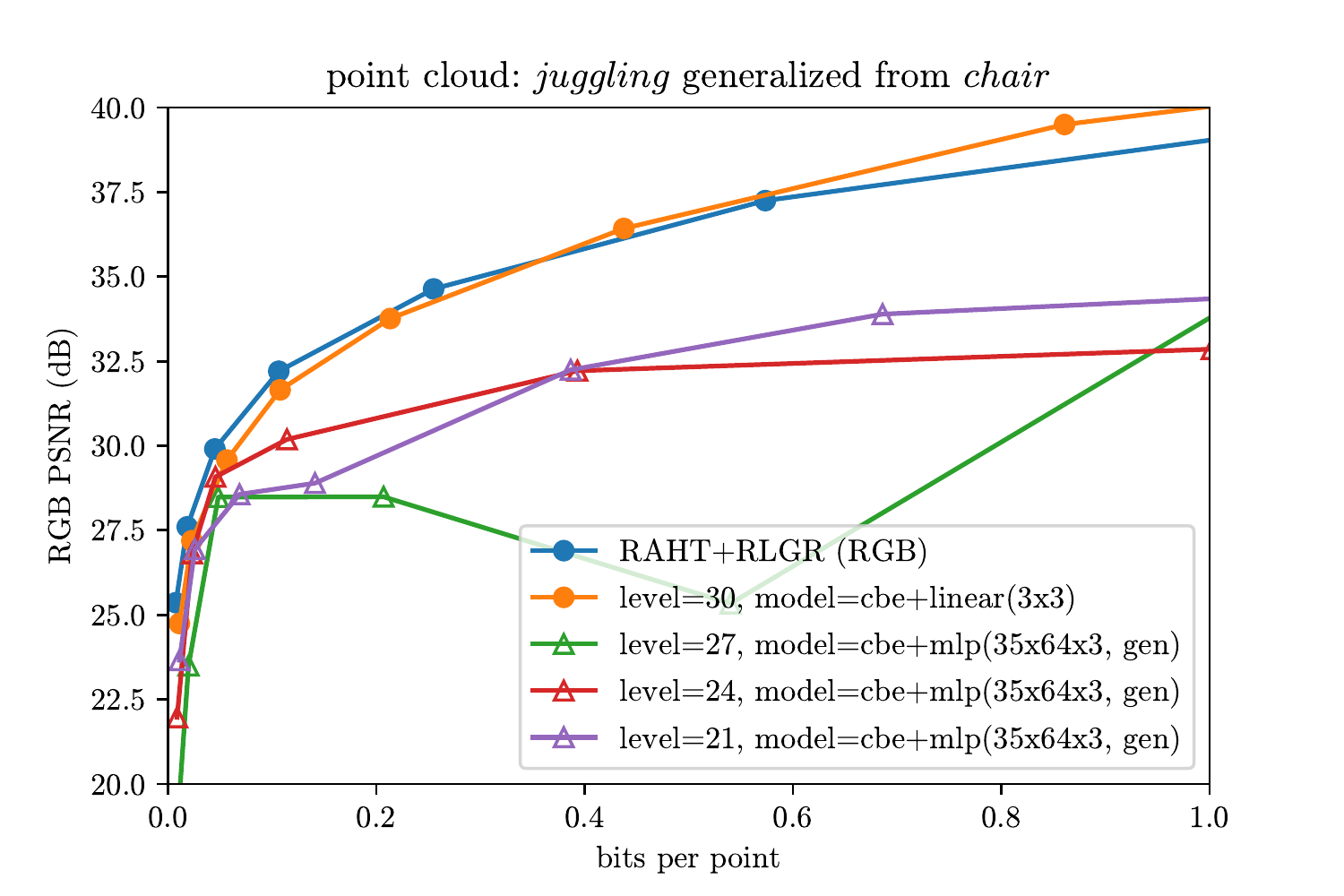}
    \includegraphics[width=0.29\linewidth, trim=20 5 35 15, clip]{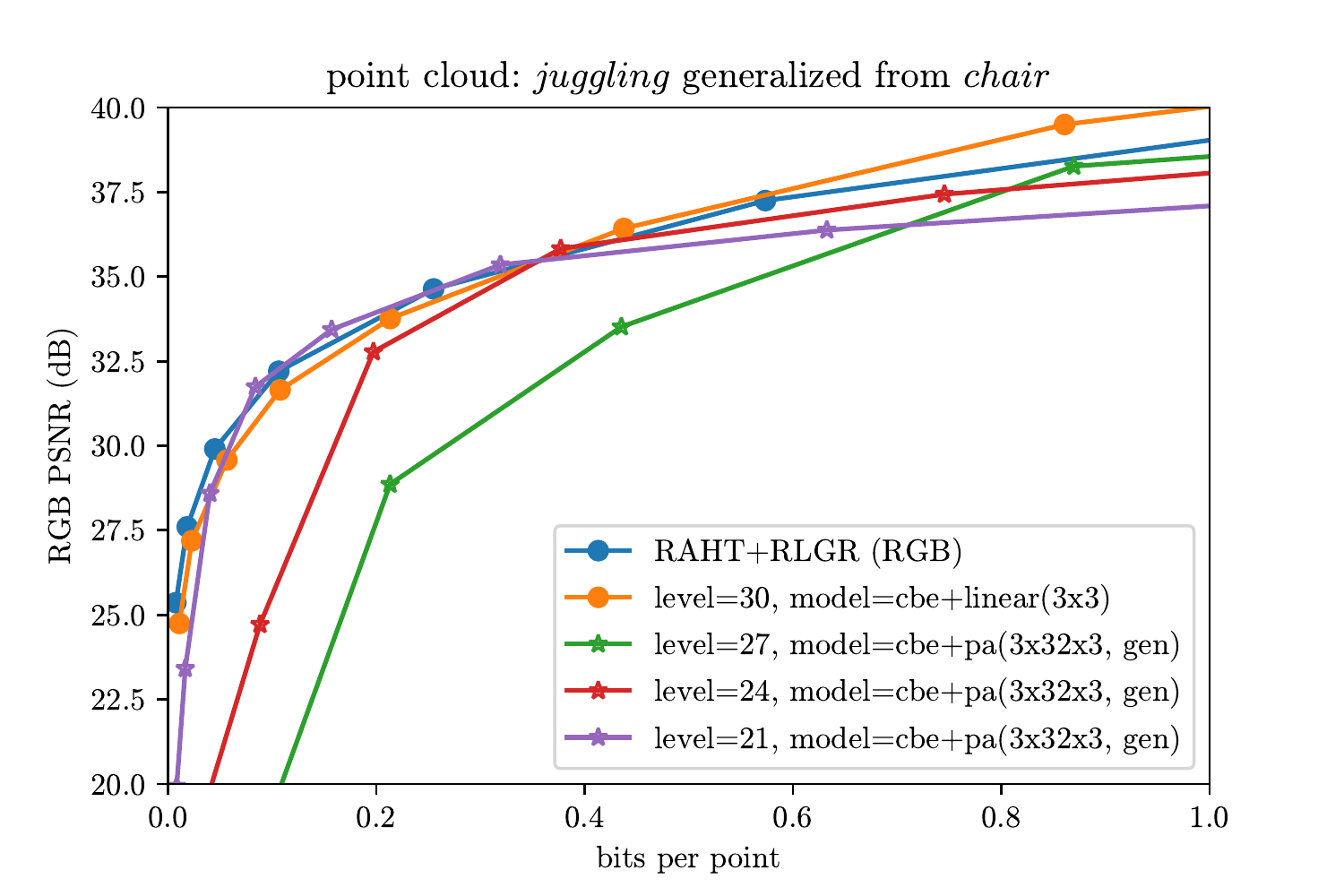}
    
    \includegraphics[width=0.29\linewidth, trim=20 5 35 15, clip]{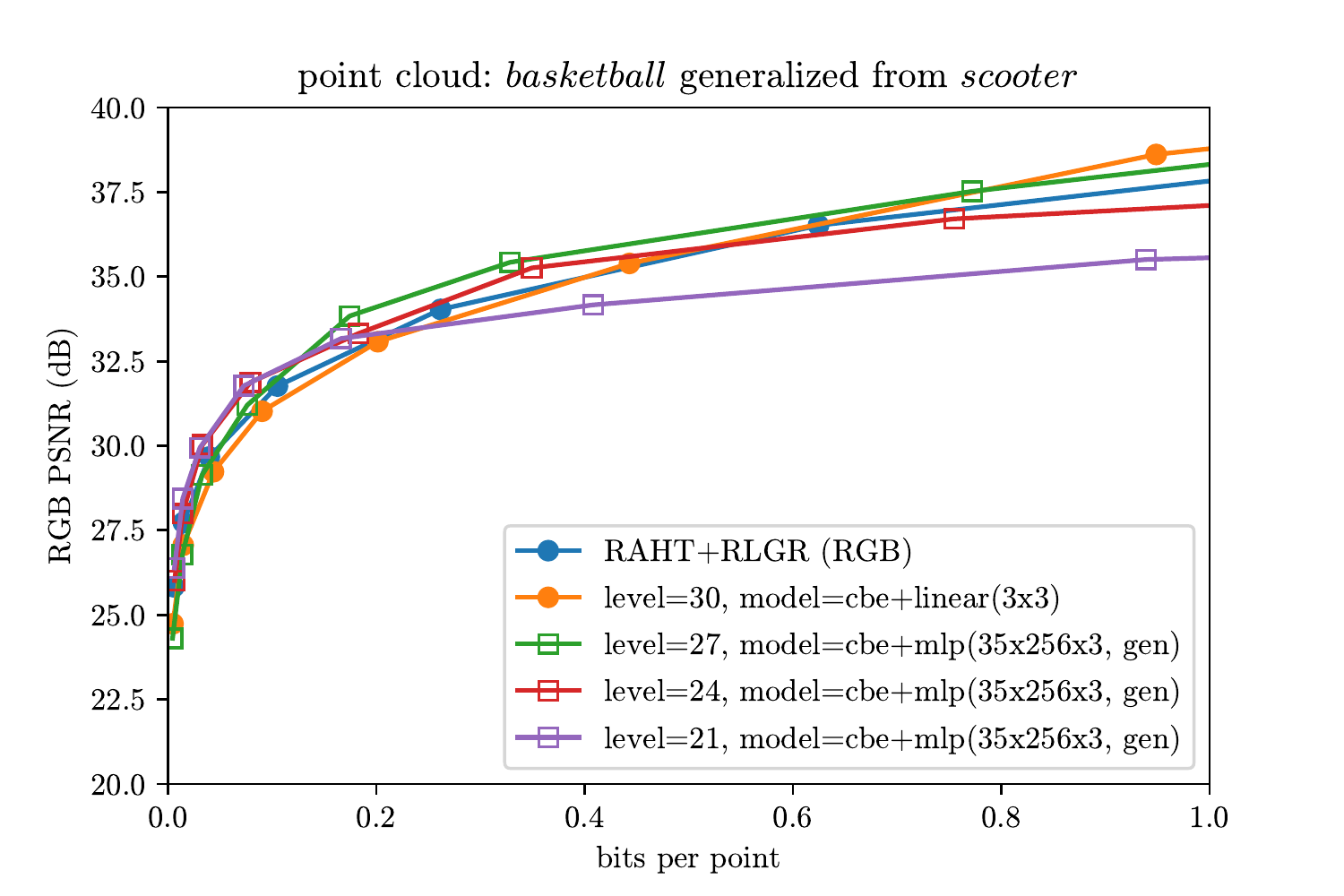}
    \includegraphics[width=0.29\linewidth, trim=20 5 35 15, clip]{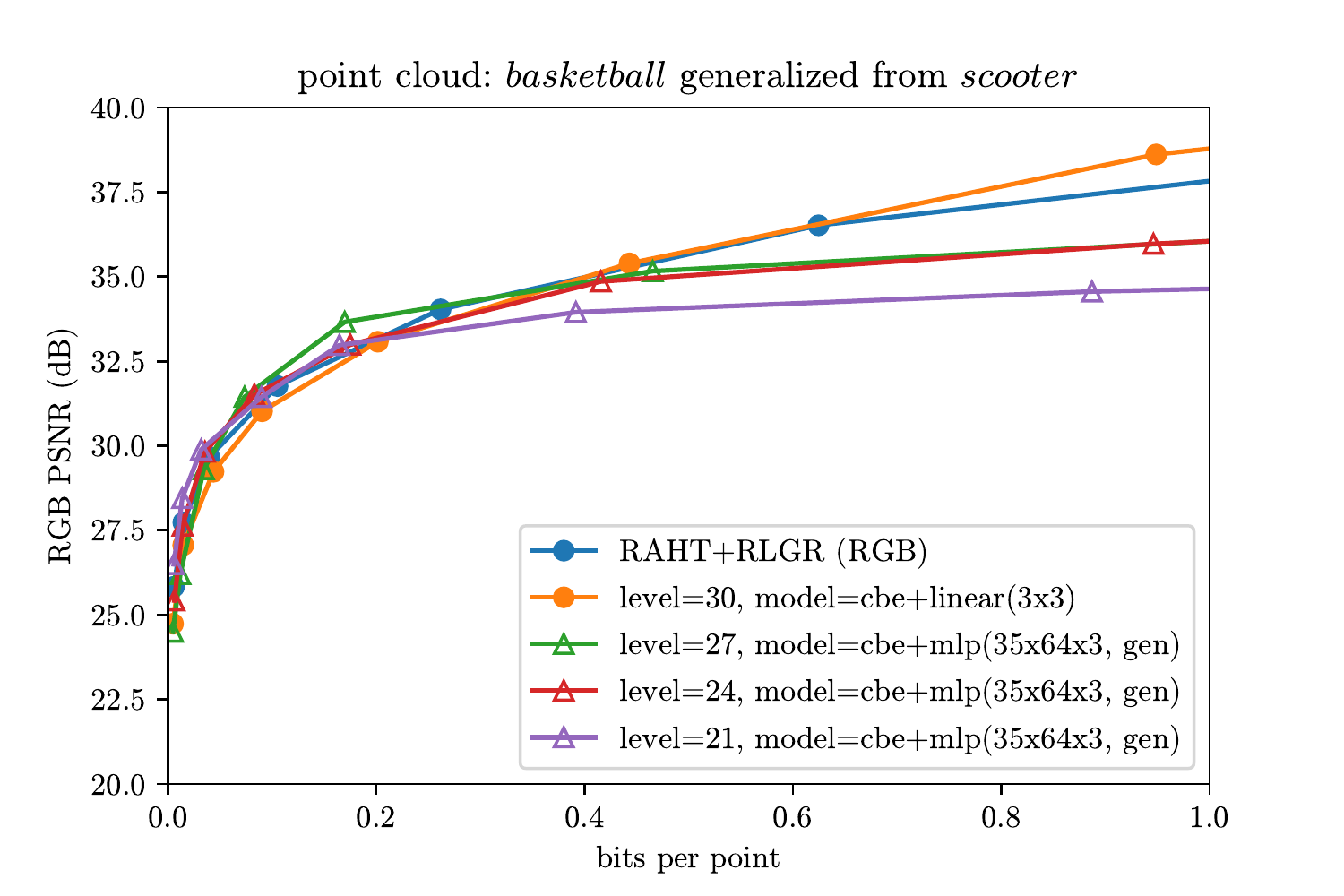}
    \includegraphics[width=0.29\linewidth, trim=20 5 35 15, clip]{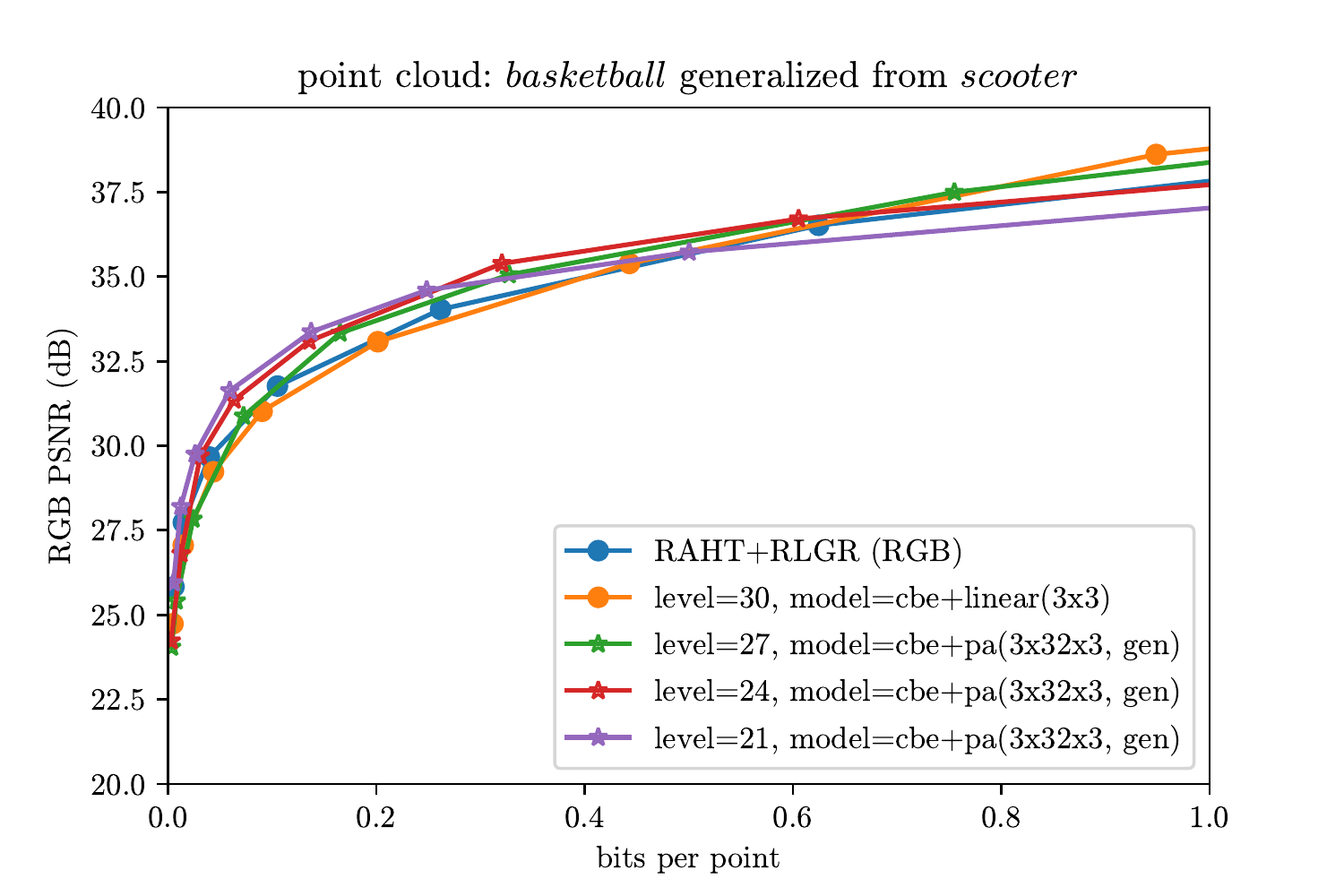}
    
    \includegraphics[width=0.29\linewidth, trim=20 5 35 15, clip]{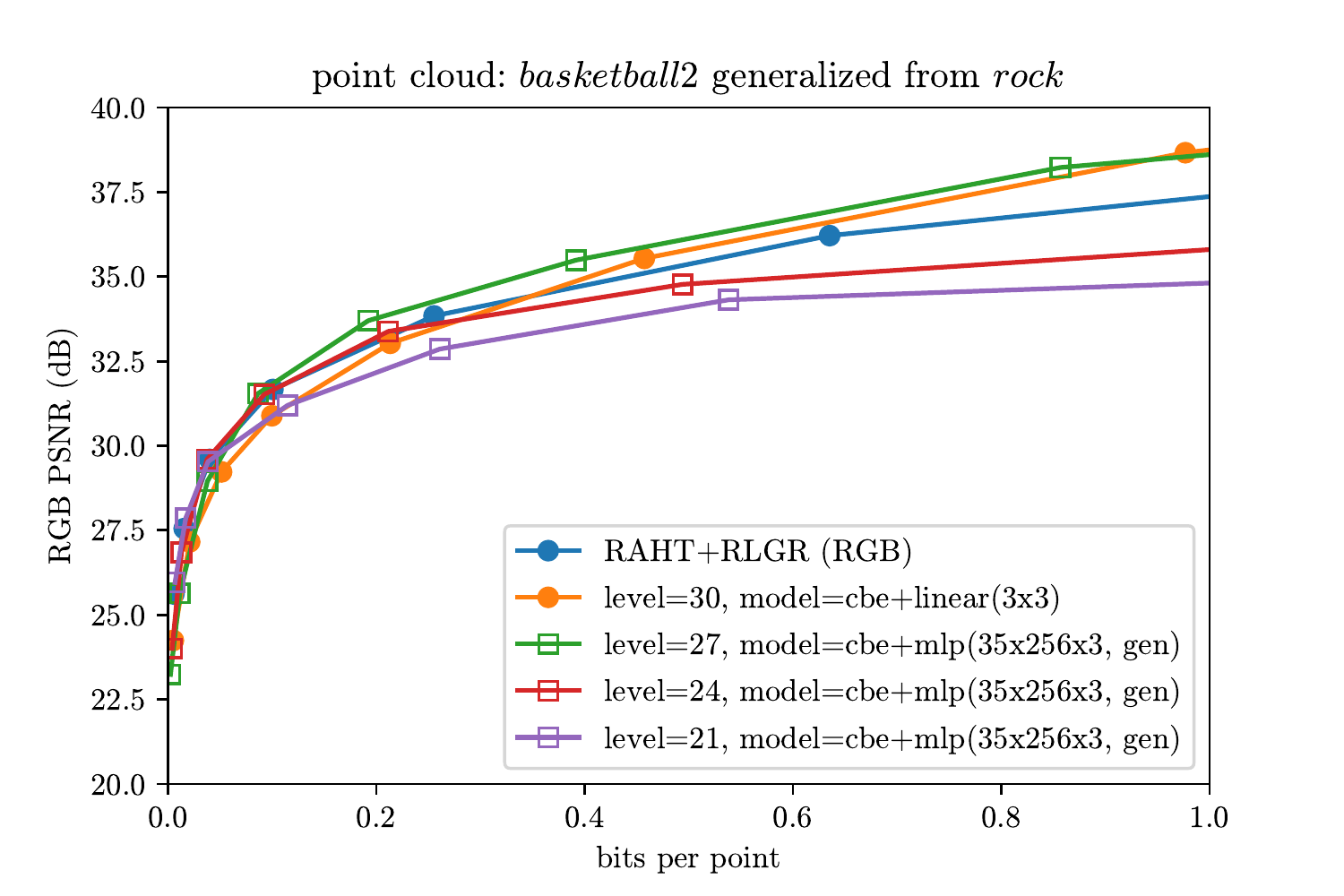}
    \includegraphics[width=0.29\linewidth, trim=20 5 35 15, clip]{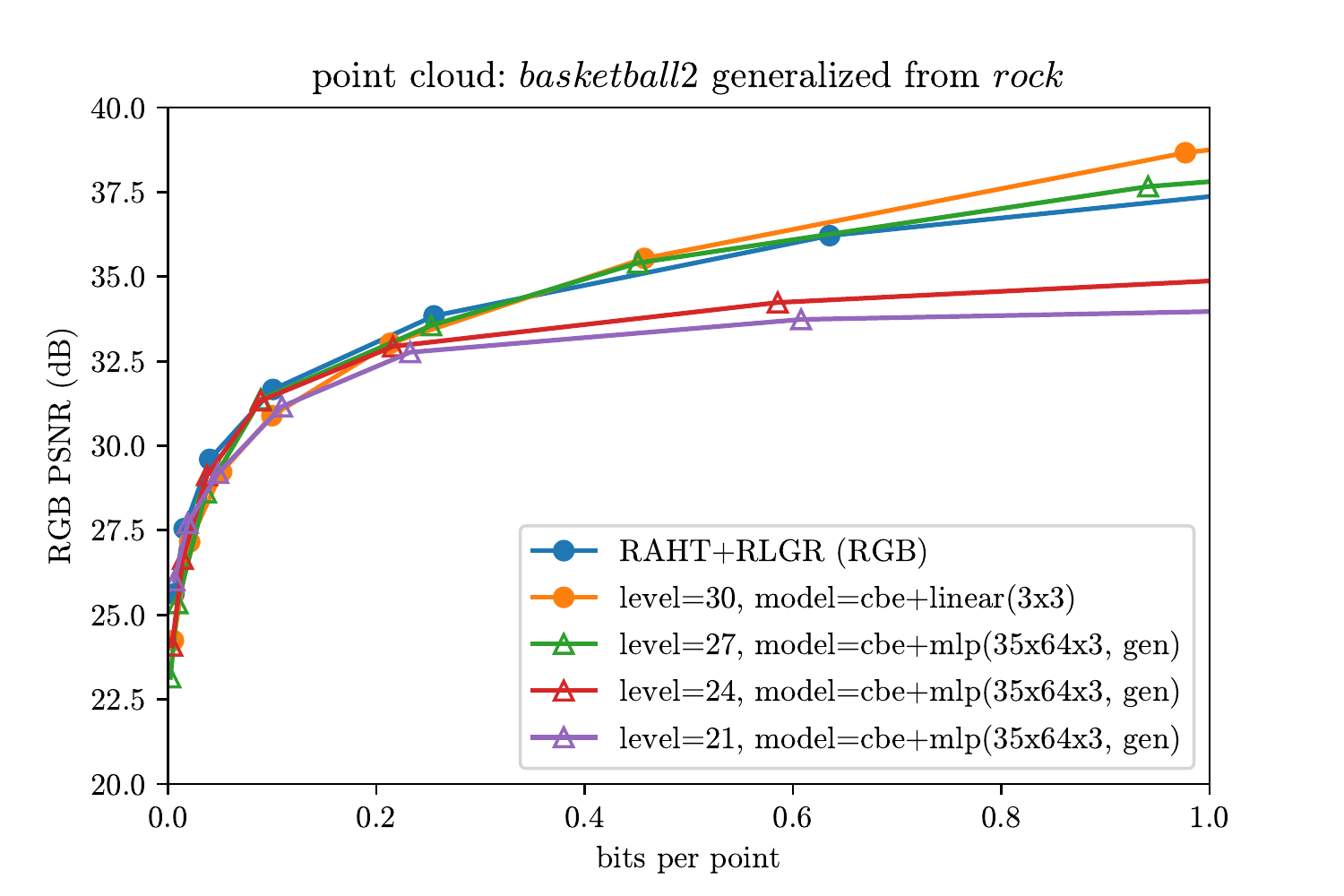}
    \includegraphics[width=0.29\linewidth, trim=20 5 35 15, clip]{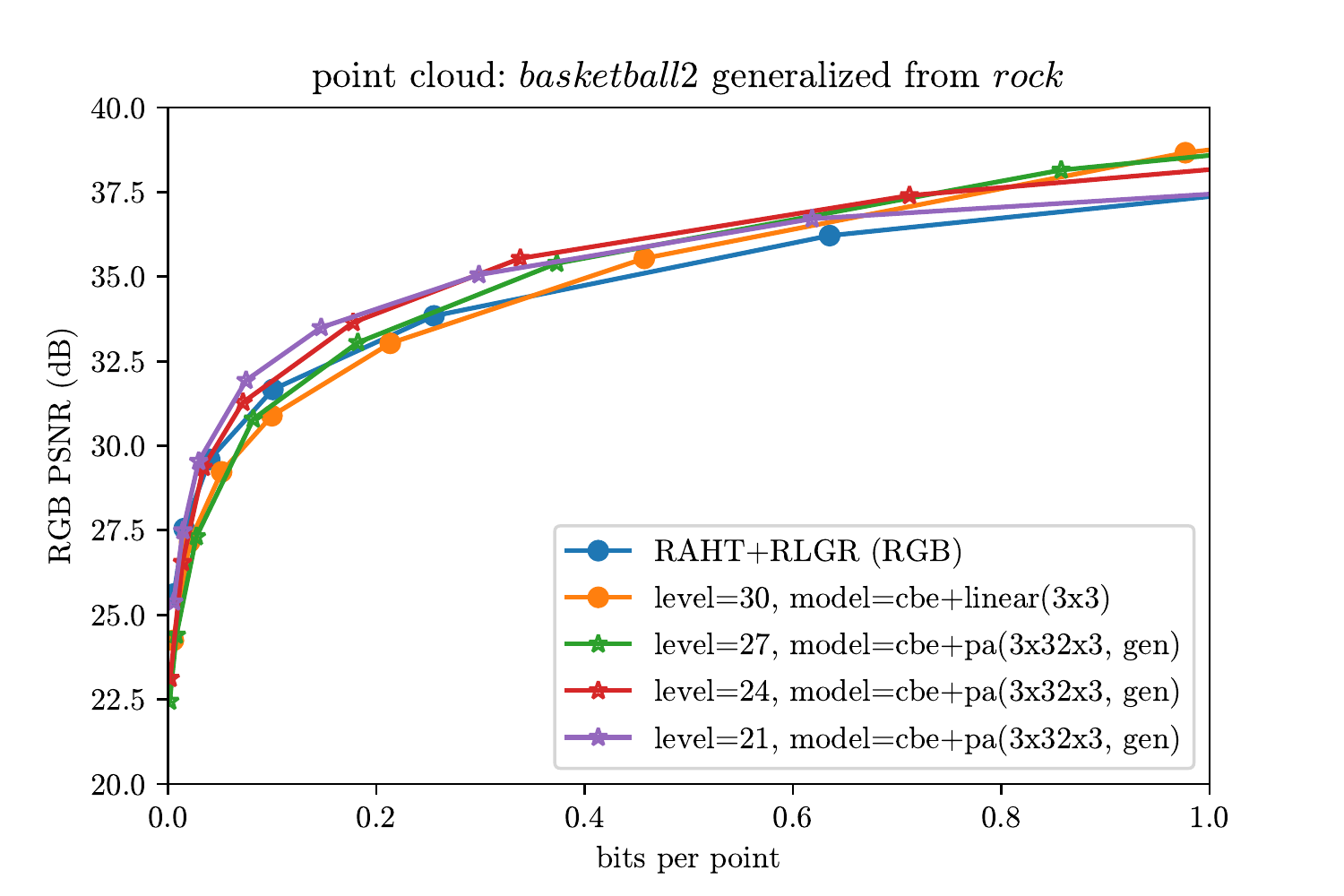}
    
    \includegraphics[width=0.29\linewidth, trim=20 5 35 15, clip]{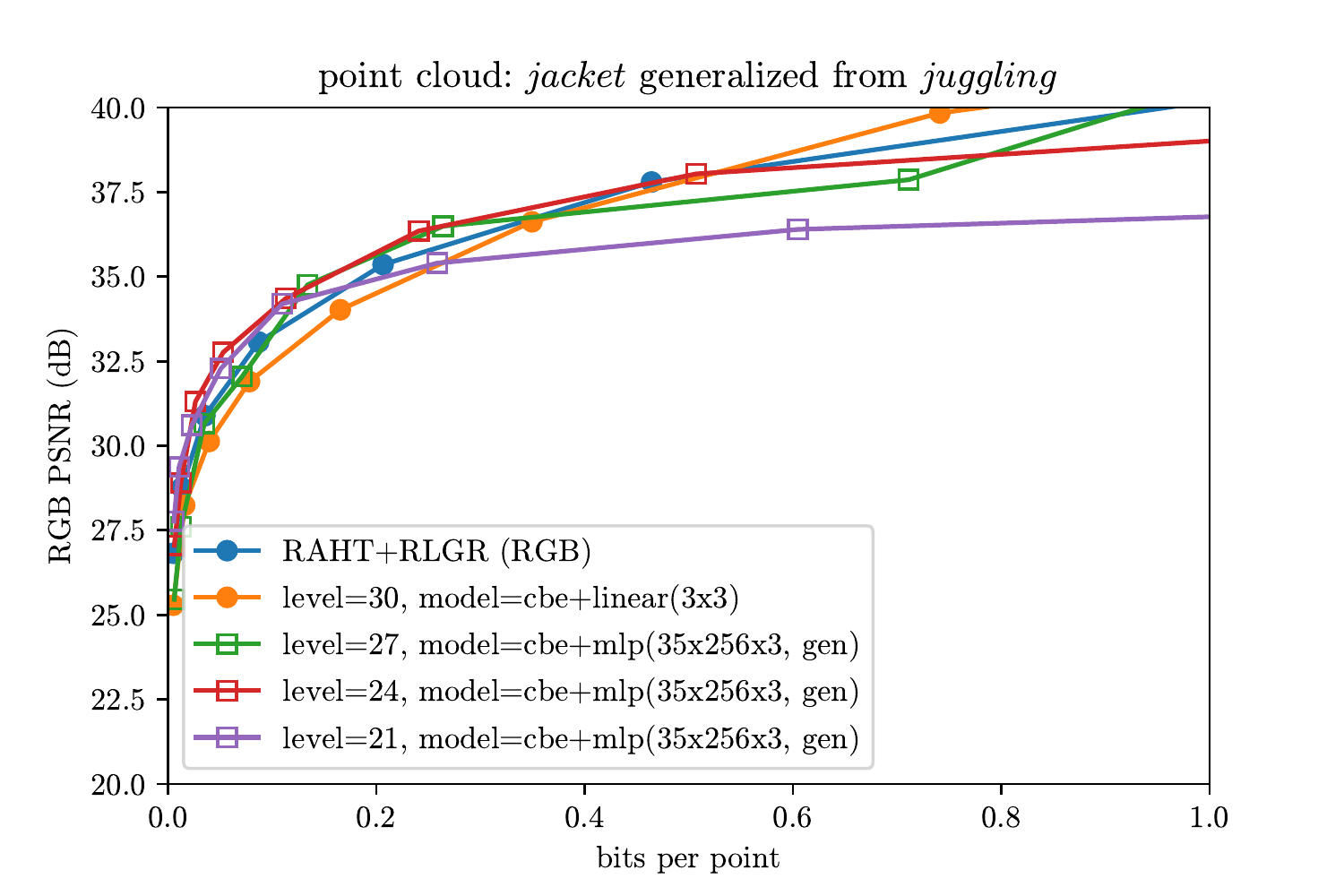}
    \includegraphics[width=0.29\linewidth, trim=20 5 35 15, clip]{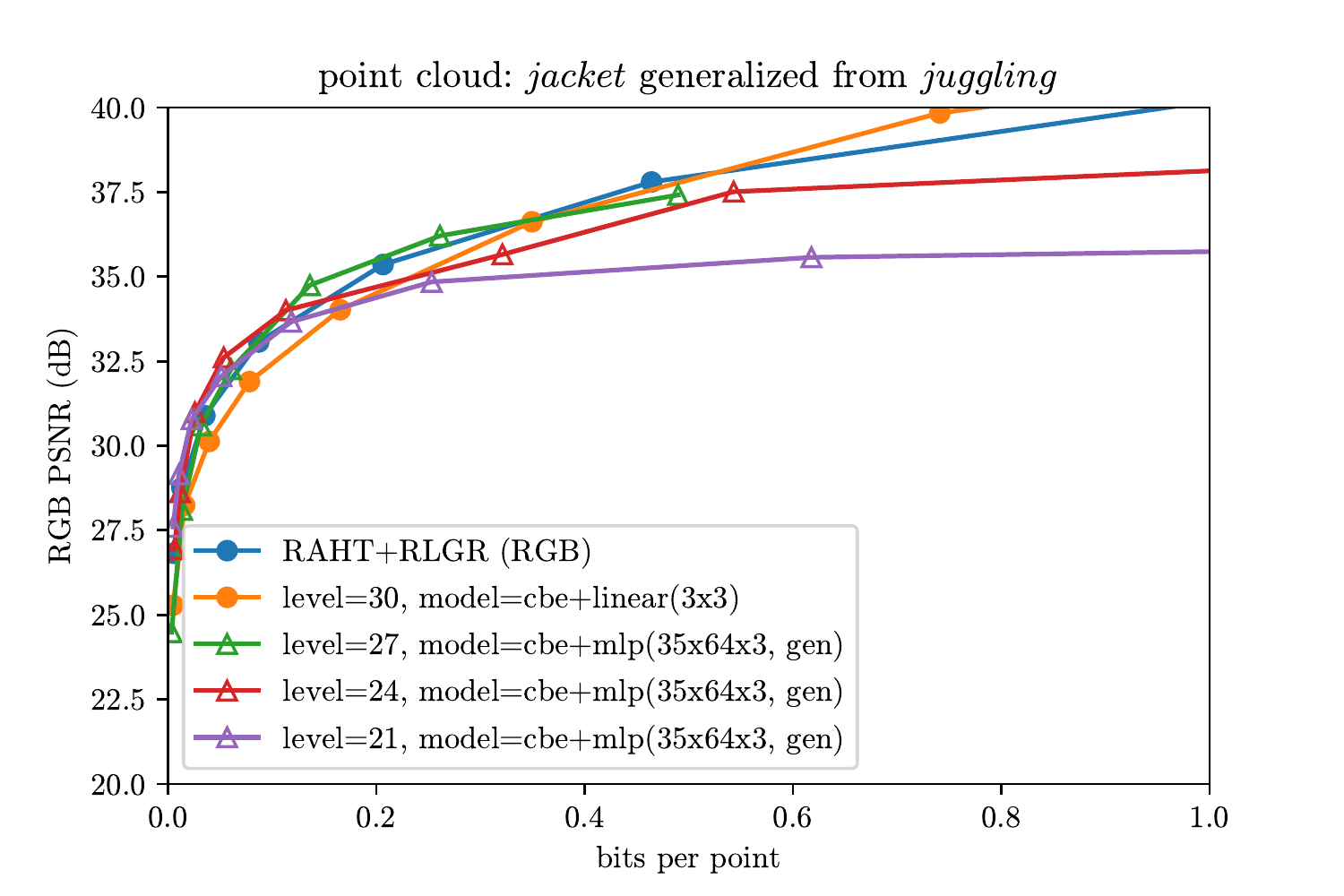}
    \includegraphics[width=0.29\linewidth, trim=20 5 35 15, clip]{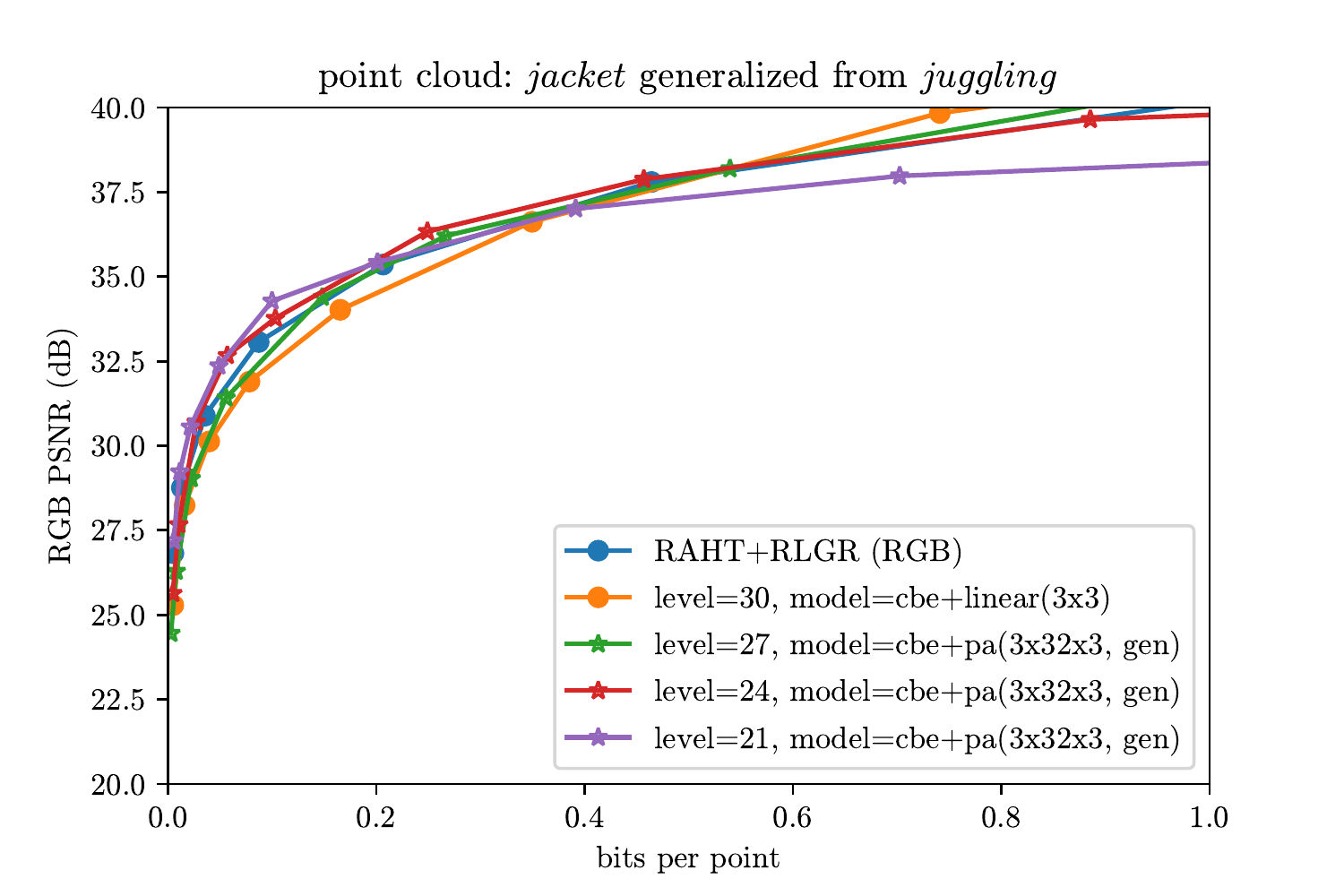}
    
    \caption{Coordinate Based Networks with generalization, by network.  Each row is a different point cloud.  Left, middle, right columns each show levels 27, 24, and 21, along with baselines, for CBNs {\em mlp(35x256x3)}, {\em mlp(35x64x3)}, and {\em pa(3x32x3)}.  See \cref{fig:cbns_gen} (bottom) for point cloud {\em rock}.}
    \label{fig:cbns_by_network_gen_supp}
\end{figure*}

\begin{figure*}
    \centering
    \includegraphics[width=0.33\linewidth, trim=20 5 35 15, clip]{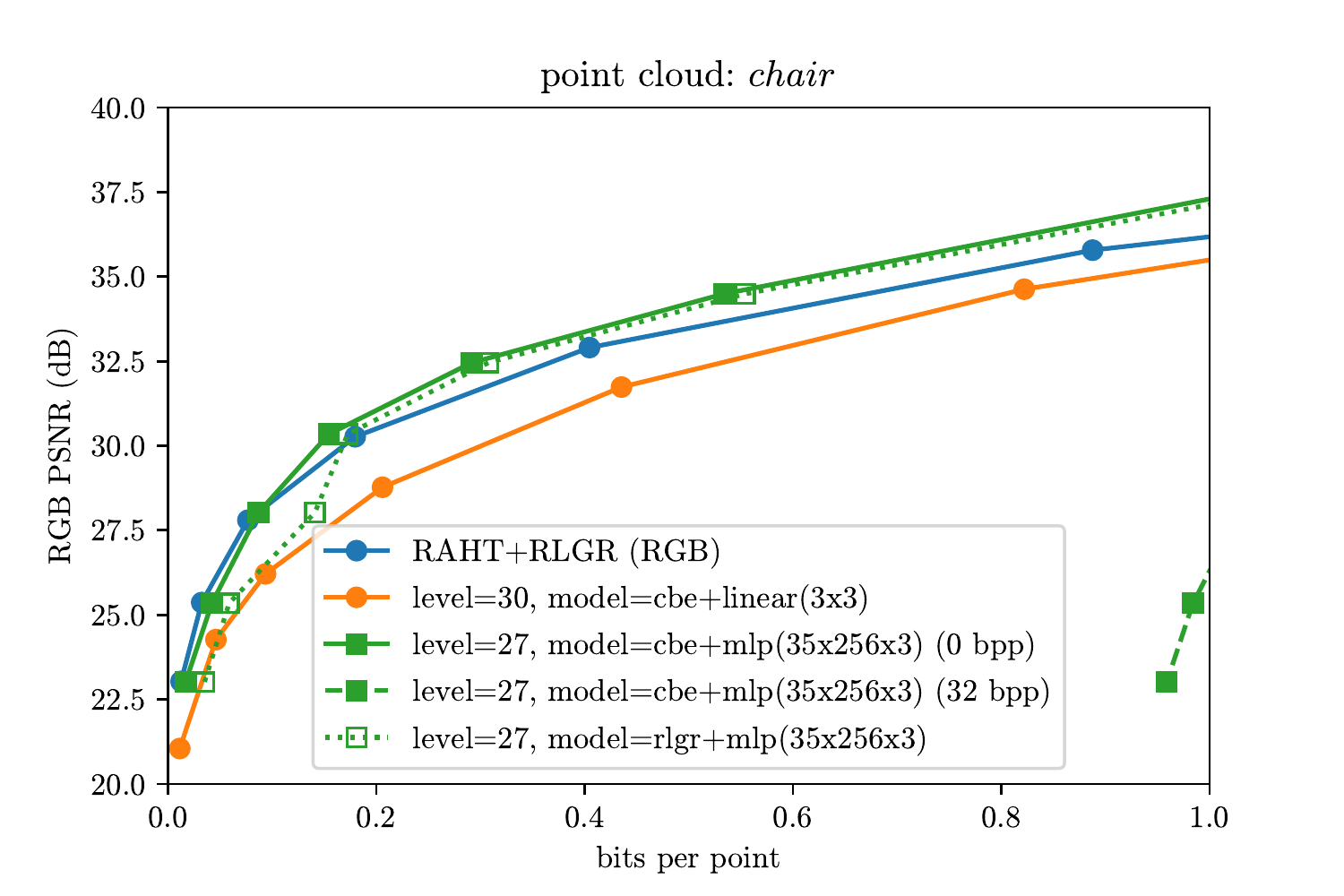}
    \includegraphics[width=0.33\linewidth, trim=20 5 35 15, clip]{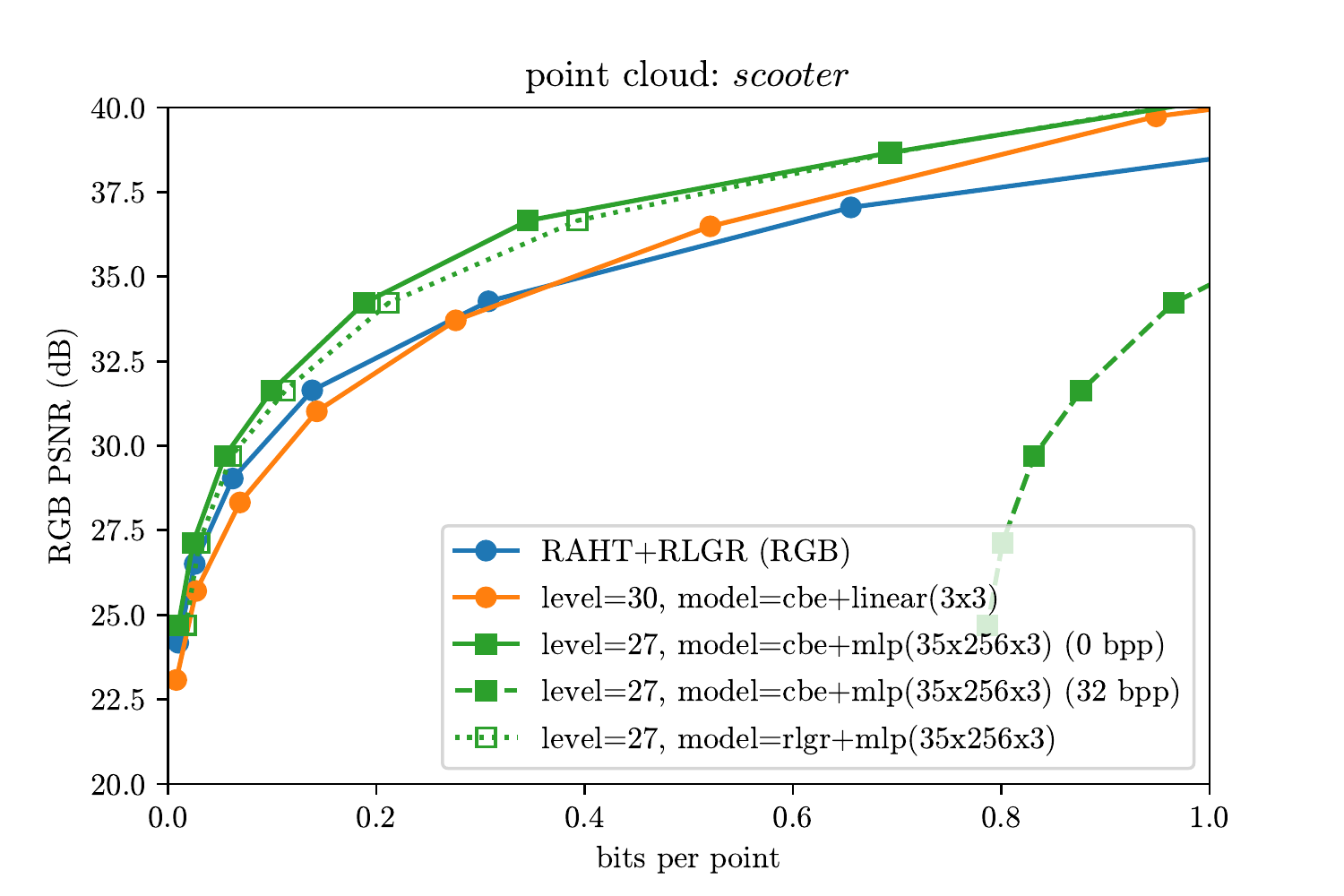}
    \includegraphics[width=0.33\linewidth, trim=20 5 35 15, clip]{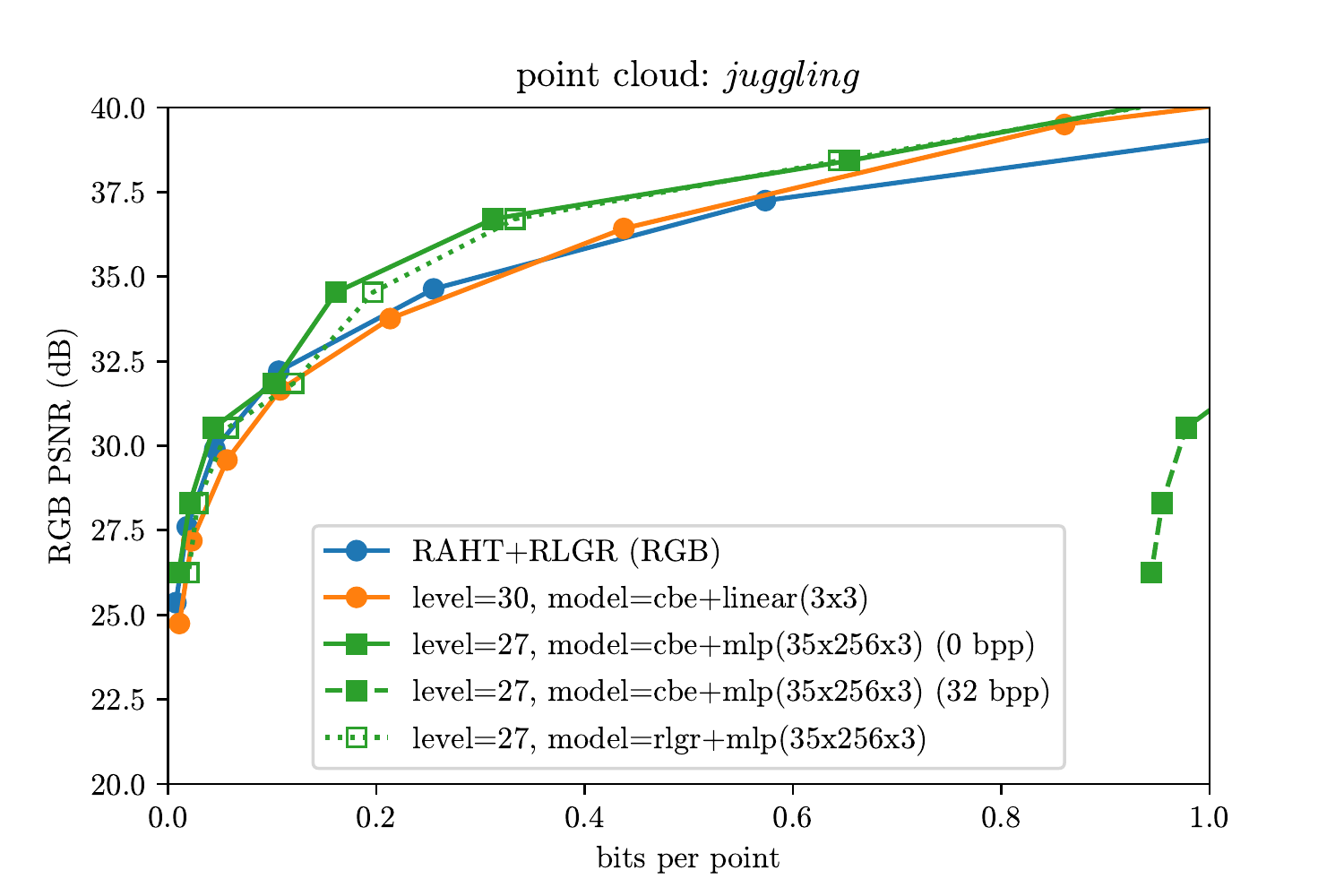}
    \includegraphics[width=0.33\linewidth, trim=20 5 35 15, clip]{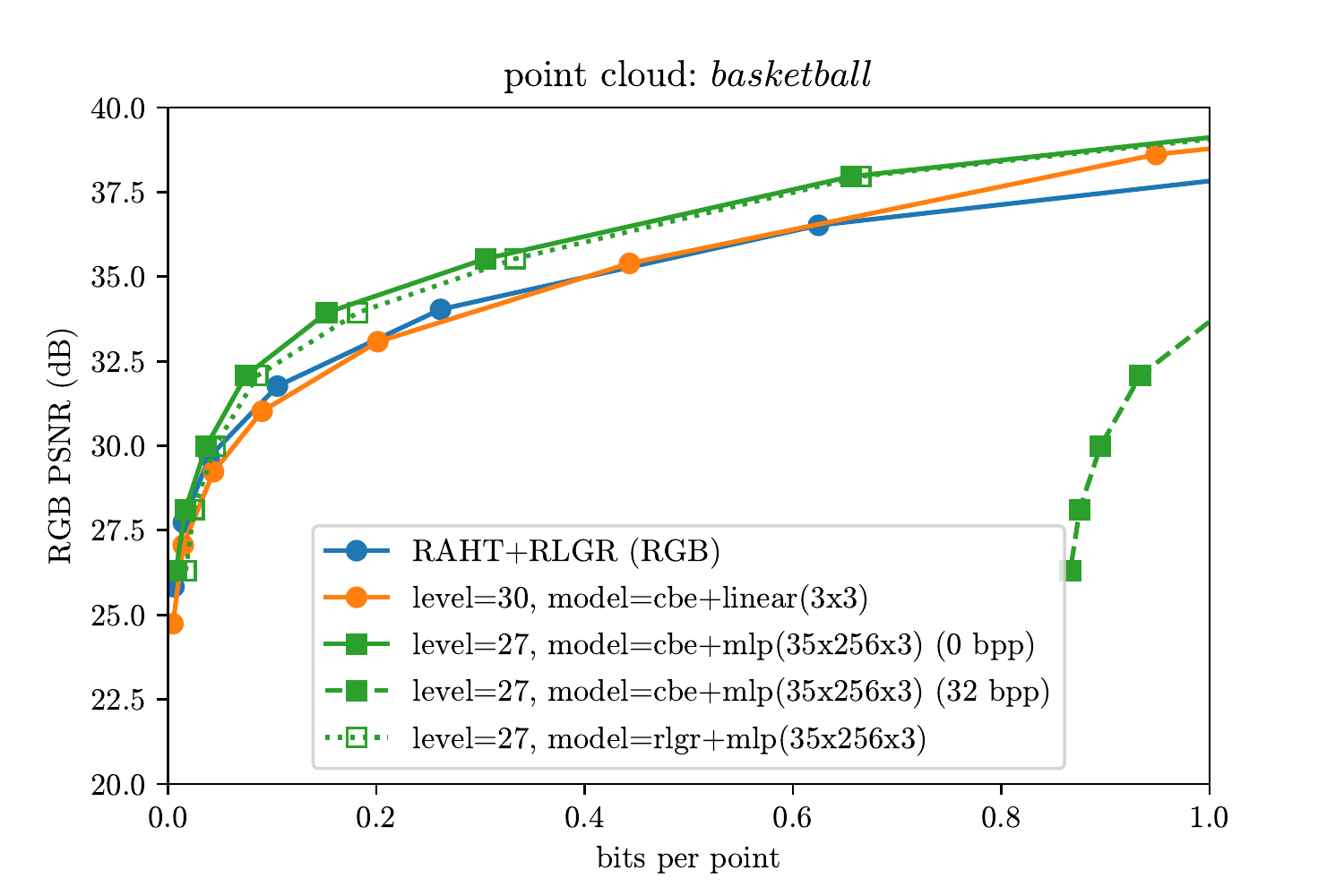}
    \includegraphics[width=0.33\linewidth, trim=20 5 35 15, clip]{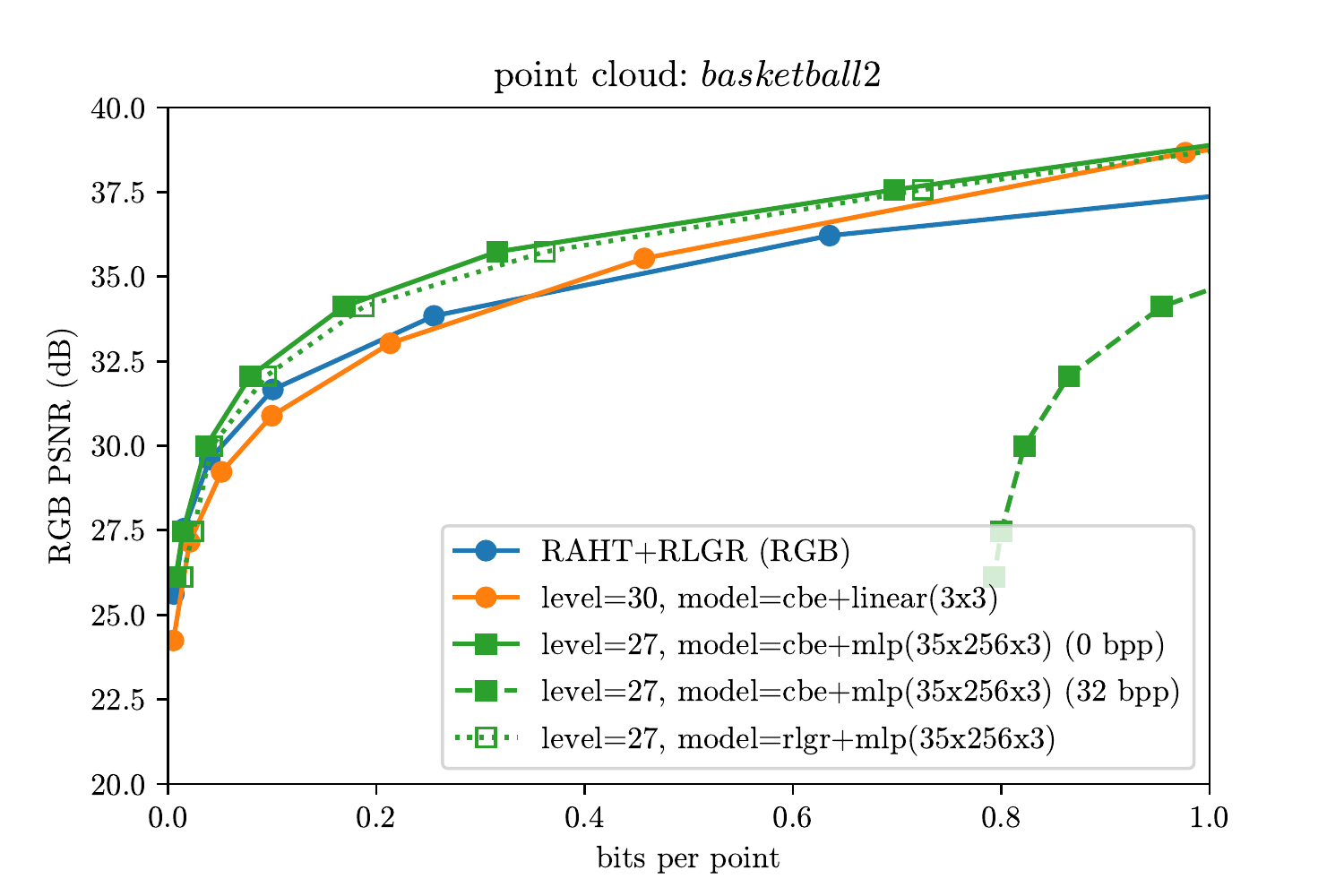}
    \includegraphics[width=0.33\linewidth, trim=20 5 35 15, clip]{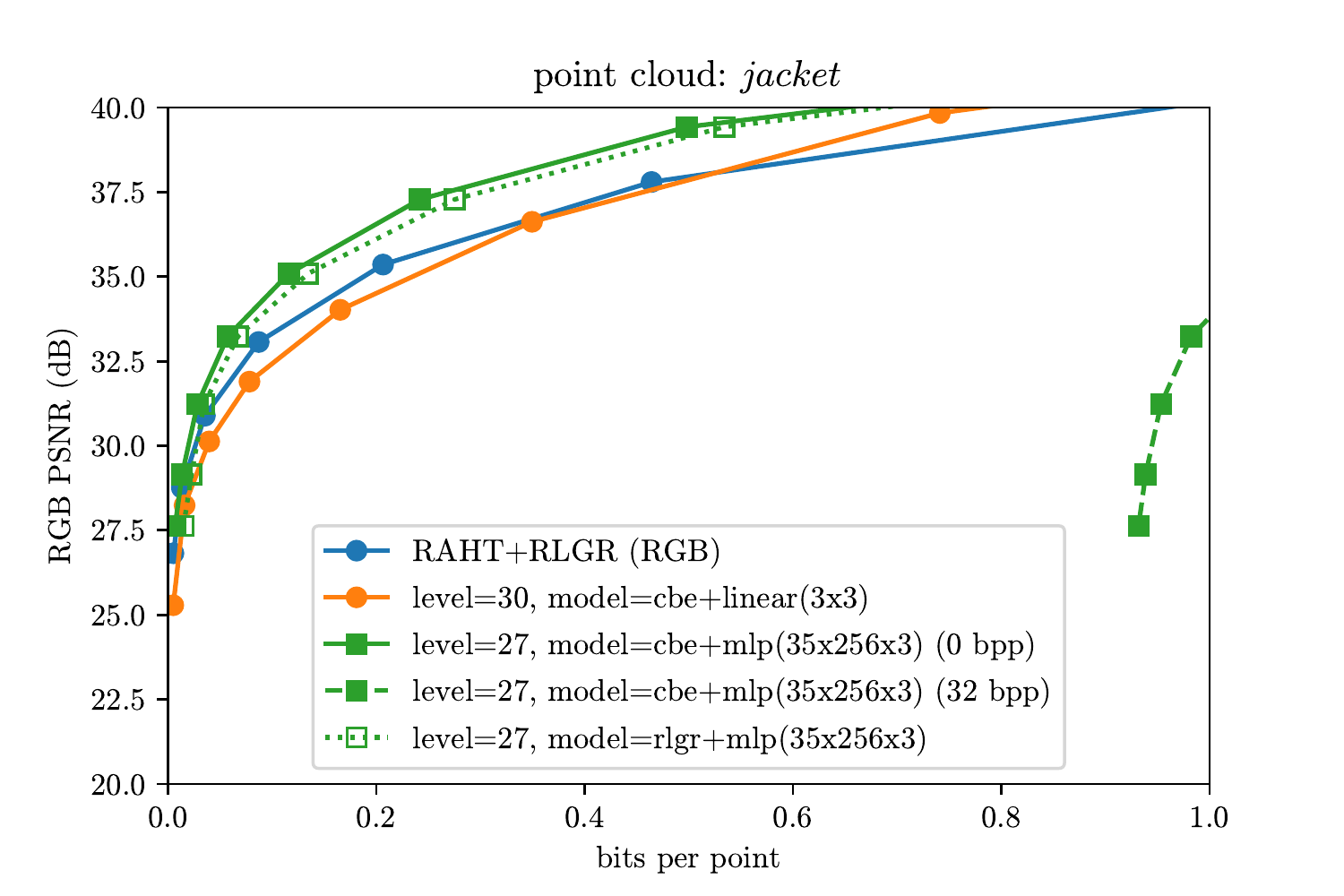}
    \caption{Side information for entropy model.  Sending 32 bits per parameter for the {\em cbe} entropy model would reduce RD performance from solid to dashed green lines.  But the backward-adaptive {\em RLGR} entropy coder (dotted, unfilled) obviates the need to send side information with almost no loss in performance.  See \cref{fig:sideinfo_cbe_vs_rlgr} for point cloud {\em rock}.}
    \label{fig:sideinfo_cbe_vs_rlgr_supp}
\end{figure*}

\begin{figure*}
    \centering
    \includegraphics[width=0.29\linewidth, trim=20 5 35 15, clip]{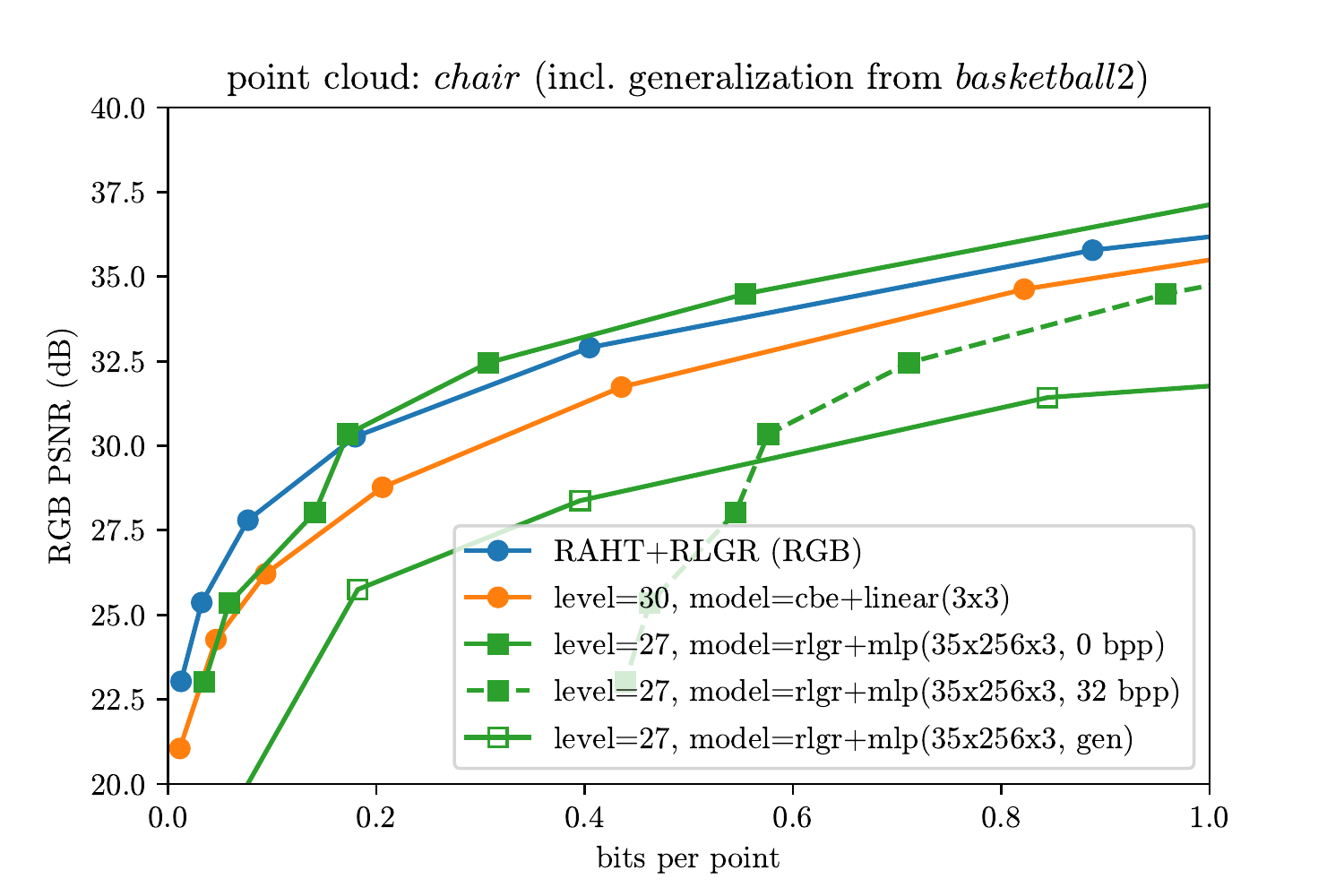}
    \includegraphics[width=0.29\linewidth, trim=20 5 35 15, clip]{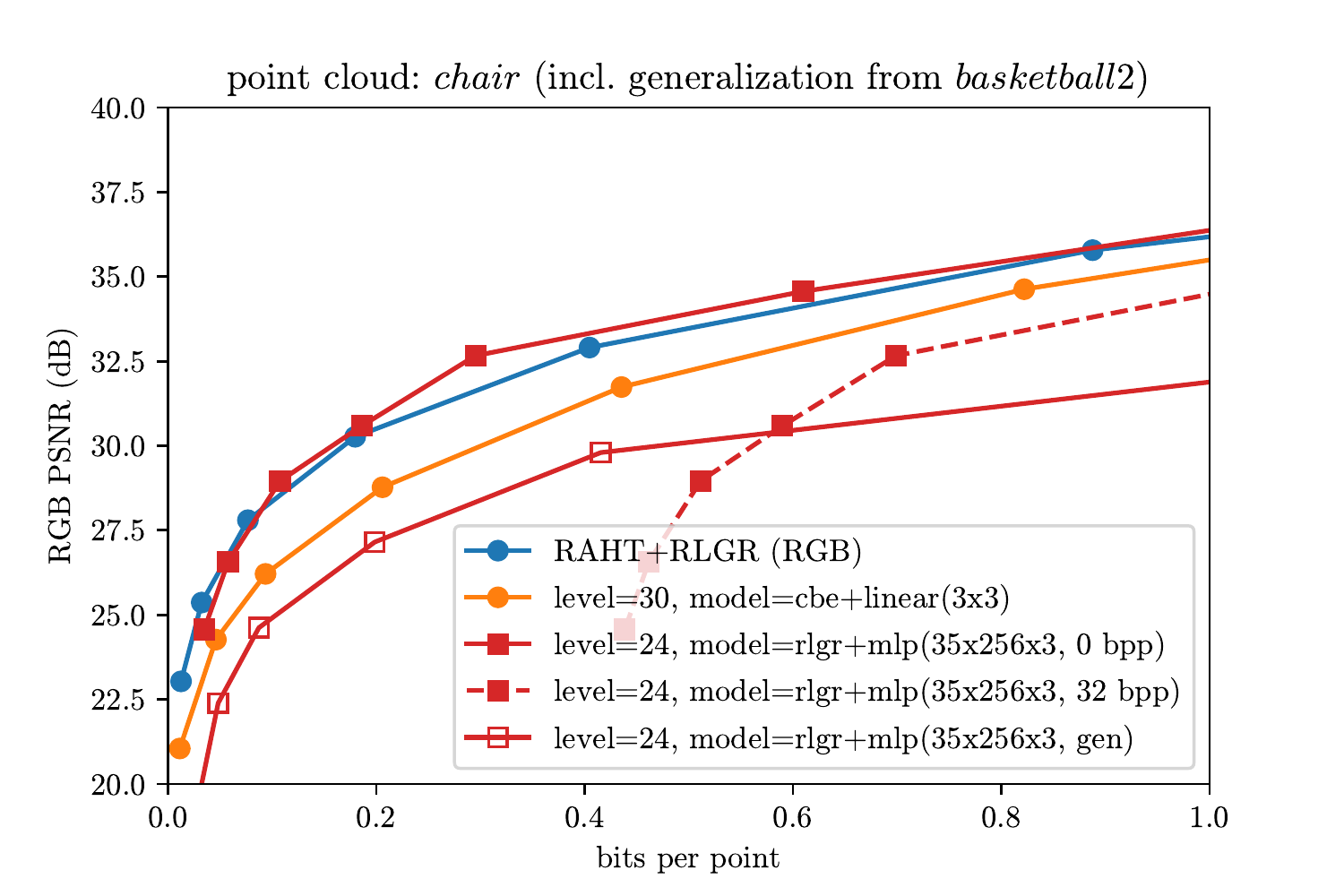}
    \includegraphics[width=0.29\linewidth, trim=20 5 35 15, clip]{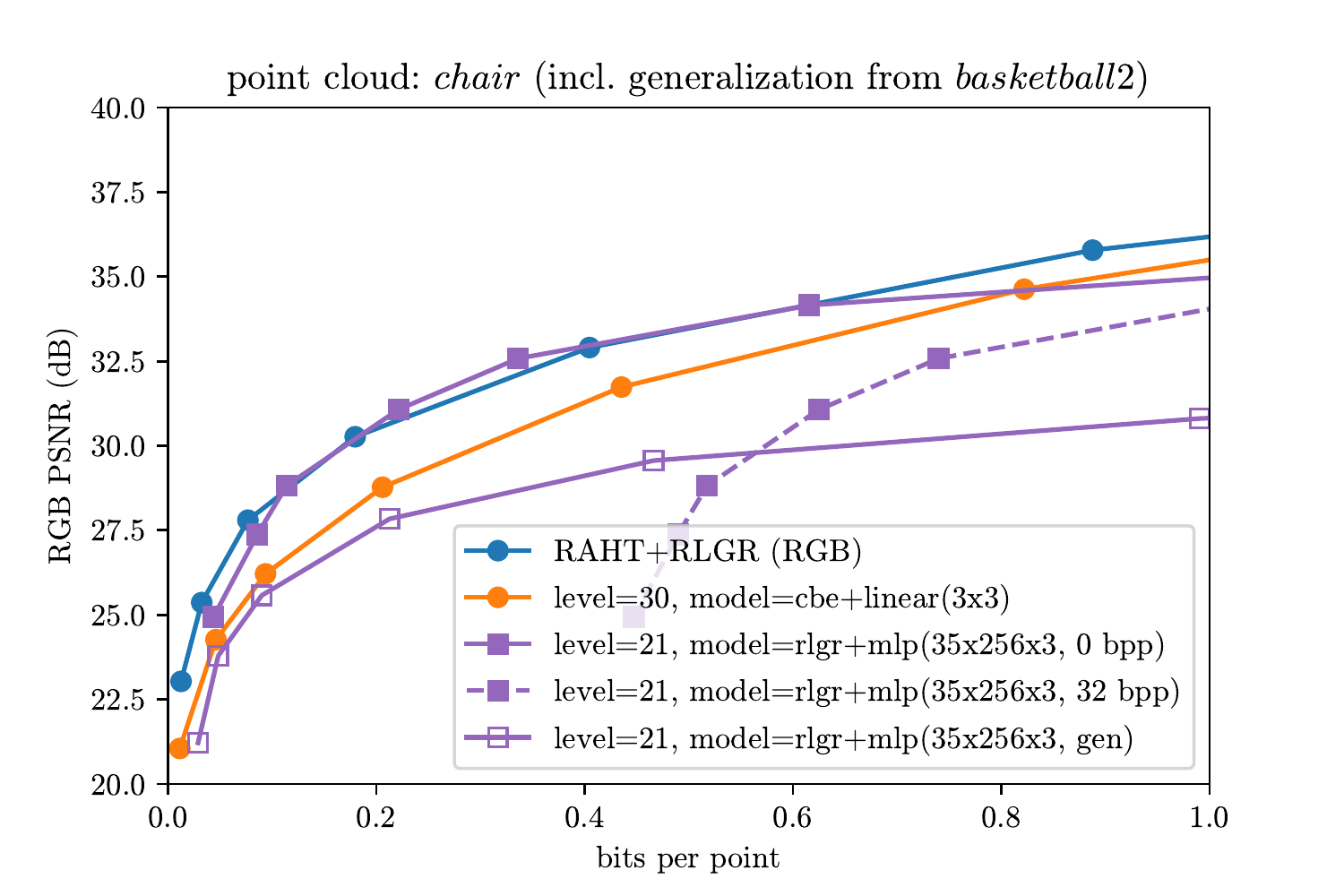}
    
    \includegraphics[width=0.29\linewidth, trim=20 5 35 15, clip]{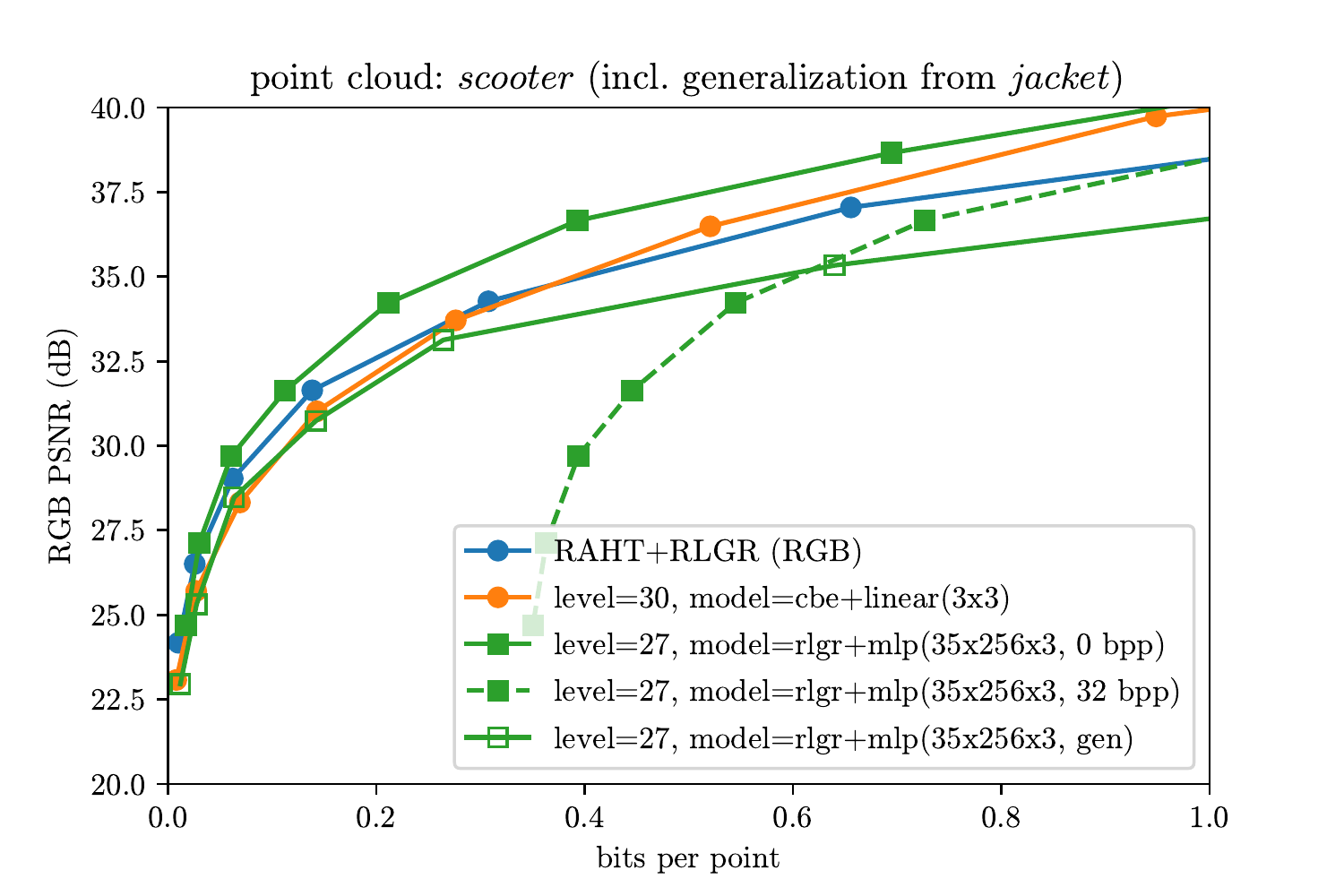}
    \includegraphics[width=0.29\linewidth, trim=20 5 35 15, clip]{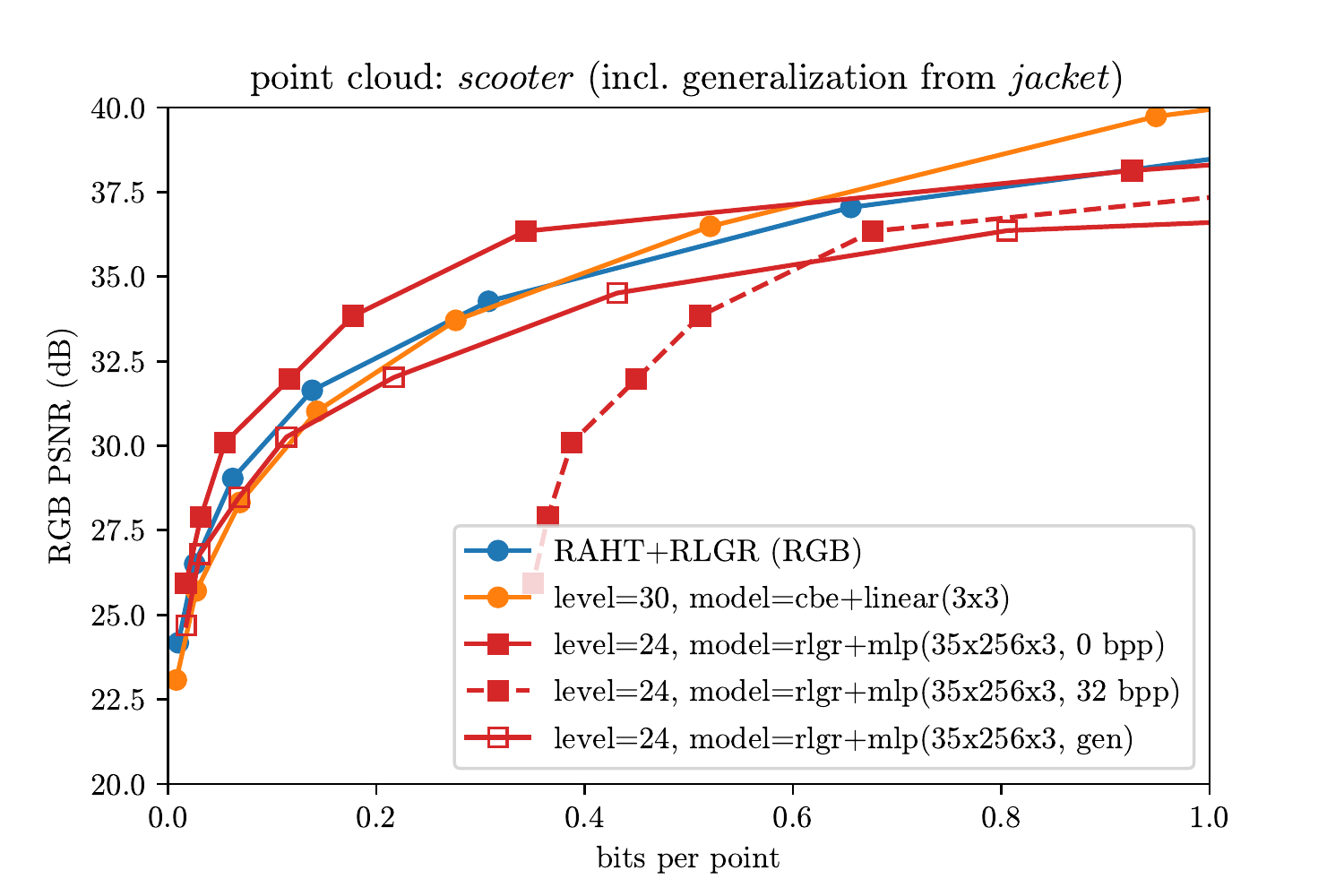}
    \includegraphics[width=0.29\linewidth, trim=20 5 35 15, clip]{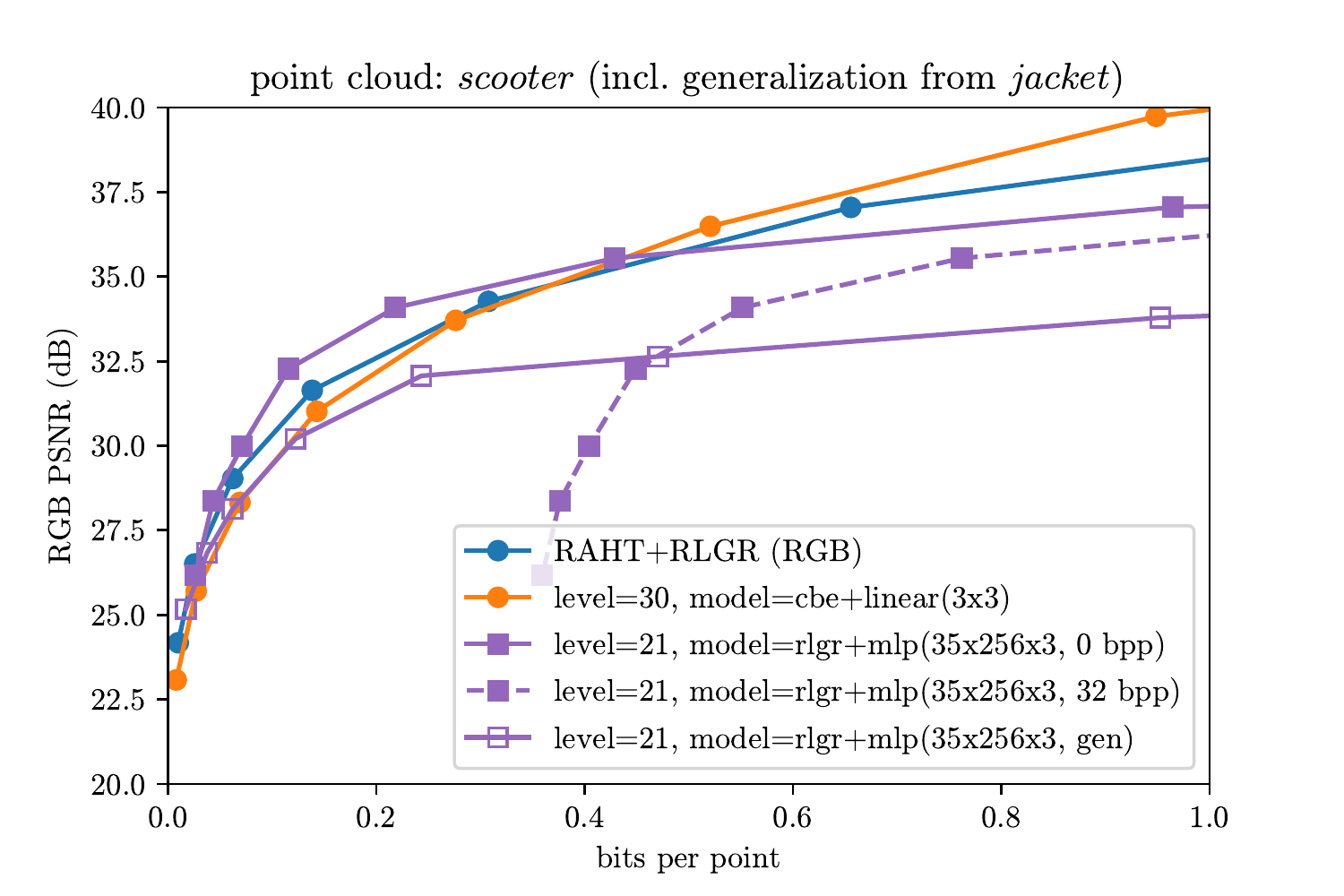}
    
    \includegraphics[width=0.29\linewidth, trim=20 5 35 15, clip]{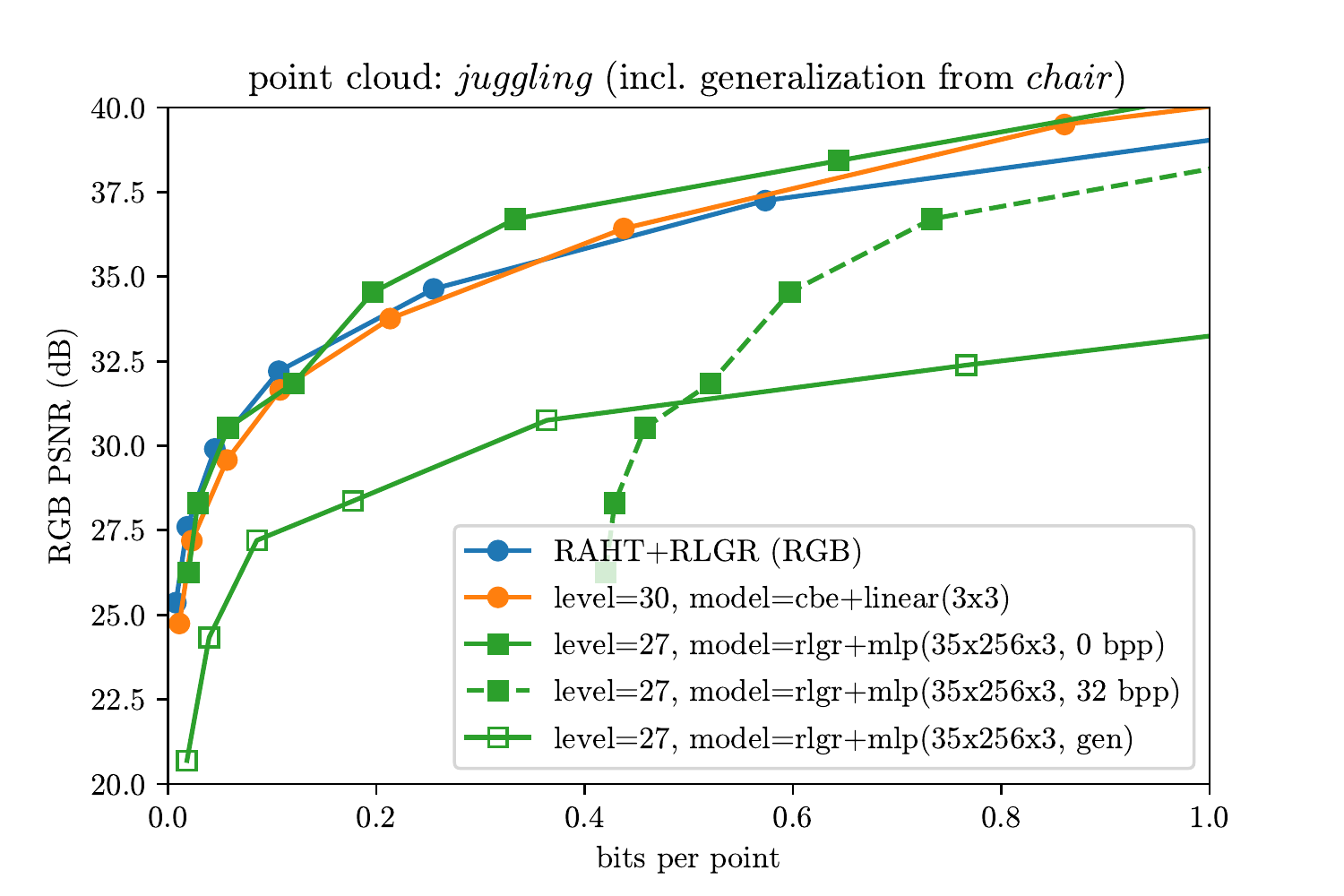}
    \includegraphics[width=0.29\linewidth, trim=20 5 35 15, clip]{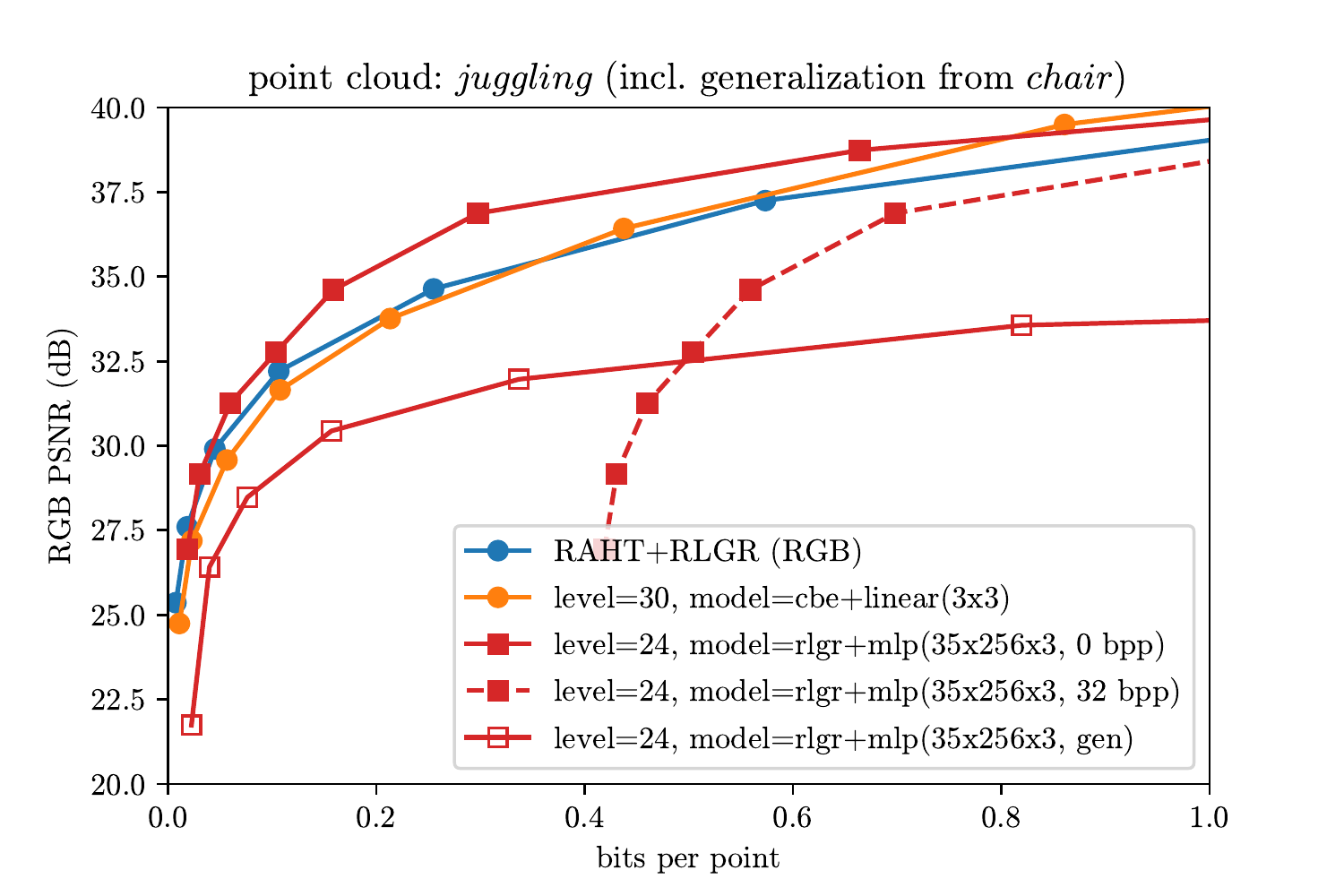}
    \includegraphics[width=0.29\linewidth, trim=20 5 35 15, clip]{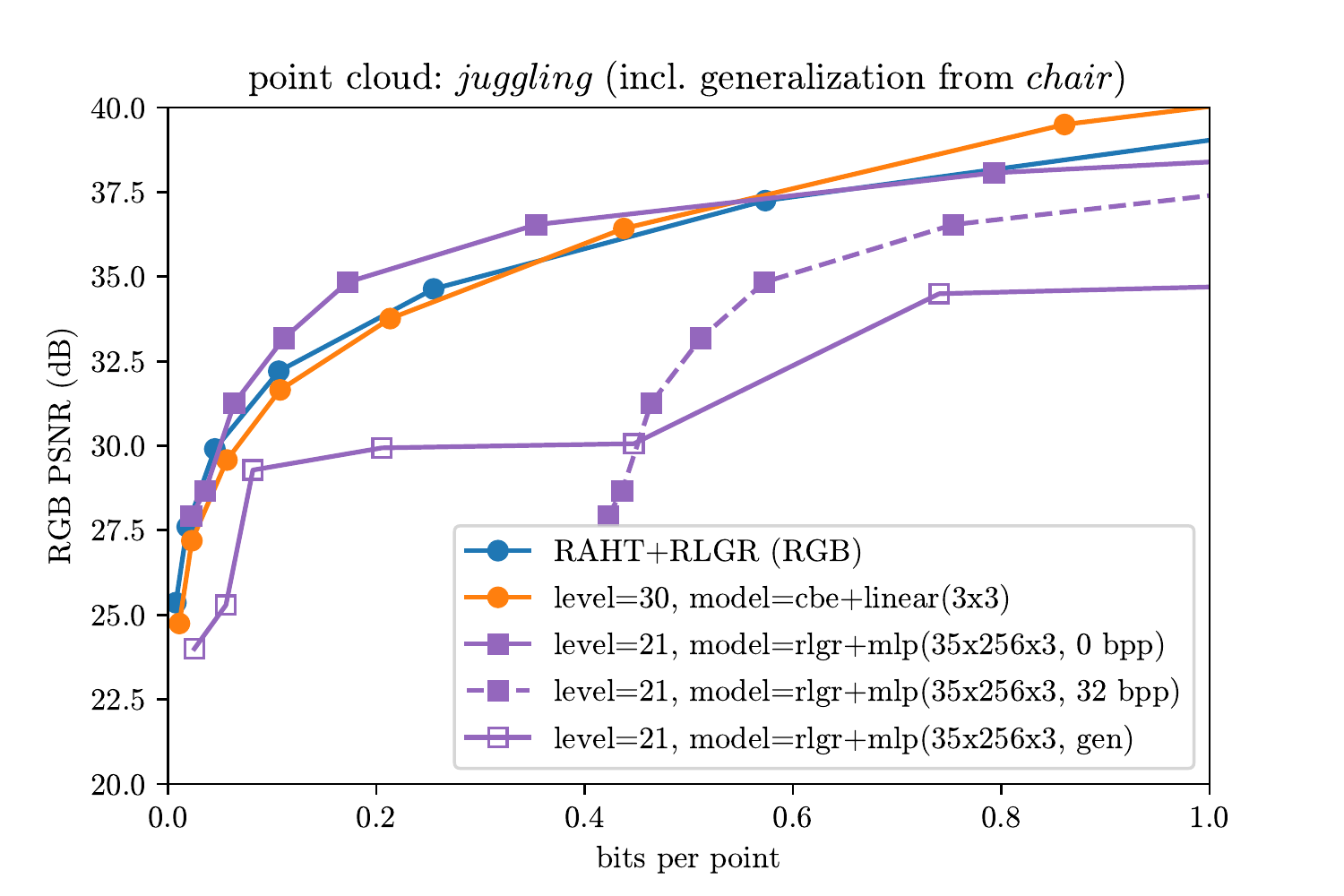}
    
    \includegraphics[width=0.29\linewidth, trim=20 5 35 15, clip]{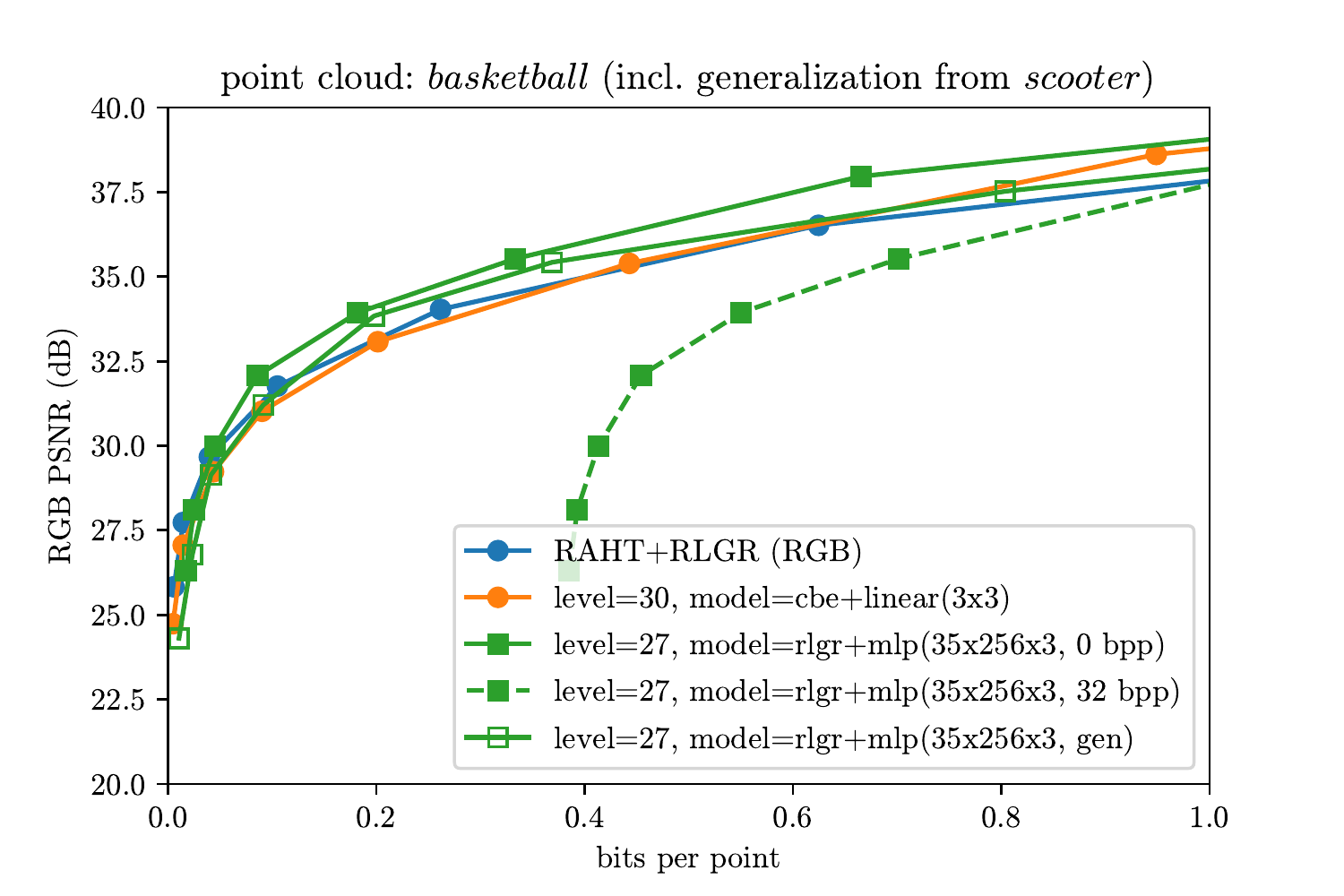}
    \includegraphics[width=0.29\linewidth, trim=20 5 35 15, clip]{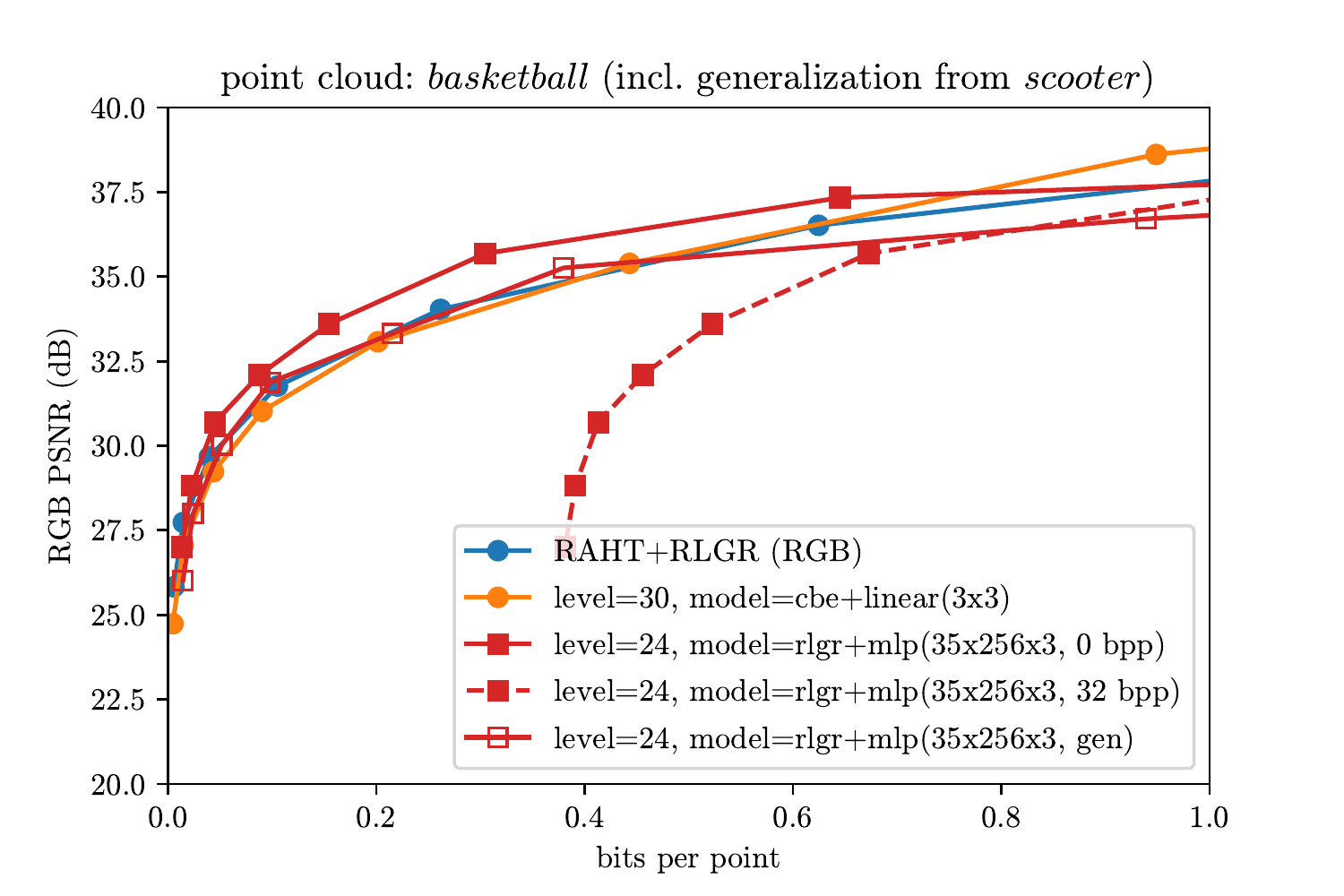}
    \includegraphics[width=0.29\linewidth, trim=20 5 35 15, clip]{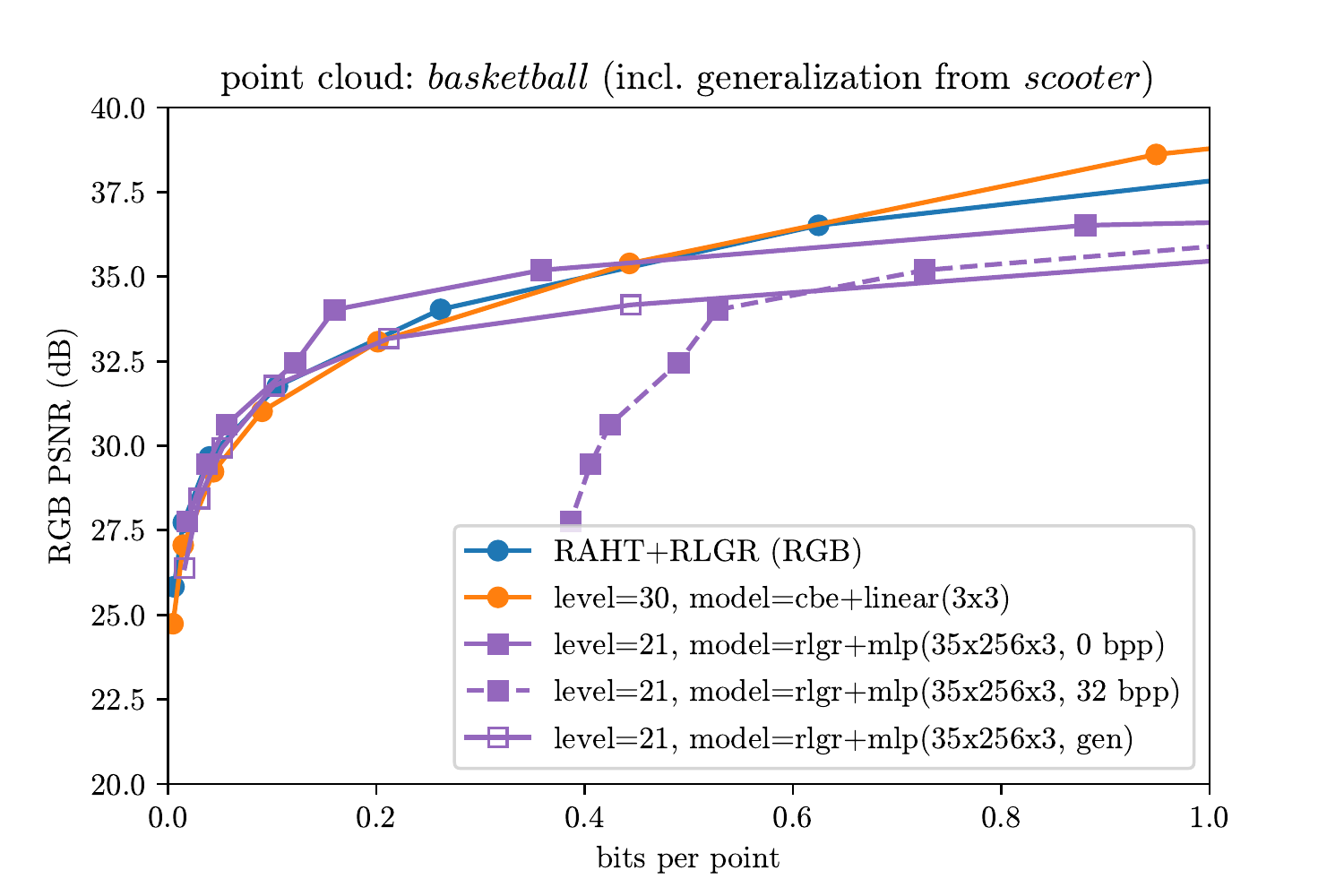}
    
    \includegraphics[width=0.29\linewidth, trim=20 5 35 15, clip]{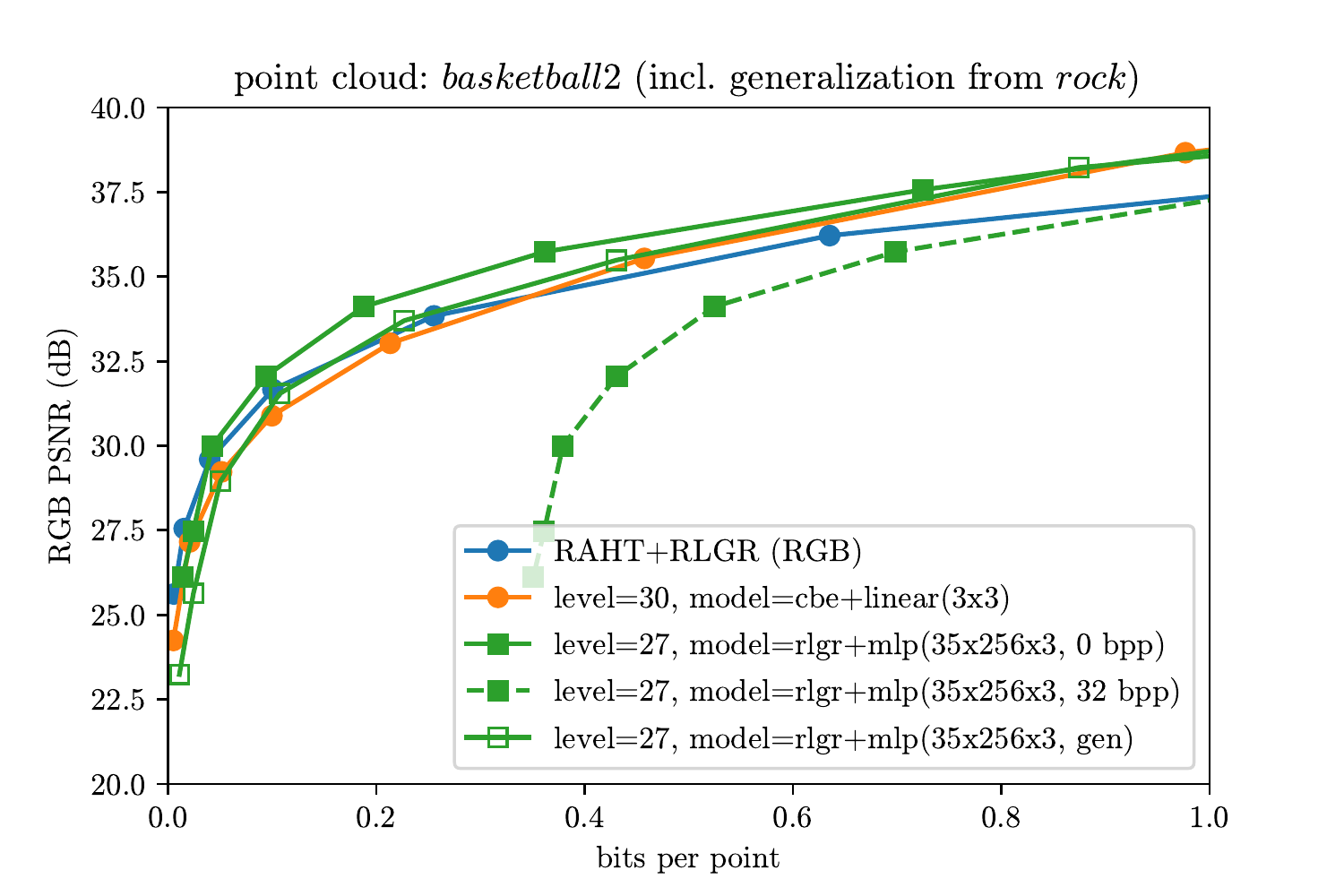}
    \includegraphics[width=0.29\linewidth, trim=20 5 35 15, clip]{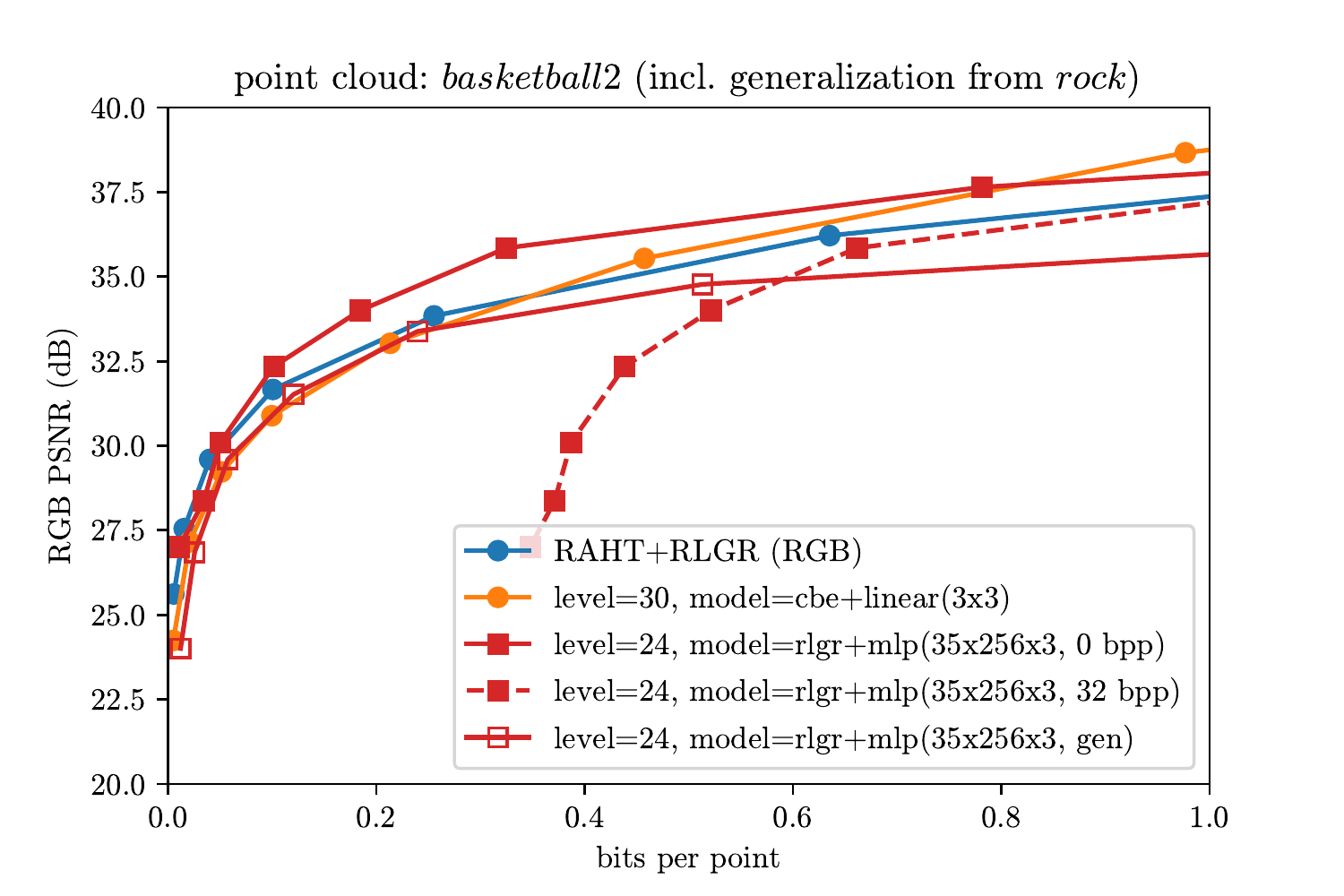}
    \includegraphics[width=0.29\linewidth, trim=20 5 35 15, clip]{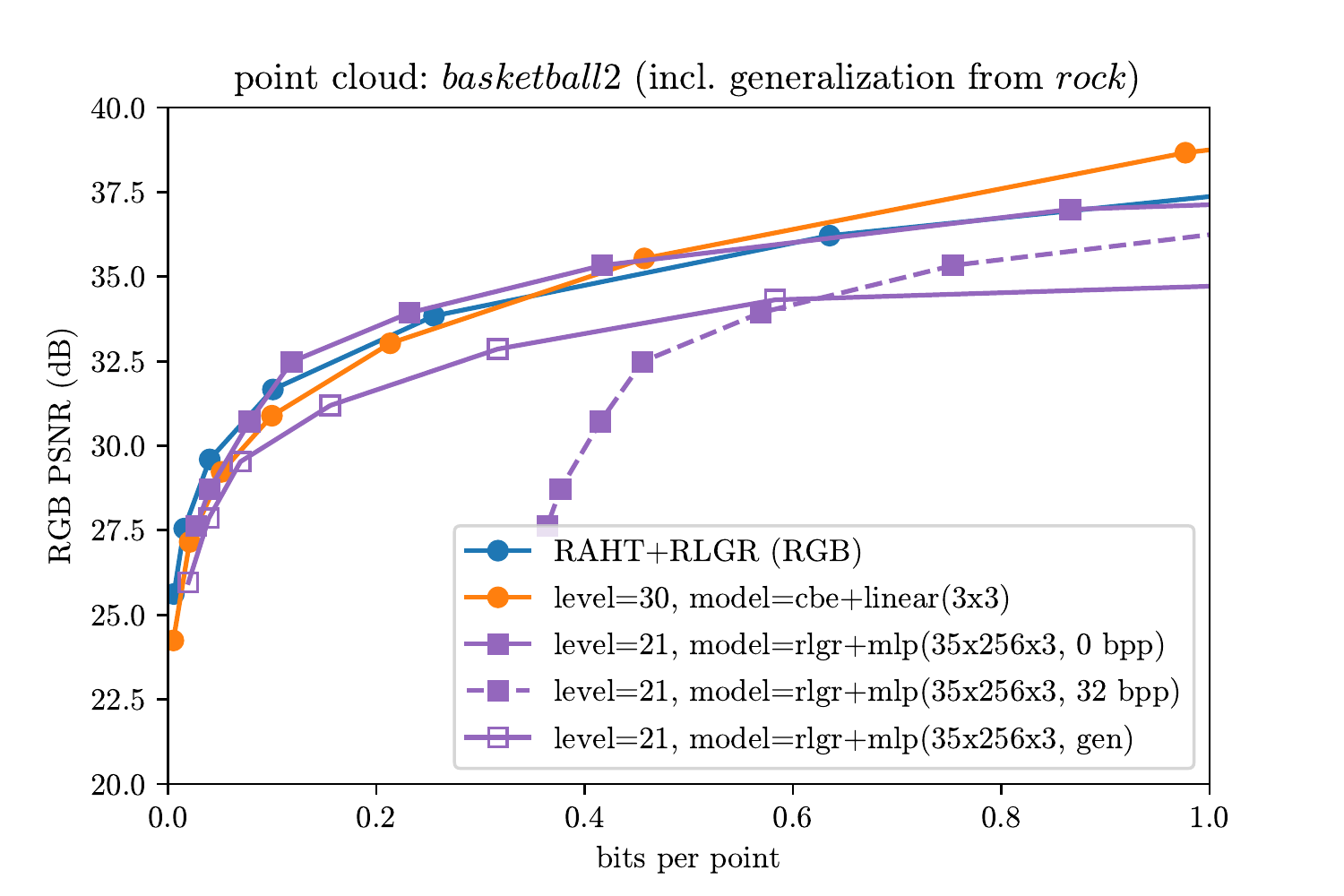}
    
    \includegraphics[width=0.29\linewidth, trim=20 5 35 15, clip]{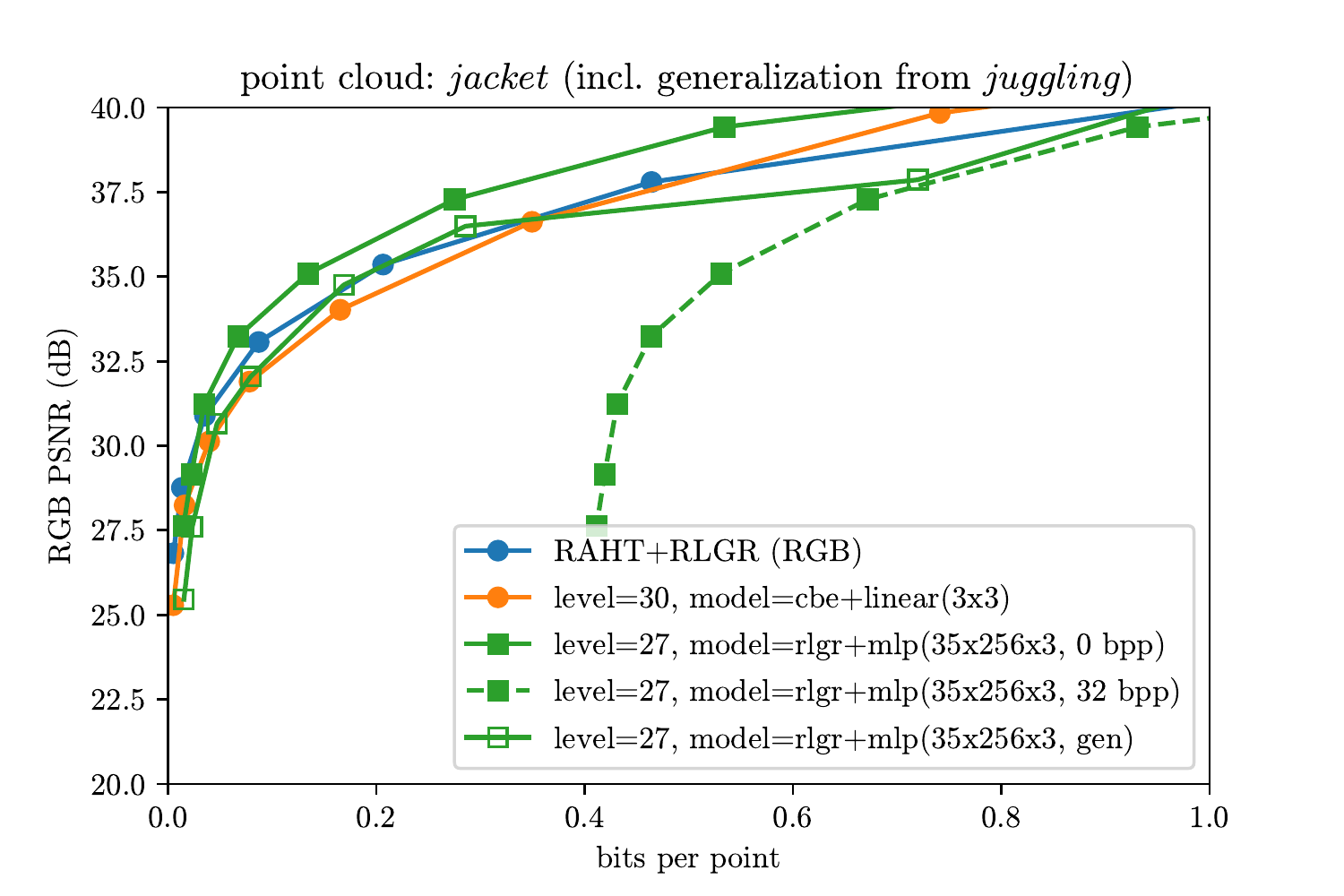}
    \includegraphics[width=0.29\linewidth, trim=20 5 35 15, clip]{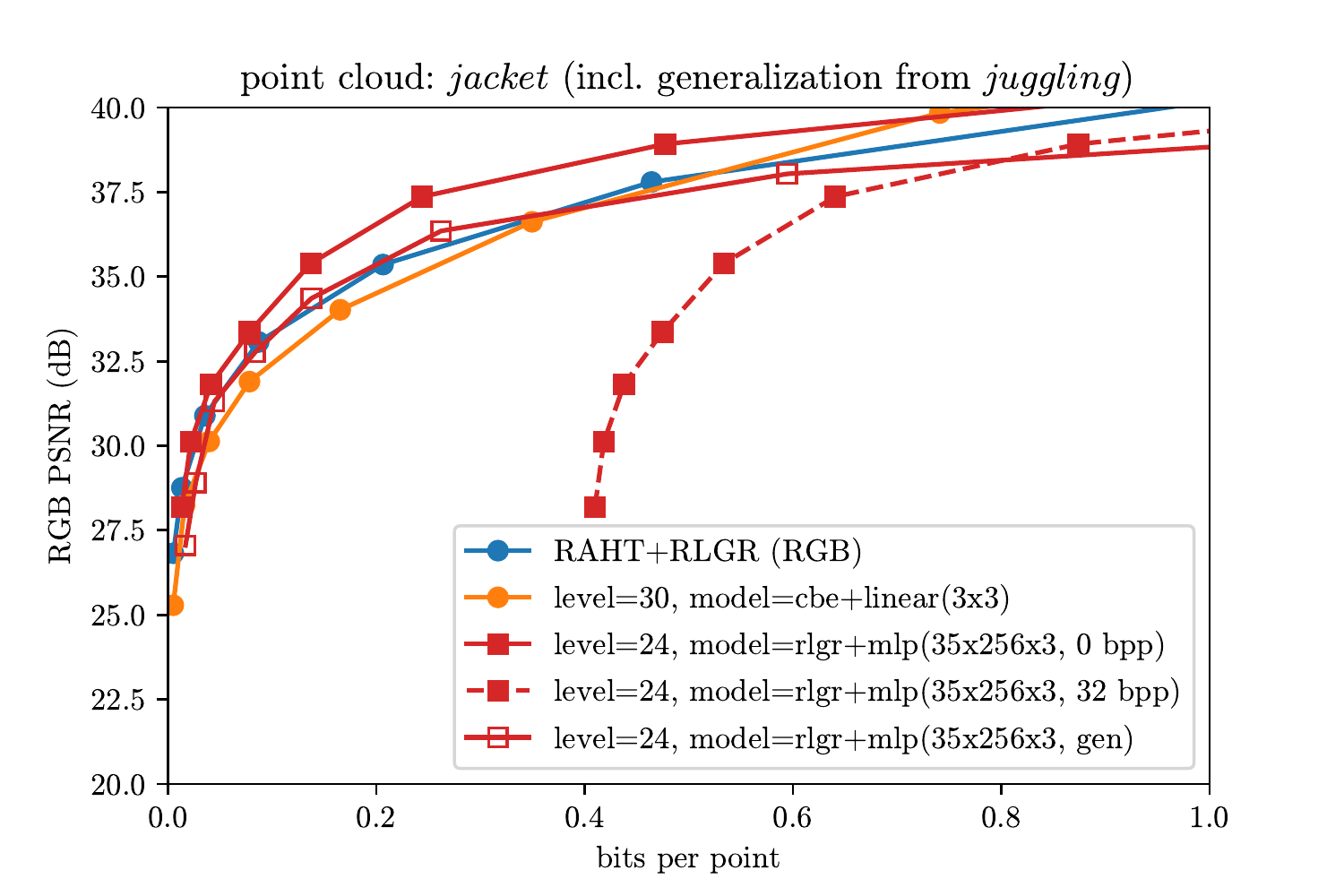}
    \includegraphics[width=0.29\linewidth, trim=20 5 35 15, clip]{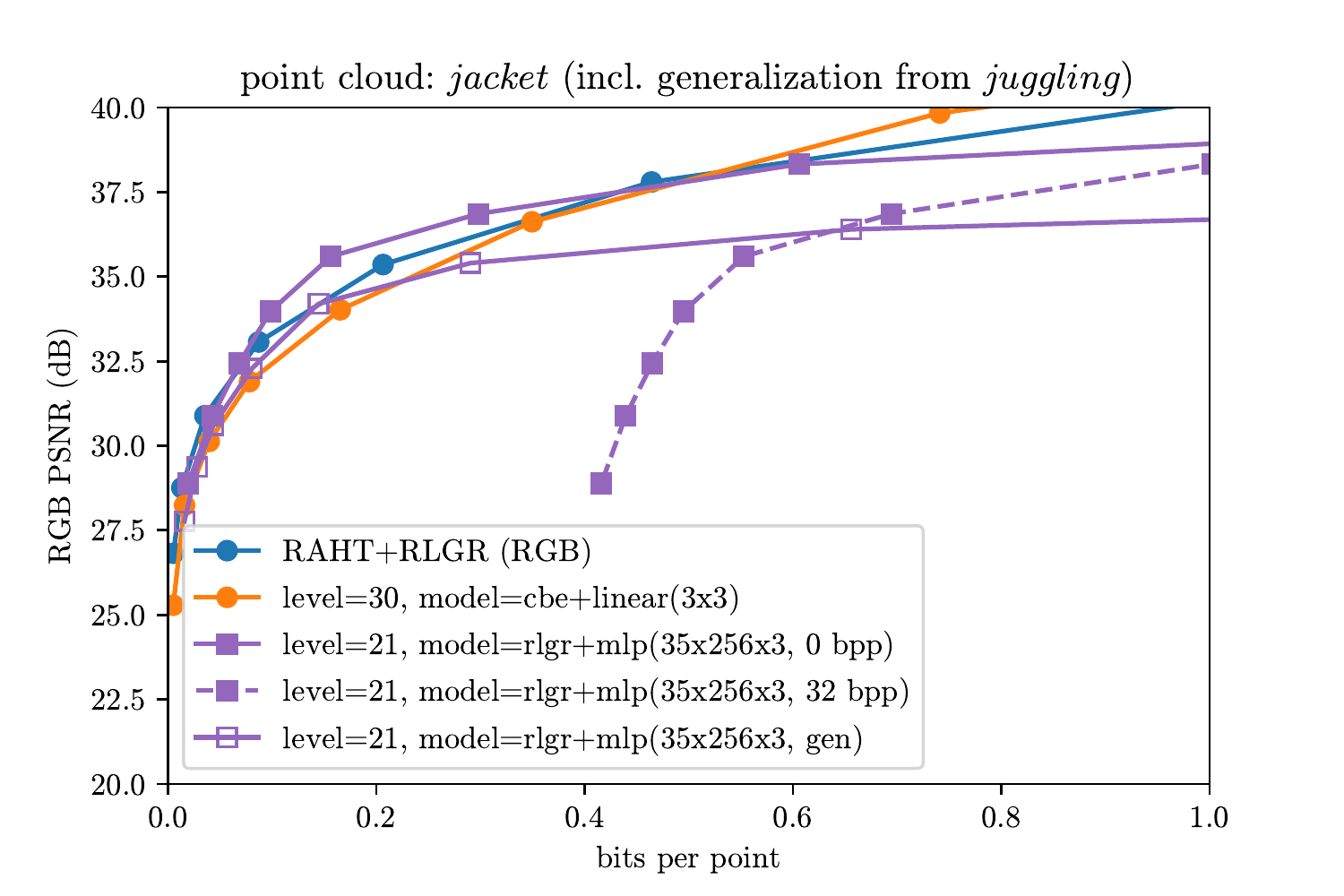}
    
    \caption{Effect of side information for coordinate based network {\em mlp(35x256x3)} at levels 27 (left), 24 (middle), and 21 (right).  Each row is a different point cloud.  See \cref{fig:sideinfo_mlp256} point cloud {\em rock}.}
    \label{fig:sideinfo_mlp256_other}
\end{figure*}

\begin{figure*}
    \centering
    \includegraphics[width=0.29\linewidth, trim=20 5 35 15, clip]{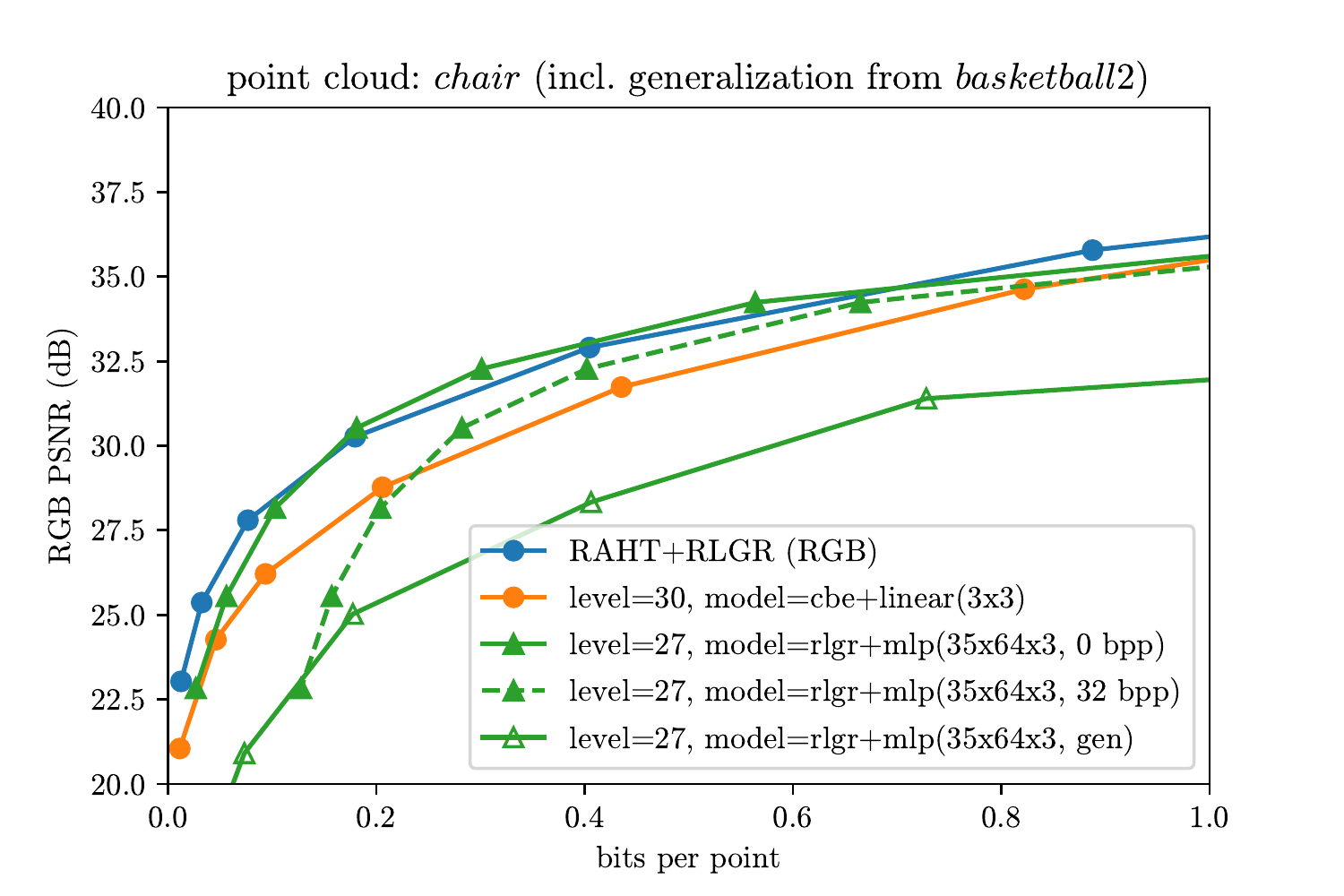}
    \includegraphics[width=0.29\linewidth, trim=20 5 35 15, clip]{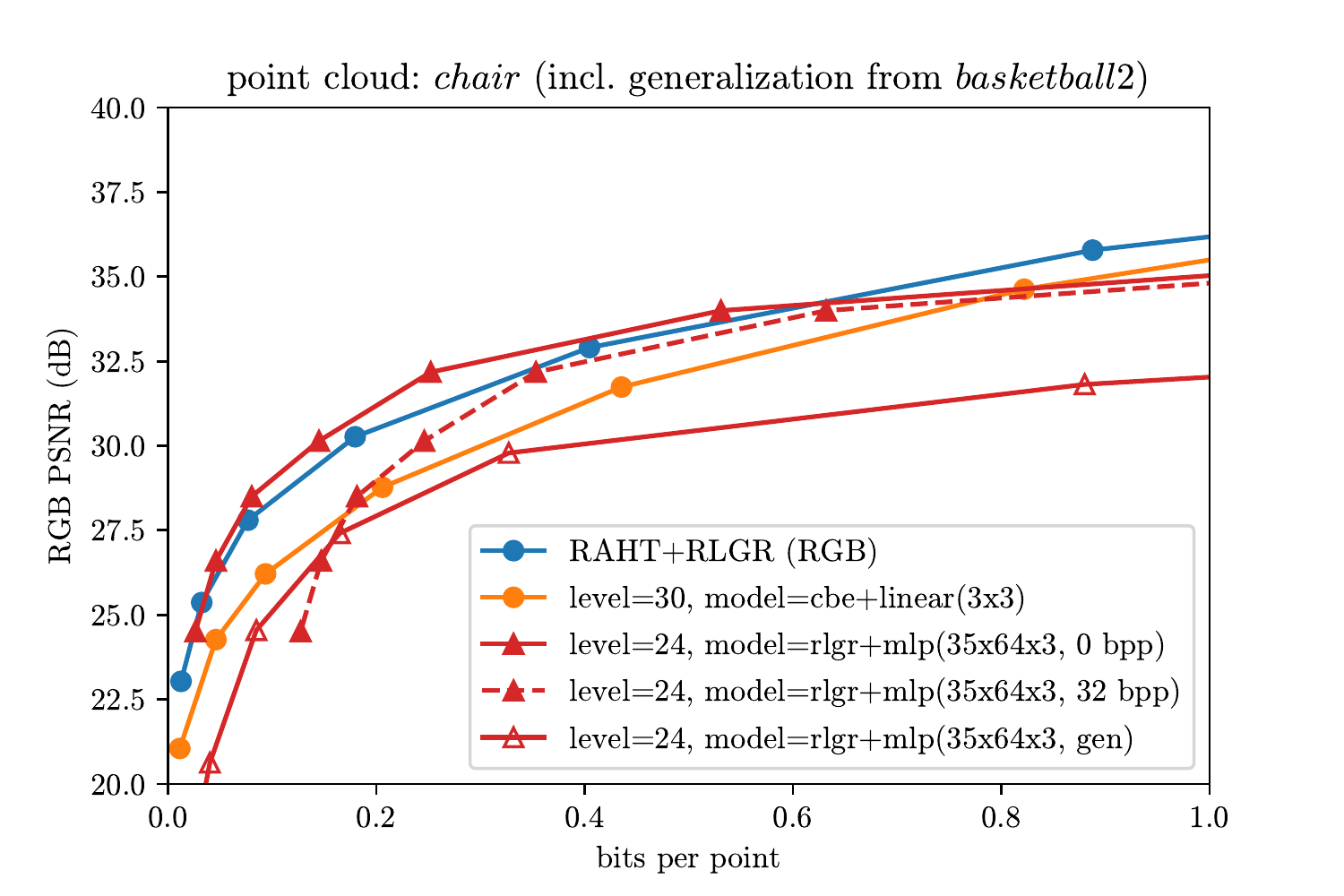}
    \includegraphics[width=0.29\linewidth, trim=20 5 35 15, clip]{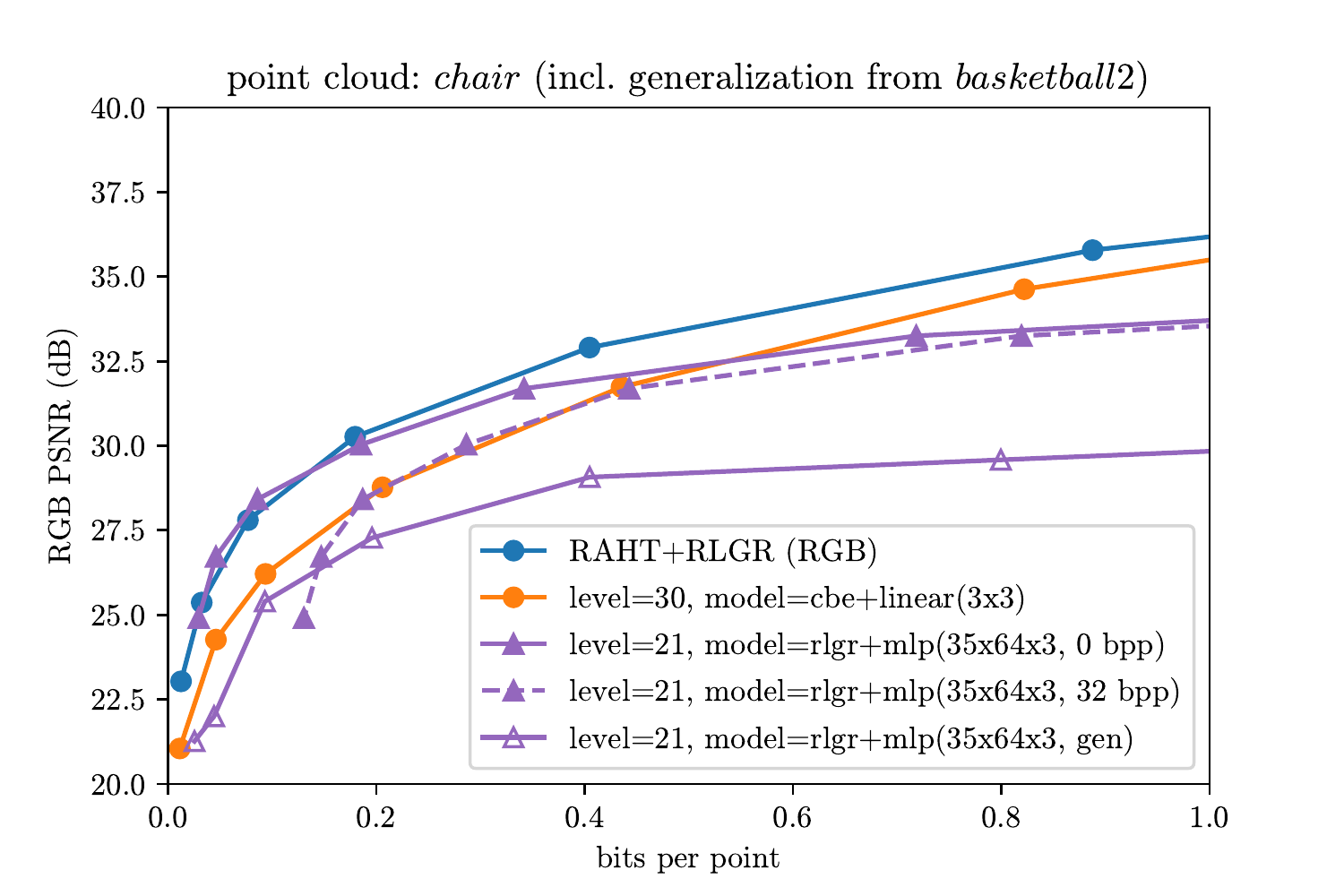}
    
    \includegraphics[width=0.29\linewidth, trim=20 5 35 15, clip]{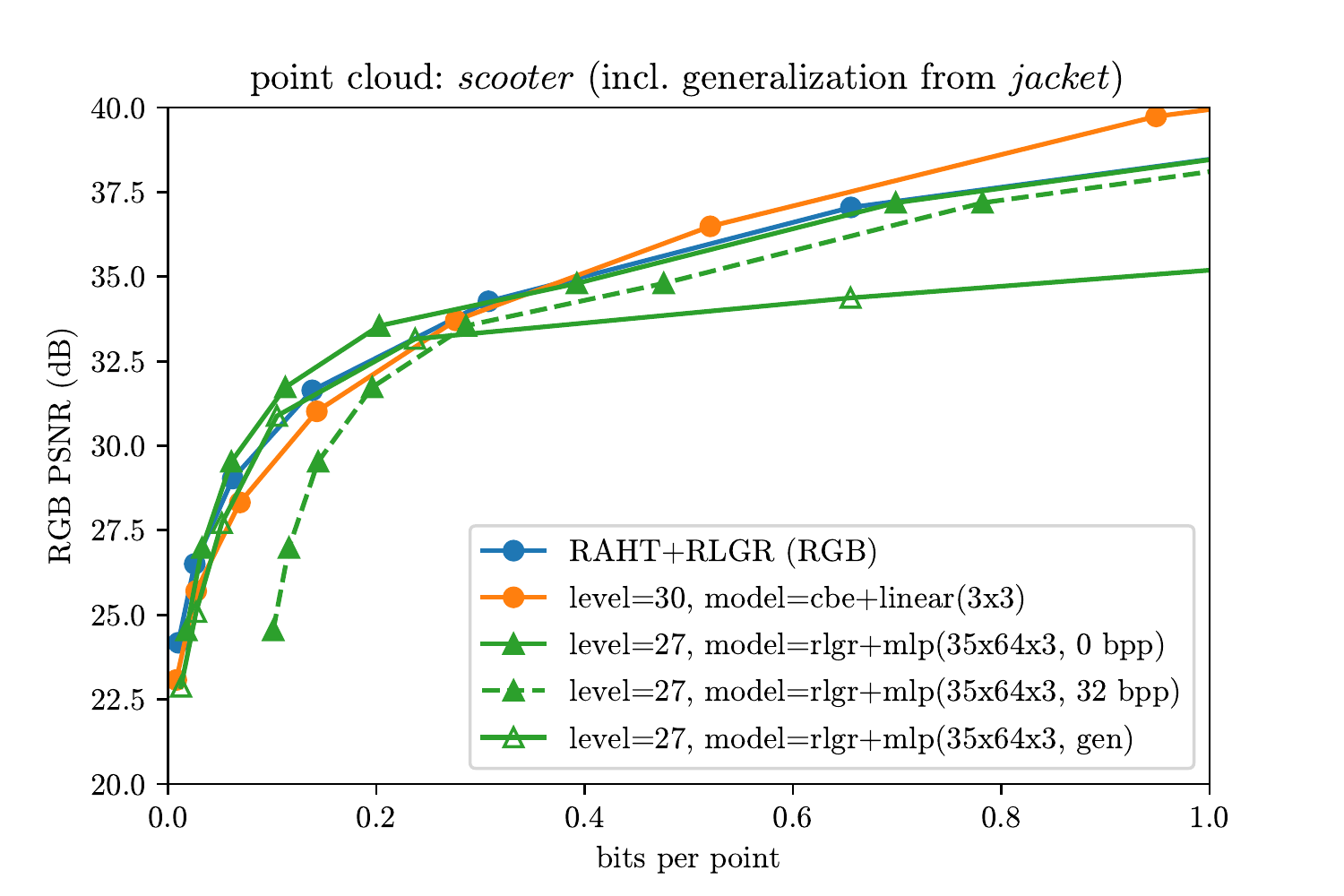}
    \includegraphics[width=0.29\linewidth, trim=20 5 35 15, clip]{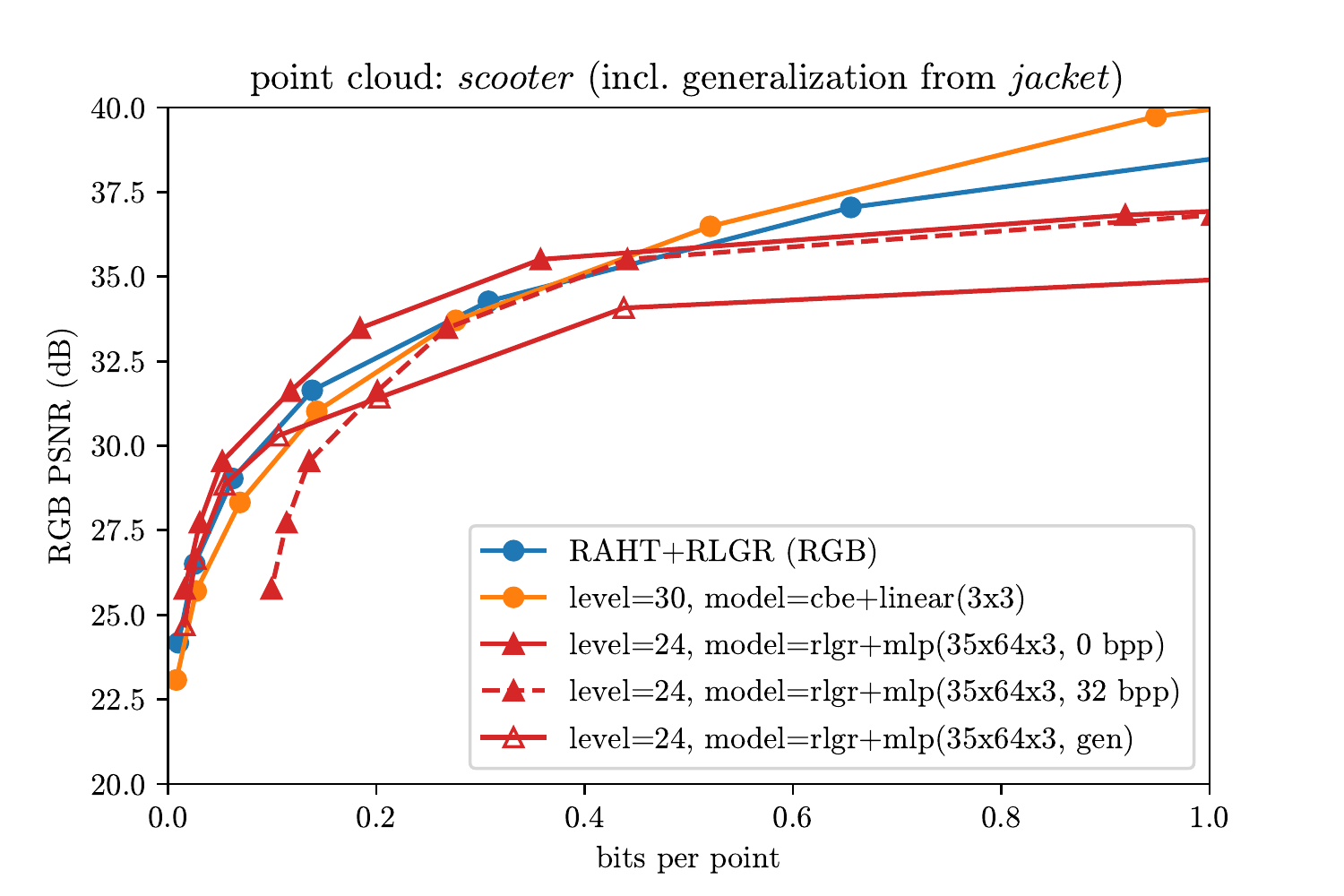}
    \includegraphics[width=0.29\linewidth, trim=20 5 35 15, clip]{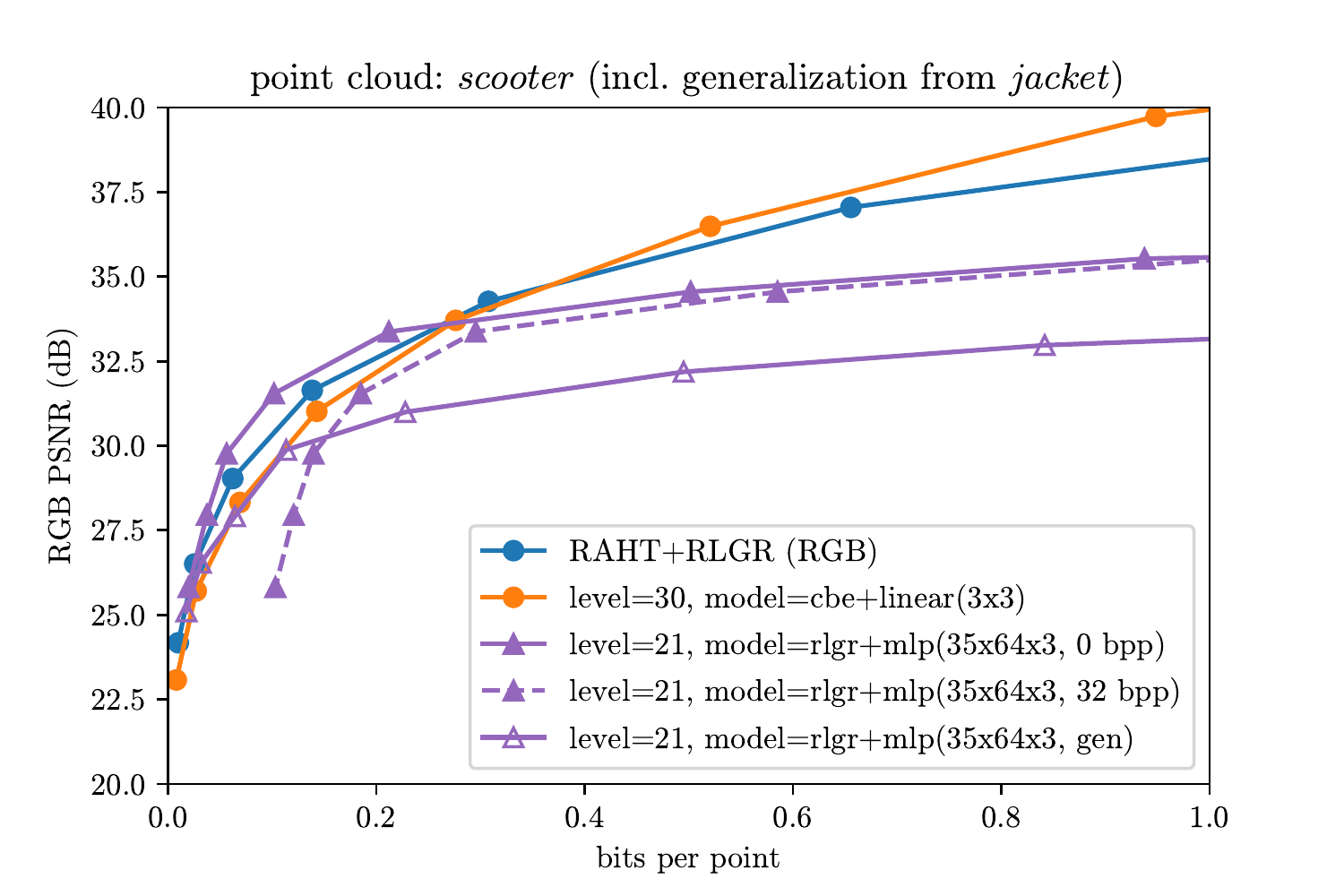}
    
    \includegraphics[width=0.29\linewidth, trim=20 5 35 15, clip]{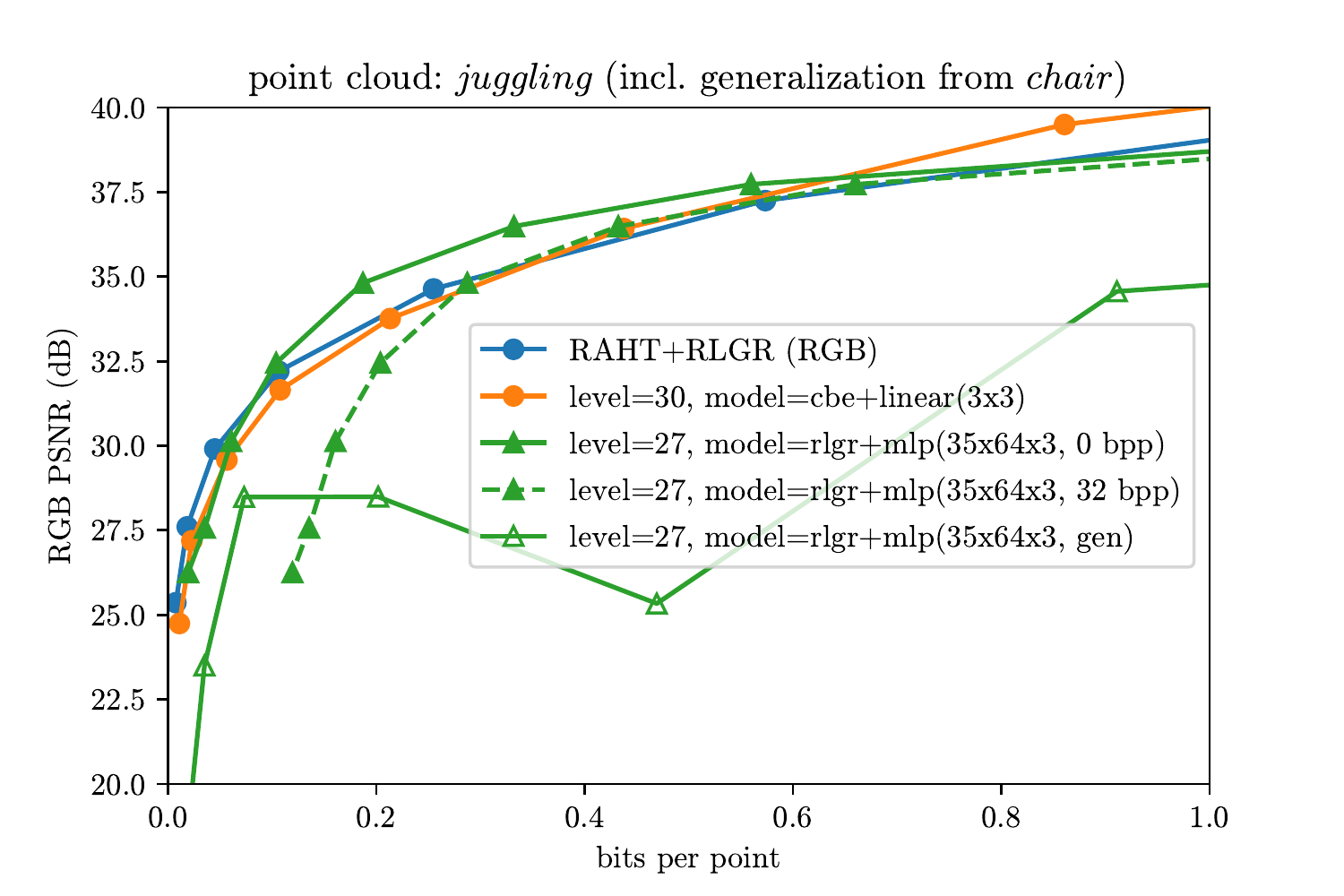}
    \includegraphics[width=0.29\linewidth, trim=20 5 35 15, clip]{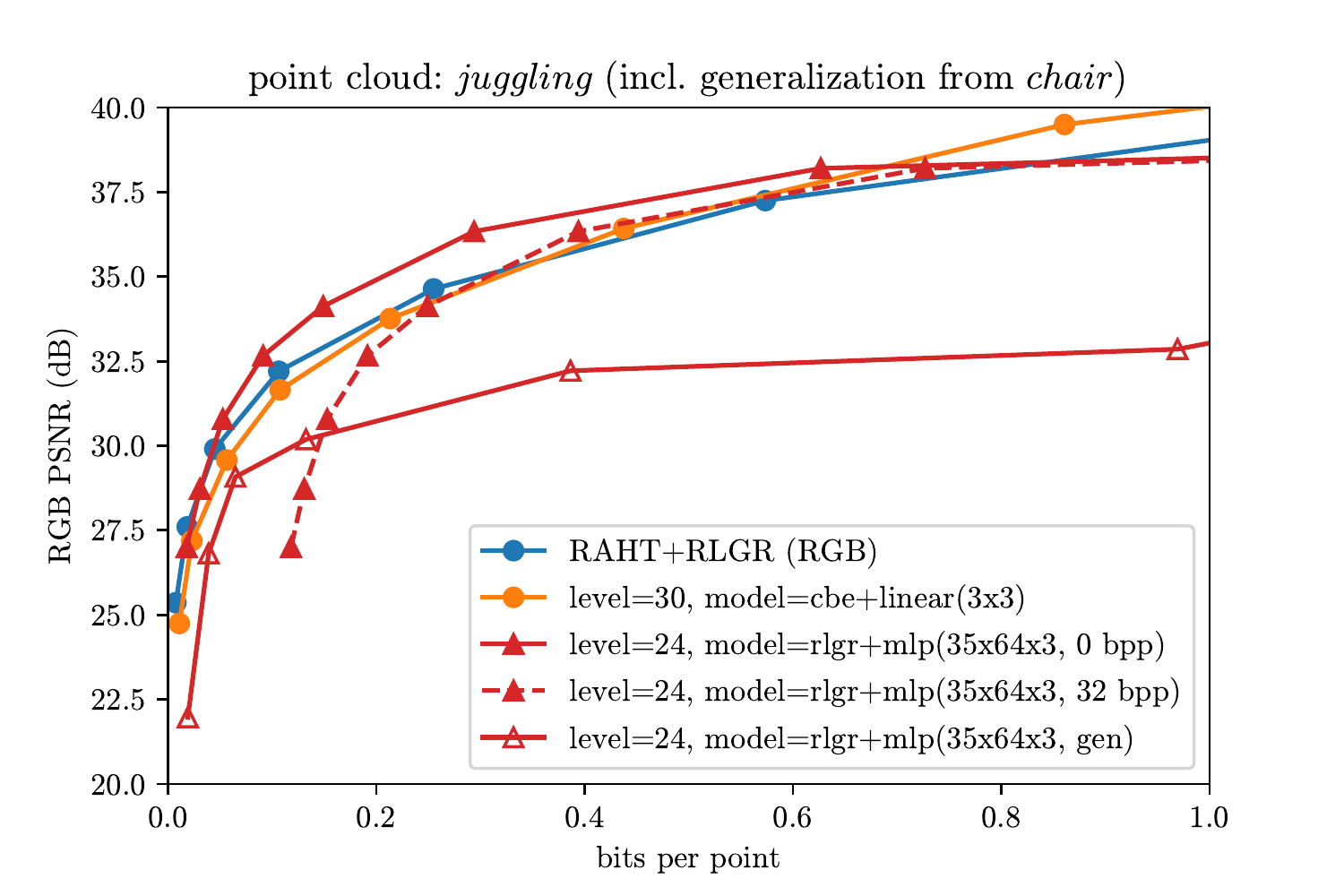}
    \includegraphics[width=0.29\linewidth, trim=20 5 35 15, clip]{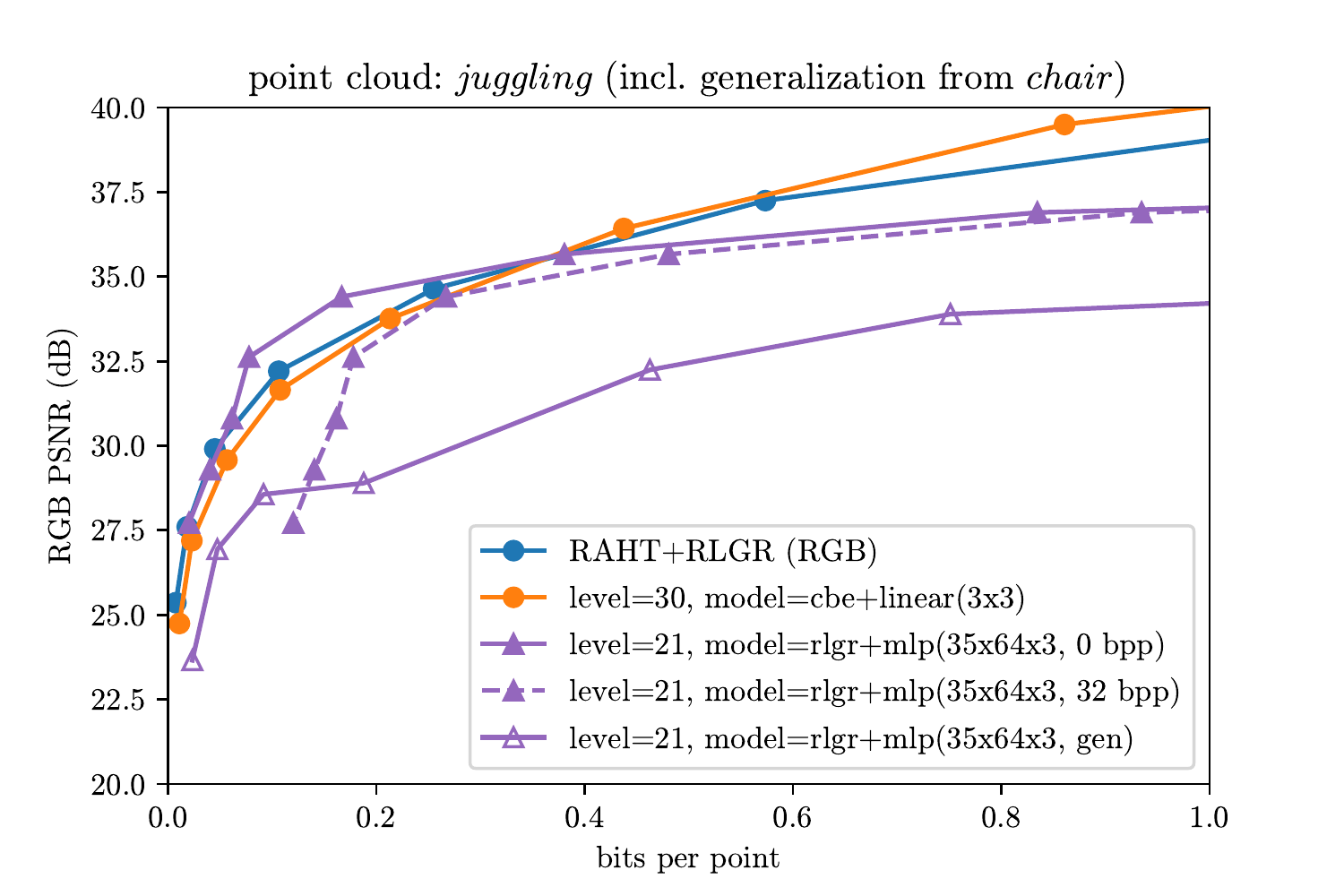}
    
    \includegraphics[width=0.29\linewidth, trim=20 5 35 15, clip]{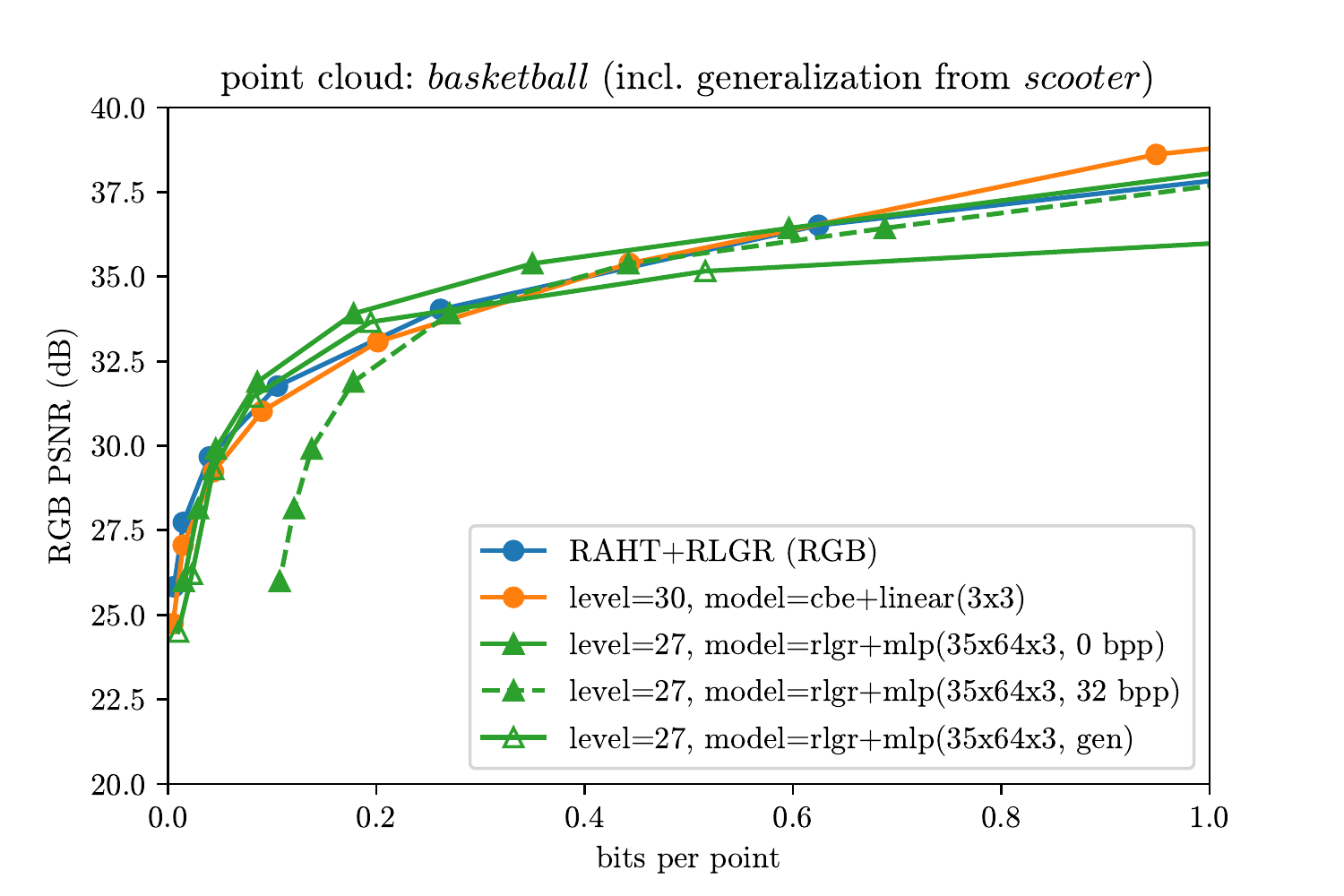}
    \includegraphics[width=0.29\linewidth, trim=20 5 35 15, clip]{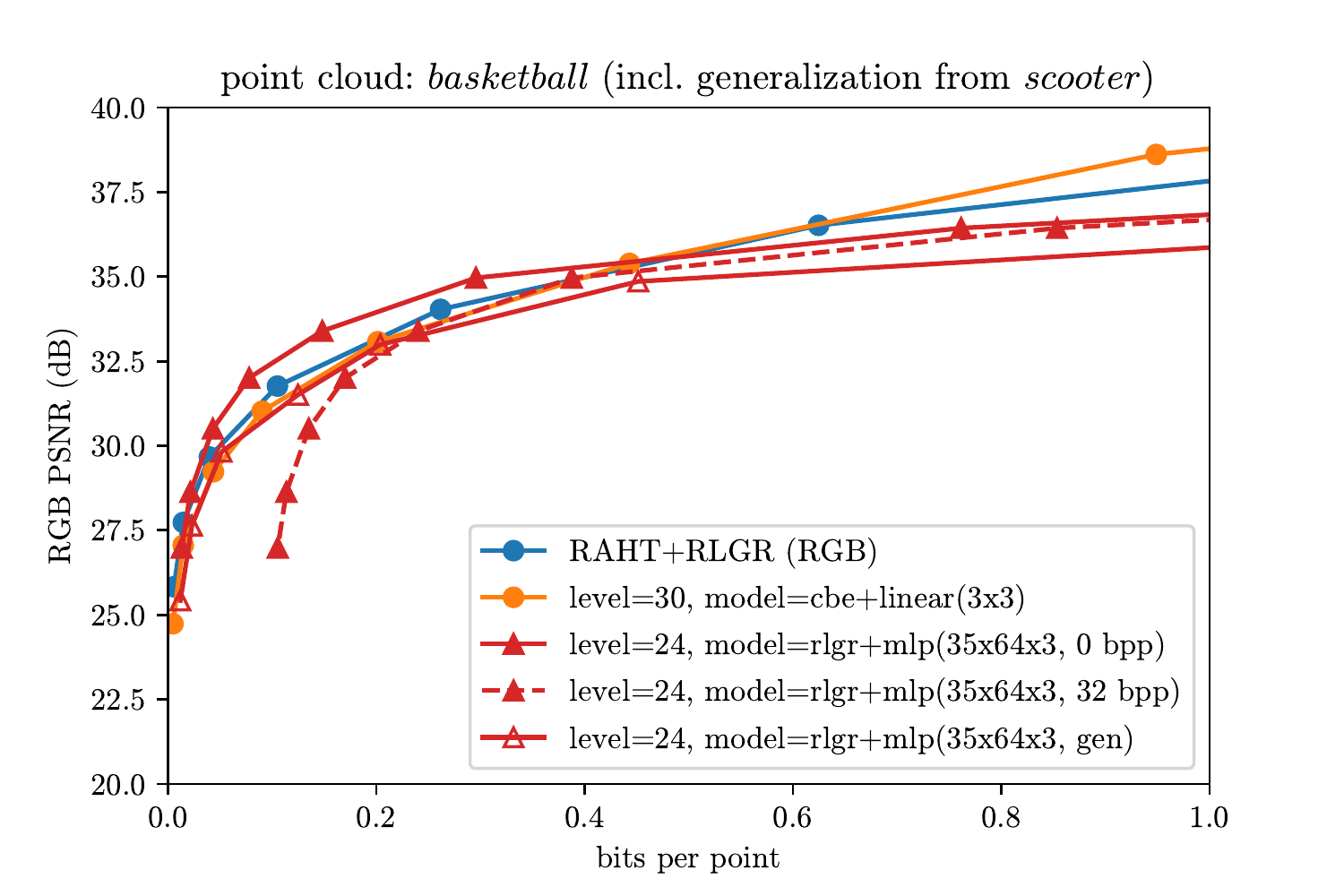}
    \includegraphics[width=0.29\linewidth, trim=20 5 35 15, clip]{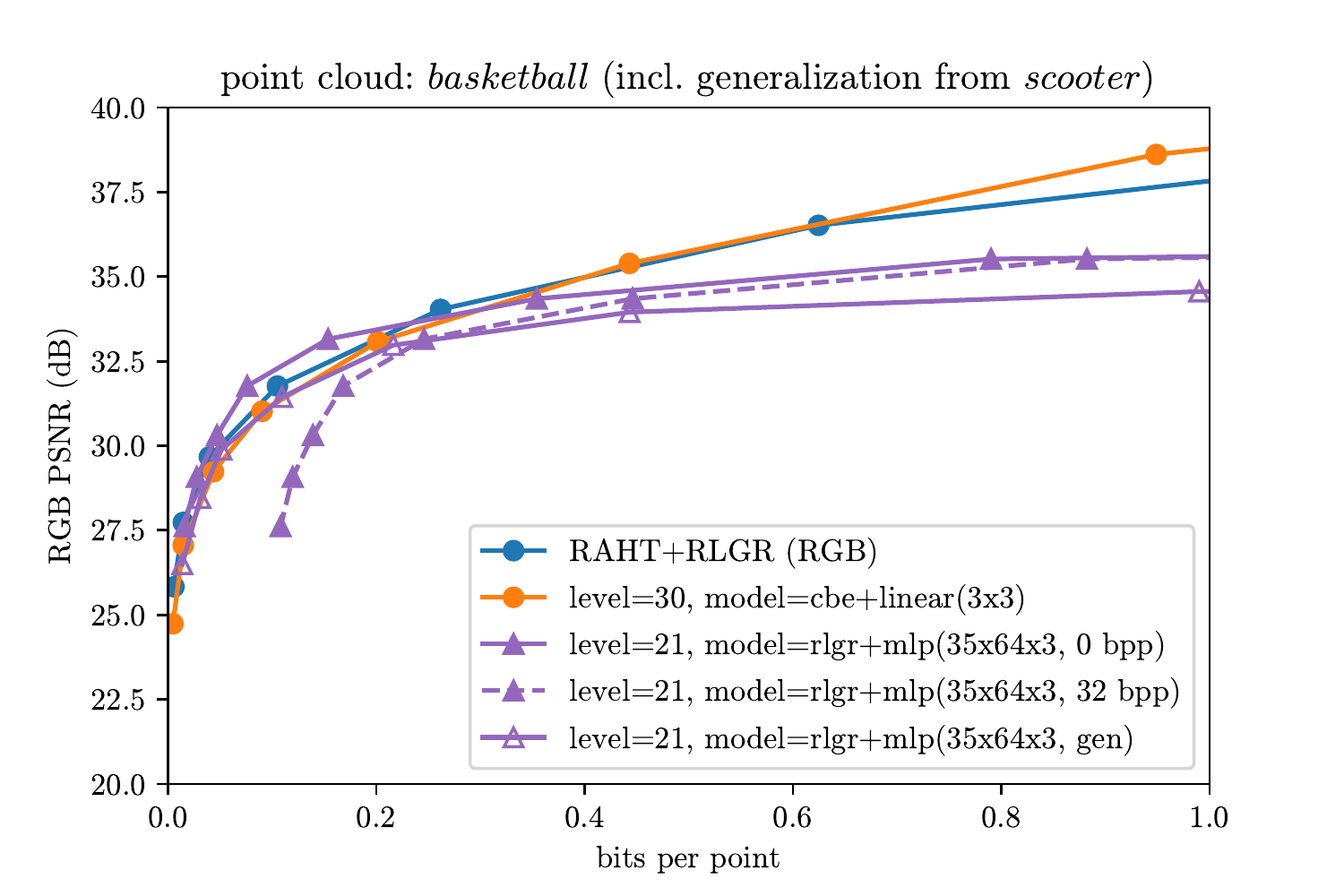}
    
    \includegraphics[width=0.29\linewidth, trim=20 5 35 15, clip]{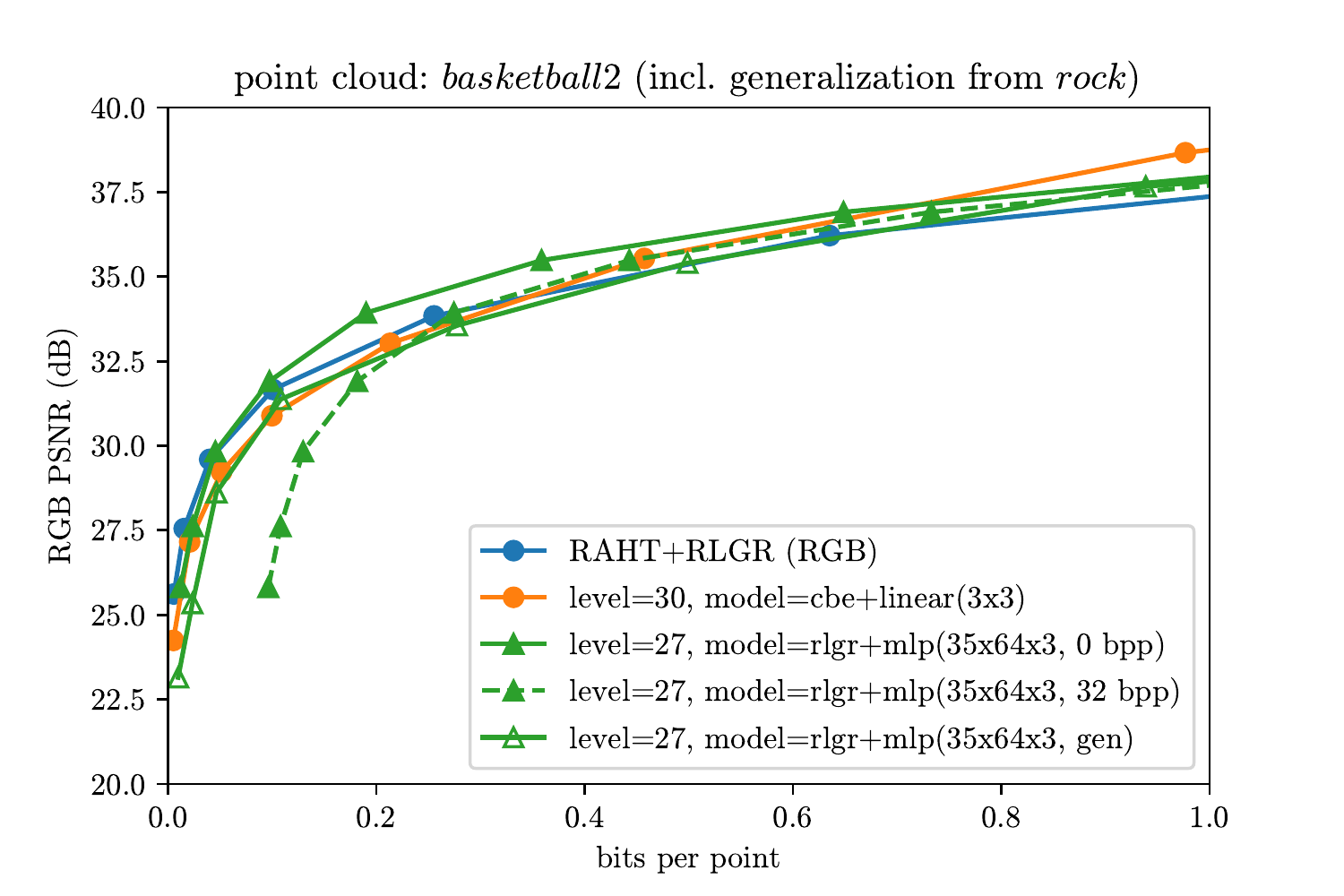}
    \includegraphics[width=0.29\linewidth, trim=20 5 35 15, clip]{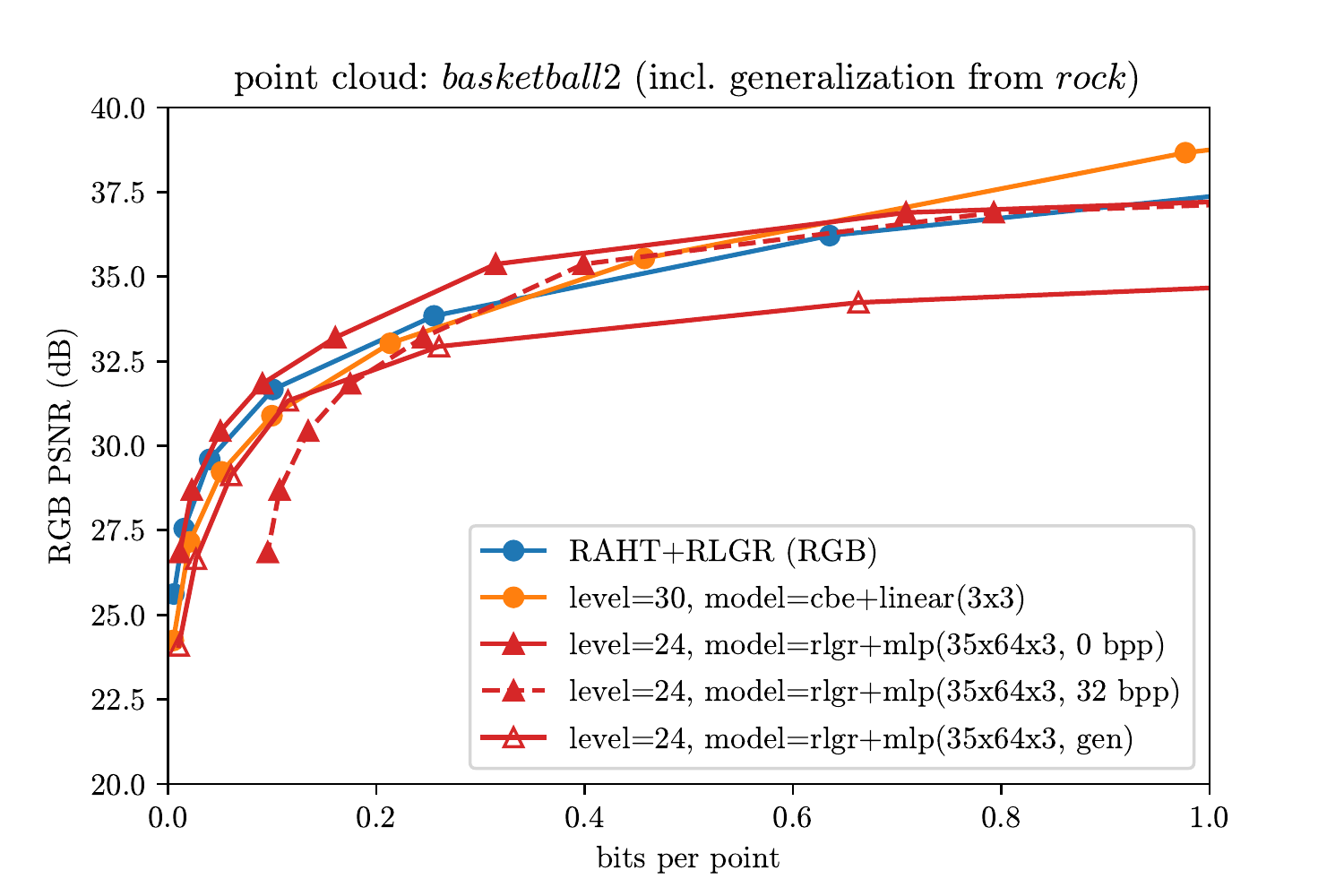}
    \includegraphics[width=0.29\linewidth, trim=20 5 35 15, clip]{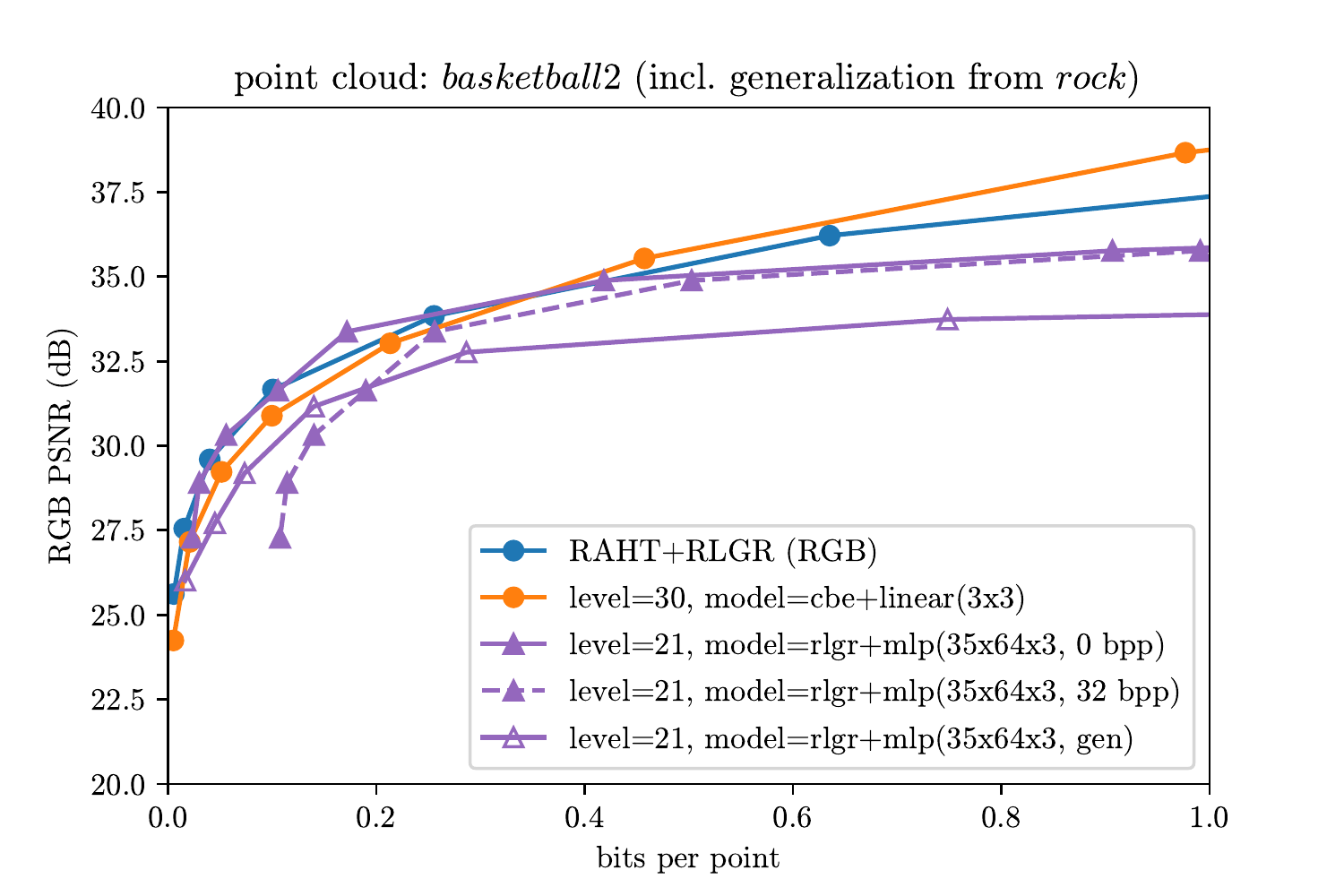}
    
    \includegraphics[width=0.29\linewidth, trim=20 5 35 15, clip]{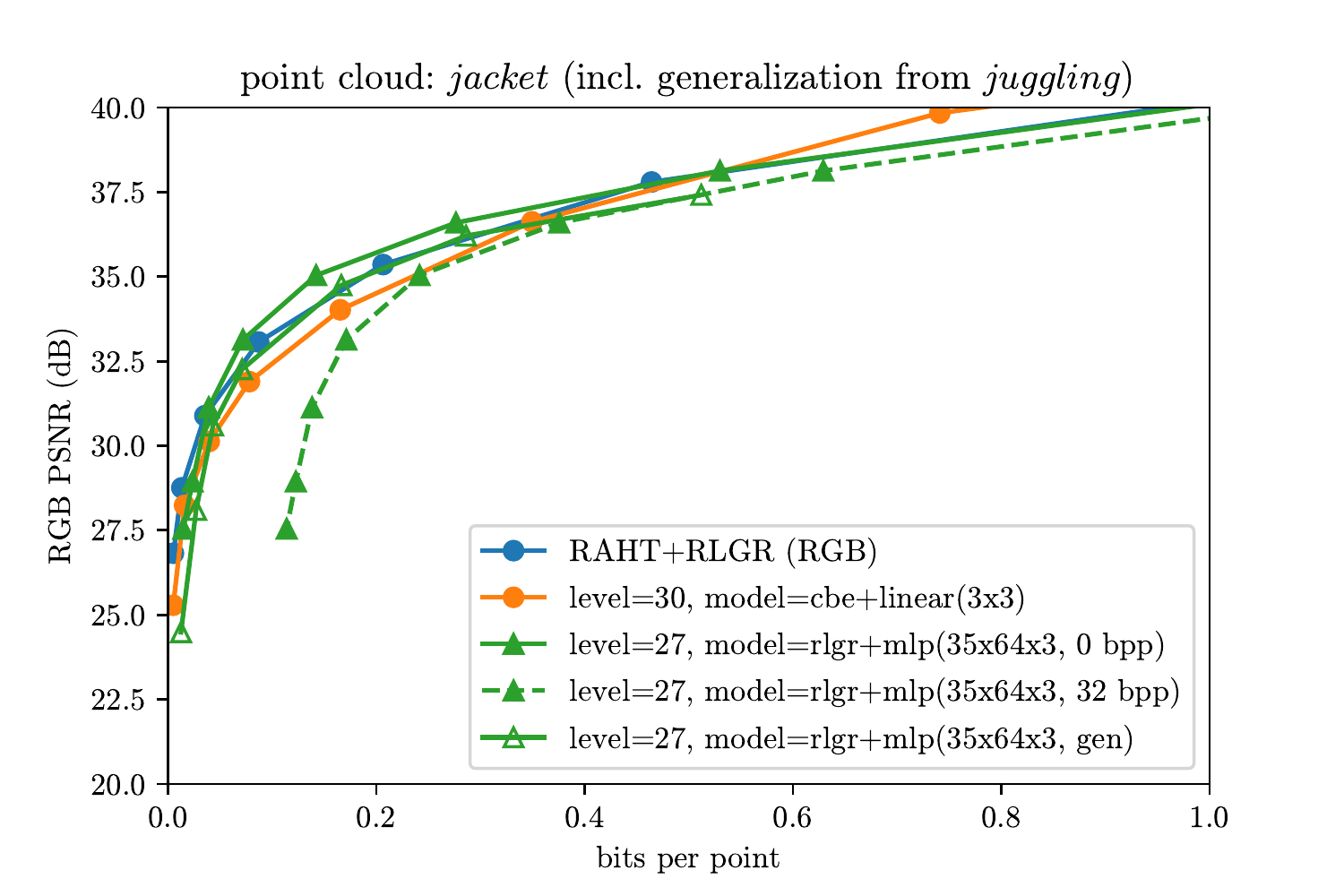}
    \includegraphics[width=0.29\linewidth, trim=20 5 35 15, clip]{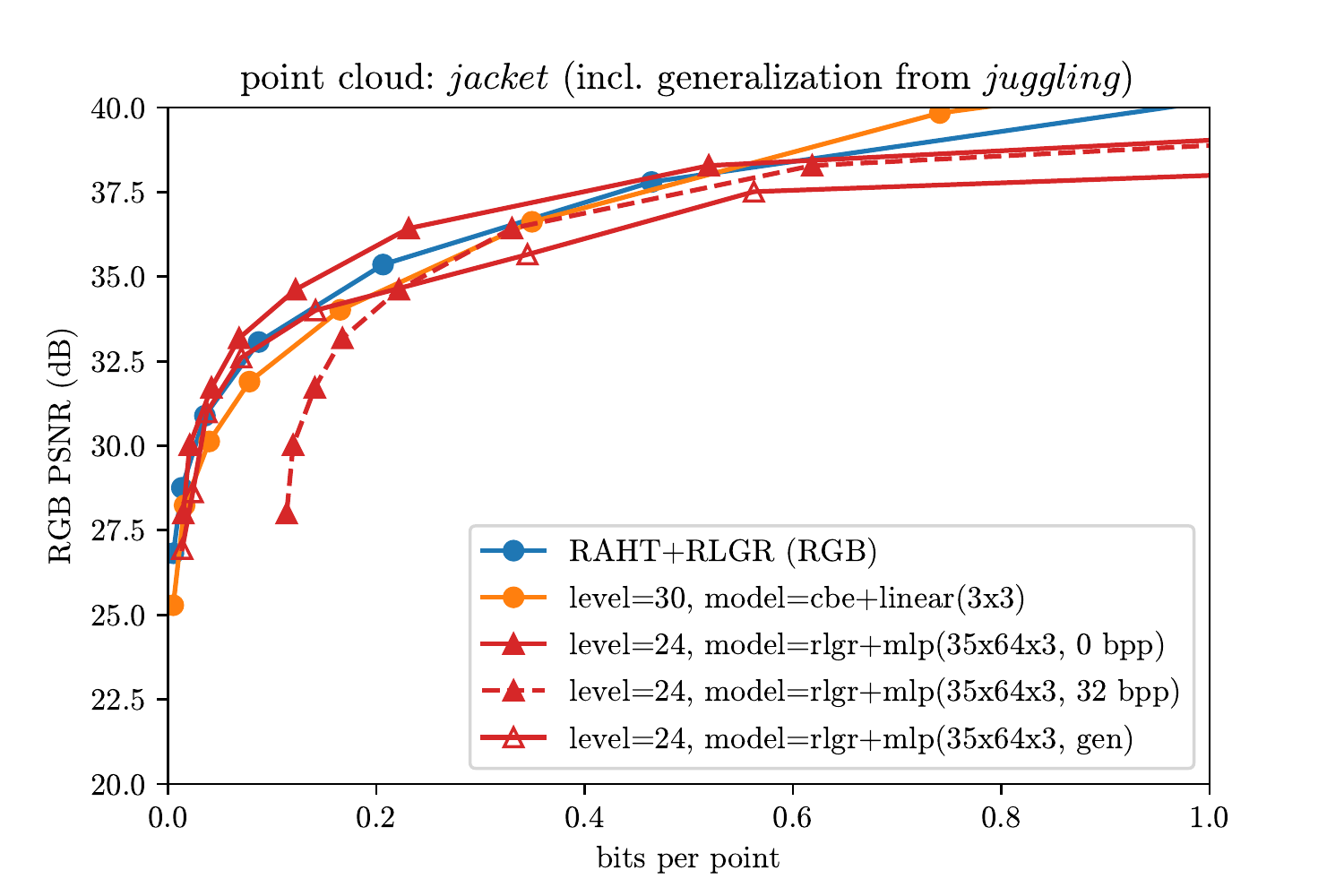}
    \includegraphics[width=0.29\linewidth, trim=20 5 35 15, clip]{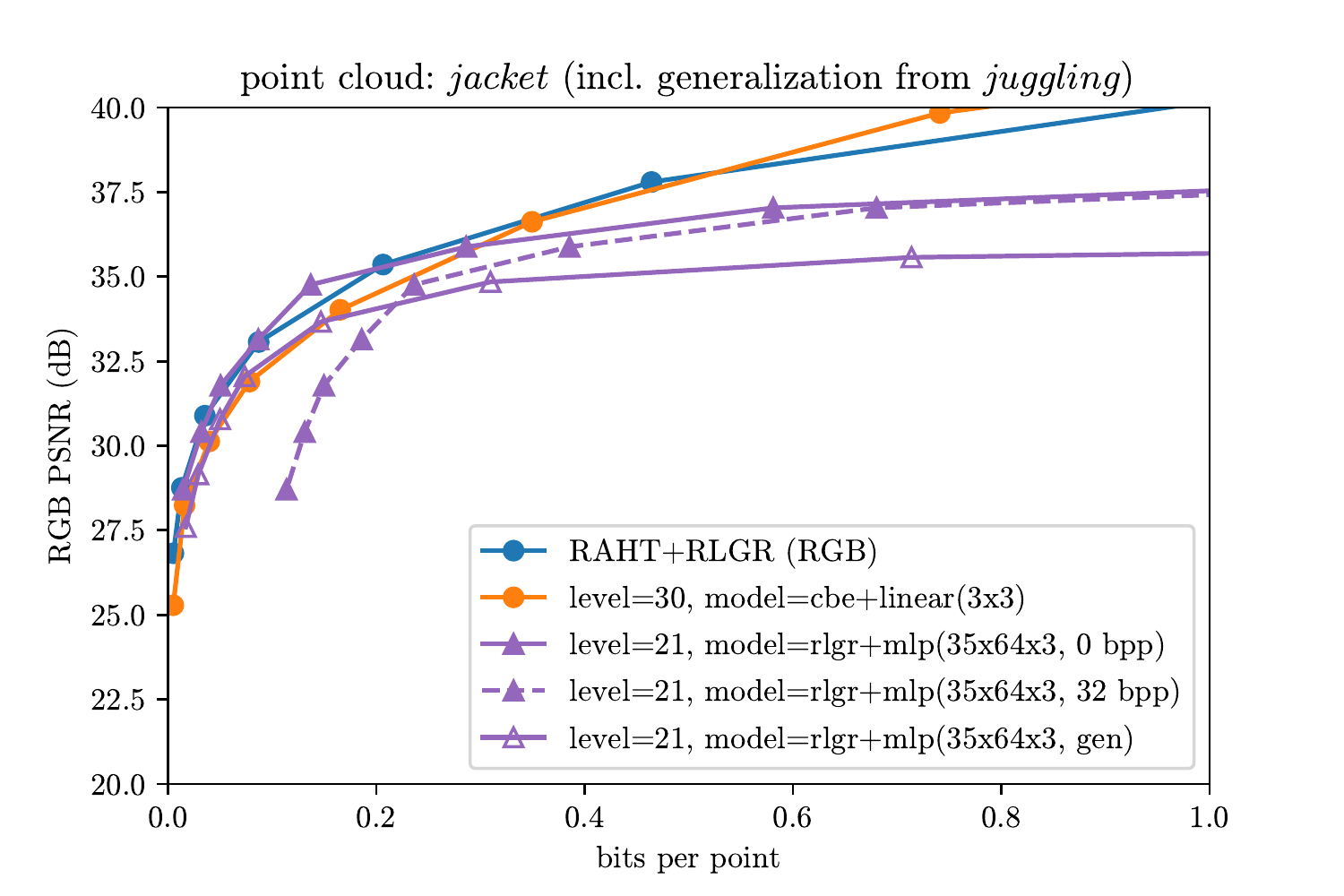}
    
    \caption{Effect of side information for coordinate based network {\em mlp(35x64x3)} at levels 27 (left), 24 (middle), and 21 (right).  Each row is a different point cloud.  See \cref{fig:sideinfo_mlp64_and_pa} (top) for point cloud {\em rock}.}
    \label{fig:sideinfo_mlp64_other}
\end{figure*}

\begin{figure*}
    \centering
    \includegraphics[width=0.29\linewidth, trim=20 5 35 15, clip]{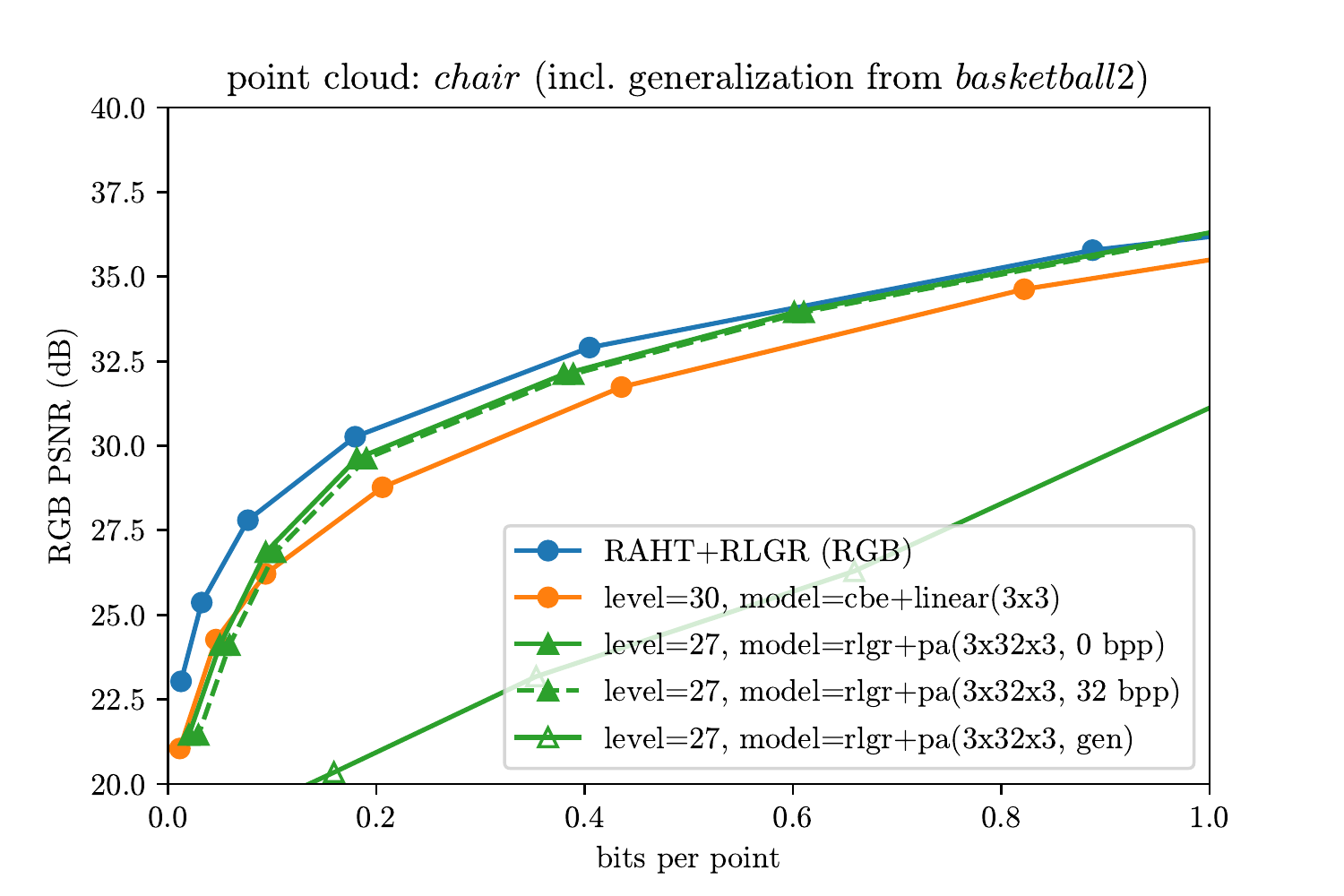}
    \includegraphics[width=0.29\linewidth, trim=20 5 35 15, clip]{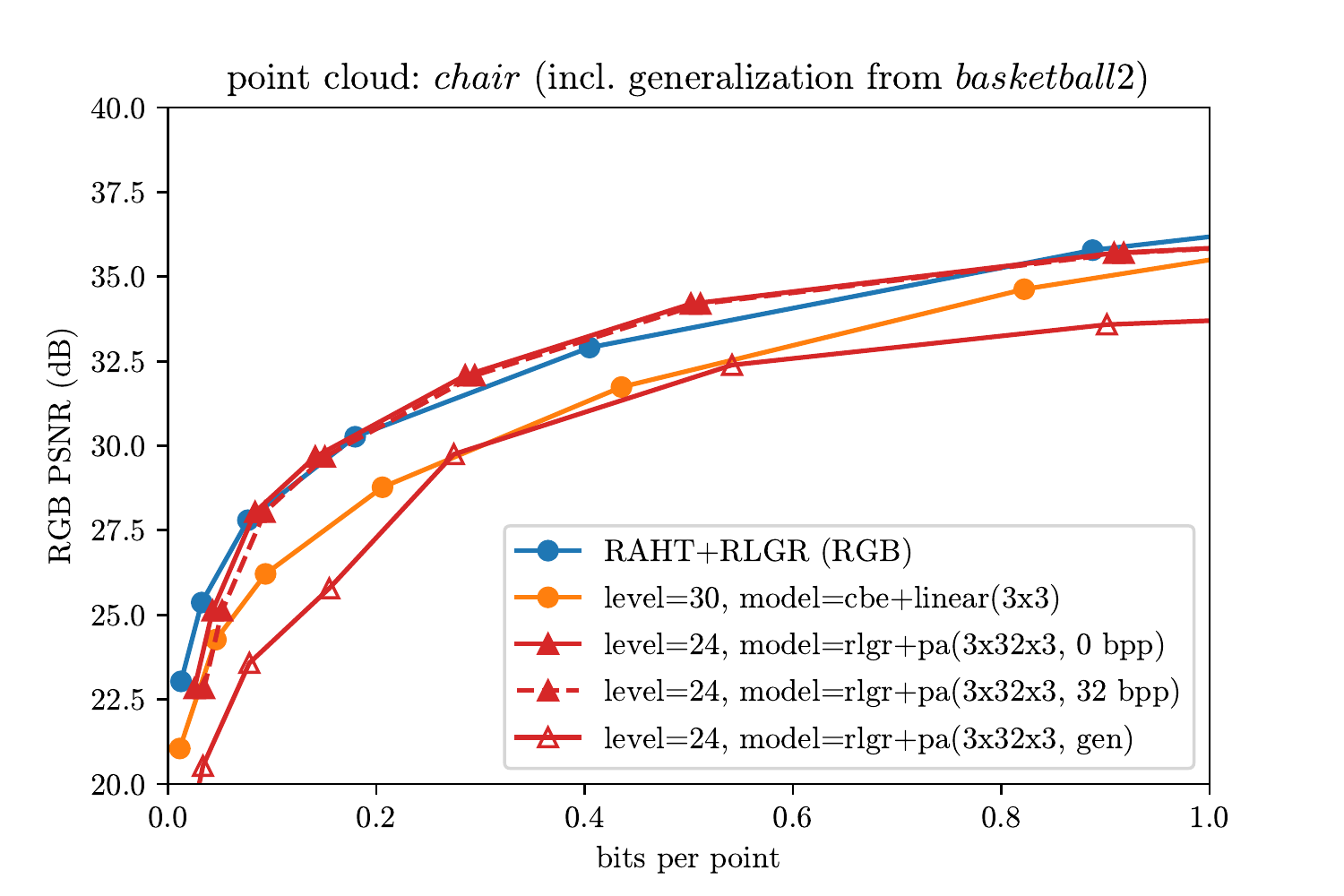}
    \includegraphics[width=0.29\linewidth, trim=20 5 35 15, clip]{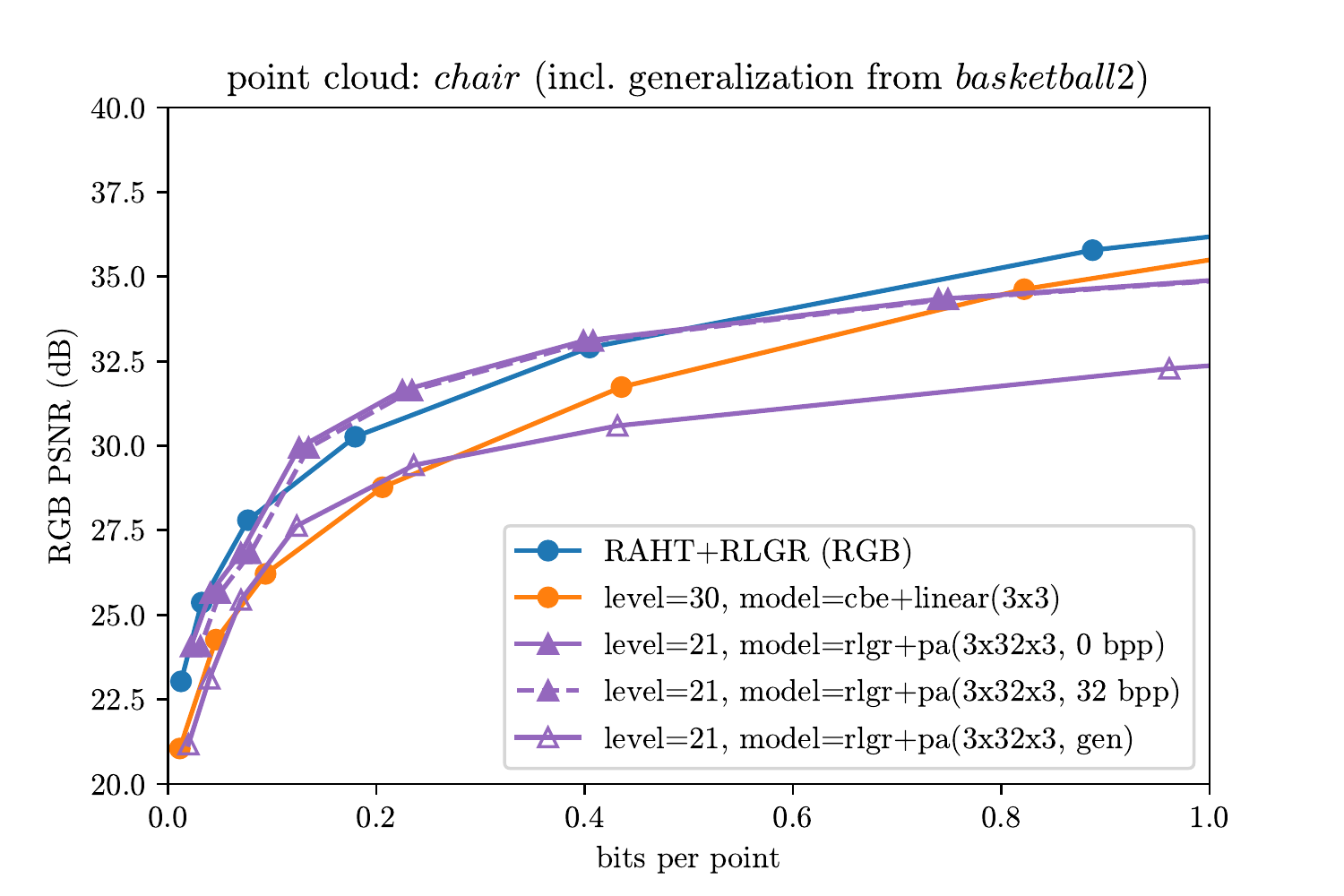}
    
    \includegraphics[width=0.29\linewidth, trim=20 5 35 15, clip]{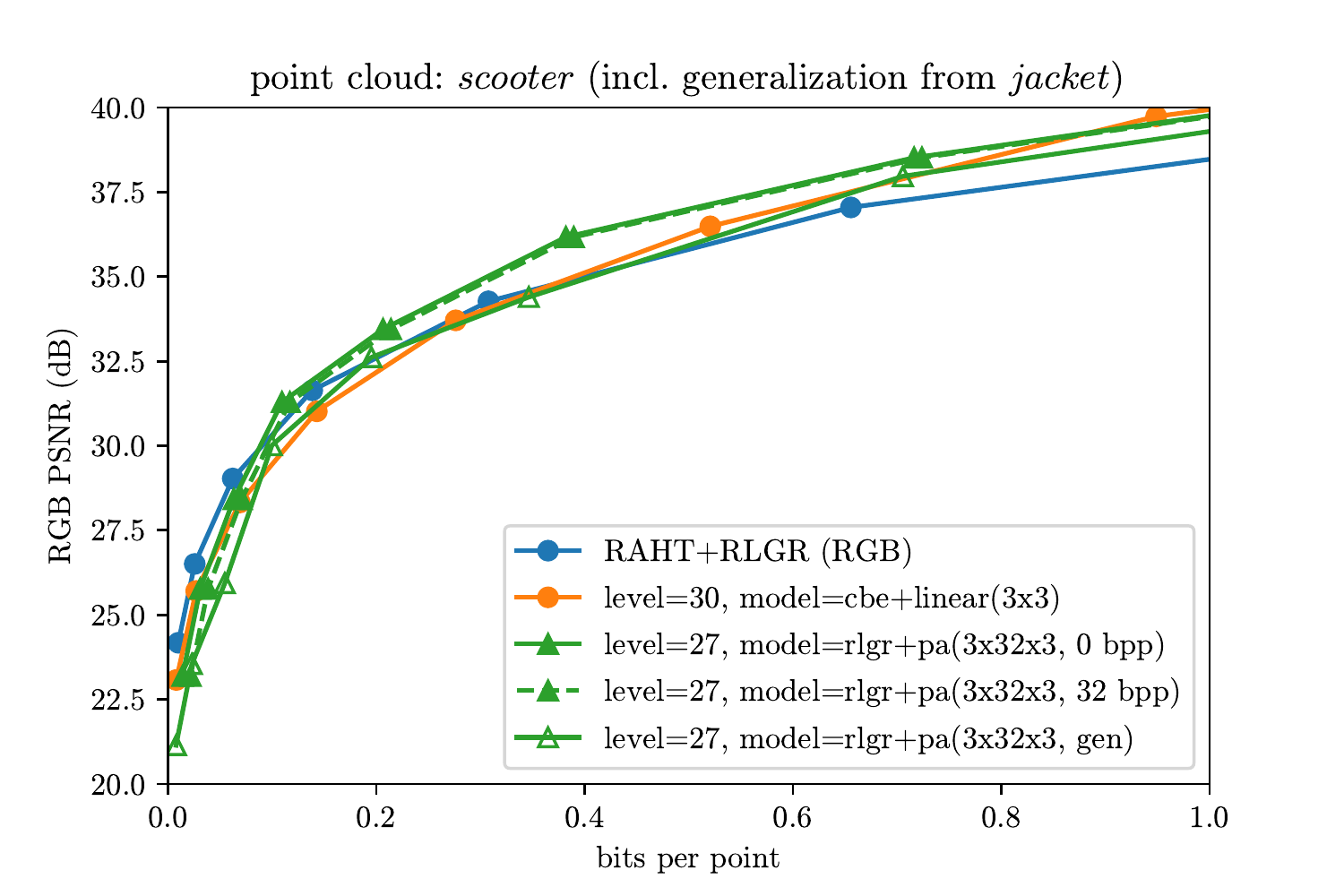}
    \includegraphics[width=0.29\linewidth, trim=20 5 35 15, clip]{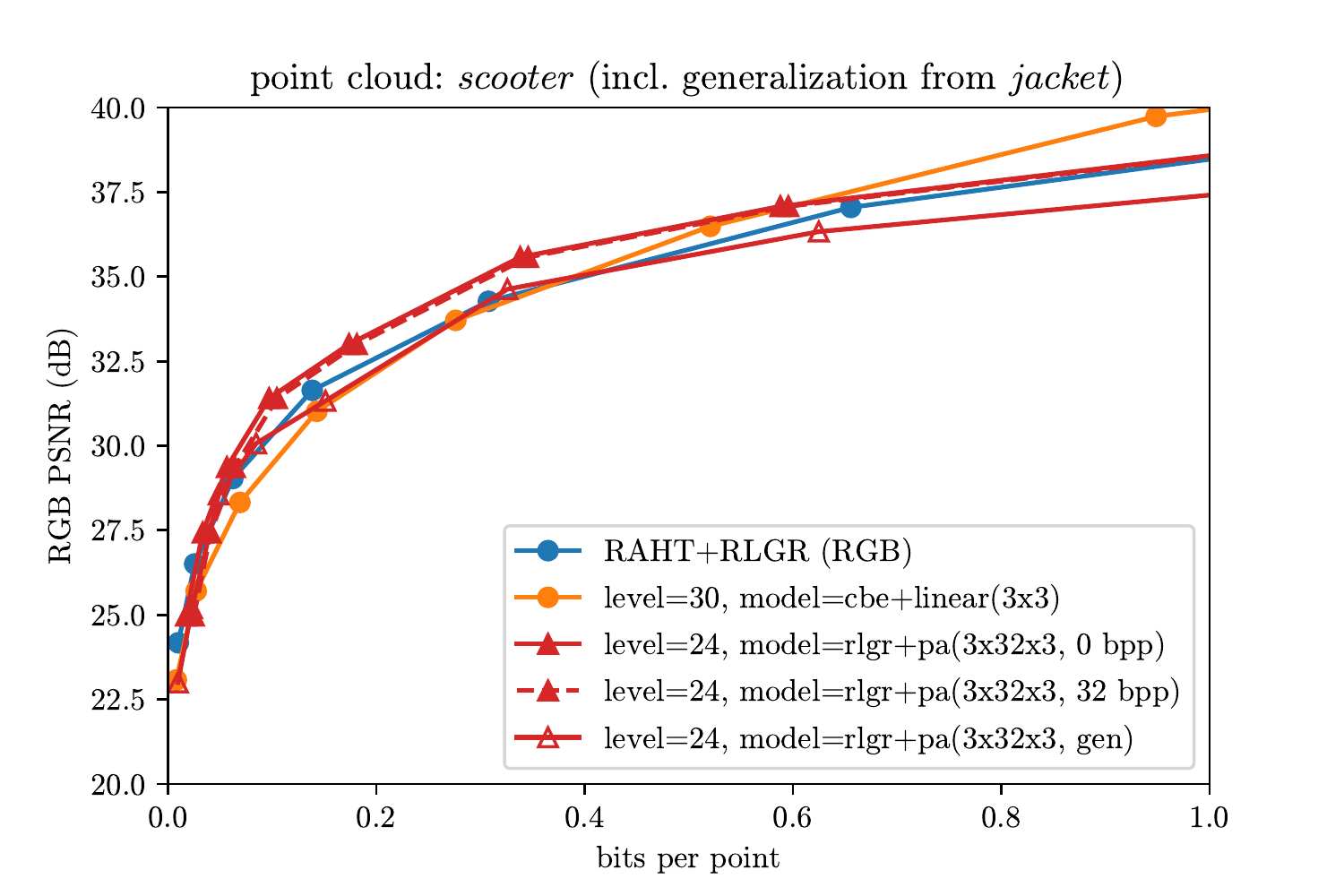}
    \includegraphics[width=0.29\linewidth, trim=20 5 35 15, clip]{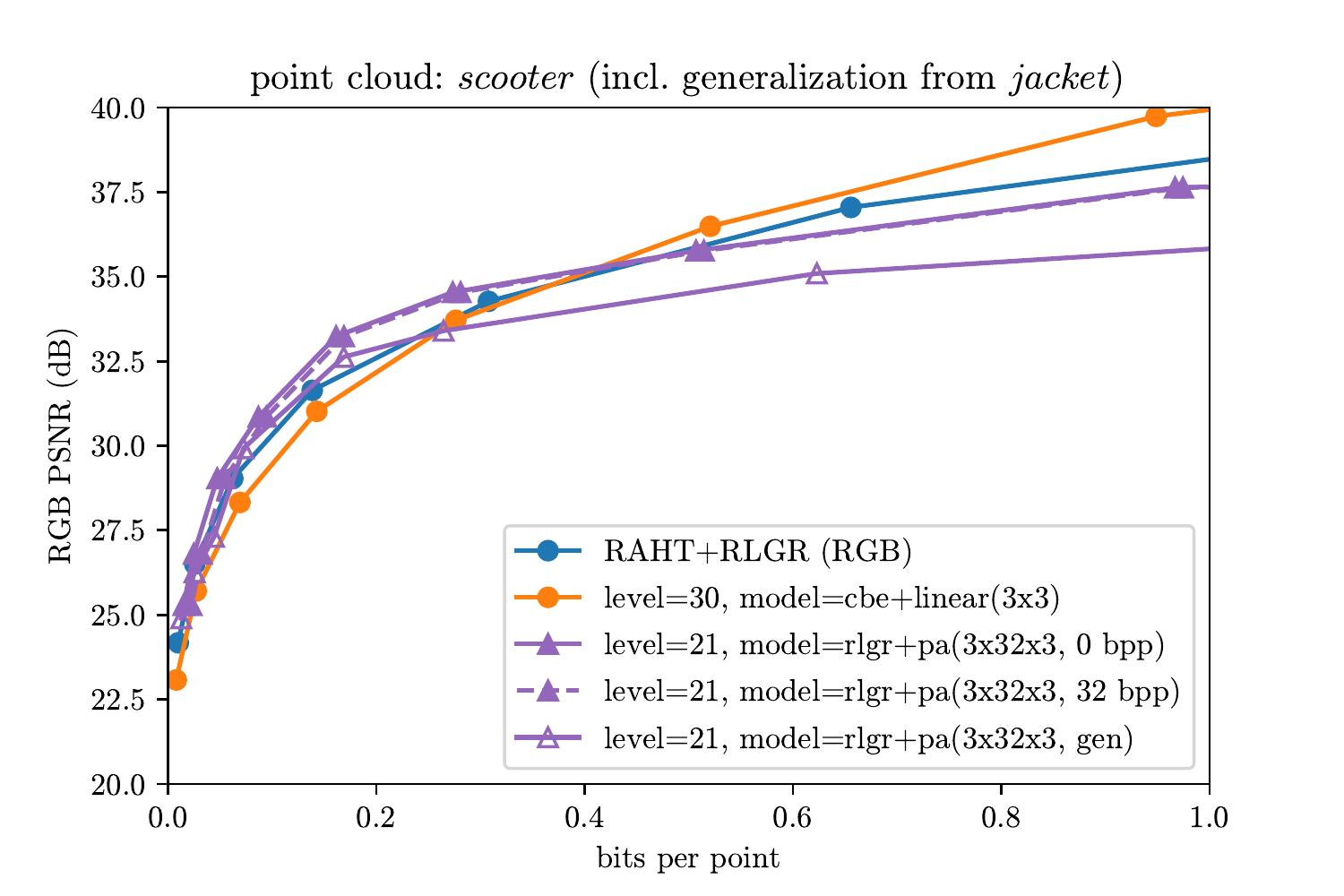}
    
    \includegraphics[width=0.29\linewidth, trim=20 5 35 15, clip]{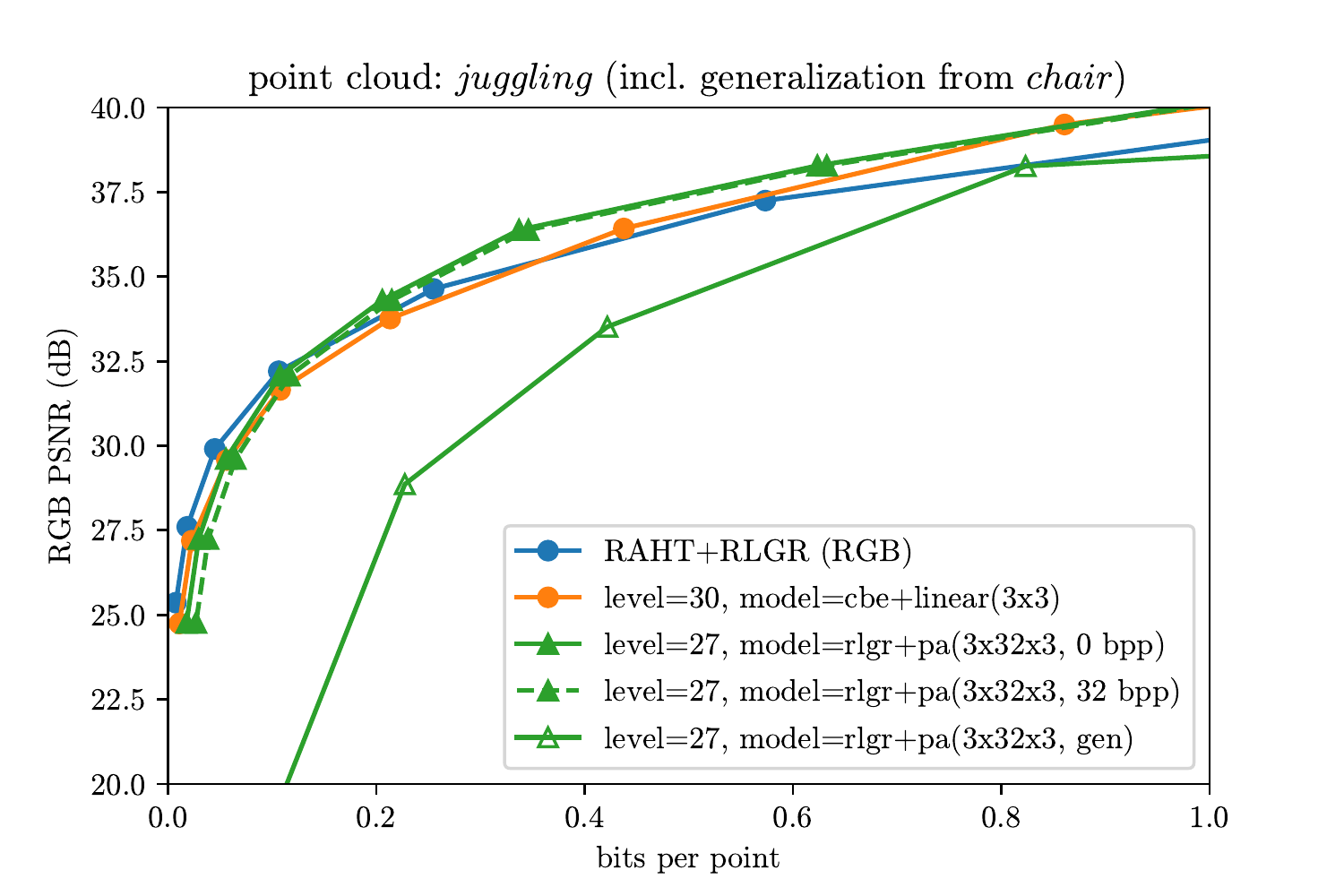}
    \includegraphics[width=0.29\linewidth, trim=20 5 35 15, clip]{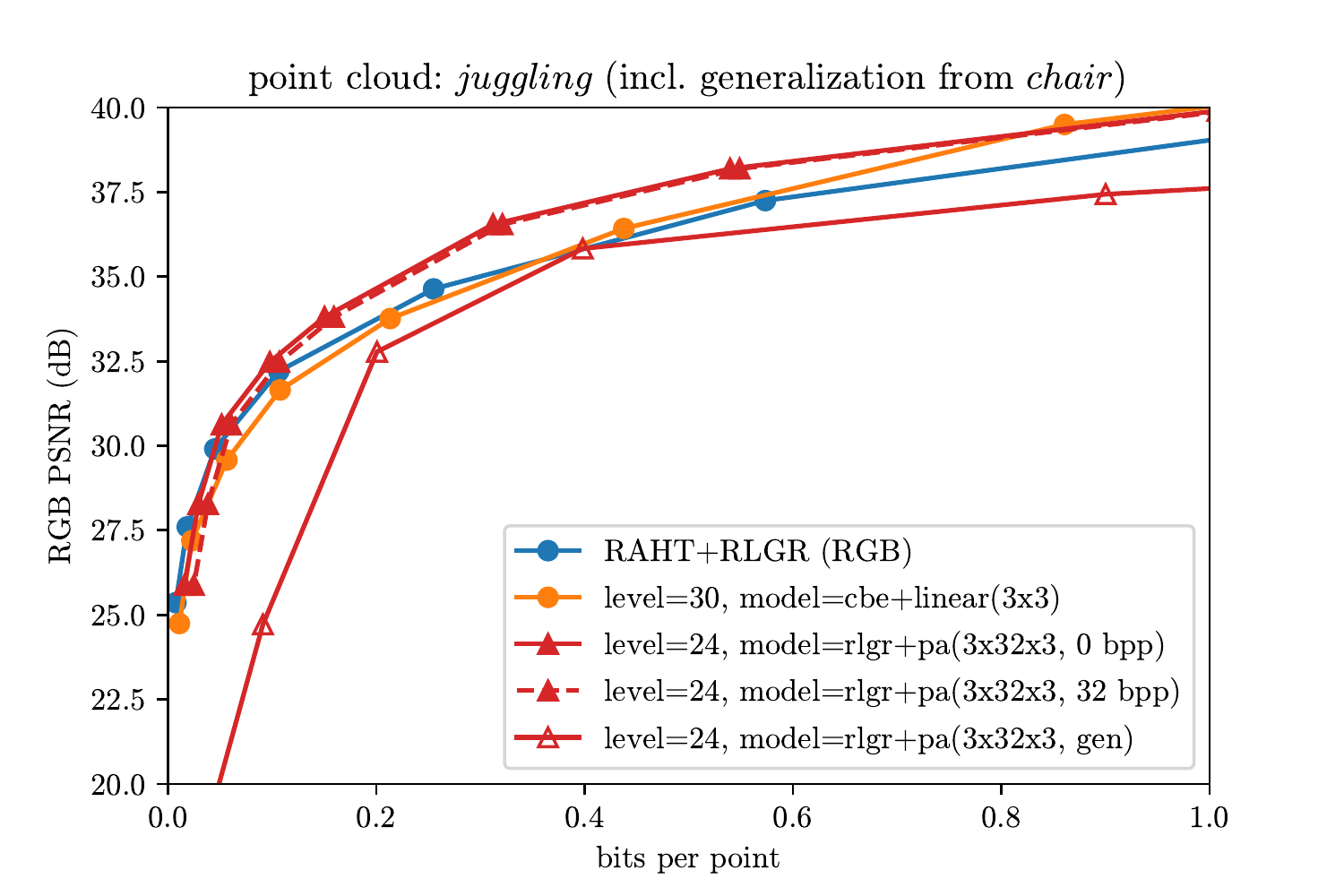}
    \includegraphics[width=0.29\linewidth, trim=20 5 35 15, clip]{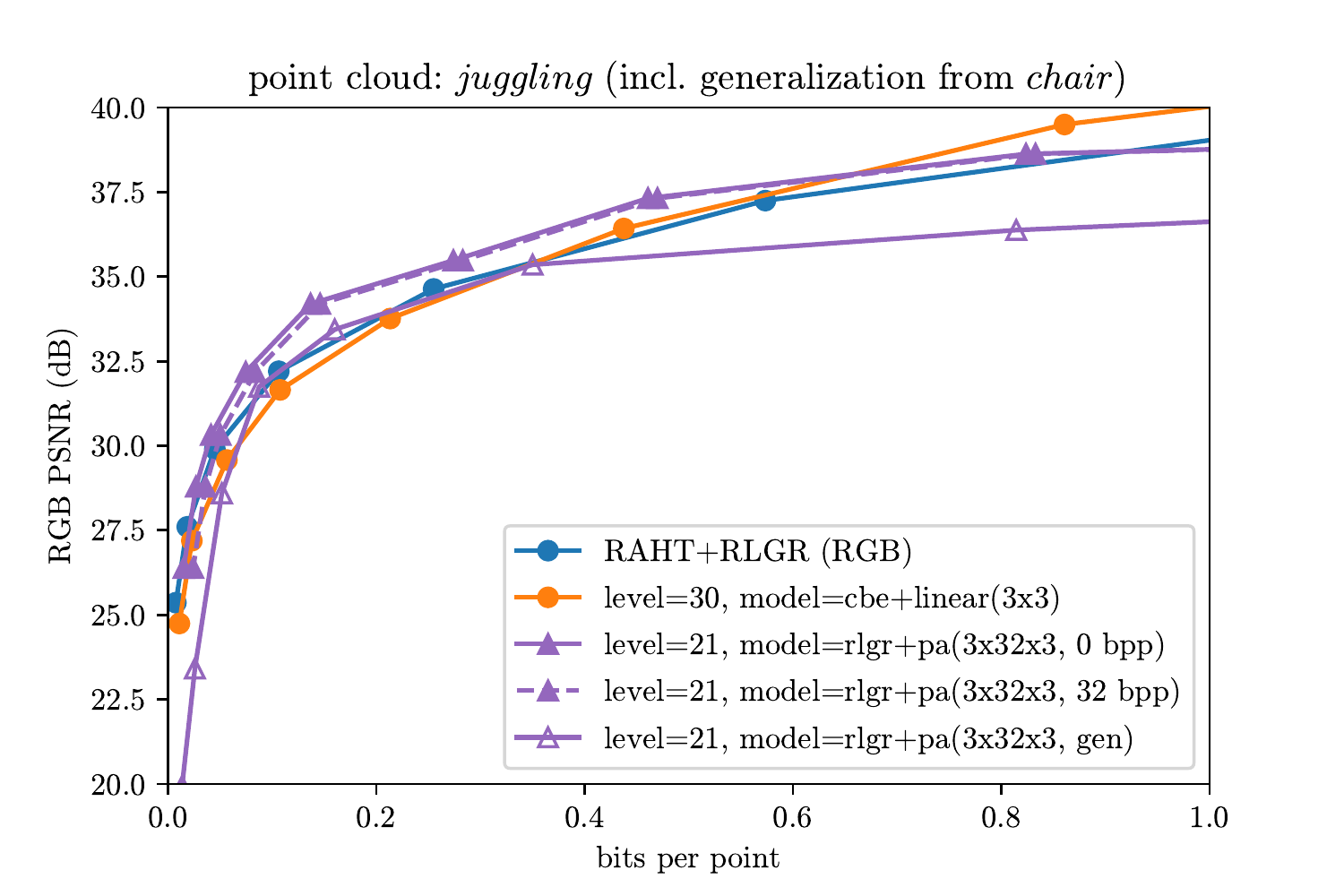}
    
    \includegraphics[width=0.29\linewidth, trim=20 5 35 15, clip]{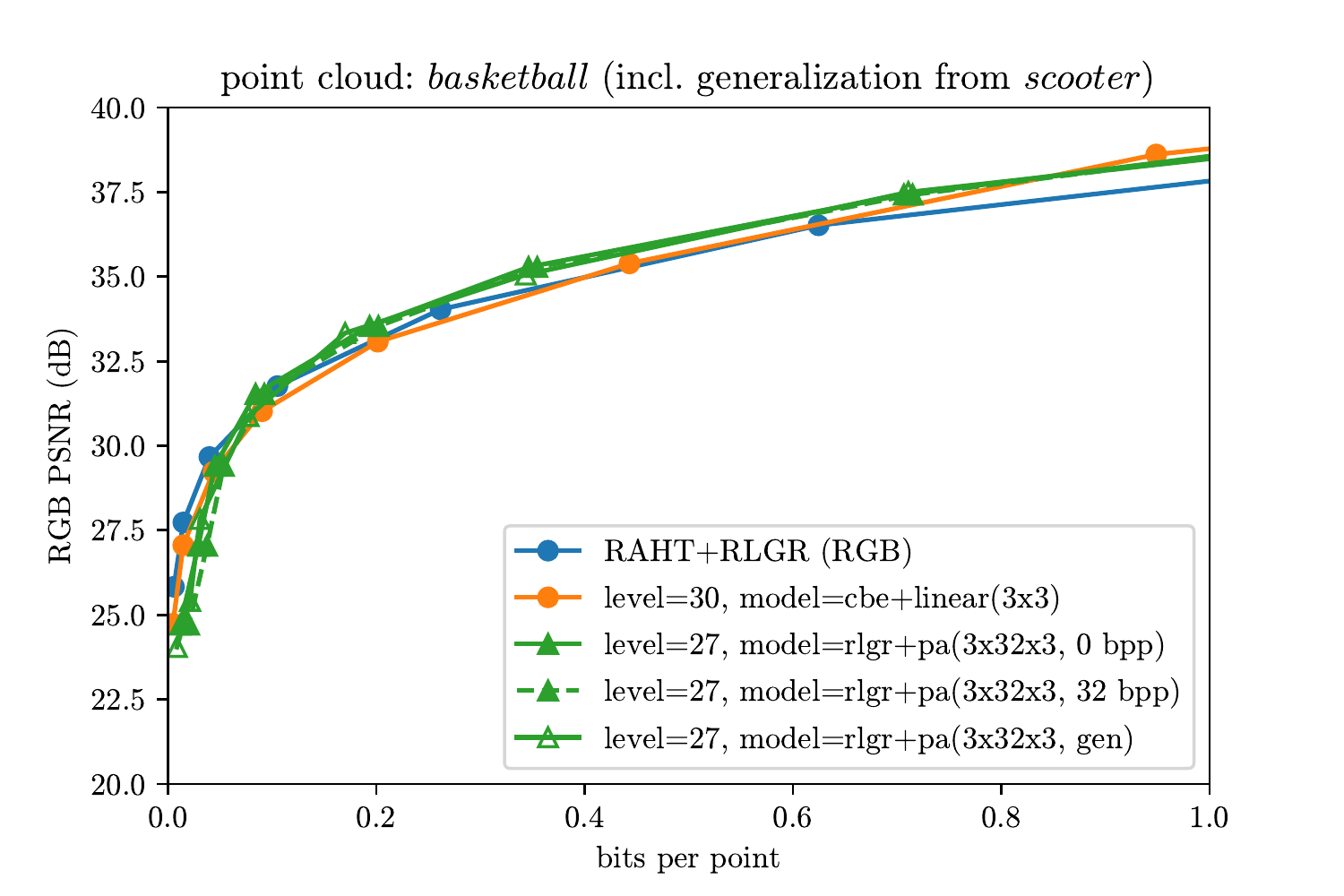}
    \includegraphics[width=0.29\linewidth, trim=20 5 35 15, clip]{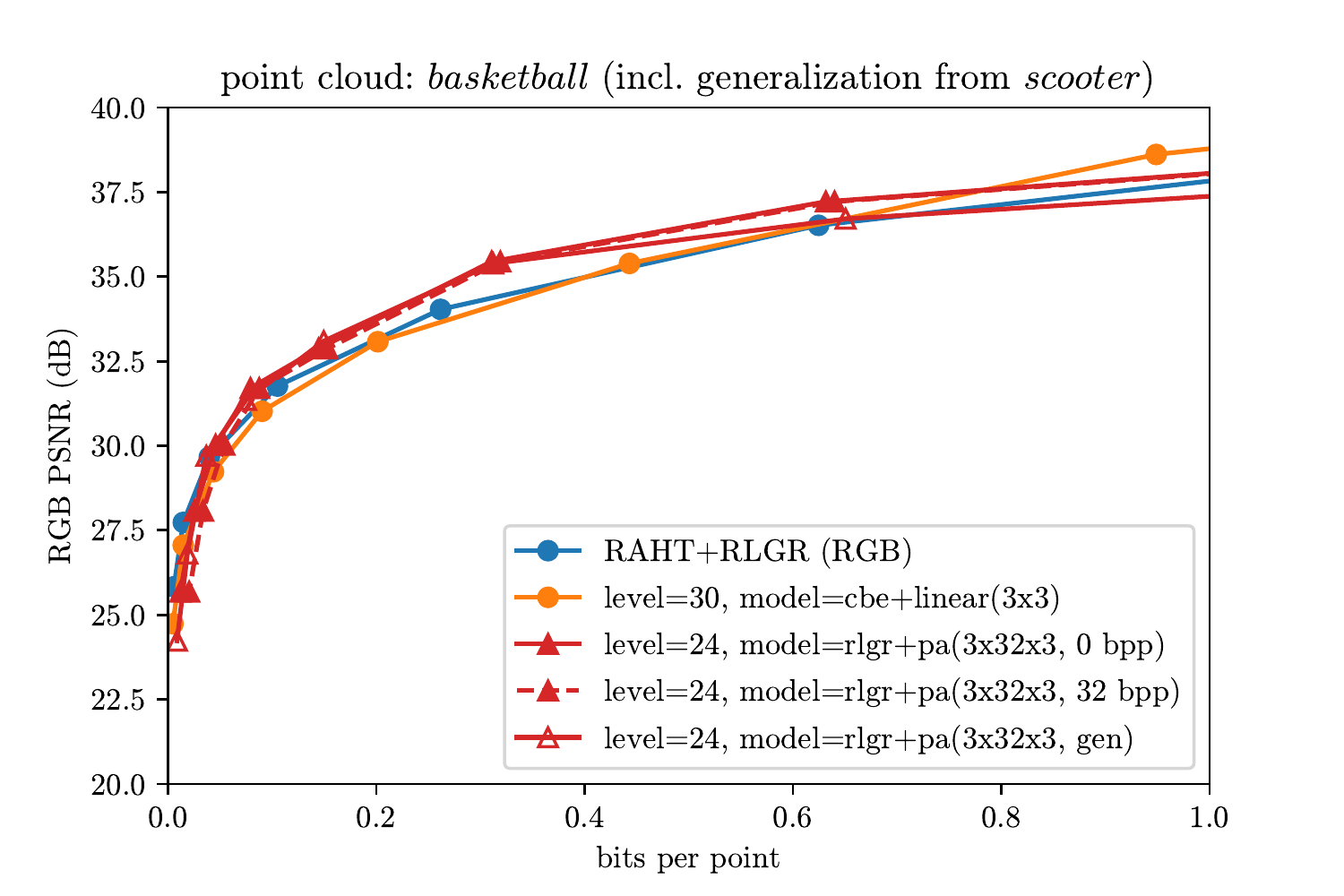}
    \includegraphics[width=0.29\linewidth, trim=20 5 35 15, clip]{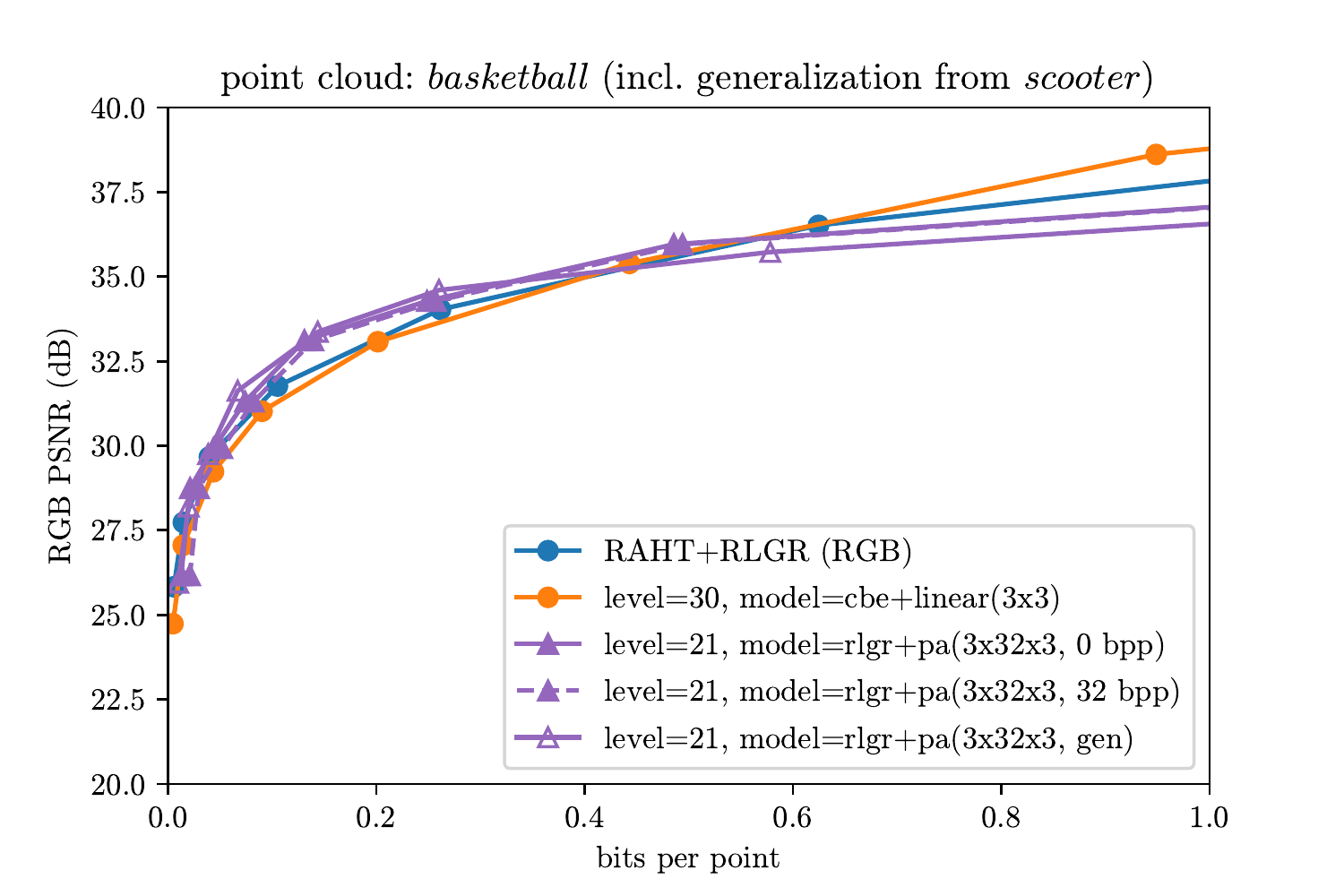}
    
    \includegraphics[width=0.29\linewidth, trim=20 5 35 15, clip]{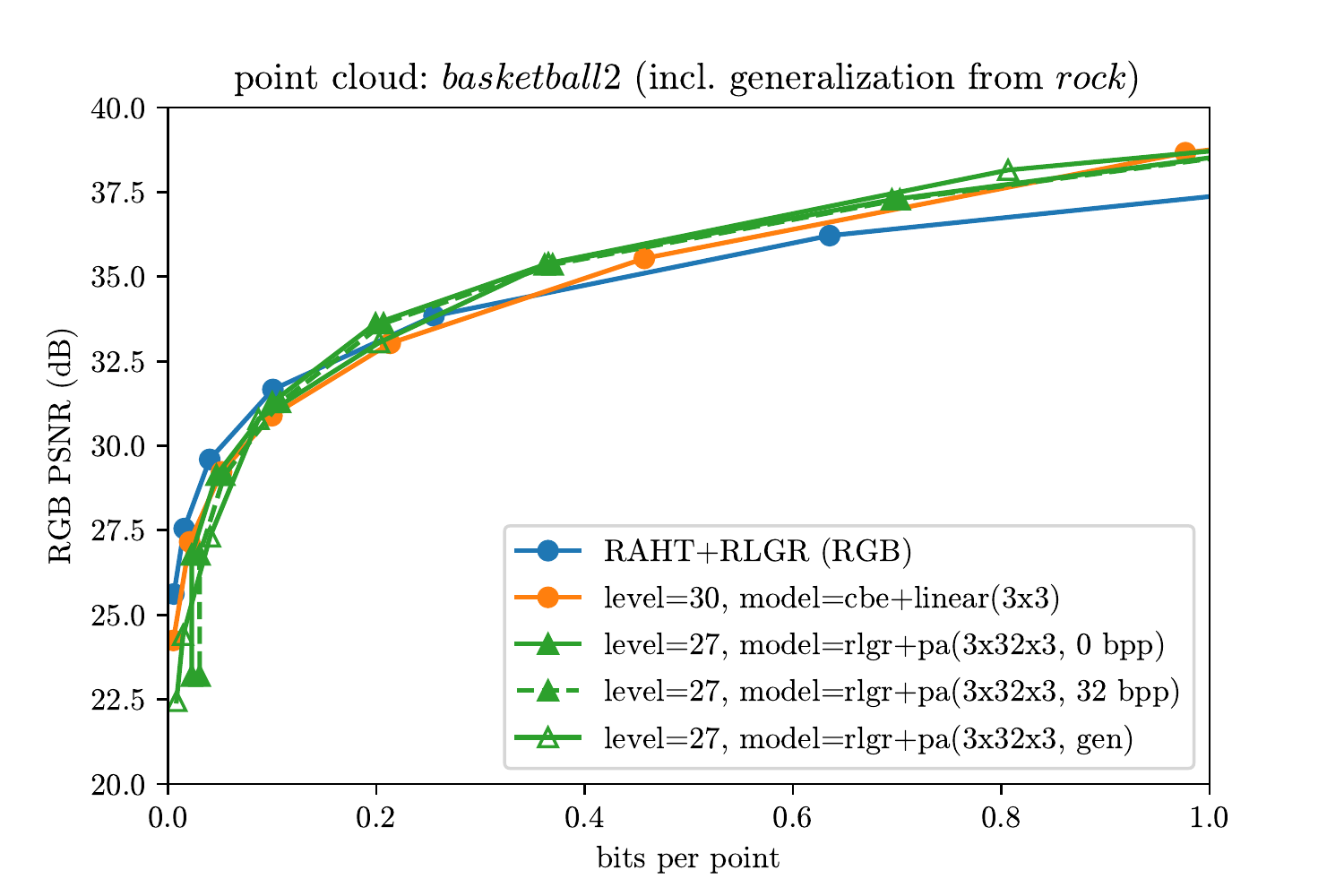}
    \includegraphics[width=0.29\linewidth, trim=20 5 35 15, clip]{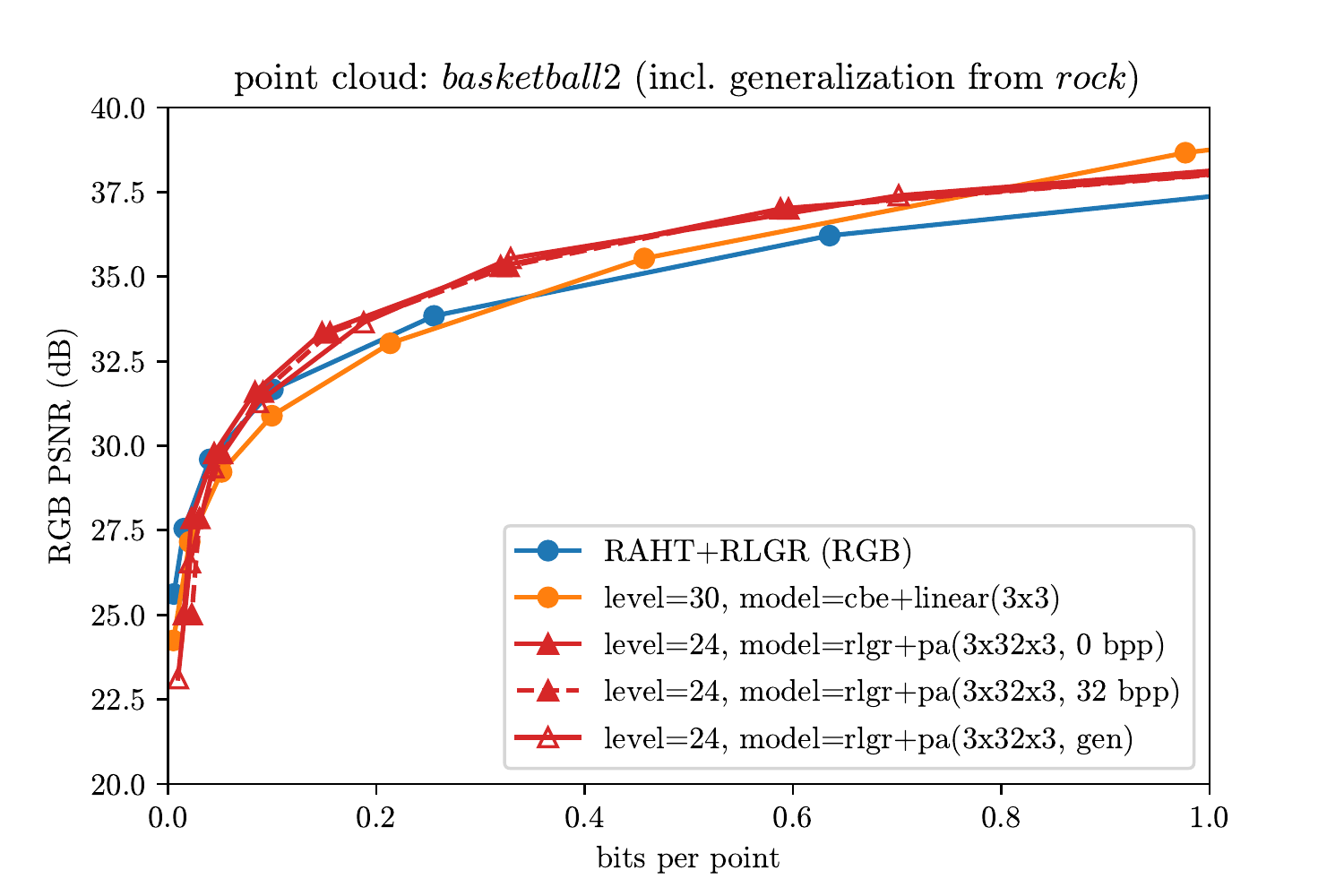}
    \includegraphics[width=0.29\linewidth, trim=20 5 35 15, clip]{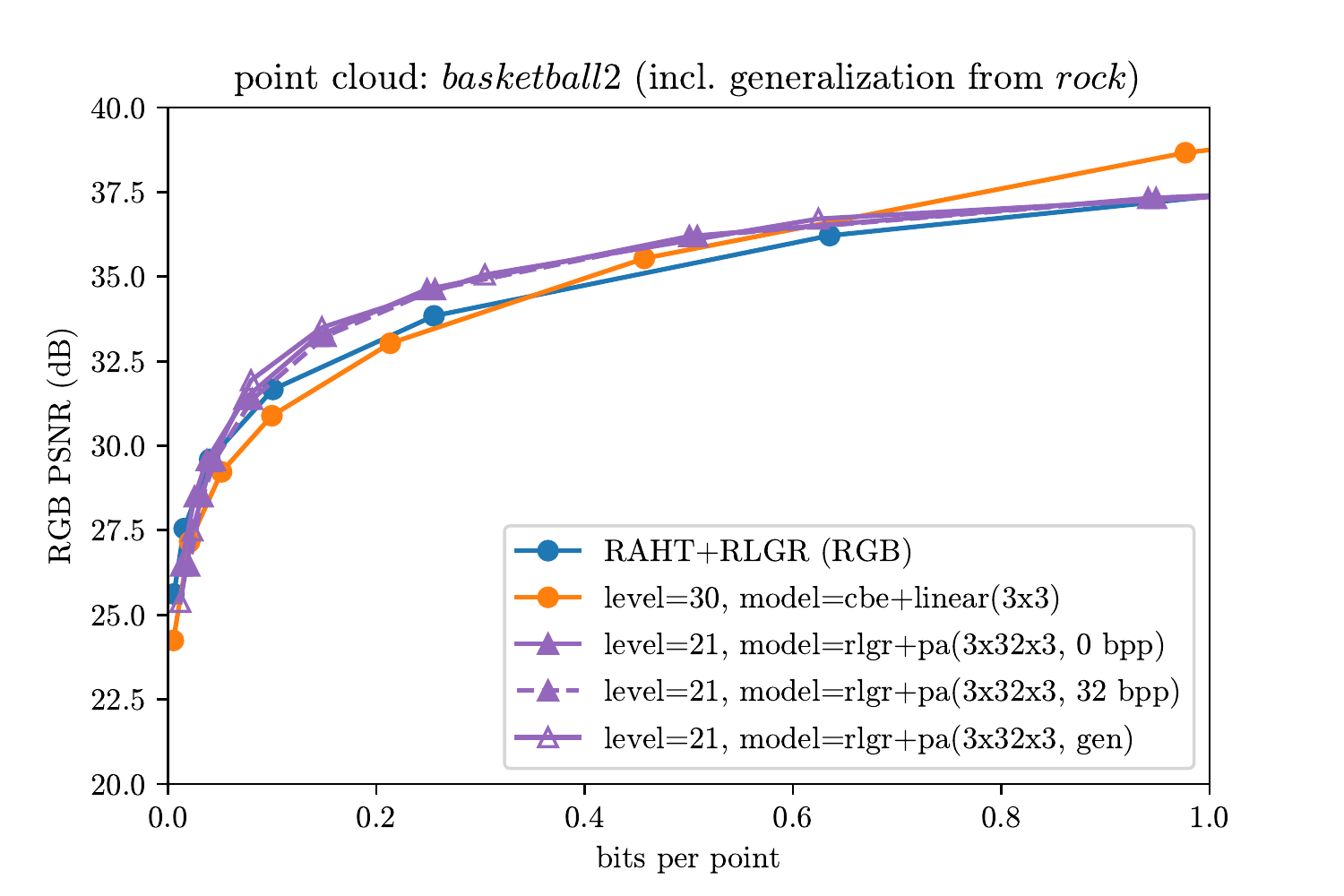}
    
    \includegraphics[width=0.29\linewidth, trim=20 5 35 15, clip]{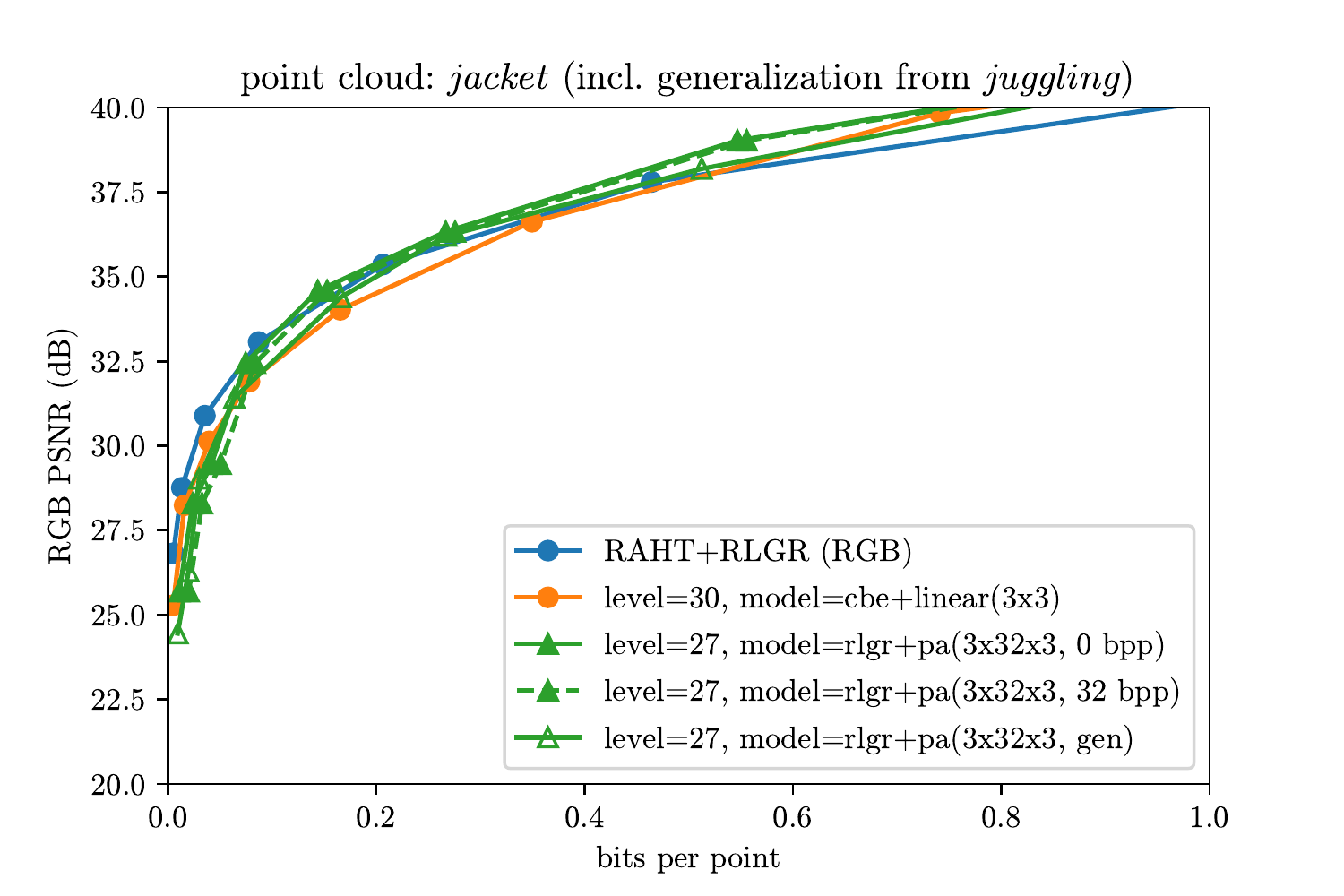}
    \includegraphics[width=0.29\linewidth, trim=20 5 35 15, clip]{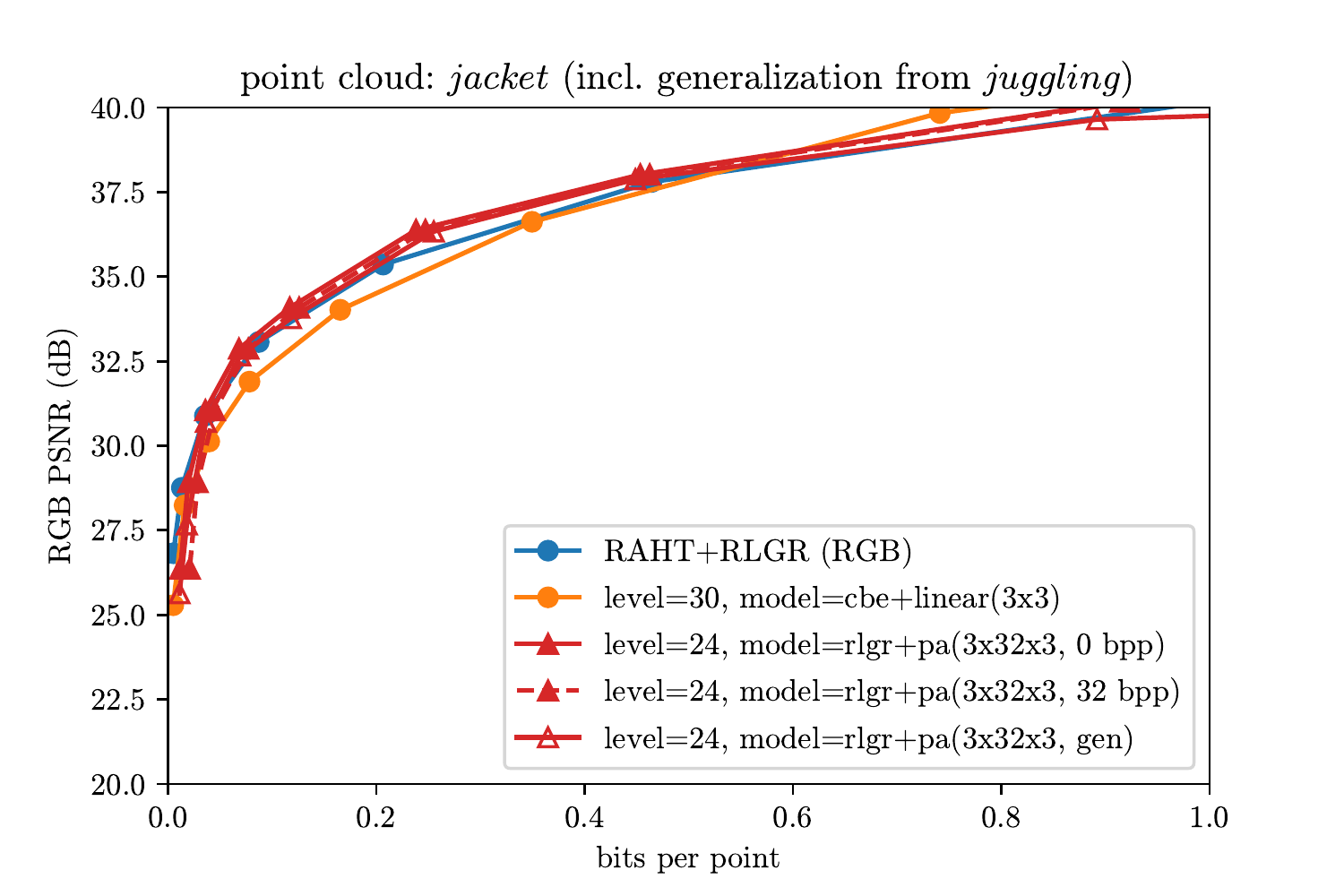}
    \includegraphics[width=0.29\linewidth, trim=20 5 35 15, clip]{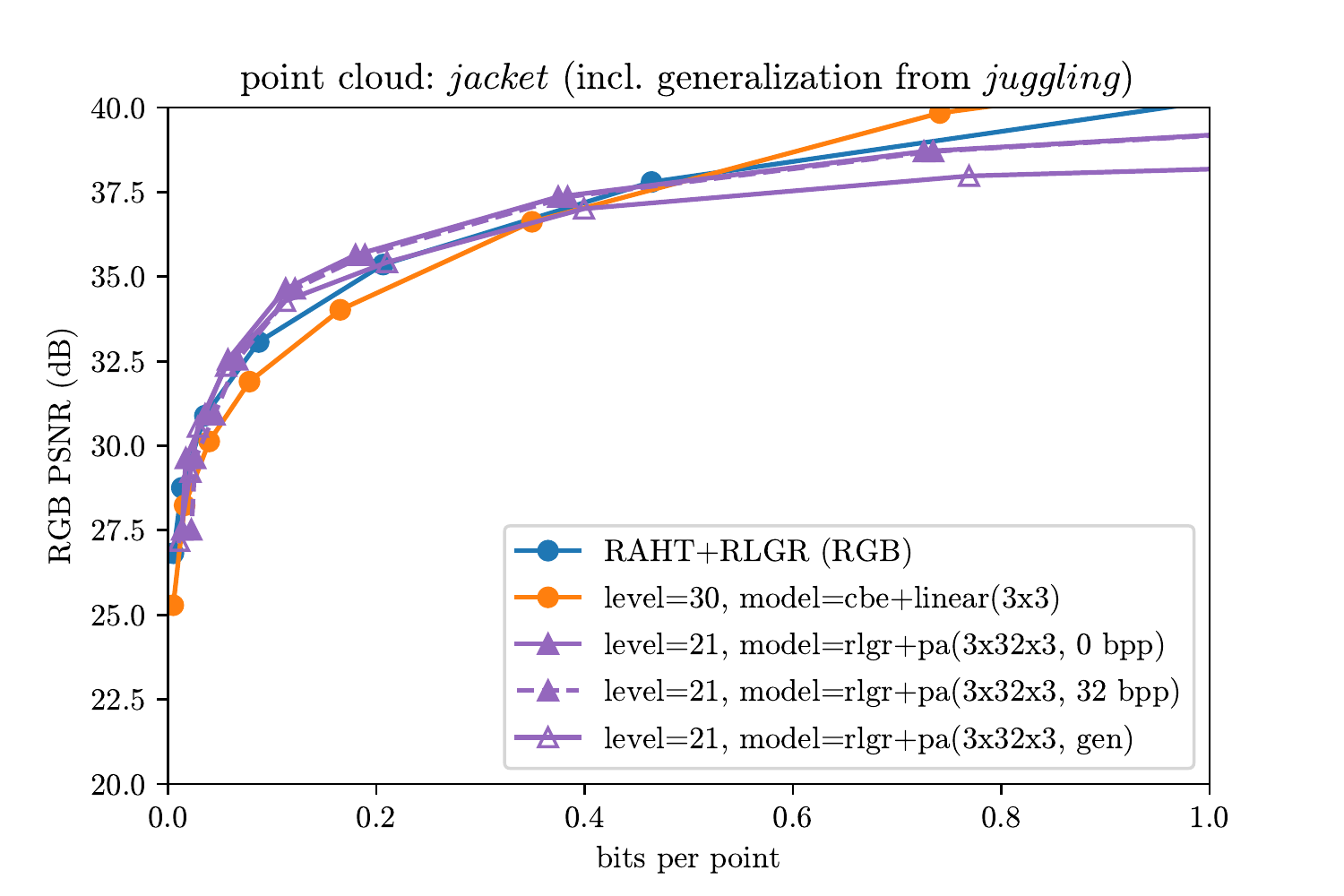}
    
    \caption{Effect of side information for coordinate based network {\em pa(3x32x3)} at levels 27 (left), 24 (middle), and 21 (right).  Each row is a different point cloud.  See \cref{fig:sideinfo_mlp64_and_pa} (bottom) for point cloud {\em rock}.}
    \label{fig:sideinfo_pa_other}
\end{figure*}

\begin{figure*}
    \centering
    \begin{minipage}{0.30\textwidth}
    \centering\small
    point cloud: {\em chair} \\
    \includegraphics[width=1.0\linewidth, trim=20 15 35 32, clip]{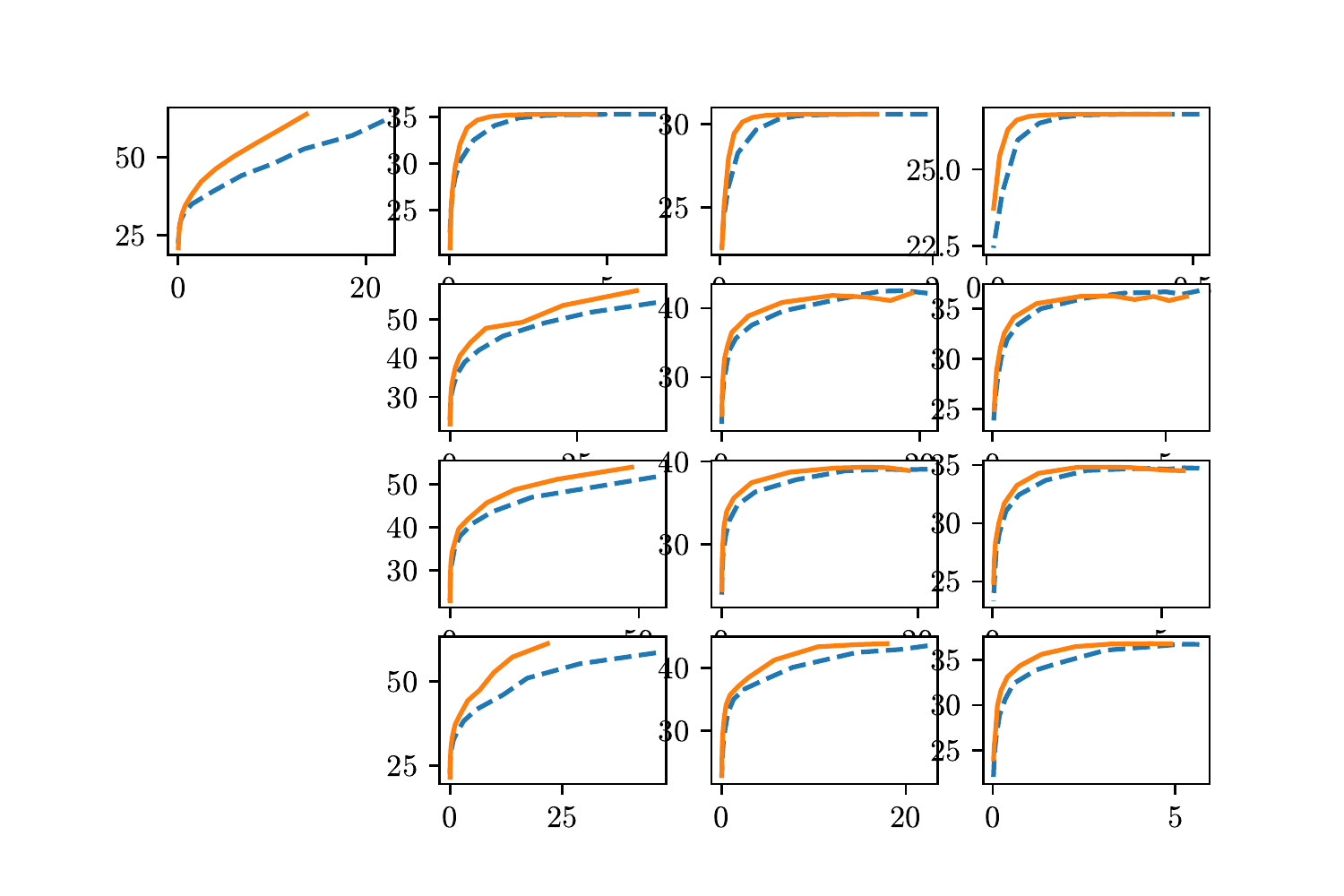}
    \end{minipage}
    \begin{minipage}{0.30\textwidth}
    \centering\small
    point cloud: {\em scooter} \\
    \includegraphics[width=1.0\linewidth, trim=20 15 35 32, clip]{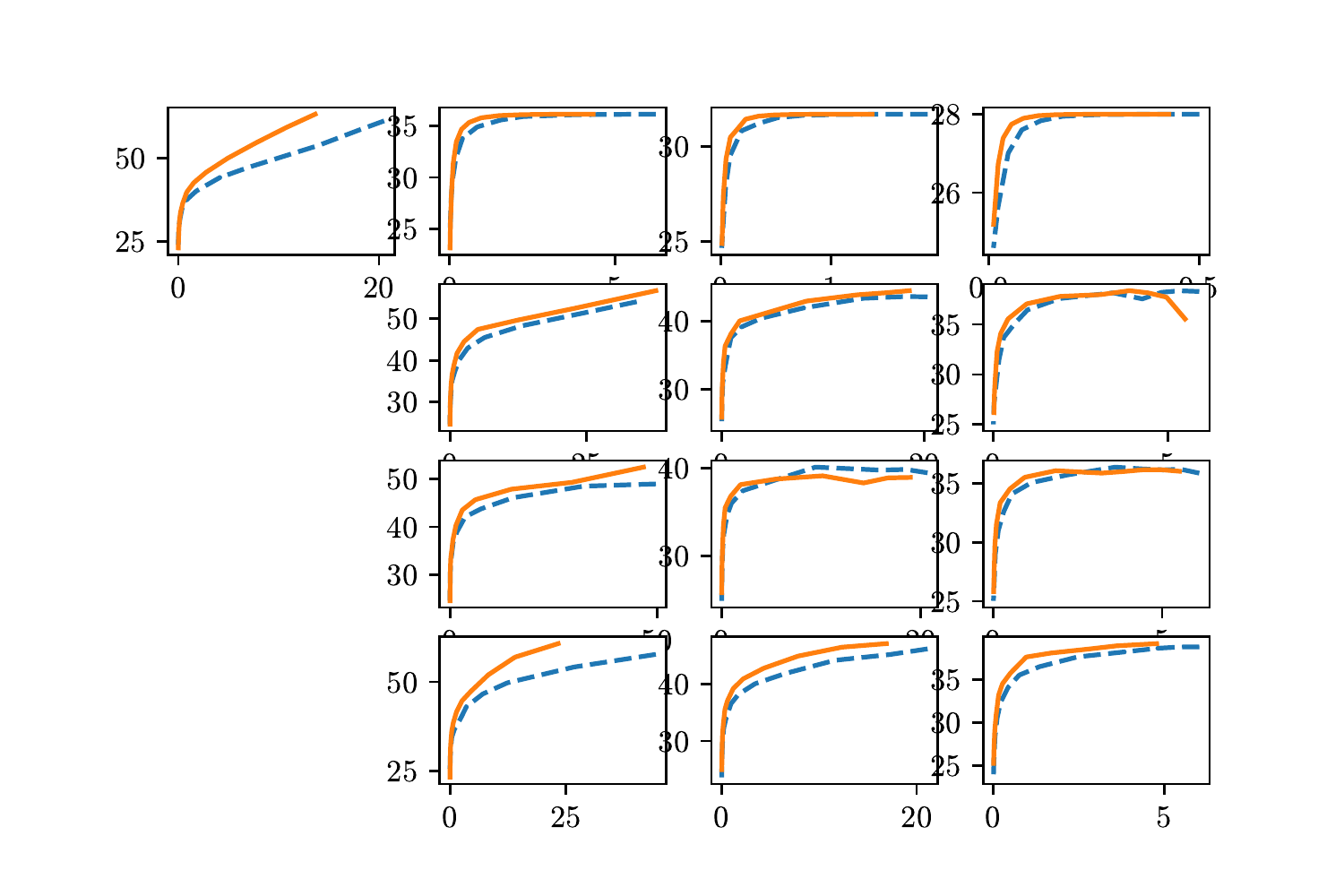}
    \end{minipage}
    \begin{minipage}{0.30\textwidth}
    \centering\small
    point cloud: {\em juggling} \\
    \includegraphics[width=1.0\linewidth, trim=20 15 35 32, clip]{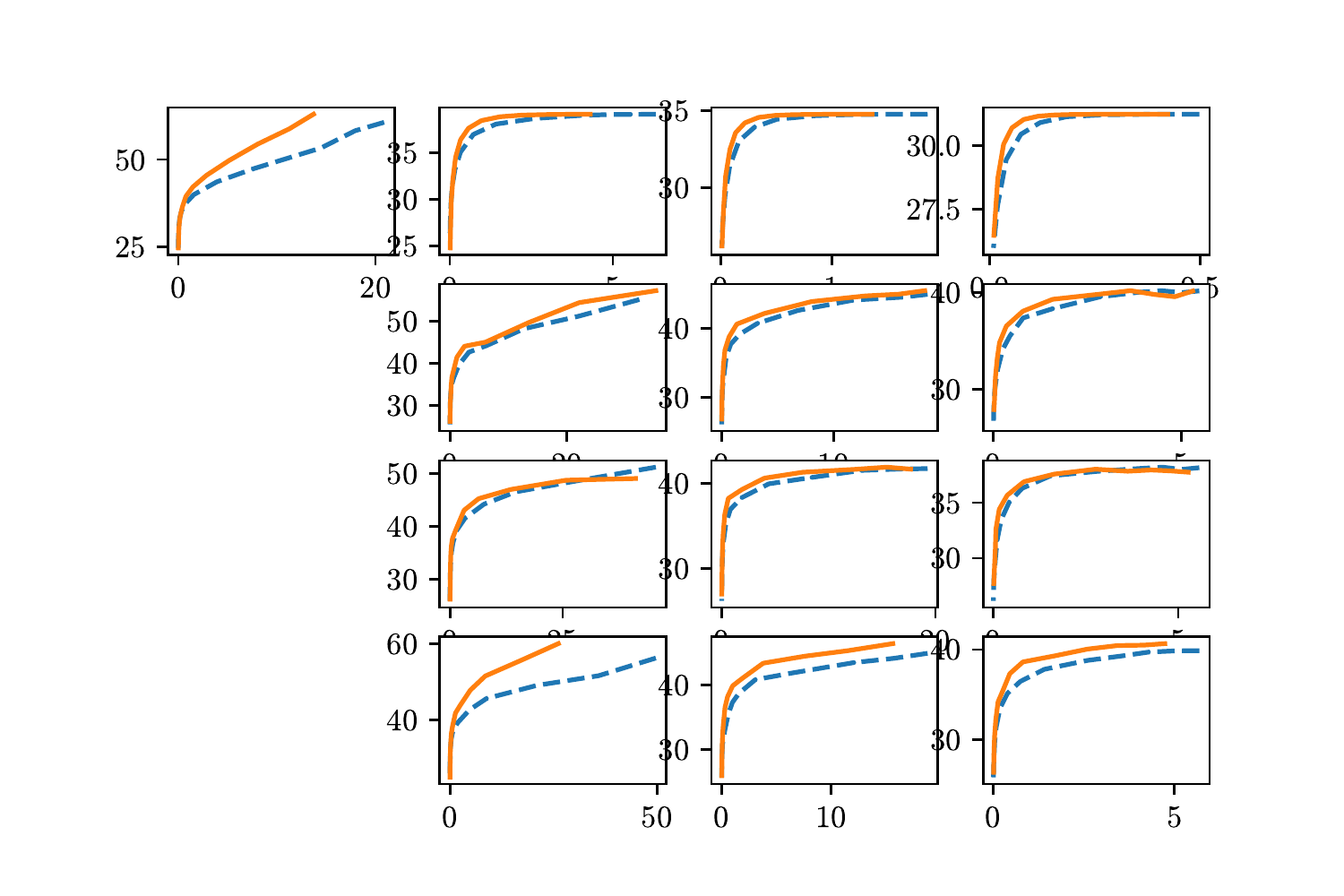}
    \end{minipage}
    \\[2ex]
    \begin{minipage}{0.30\textwidth}
    \centering\small
    point cloud: {\em basketball} \\
    \includegraphics[width=1.0\linewidth, trim=20 15 35 32, clip]{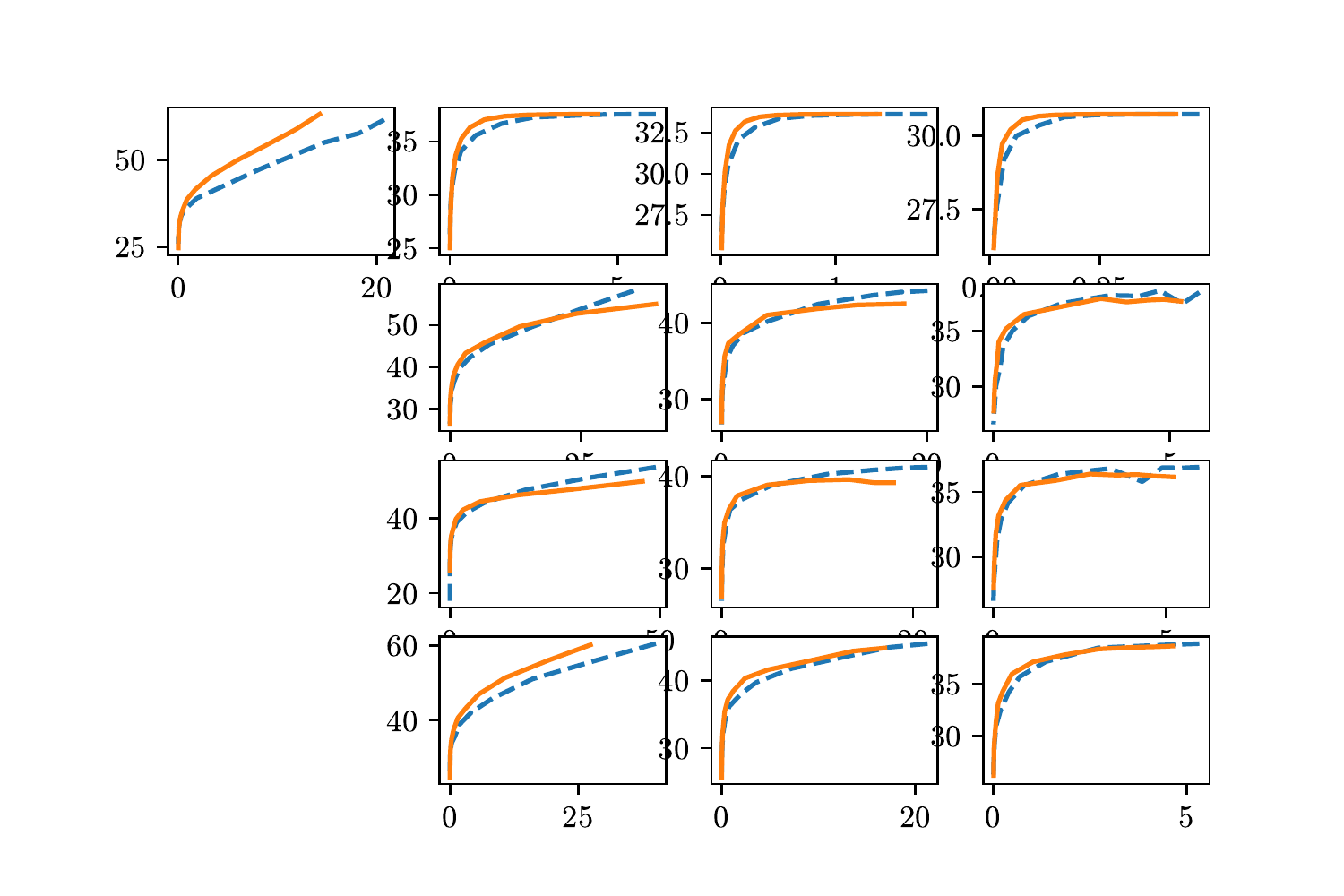}
    \end{minipage}
    \begin{minipage}{0.30\textwidth}
    \centering\small
    point cloud: {\em basketball2} \\
    \includegraphics[width=1.0\linewidth, trim=20 15 35 32, clip]{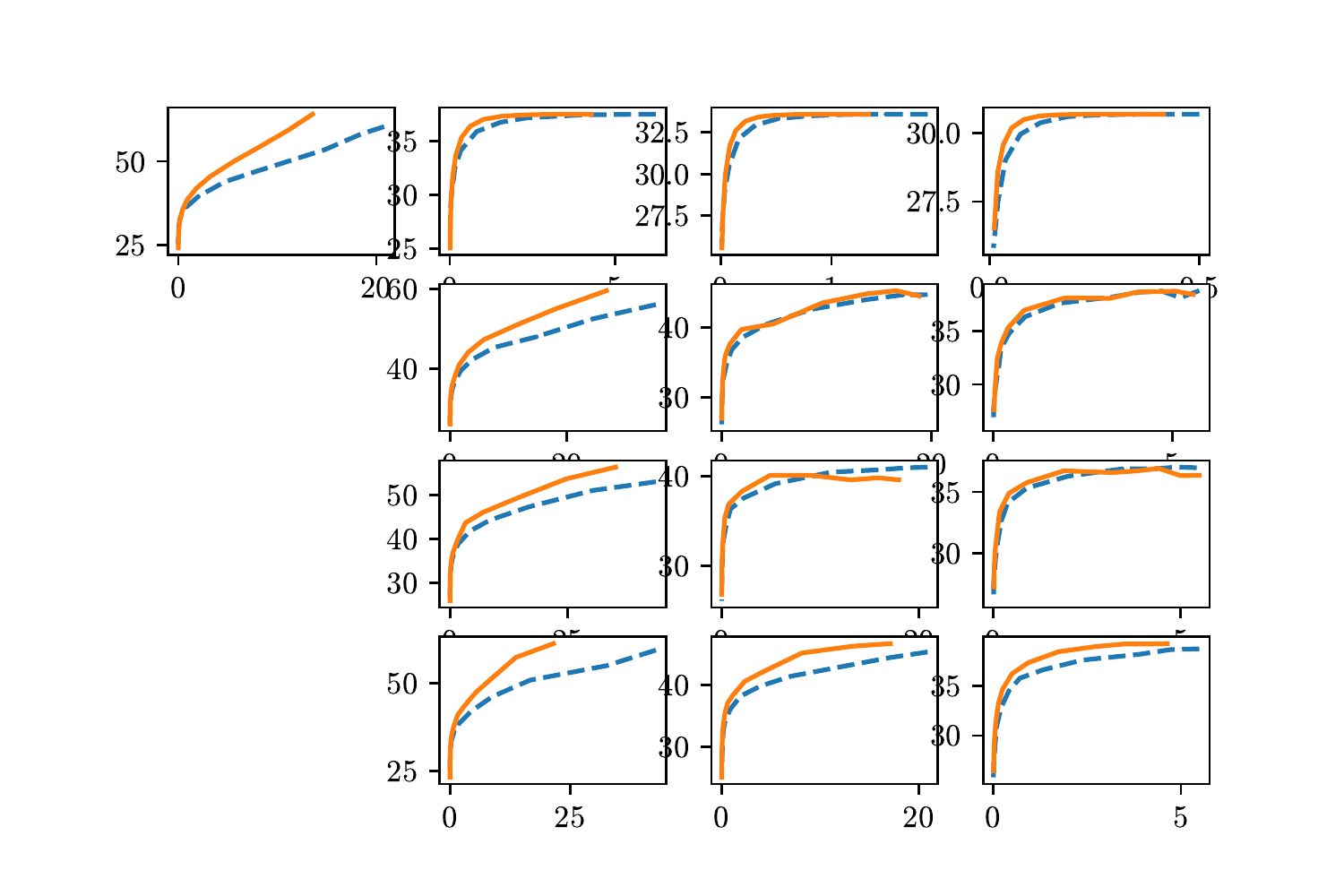}
    \end{minipage}
    \begin{minipage}{0.30\textwidth}
    \centering\small
    point cloud: {\em jacket} \\
    \includegraphics[width=1.0\linewidth, trim=20 15 35 32, clip]{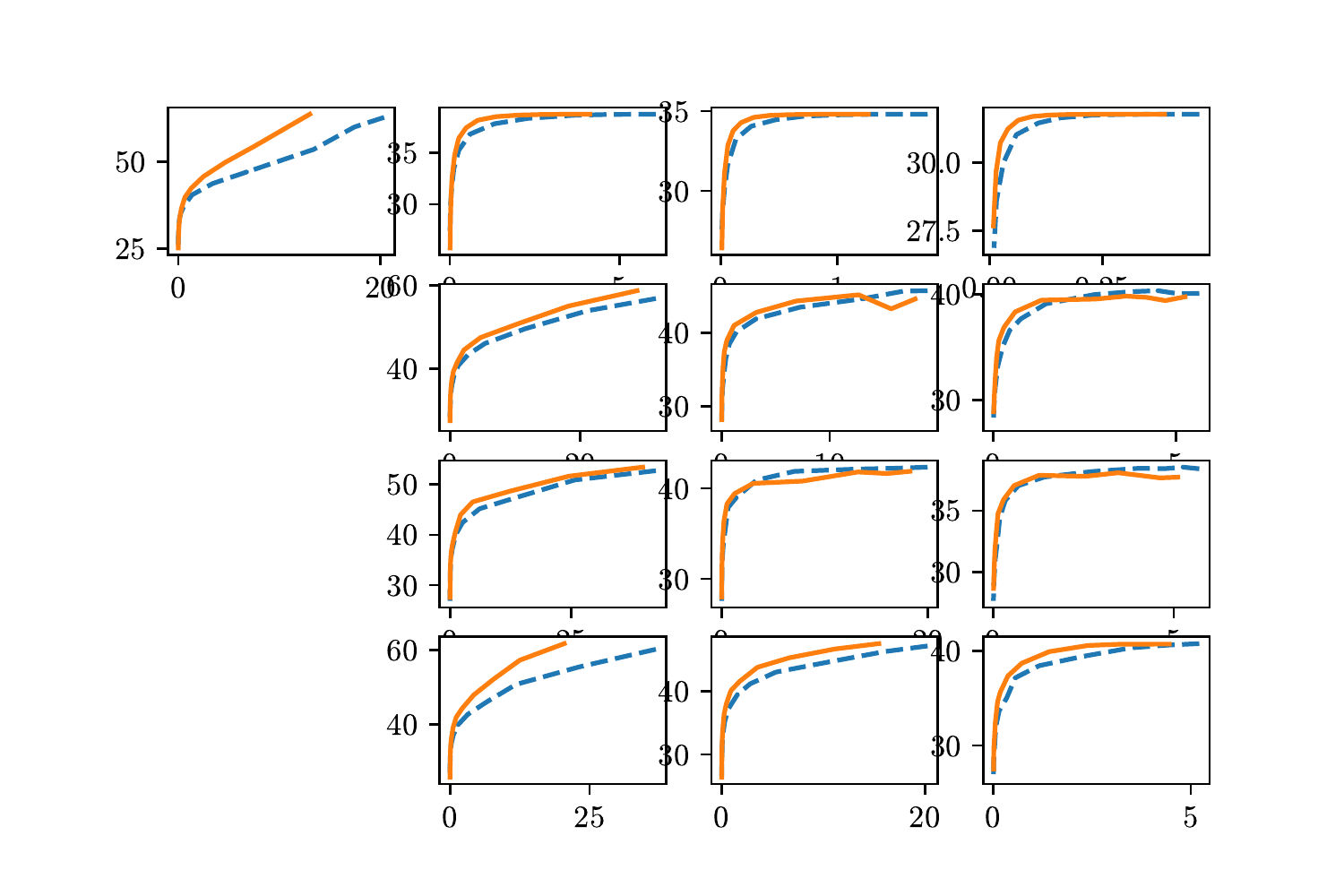}
    \end{minipage}
    \caption{RD performance (RGB PSNR vs.\ bit rate) improvement due to normalization, corresponding to entries in \cref{tab:normalization_other}.  See \cref{fig:normalization} for point cloud {\em rock}.}
    \label{fig:normalization_other}
\end{figure*}

\begin{figure*}
    \centering
    \includegraphics[width=0.29\linewidth, trim=20 5 35 15, clip]{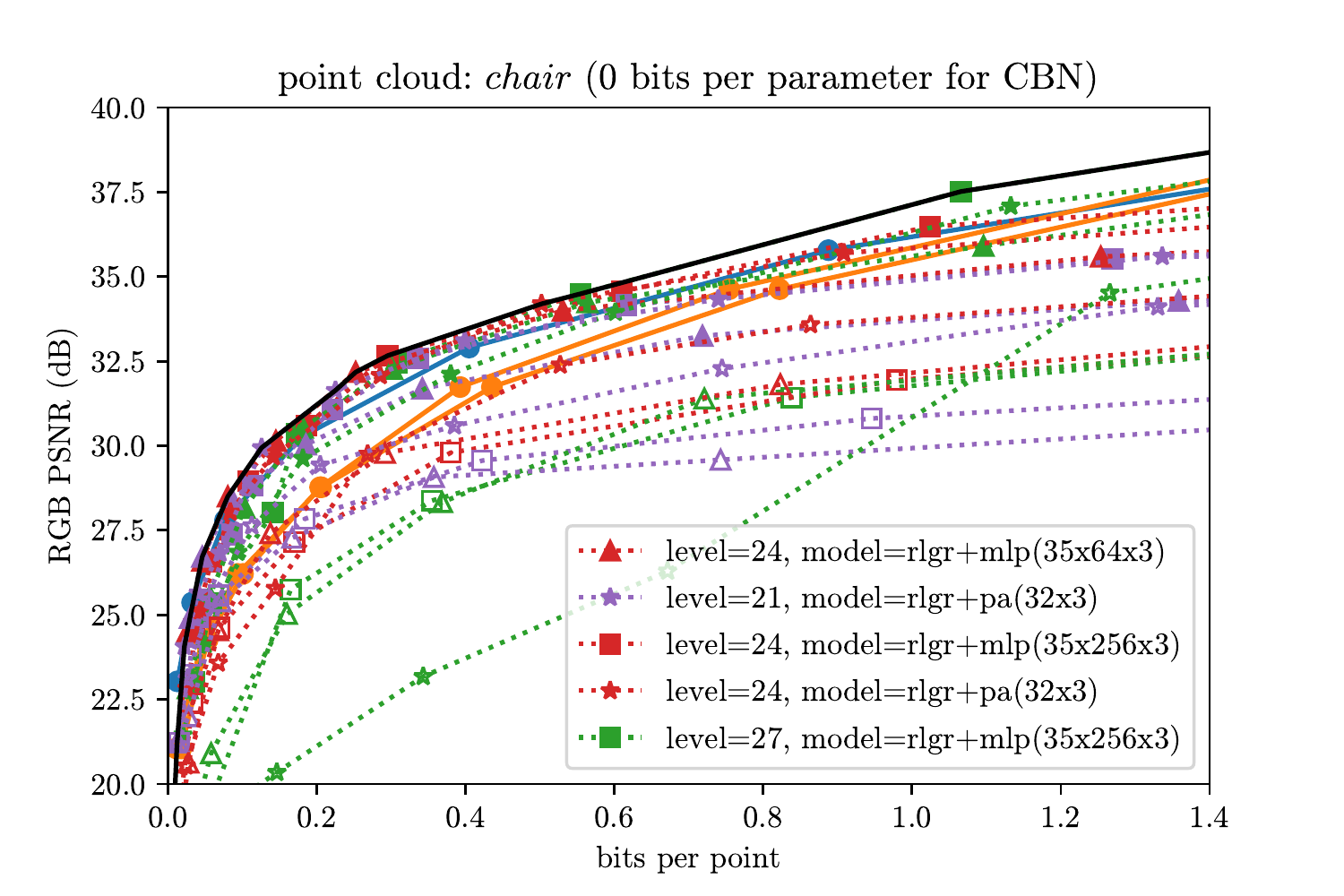}
    \includegraphics[width=0.29\linewidth, trim=20 5 35 15, clip]{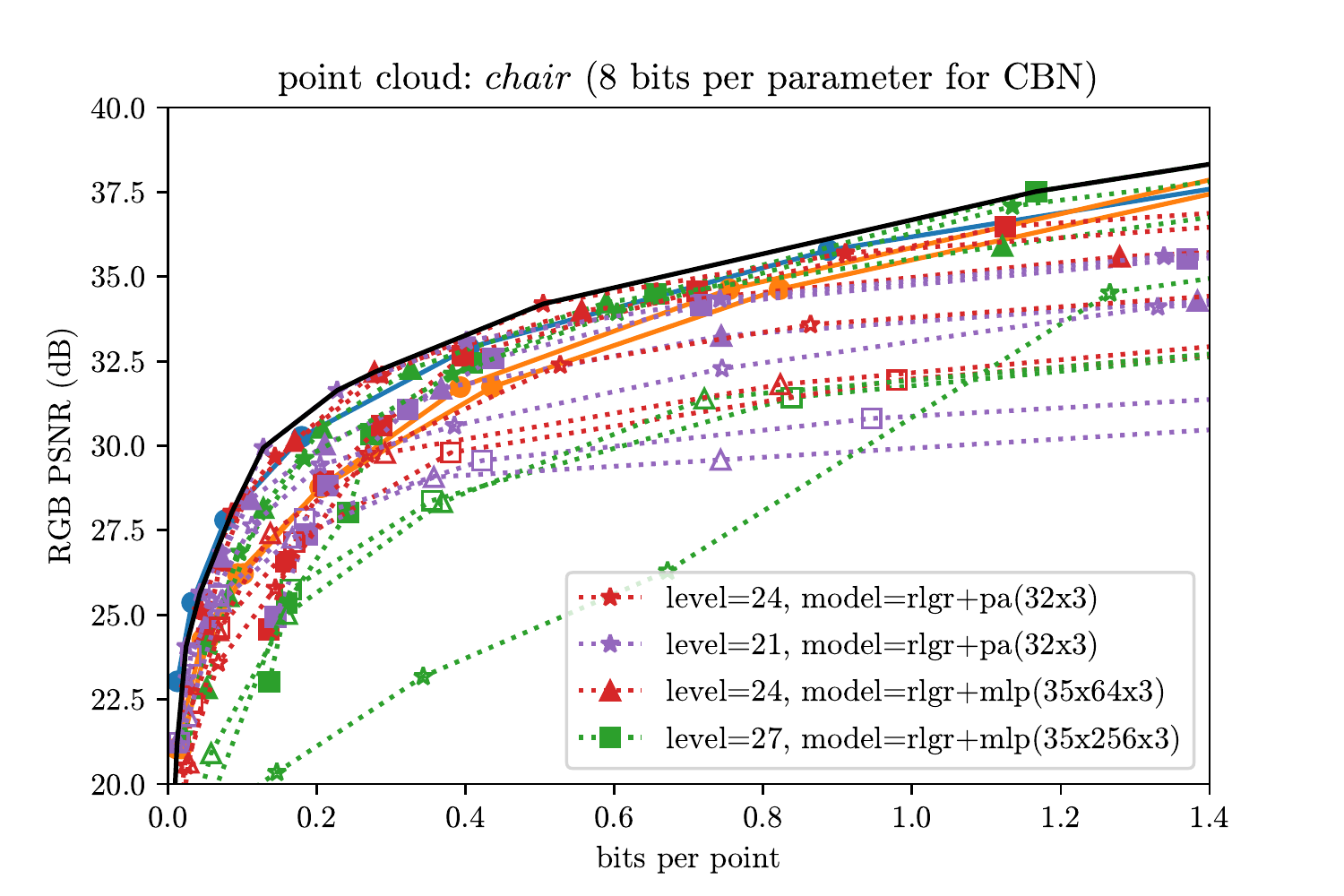}
    \includegraphics[width=0.29\linewidth, trim=20 5 35 15, clip]{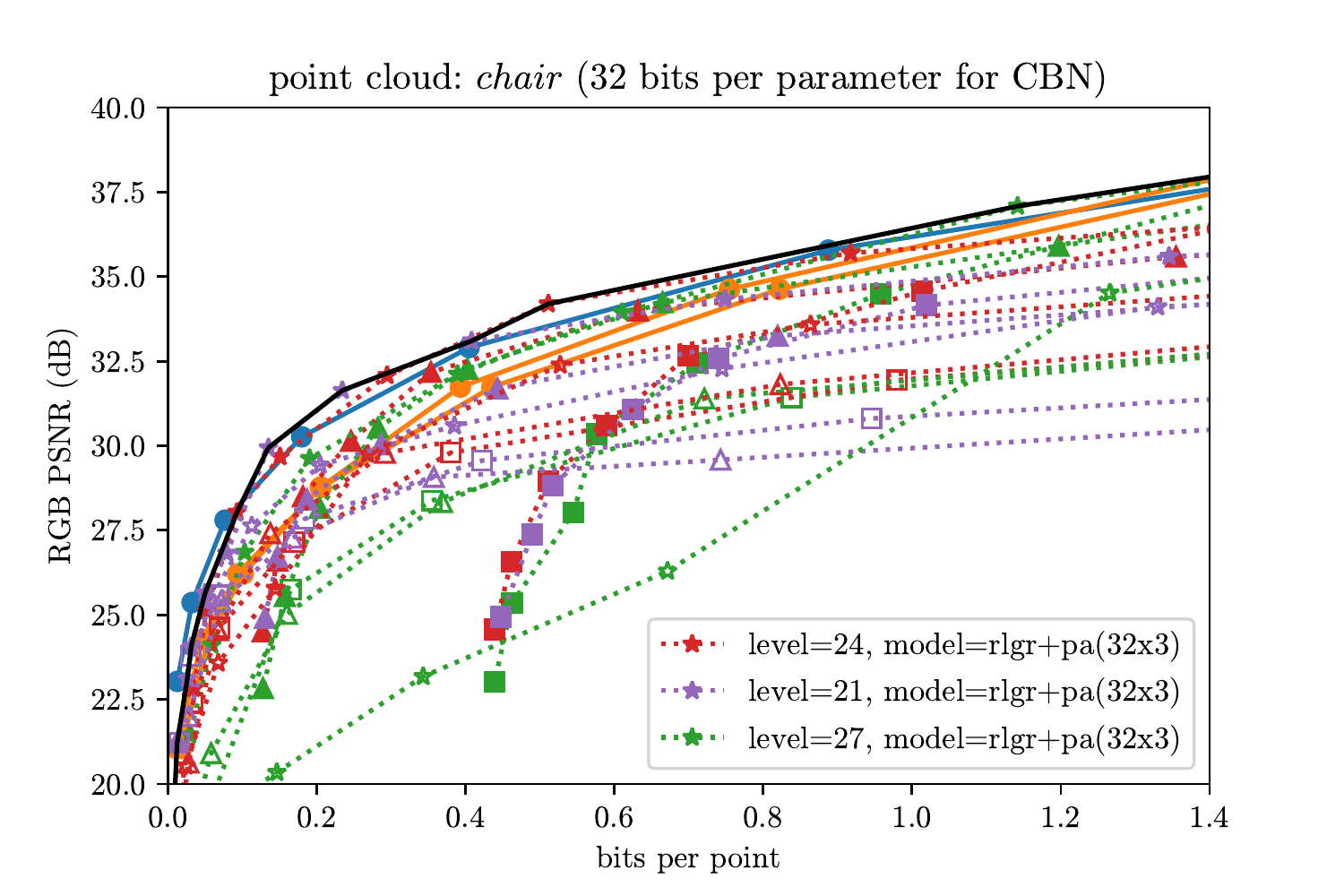}
    
    \includegraphics[width=0.29\linewidth, trim=20 5 35 15, clip]{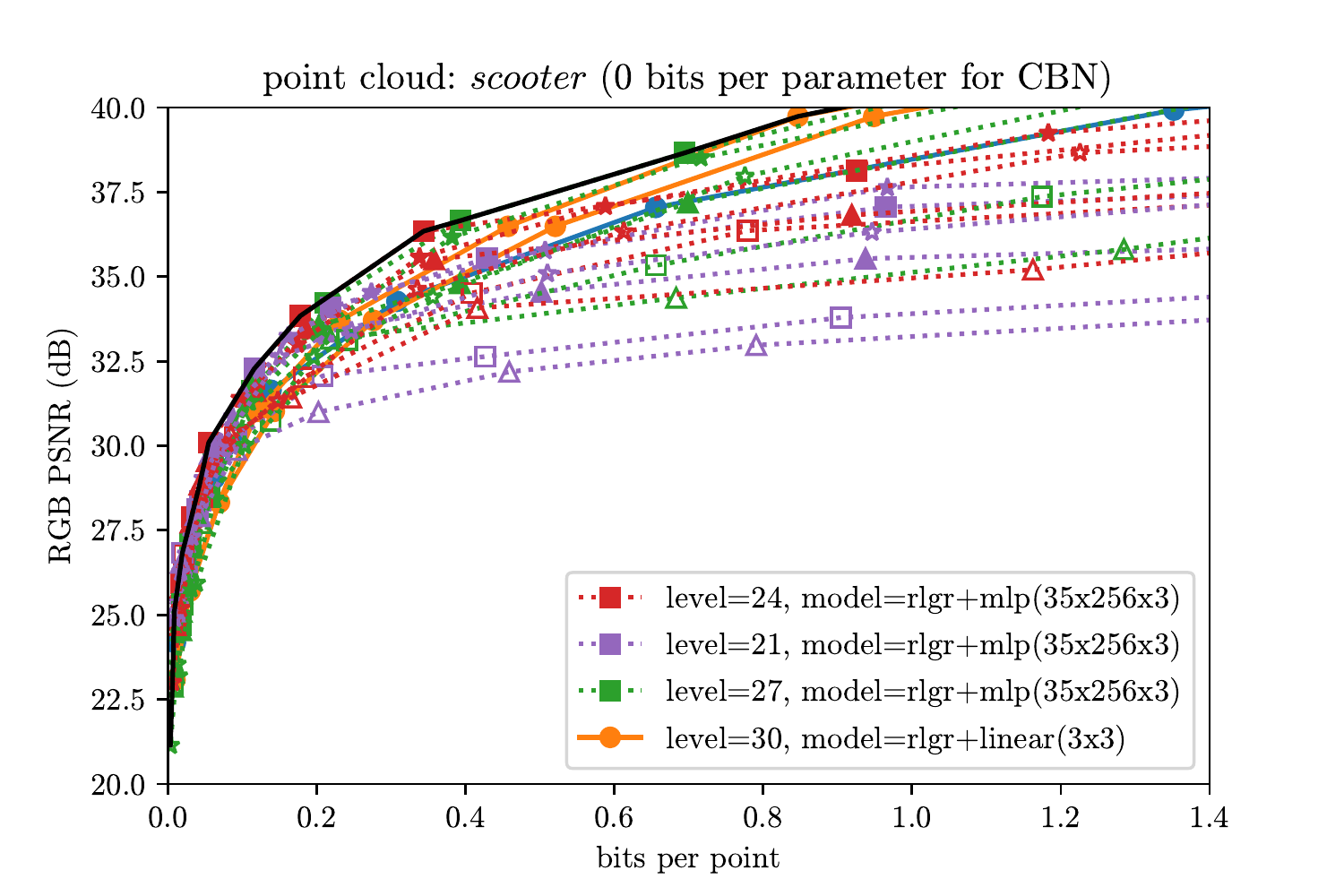}
    \includegraphics[width=0.29\linewidth, trim=20 5 35 15, clip]{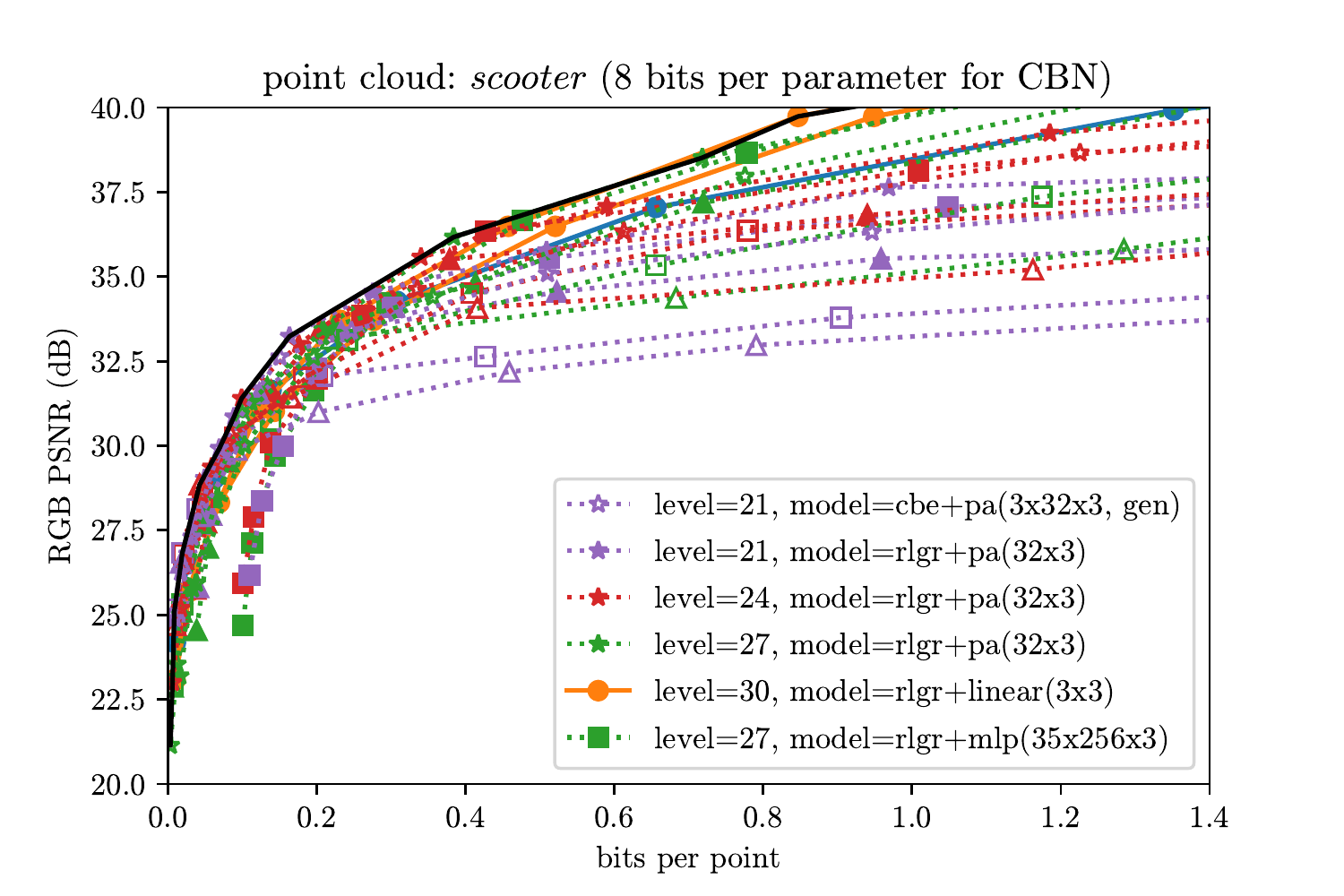}
    \includegraphics[width=0.29\linewidth, trim=20 5 35 15, clip]{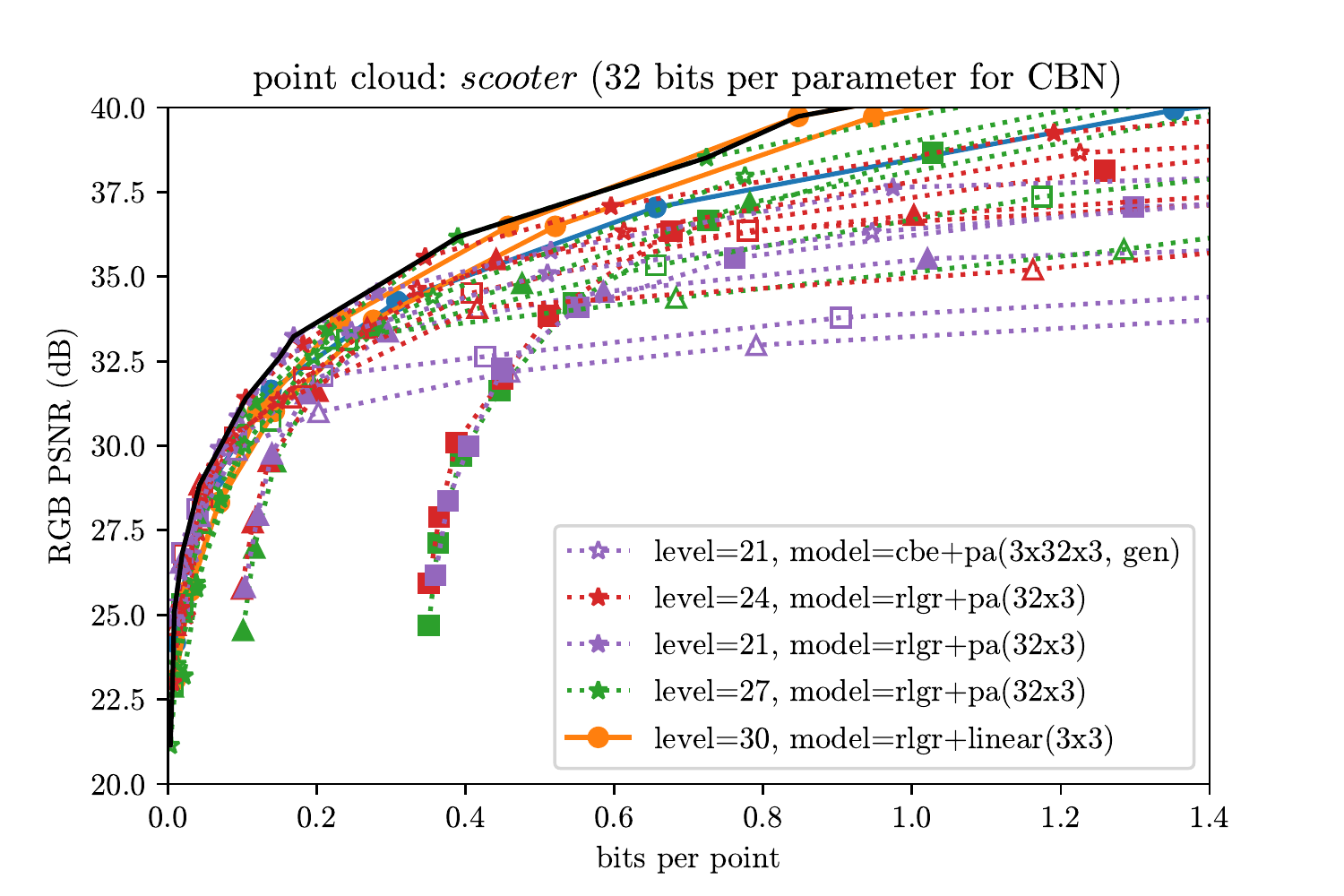}
    
    \includegraphics[width=0.29\linewidth, trim=20 5 35 15, clip]{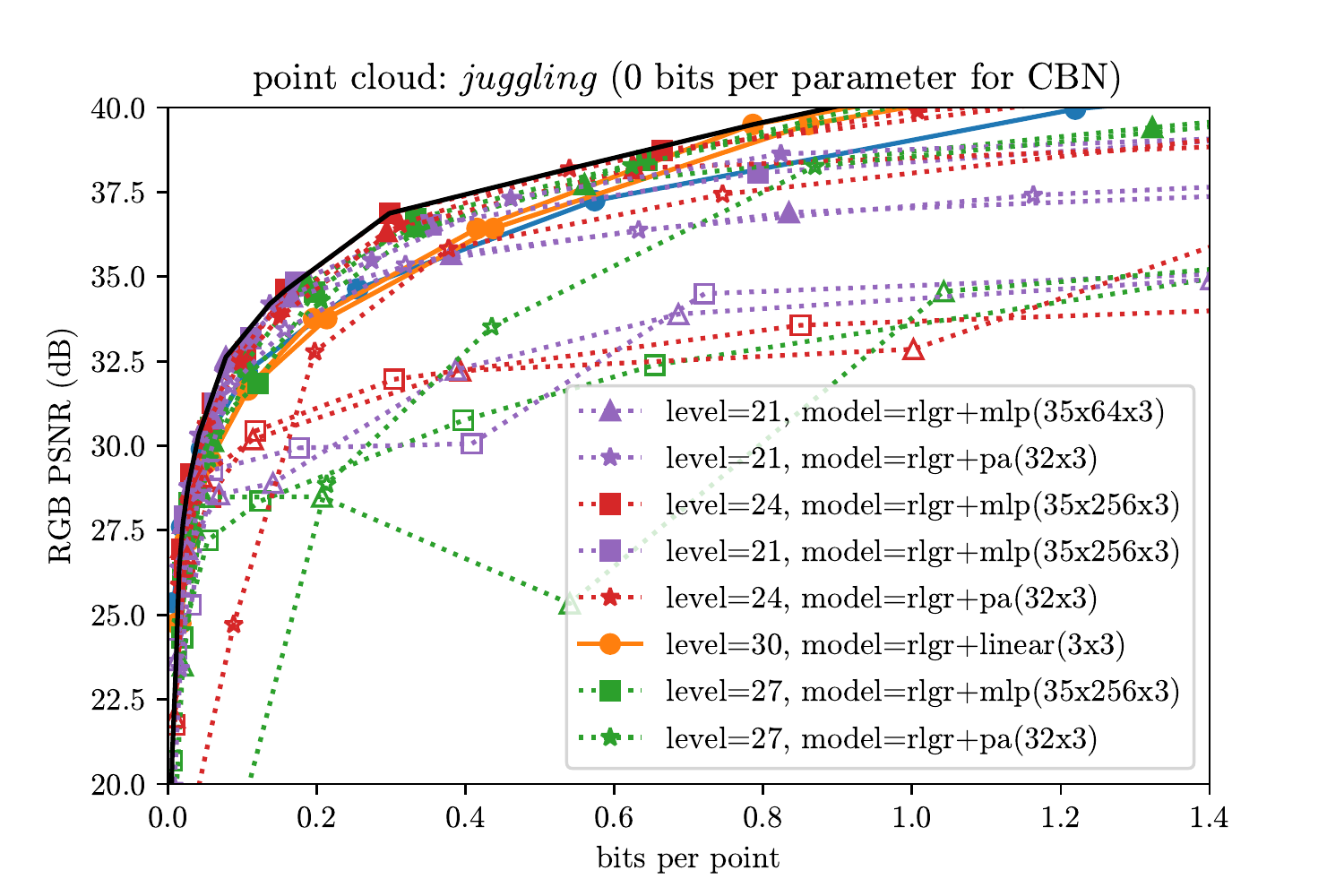}
    \includegraphics[width=0.29\linewidth, trim=20 5 35 15, clip]{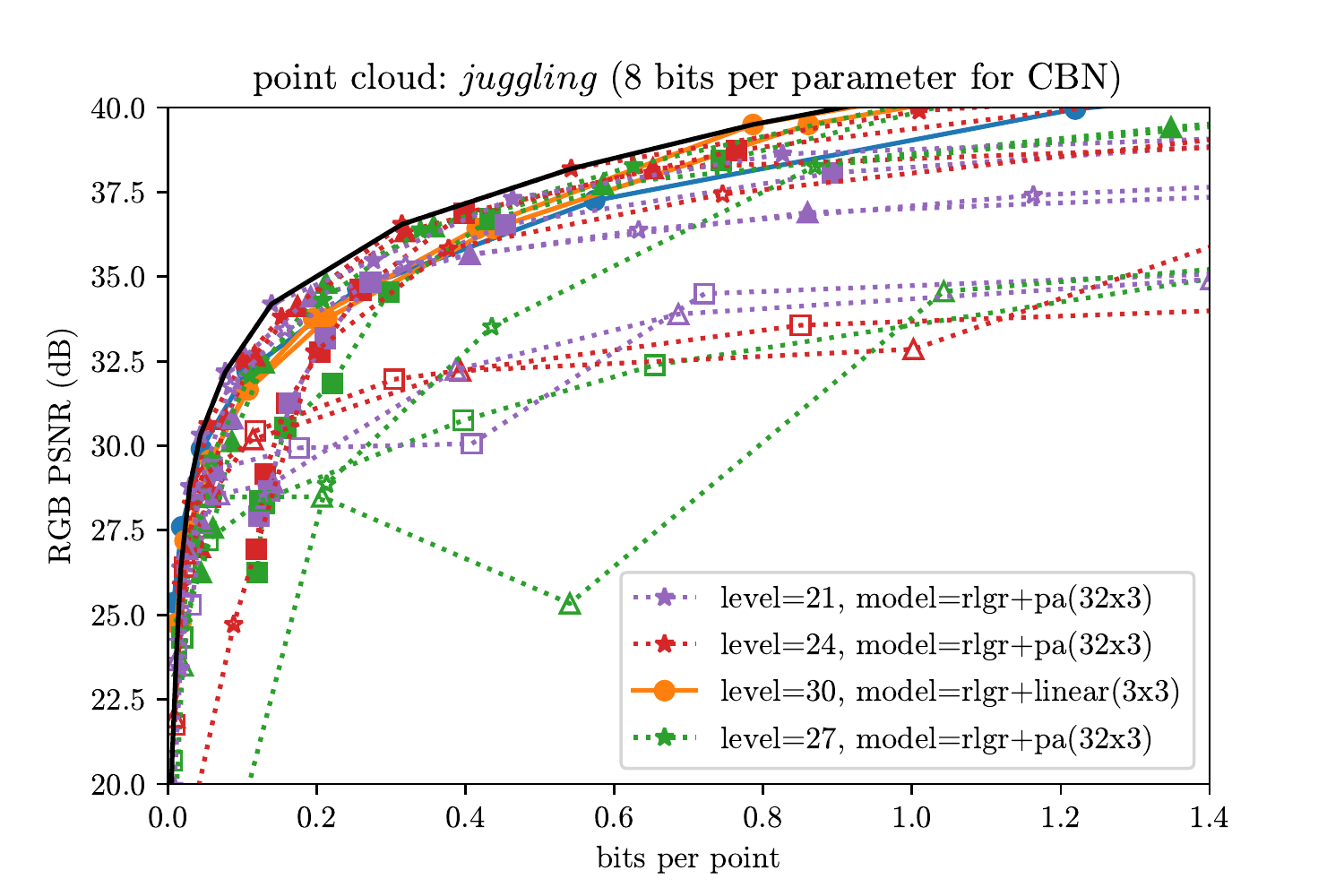}
    \includegraphics[width=0.29\linewidth, trim=20 5 35 15, clip]{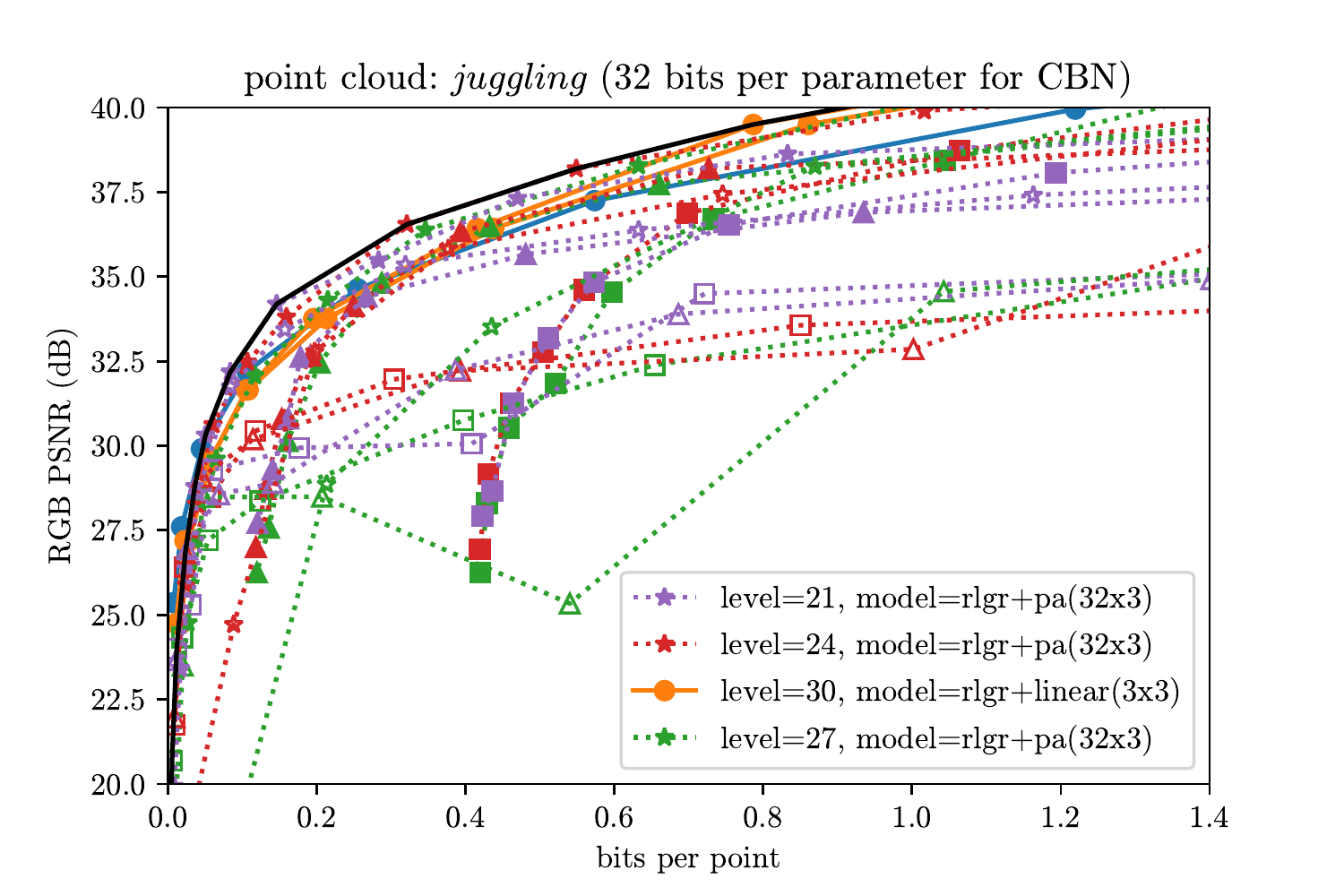}
    
    \includegraphics[width=0.29\linewidth, trim=20 5 35 15, clip]{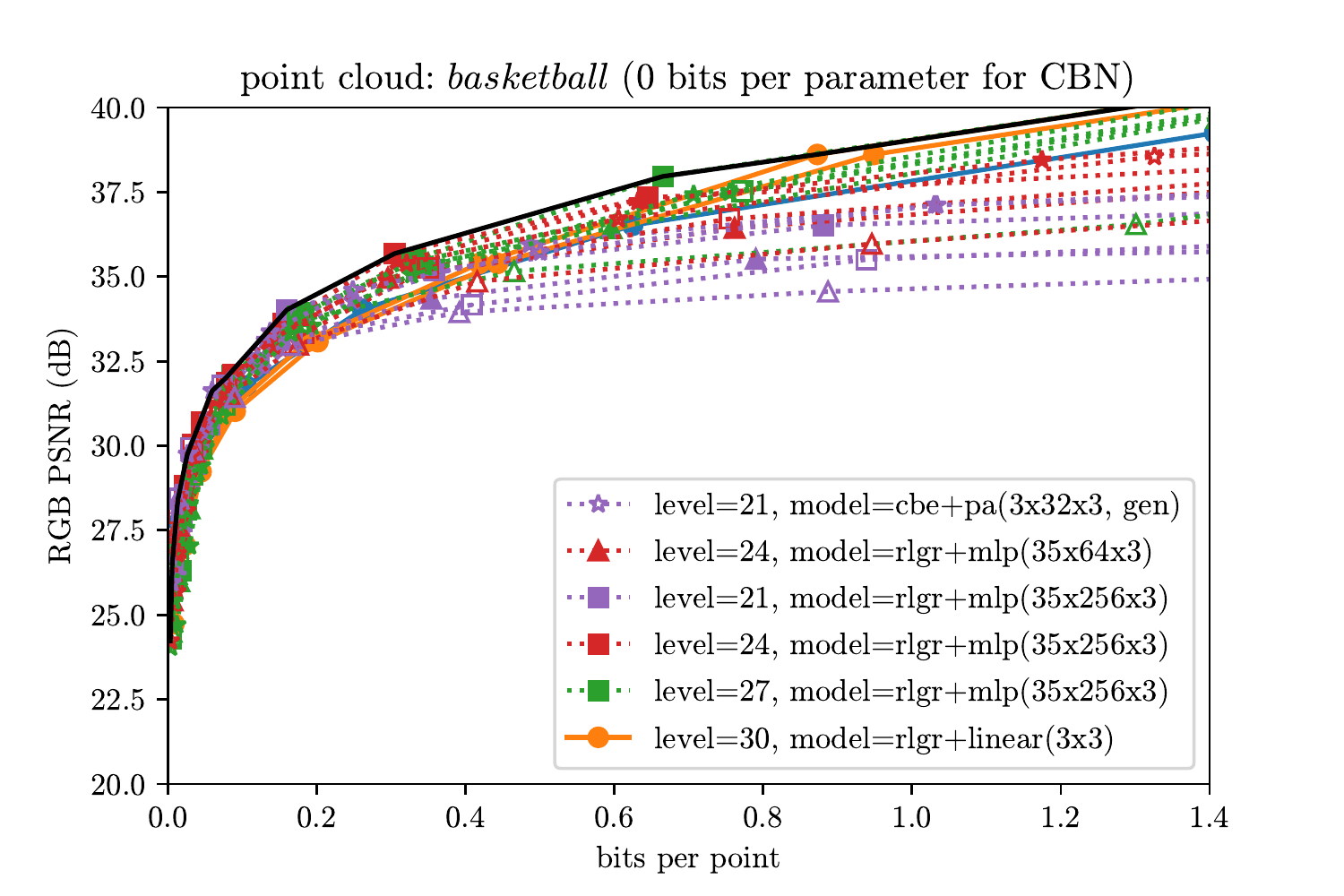}
    \includegraphics[width=0.29\linewidth, trim=20 5 35 15, clip]{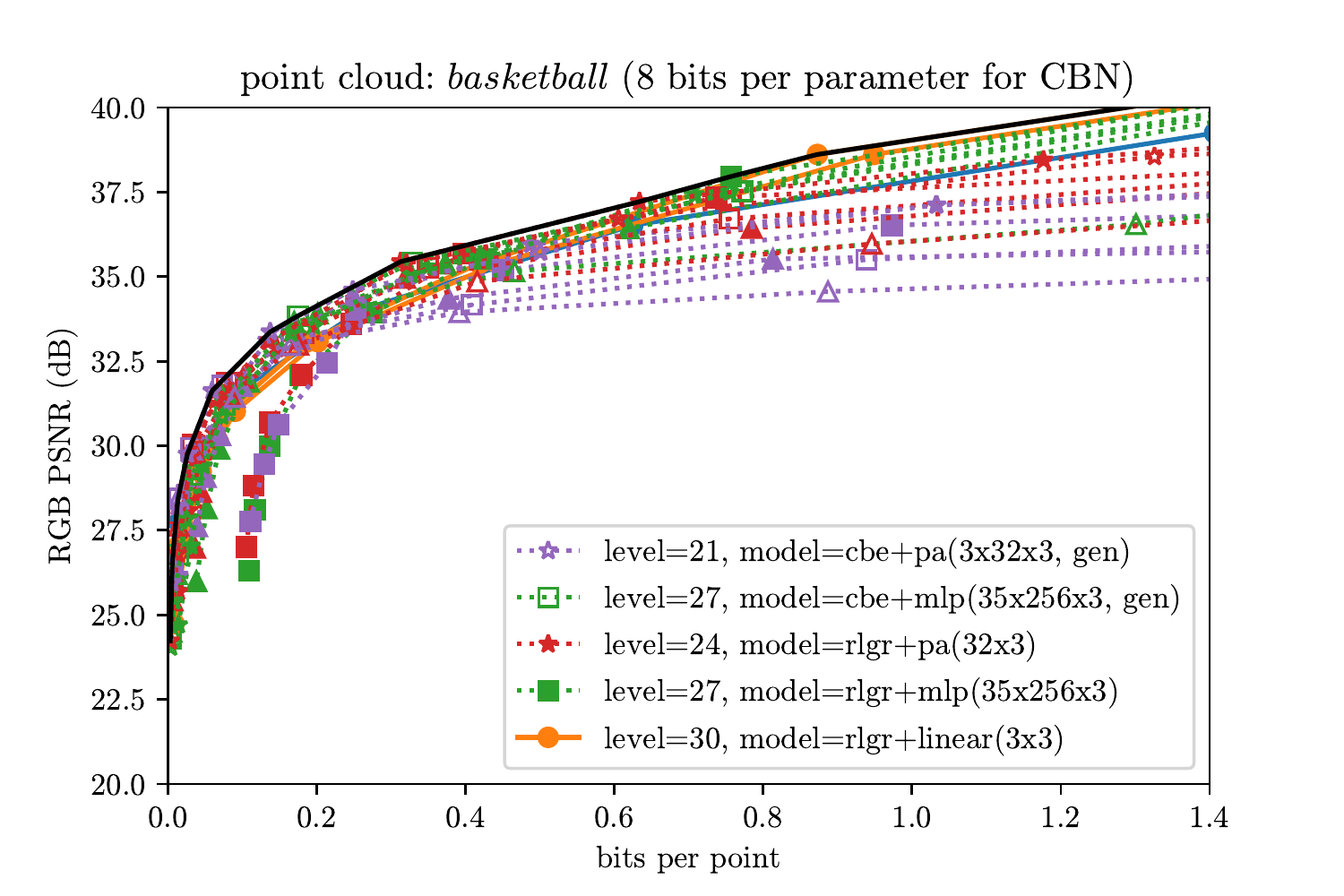}
    \includegraphics[width=0.29\linewidth, trim=20 5 35 15, clip]{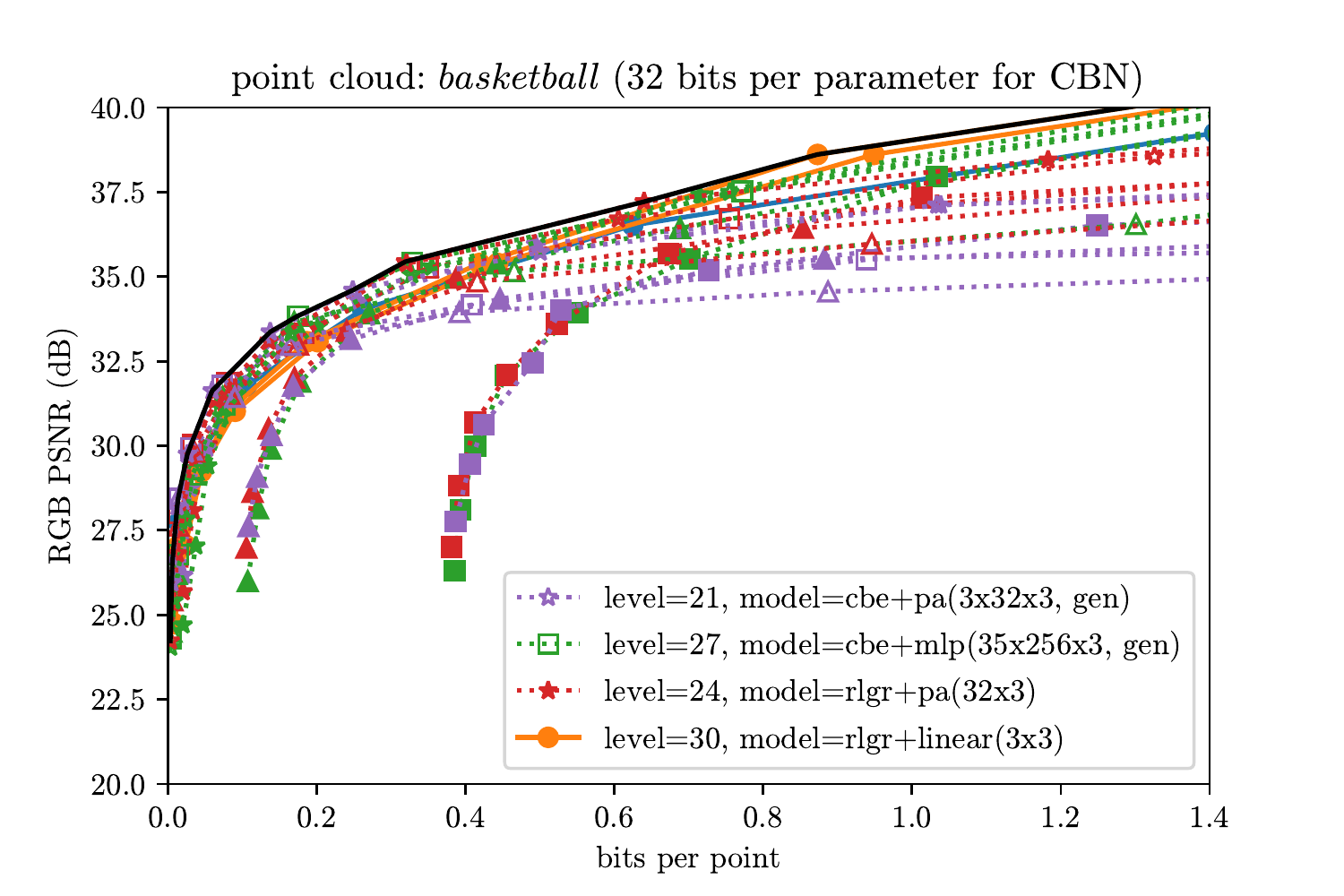}
    
    \includegraphics[width=0.29\linewidth, trim=20 5 35 15, clip]{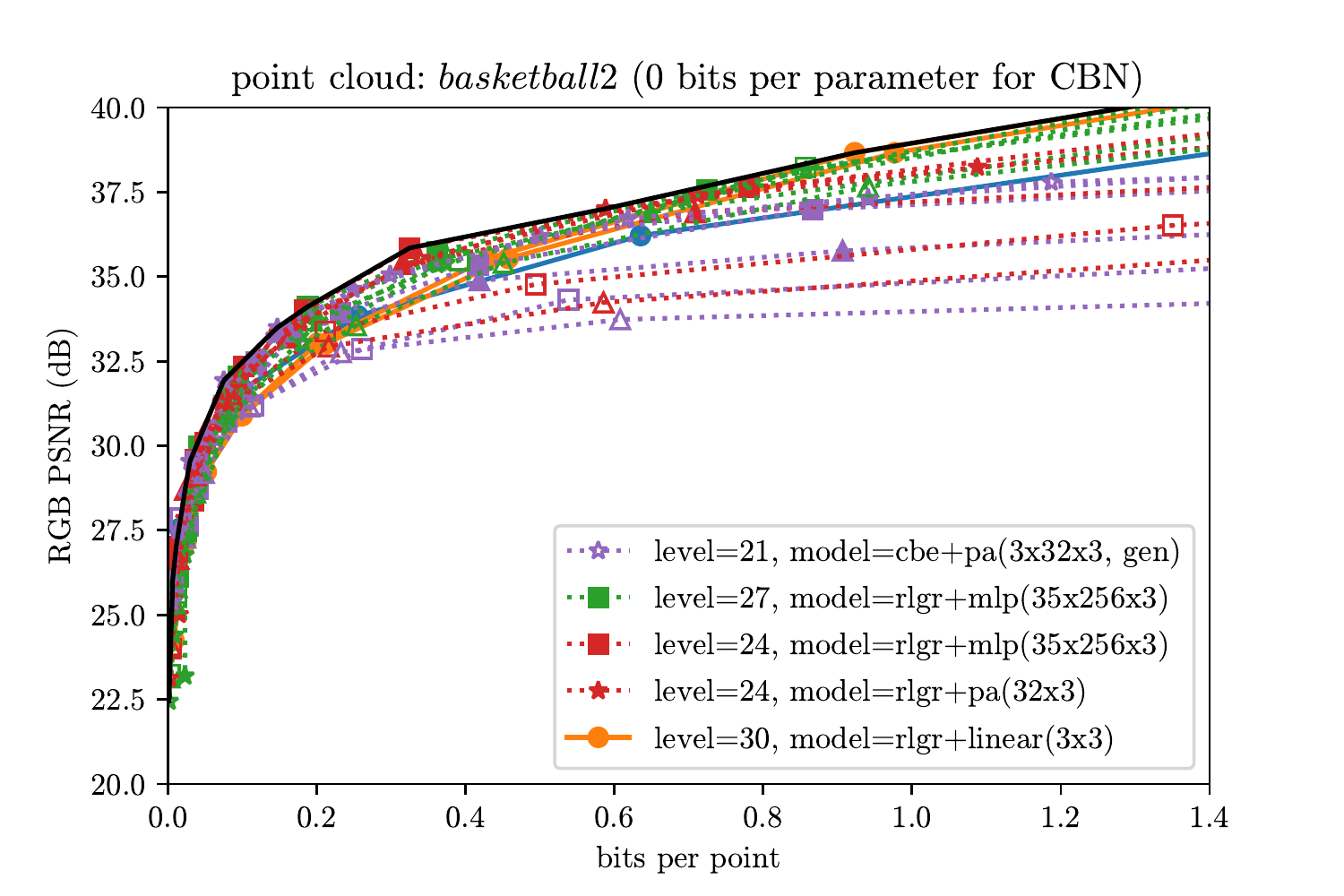}
    \includegraphics[width=0.29\linewidth, trim=20 5 35 15, clip]{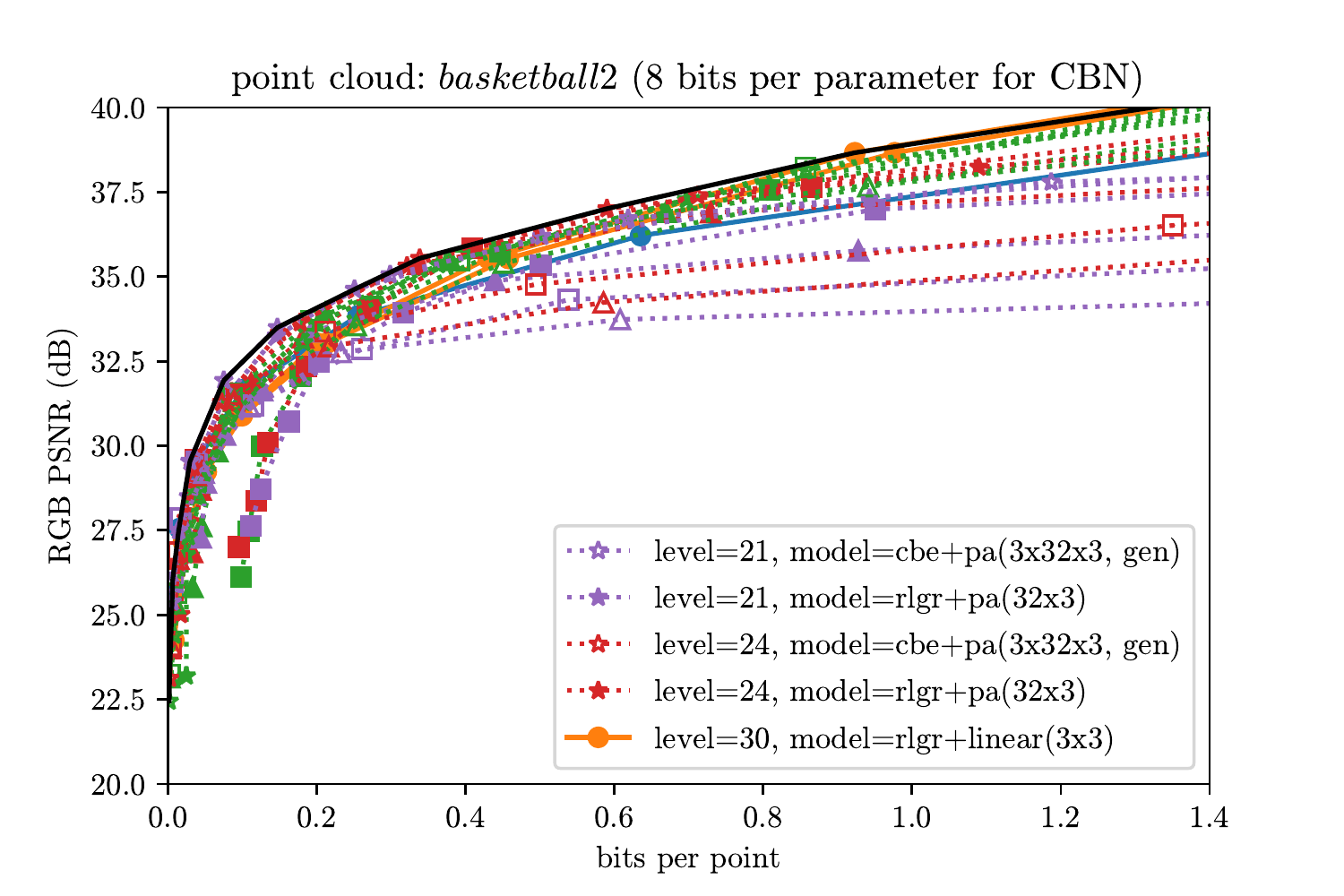}
    \includegraphics[width=0.29\linewidth, trim=20 5 35 15, clip]{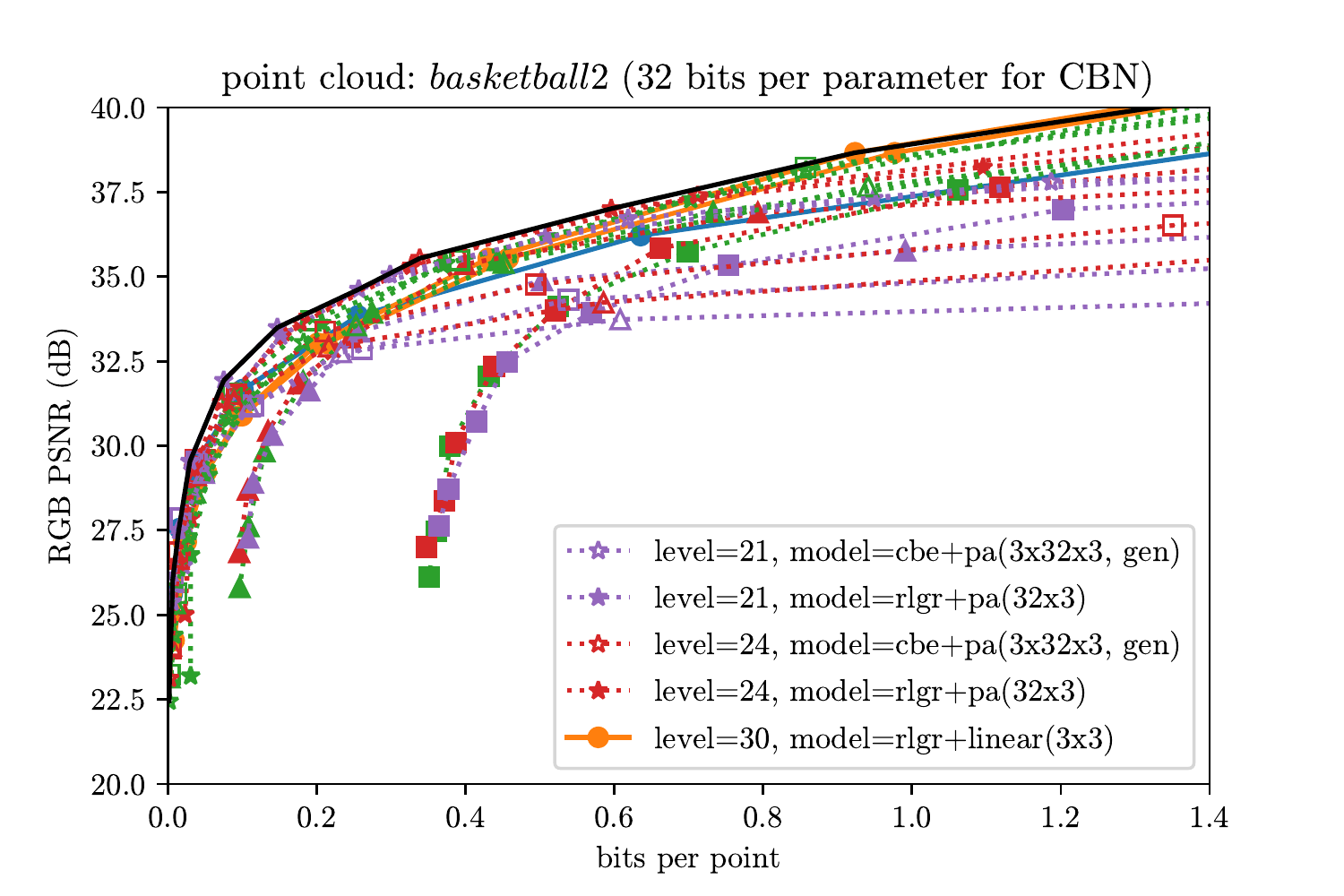}
    
    \includegraphics[width=0.29\linewidth, trim=20 5 35 15, clip]{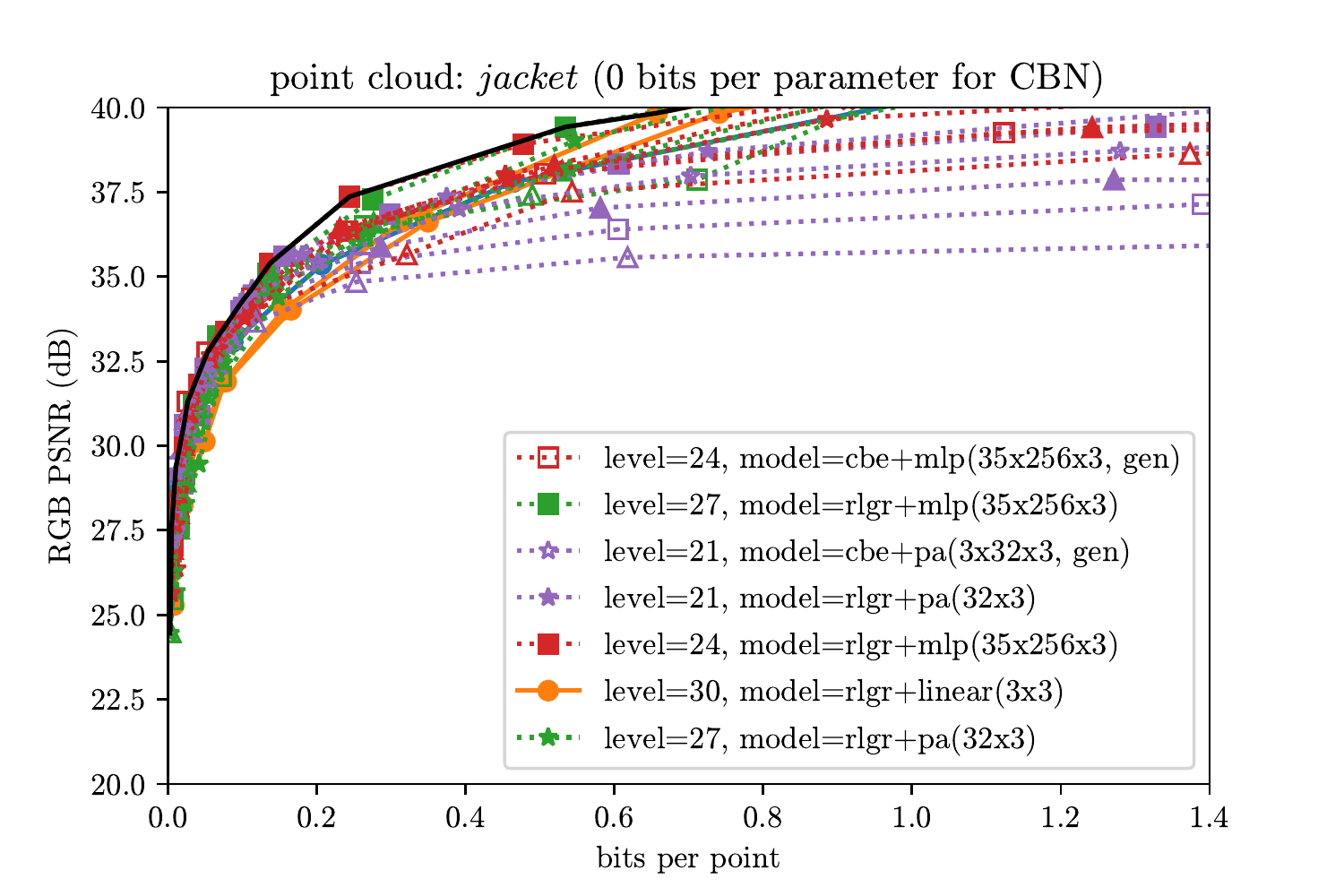}
    \includegraphics[width=0.29\linewidth, trim=20 5 35 15, clip]{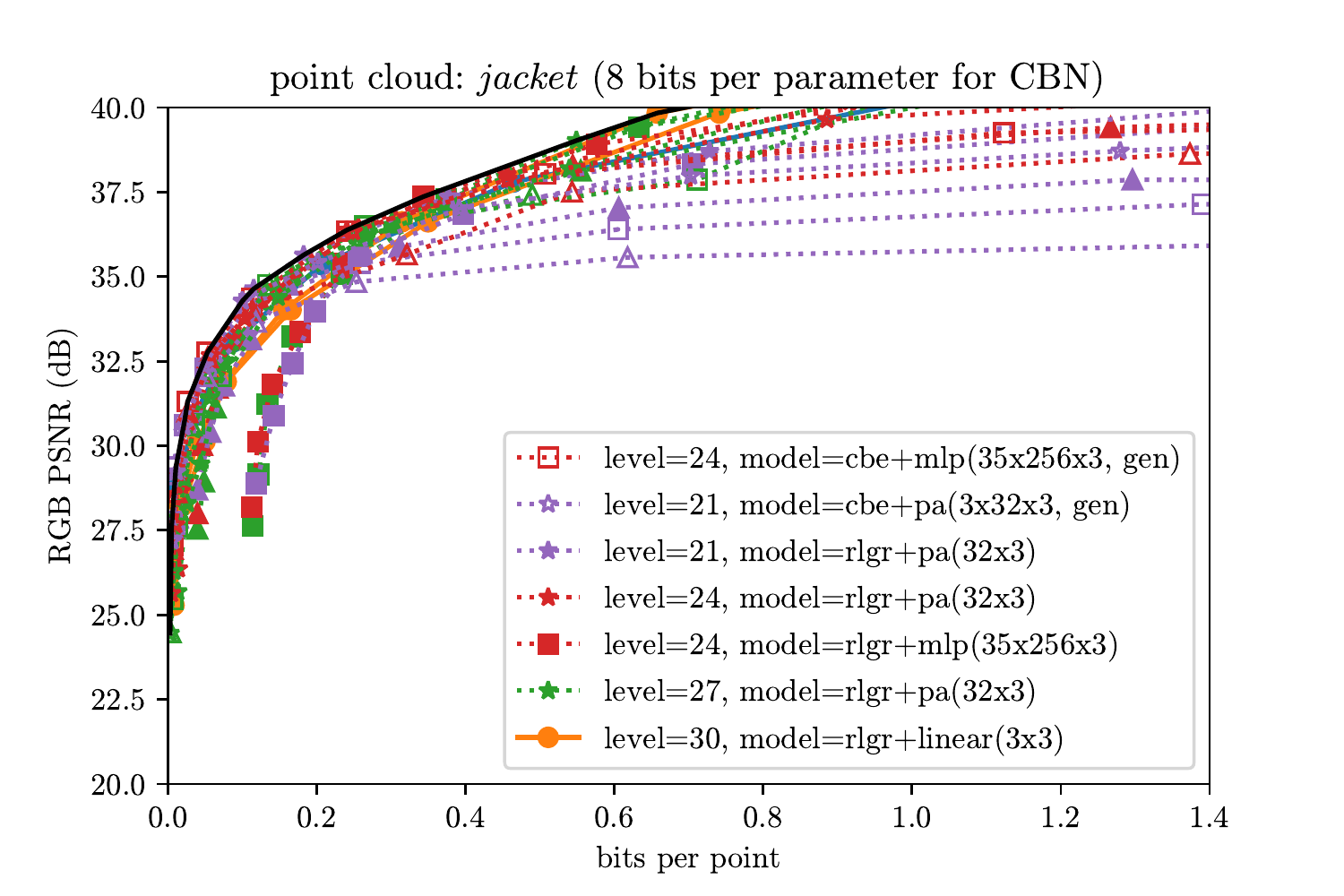}
    \includegraphics[width=0.29\linewidth, trim=20 5 35 15, clip]{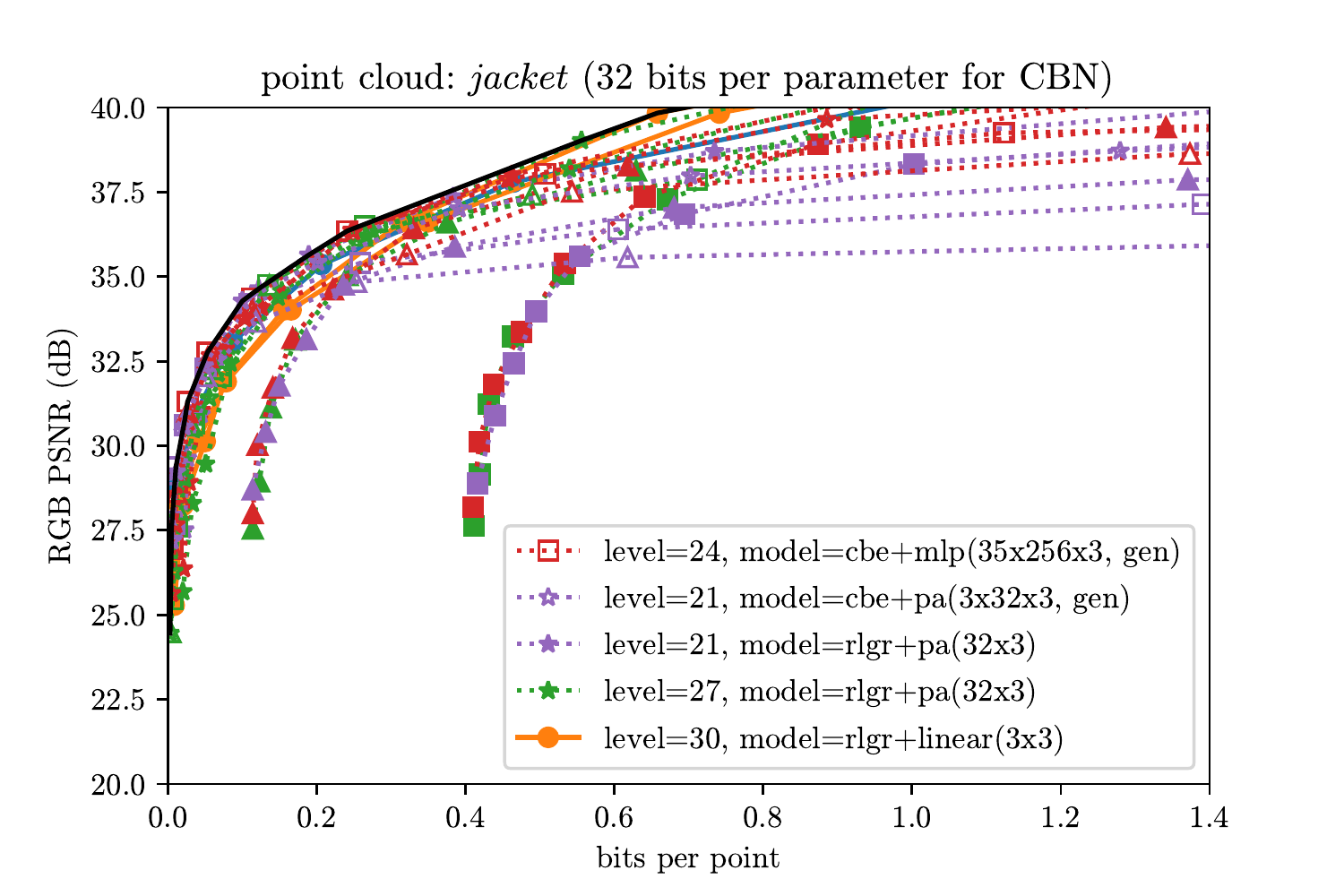}
    
    \caption{Convex hull (solid black line) of RD performances of all CBN configurations across all levels, including side information using 0 (left), 8 (middle), and 32 (right) bits per CBN parameter.  Each row is a different point cloud.  See \cref{fig:convhull} for point cloud {\em rock}.}
    \label{fig:convhull_other}
\end{figure*}

\Cref{fig:baselines_supp}, \cref{fig:cbns_by_level_supp}, \cref{fig:cbns_by_network_supp}, \cref{fig:cbns_by_level_gen_supp} \& \cref{fig:cbns_by_network_gen_supp}, \cref{fig:sideinfo_cbe_vs_rlgr_supp}, \cref{fig:sideinfo_mlp256_other}, \cref{fig:sideinfo_mlp64_other} \& \cref{fig:sideinfo_pa_other}, \cref{tab:normalization_other}, \cref{fig:normalization_other}, and \cref{fig:convhull_other} show results for point clouds {\em chair}, {\em scooter}, {\em juggling}, {\em basketball}, {\em basketball2}, and {\em jacket} corresponding respectively to \cref{fig:baselines}, \cref{fig:cbns_by_level}, \cref{fig:cbns_by_network}, \cref{fig:cbns_gen}, \cref{fig:sideinfo_cbe_vs_rlgr}, \cref{fig:sideinfo_mlp256}, \cref{fig:sideinfo_mlp64_and_pa}, \cref{tab:normalization}, \cref{fig:normalization}, and \cref{fig:convhull} for {\em rock}.

\begin{table}
    \centering
    \begin{tabular}{|c|cccc|} \hline
    {\em rock} & \multicolumn{4}{c|}{level} \\
    CBN & 30 & 27 & 24 & 21 \\ \hline
    linear(3x3) & -36.5\% & -26.1\% & -29.6\% & -35.2\% \\
mlp(35x256x3) & N/A & -35.8\% & -39.4\% & -28.7\% \\
mlp(35x64x3) & N/A & -32.7\% & -34.4\% & -27.8\% \\
pa(3x32x3) & N/A & -41.6\% & -39.8\% & -37.5\%
    \\ \hline
    \end{tabular}
    
    \begin{tabular}{|c|cccc|} \hline
    {\em chair} & \multicolumn{4}{c|}{level} \\
    CBN & 30 & 27 & 24 & 21 \\ \hline
    linear(3x3) & -36.8\% & -26.5\% & -34.1\% & -47.4\% \\
mlp(35x256x3) & N/A & -34.8\% & -39.5\% & -28.0\% \\
mlp(35x64x3) & N/A & -34.3\% & -47.5\% & -38.4\% \\
pa(3x32x3) & N/A & -45.2\% & -44.7\% & -45.6\%

    \\ \hline
    \end{tabular}
    
    \begin{tabular}{|c|cccc|} \hline
    {\em scooter} & \multicolumn{4}{c|}{level} \\
    CBN & 30 & 27 & 24 & 21 \\ \hline
    linear(3x3) & -32.6\% & -19.5\% & -34.6\% & -41.8\% \\
mlp(35x256x3) & N/A & -32.1\% & -38.1\% & -20.6\% \\
mlp(35x64x3) & N/A & -27.1\% & -30.2\% & -39.1\% \\
pa(3x32x3) & N/A & -41.9\% & -42.2\% & -40.6\%

    \\ \hline
    \end{tabular}
    
    \begin{tabular}{|c|cccc|} \hline
    {\em juggling} & \multicolumn{4}{c|}{level} \\
    CBN & 30 & 27 & 24 & 21 \\ \hline
    linear(3x3) & -26.9\% & -8.5\% & -23.4\% & -32.4\% \\
mlp(35x256x3) & N/A & -22.7\% & -35.0\% & -27.7\% \\
mlp(35x64x3) & N/A & -16.3\% & -35.2\% & -30.5\% \\
pa(3x32x3) & N/A & -47.7\% & -47.6\% & -39.1\%

    \\ \hline
    \end{tabular}
    
    \begin{tabular}{|c|cccc|} \hline
    {\em basketball} & \multicolumn{4}{c|}{level} \\
    CBN & 30 & 27 & 24 & 21 \\ \hline
    linear(3x3) & -30.0\% & -19.7\% & -27.7\% & -28.9\% \\
mlp(35x256x3) & N/A & -20.9\% & -29.2\% & -42.3\% \\
mlp(35x64x3) & N/A & -5.8\% & -28.4\% & -32.8\% \\
pa(3x32x3) & N/A & -31.1\% & -27.7\% & -24.5\%

    \\ \hline
    \end{tabular}
    
    \begin{tabular}{|c|cccc|} \hline
    {\em basketball2} & \multicolumn{4}{c|}{level} \\
    CBN & 30 & 27 & 24 & 21 \\ \hline
    linear(3x3) & -29.2\% & -14.6\% & -20.8\% & -36.5\% \\
mlp(35x256x3) & N/A & -34.5\% & -24.0\% & -15.9\% \\
mlp(35x64x3) & N/A & -28.7\% & -27.8\% & -24.1\% \\
pa(3x32x3) & N/A & -41.2\% & -42.5\% & -41.7\%

    \\ \hline
    \end{tabular}
    
    \begin{tabular}{|c|cccc|} \hline
    {\em jacket} & \multicolumn{4}{c|}{level} \\
    CBN & 30 & 27 & 24 & 21 \\ \hline
    linear(3x3) & -29.3\% & -15.2\% & -27.7\% & -41.9\% \\
mlp(35x256x3) & N/A & -28.8\% & -35.1\% & -28.4\% \\
mlp(35x64x3) & N/A & -22.0\% & -21.3\% & -25.1\% \\
pa(3x32x3) & N/A & -39.1\% & -38.0\% & -41.7\%

    \\ \hline
    \end{tabular}
    
    \caption{BD-Rate reductions due to normalization, for each point cloud.  See \cref{tab:normalization} for aggregate results.}
    \label{tab:normalization_other}
\end{table}

\end{document}